%% file: paper.tex
\documentclass{aa}

\usepackage{graphicx,xcolor}
\usepackage{txfonts,color,epstopdf}
\usepackage[normalem]{ulem}
\usepackage{tikz}
\usepackage{float}
\usepackage{natbib}
\bibpunct{(}{)}{;}{a}{}{,}

\newcommand{\Msun}{\,\mathrm{M}_\odot}
\newcommand{\Mdot}{\,\mathrm{M}_\odot\,\mathrm{yr}^{-1}}
\newcommand{\Mrun}{M_\mathrm{\ast,run}}
\newcommand{\Myr}{\,\mathrm{Myr}}

\newcommand{\MWDf}{M_\mathrm{WD,f}}
\newcommand{\MWd}{M_\mathrm{\star,run}}

\newcommand{\kms}{\,$\mathrm{km}\,\mathrm{s}^{-1}$}
\newcommand{\mkms}{\,\mathrm{km}\,\mathrm{s}^{-1}}
\newcommand{\kkpc}{$\mathrm{kpc}^{-3}$}
\newcommand{\kpc}{\,kpc }
\newcommand{\mkpc}{\,\mathrm{kpc}}
\newcommand{\sub}[1]{_\mathrm{#1}}
\newcommand{\magten}[1]{\cdot10^{#1}}
\newcommand{\mas}{\,\mathrm{mas}}
\newcommand{\masyr}{\,\mathrm{mas}\,\mathrm{yr}^{-1}}
\newcommand{\pyr}{\,\mathrm{yr}^{-1}}

\begin{document}

   \title{Properties and applications of a predicted population of runaway He-sdO/B stars ejected from single degenerate He-donor SNe}

   \author{P. Neunteufel
        \inst{1}
        \and
        H. Preece
        \inst{1}
        \and
        M. Kruckow
        \inst{2}
        \and
        S. Geier
        \inst{3}
        \and
        A.S. Hamers
        \inst{1}
        \and
        S. Justham
        \inst{1,4,5,6}
        \and
        Ph. Podsiadlowski
        \inst{7}
   }

\institute{\inst{1}Max Planck Institut f\"ur Astrophysik,
	Karl-Schwarzschild-Straße 1, 85748 Garching bei M\"unchen, Germany\\
	\inst{2}Yunnan Observatories, Chinese Academy of Sciences, 
	Kunming 650011, China\\
	\inst{3}Institut f\"{u}r Physik und Astronomie, Universit\"{a}t Potsdam, 
	Haus 28, Karl-Liebknecht-Str. 24/25, D-14476 Potsdam-Golm, Germany\\
	\inst{4}Anton Pannekoek Institute of Astronomy and GRAPPA, University of Amsterdam, Science Park 904, 1098 XH Amsterdam, Netherlands \\
	\inst{5}School of Astronomy and Space Science, University of the Chinese Academy of Sciences, Beijing 100012, China \\
	\inst{6}National Astronomical Observatories, Chinese Academy of Sciences, Beijing 100012, China\\
	\inst{7}University of Oxford, St Edmund Hall, Oxford, OX1 4AR, UK \\
	\email{pneun@mpa-garching.mpg.de} }

   \date{Received (month) (day), (year); accepted (month) (day), (year)}
\abstract
   {Thermonuclear supernovae (SNe), a subset of which are the highly important SNe of Type Ia and Iax, are relatively poorly understood phenomena. One of the more promising scenarios leading up to the creation of a thermonuclear SN involves accretion of helium-rich material from a binary companion. Following the SN, the binary companion is then ejected from the location of the progenitor binary at velocities possibly large enough to unbind it from the gravitational potential of the Galaxy. Ejected companion stars should form a detectable population, if their production mechanism is not exceedingly rare.}
   { This study builds on previous works, producing the most extensive prediction of the properties of such a hypothetical population to date, taking both Chandrasekhar and non-Chandrasekhar mass events into account. These results are then used to define criteria for membership of this population and characterise putative subpopulations.}
   {This study contains $6\cdot10^{6}$ individual ejection trajectories out of the Galactic plane calculated with the stellar kinematics framework SHyRT, which are analysed with regard to their bulk observational properties. These are then put into context with the only previously identified population member US\,708 and applied to a number of other possible candidate objects.}
   {We find that two additional previously observed objects possess properties to warrant a designation as candidate objects. Characterisation of these object with respect to the predicted population finds all of them to be extreme in at least one astrometric observable. Higher mass ($>0.7\Msun$) objects should be over-represented in the observationally accessible volume, with the ratio of bound to unbound objects being an accessible observable for the determination of the dominant terminal accretor mass. We find that current observations of runaway candidates within 10\kpc support a Galactic SN rate of the order of $\sim 3\magten{-7}\pyr$ to $\sim 2\magten{-6}\pyr$, three orders of magnitude below the inferred Galactic SN Ia rate and two orders of magnitude below the formation rate of predicted He-donor progenitors.}
  {The number of currently observed population members suggests that the He-donor scenario, as suspected before, is not a dominant contributor to the number of observed SNe Ia. However, even at the low event rate suggested, we find that the majority of possibly detectable population members is still undetected. The extreme nature of current population members suggests that a still larger number of objects has simply evaded detection up to this point, hinting at a higher contribution than is currently supported by observation.}
   
   \titlerunning{Runaway sdO/B population}
\maketitle

\section{Introduction} \label{sec:introduction}
For the purposes of measuring cosmological distances, supernovae (SN) of Type Ia, in their role as standard candles, occupy a dominating position in the observational toolset \citep[leading to a Nobel Prize in 2011, see][]{PAG1999}. The usefulness of transients of this type relies heavily on the well-established relationship between the peak-luminosity light curve evolution \citep{P1993} of the event, allowing the observer to independently infer the brightness of the event from parameters accessible to observation. The observational characterisation, which includes spectral analysis \citep{BMK2012,HKRR2013} of a SN Ia or other type has to be distinguished from theoretical classification of the production mechanism. For example, in the context of the presumed production mechanism, a collapse of the iron core of a massive star (core collapse SNE, CCSNe) may result in a variety of spectral classifications depending on the state of the progenitor in terms of, for example, its mass, its chemical composition, its structure and other parameters\citep[see e.g.][]{J2012,JMS2016}. In the same vein, SNe Ia may result from a variety of different progenitor scenarios. It is generally accepted that SNe Ia result from the thermonuclear explosion of a white dwarf star (WD) - the degenerate, burnt-out remnant of a relatively low mass ($\lesssim 8 \Msun$)  main sequence (MS) star \citep[e.g.][]{W2000,CKT2018}. WDs are inert in a thermonuclear sense and, if left to their own devices, will simply cool radiatively until they reach thermal equilibrium with the surrounding medium. It is therefore further generally accepted that WDs require material to be donated from another source in order to reinitiate nuclear burning, resulting in an explosion. The process by which this reignition occurs and then proceeds through the WD is currently not sufficiently well understood \citep[see e.g.][]{HN2000, HKRR2013,LM2018,R2020}. While a number of different progenitor scenarios for SNe Ia have been proposed in recent decades, as will be elaborated on below, a clear favourite for the dominant progenitor channel has not been forthcoming. 

The likely nature of SNe Ia as accretion induced thermonuclear explosions associated with WDs puts them into a conceptual family of  thermonuclear transients which also includes classical novae \citep[see e.g.][]{JHI2006,CMS2021}, cataclysmic variables \citep[CVs, see e.g.][]{G2000} and AM Canum Venaticorum \citep[AM CVn, which have been associated with SN Ia progenitor candidate systems,][]{PHR2003,N2005,SYu2005,S2010,PYT2015} and its analogues. Additionally, a number of  other transients originating from WDs have been either proposed or classified. Among these are SNe of Type Iax \citep{LFC2003,FCC2013} and other peculiar SN Ia variants \citep{WK2011}. All of these together, such as thermonuclear events, including ordinary SNe Ia, originating from WDs which are brighter and more energetic than CVs and classical novae and for the purposes of this paper, form the family of 'thermonuclear SNe'. From this point onwards, we refer to the spectral classification of a SN of Type Ia only and exclusively as 'ordinary' SNe Ia. Attempts to identify the progenitor systems of observed individual ordinary SNe Ia have generally yielded ambiguous results.  Counter-intuitively, investigations in this direction have actually been more successful for SNe Iax than for ordinary SNe Ia, \citet{MJS2014} having identified a luminous blue source at the location of SN 2012Z, commensurate with either a massive main sequence star or a hydrogen depleted star of a mass $\sim 1.0 \Msun$.

Progenitor scenarios for thermonuclear SNe are generally accepted to fall into two distinct categories, these being the single degenerate and the double degenerate channels. In the double degenerate channel, the progenitor binary is assumed to consist of two WDs (which may be of different mass and composition). Under the influence of angular momentum loss due to gravitational wave radiation \citep[GWR][]{LL1975}, and assuming a not exceedingly large initial orbital separation, this progenitor system will become tight enough within Hubble-time for the less massive WD to fill its Roche-lobe. Then, given a mass ratio not too different from unity, it will undergo dynamically unstable Roche-lobe overflow (RLOF). Unstable ignition of either the accreting WD, both WDs individually, or an object resulting from the merger of the two components, is expected, leading to a SN  \citep[see e.g][]{W1984,TM1986,CI1986,SRH2010,PKT2012,PKT2013}. The number of candidate progenitors in this channel is surprisingly limited, with only two currently known candidates \citep[KPD1930+2752, HD 265435,][respectively]{MMN2000,PN2021}. In the single degenerate channel, the accreting (degenerate) WD is assumed to receive mass from a non-degenerate companion of some description. The donor star, according to different sub-channels, may be an ordinary main sequence star, an evolved post-main sequence star, either of which would chiefly donate hydrogen-rich material, or a hydrogen depleted star which would chiefly donate helium enriched material \citep{N1980P,N1982a,N1982b,WPH2017}. Again, a number of different ignition and explosion mechanisms have been proposed, a thorough discussion of which would extend beyond the limits of this article \citep[though see][for an extensive review]{R2020}. Between these proposed mechanisms and a lack of direct observational evidence, the structure, in terms of component masses and stellar types involved, of the candidate progenitor binaries, as well as the identity of the spectral type resulting from a particular channel and sub-channel, remain obscured. 

One feature distinguishing the single degenerate channel from the double degenerate channel is that the donor star is generally expected to survive the destruction of the accretor \citep{P2003,LPS2013,BWB2019,N2020,LZ2020,LRYH2021}. Therefore, in this channel, as the accreting WD is disrupted in a SN, the former donor star is flung away, under conservation of its pre-SN orbital angular momentum, from the centre of mass of the progenitor binary \citep[see e.g.][]{Hansen2003,JWP2009}. The velocity of this putative runaway star depends, as will be discussed below, on the star's pre-ejection mass and structure, with highest runaway velocities in the single degenerate channel associated with the helium donor sub-channel. 

In many cases, the ejected helium star will be fast enough to be unbound from the Galactic potential, in the process becoming a hypervelocity star (HVS). HVSs, of which the type under consideration here is not the only variant, have attracted considerable interest over the years. Specifically, higher mass and main sequence candidates are expected to form during interaction of a binary star with a supermassive black hole \citep[][]{H1988,B2015,HP2017}. Notably, \citet{KBD2020} reported on the discovery of an object originating at the Galactic centre. HVS of Galactic centre origin have been utilised in probing the Galactic matter distribution \citep{YM2007} and the environment of the Galactic centre \citep{BLK2018,GM2020,EMR2021}. This present work should be considered in the same line as others focused on the kinematic and population properties of such objects \citep[e.g.][]{BGK2009,LZY2010,RKS2014}, though, unlike these, considering ejected He-stars of SN origin \citep{GFZ2015,VNK2017,SBG2018}. We mention that origins as former companion stars in core collapse SNe have also been considered \citep{RZM2019,ERR2020}. In any case, HVS stars of all masses and configurations are a topic of active research \citep{RIH2020,LLL2020} and observational campaigns focused on their detection, especially in the era of the Gaia Instrument \citep{Gaia-mission2016}, are ongoing \citep{IKH2018,BKB2018,RIH2020,M2021,IDHR2021} and not limited to the Galaxy \citep{EMRB2021}. Most recently, ejection of high velocity stars from dynamical disruption of triple systems has also been considered \citep{HPT2021}. We note that HVS are, conceptually, a variant of runaway stars - stars with a velocity excess over their direct association - as first considered by \citet{B1961}.
In this study we focus on the single degenerate channel, with a particular focus on the helium donor sub-channel (Hereafter: SD-He channel). Specifically, we attempt to resolve the following: firstly, the structure and distribution of a hypothetical population of runaway stars originating from the SD-He channel; secondly, how this structure depends on initial assumptions on the parameter space, specifically with respect to terminal accretor mass; thirdly, whether there are any candidate members of this population; fourthly, which conclusions on the structure of the progenitor binary can be drawn from the observational record; lastly, the theoretically predicted occurrence rates current observational record supports for this channel.

This paper is organised as follows: In Sec.\,\ref{sec:context} we expound on the context of this study and motivate our work. In Sec.\,\ref{sec:assumptions}, we discuss our physical assumptions and explain our numerical methodology. In Sec.\,\ref{sec:results} we summarise the kinematic properties of the population. In Sec.\,\ref{sec:qualitative properties} we use the results presented in the previous section to qualitatively define criteria for adherence to this population and discuss possible sub-divisions. In Sec.\,\ref{sec:quantitative_properties} we discuss the sensitivity of population parameters to assumed initial conditions and define corresponding observables, broadly taking into account constraints on the observability of population members. In Sec.\,\ref{sec:candidates} we attempt to identify possible observed members of our predicted population and categorise them in their expected subpopulations. We comment on astrometric parameters of our population in Sec.\,\ref{sec:astrometry}. Using the identified members as well as the predictions resulting from our simulations, we estimate the event rate for SD-He ejections of US 708-like stars as supported by the currently available observational record as well as calculate the number of detections necessary to support these ejections as significant contributors to the Galactic SN rate in Sec.\,\ref{sec:rates}. We conclude and summarise our study in Sec.\ref{sec:conclusions}.

\section{Motivation and context} \label{sec:context}

\subsection{Thermonuclear explosions of WD stars put into context}
As stated above, this article is chiefly concerned with survivors of the single-degenerate helium donor channel for thermonuclear SNe. In recent years, three major channels and a number of minor ones (in the sense of community interest) have emerged: firstly, the double degenerate channel, which has already been commented upon, which involves a progenitor system composed of two WD stars being driven to RLOF by emission of GWR. This channel is usually not supposed to produce a surviving remnant \citep{W1984}, though exceptions leading to the ejection of a fast runaway have been proposed (see discussion in Sec.\,\ref{ssec:sub_ch});
 secondly, the hydrogen donor channel (also: classical mechanism) where the accreting WD receives mainly hydrogen-rich material from a companion star in on the main sequence or in a red giant configuration. In this scenario, the accreting WD is usually assumed to reach the Chandrasekhar mass, leading to carbon ignition at its core \citep[see reviews by][]{HN2000,WH2012,R2020}. Due to the size of the assumed donor star, this mechanism is incapable of producing a high-velocity runaway. Thirdly, the donor star is assumed to be a hydrogen-depleted helium star. The transferred material will thus be helium enriched. As shown by \citet{HK2004,PTY2014} and others, depending on the mass transfer rate (and assumed input physics, such as rotation), the transferred material will be accumulated quiescently ($\dot{M}\lesssim3\magten{-8}\Mdot$), undergo violent flashes ($3\magten{-8}\Mdot \lesssim \dot{M} \lesssim 1\magten{-7}\Mdot$), weak flashes ($1\magten{-7}\Mdot \lesssim \dot{M} \lesssim 5\magten{-6}\Mdot$), or steady burning ($\dot{M}\gtrsim 5\magten{-6}\Mdot$). If sufficient material is accumulated, the WD's mass may grow to the Chandrasekhar mass, eventually undergoing detonation via the classical mechanism or, if helium ignition conditions are reached, produce a He-detonation, igniting the CO core at the He-CO boundary. The latter may happen at terminal accretor masses different from the Chandrasekhar mass. 

\subsection{Chandrasekhar mass explosions}
For the purposes of this study, we need to distinguish the classical mechanism, and Chandrasekhar mass double detonations. The classical mechanism assumes that a WD accretes sufficient (hydrogen-rich) material from a companion to approach the Chandrasekhar mass. Close to the Chandrasekhar mass, the degenerate electron gas will approach the relativistic limit, leading to a change in the equation of state. The consequent increase in density precipitates thermonuclear ignition and explosion of the WD \citep{HF1960, A1969, HW1969}.
Detailed modelling of the explosion has seen significant advances over the years \citep[see e.g.][]{NTY1984, HN2000}, with a number of different mechanisms having been proposed, but none having been identified conclusively as dominant. Further questions persist as to the ability of a system of this structure (i.e. a WD + hydrogen-rich companion) to provide the necessary conditions for the WD to grow to the Chandrasekhar mass and their contribution to the total SN Ia rate has been estimated to be $<5\%$ by \citet{GB2010}, but see for example \citet{D2010}. However, since, for hydrogen-rich MS stars, expected ejection velocities lie well below the velocity required to unbind the ejected star from the Galactic potential, the classical mechanism is not considered in this study. 

Instead, for our purposes, we consider Chandrasekhar mass explosions to result from the same mechanism as sub-Chandrasekhar mass explosion (as discussed in greater detail in Sec.\,\ref{ssec:sub_ch}) and in the same type of system, such as He-star+WD binaries, likewise producing a runaway He-star. The possibility of a near-Chandrasekhar mass double detonation is usually predicated on inclusion of microphysics, such as magnetic torques, as proposed by \citet{S2002} and modelled by \citet{NYL2017,NYL2019}. However, Chandrasekhar mass explosions originating from He-accretion have also been connected to SNe of Type Iax and have been predicted to leave a partially burnt remnant \citep{GKO2005,BBT2004,PWF2007,FCF2009,KFS2013,FKS2014,KOP2015,MKS2016,LCB2021}.

\subsection{Sub-Chandrasekhar mass explosions} \label{ssec:sub_ch}
Faced with difficulty in explaining how the future SN progenitor (i.e. the accreting WD) can accumulate sufficient mass to reach the Chandrasekhar mass, the question arises whether the star is actually required to reach it, leading to any number of sub-Chandrasekhar mass scenarios. The most well know of these was originally proposed in the 1980s by \cite{N1980P,N1982b}, but has consistently generated interest in the intervening years \citep[e.g.][]{L1990,LA1995,YL2004a,YL2004b,FHR2007,SRH2010,WK2011,NYL2016,NYL2017,GCD2018,NYL2019,GC2021,GCL2021,SBTK2021}. This mechanism, known as double detonation mechanism achieves detonation of the WD's CO core by treating an accumulated, quiescent helium layer, donated by a hydrogen depleted companion, as a thermonuclear fuse. Here, compressive heating, additional nuclear heating via the NCO-effect \citep{HNAK1986} and (depending on included microphysics, viscous heating), precipitate ignition and, under certain circumstances, detonation of the accumulated helium layer (the first detonation). The first detonation then propagates into the CO-core, leading to explosive carbon ignition (the second detonation). The amount of helium needed for the first detonation to occur is being debated in literature but is generally thought, depending on included microphysics and the temperature and mass of the accreting body, to lie in the range of $0.05\Msun - 0.4\Msun$. The minimum mass of the WD (prior to the terminal accretion episode) has been argued to be as low as $0.75\Msun$ \citep{LA1995}.
One peculiarity of progenitors in this channel is the requirement for the gravitational wave merger timescale to be short enough for the He-donor, most likely a subdwarf B or subdwarf O star \citep{H2009,H2016,PTJ2018,PJT2019}, to fill its Roche lobe while still in its core-He burning phase (which is expected to last for $\sim 100\Myr$). This requires the binary at formation of the terminal configuration of the system to be compact. Compactness is customarily argued to be a result of the formation of either the subdwarf BO star or the WD in a common envelope \citep{HP2004,HPM2002,HPM2003,JPH2011,BK2021,KND2021}. Only two  candidate progenitor systems in this channel have been characterised so far. One is CD-30$^{\circ}$11223 \citep{VKO2012} and, potentially and very recently, PTF1J2238+7430 \citep{KB2021}.

Another proposed mechanism in this category bears mentioning: The so-called prompt double detonation or dynamically driven double degenerate double detonation mechanism. Here, turbulent ignition of a thin $<0.05~M_\odot$ helium layer, accreted during the dynamical disruption of a degenerate companion, for example a He-WD or hybrid HeCO WD \citep[e.g.][]{PZ2021}, ignites a secondary detonation of the accretor's CO core \citep{PKT2013}. This mechanism is especially important to keep in mind in the context of this study, as one of the candidates (LP 398-9, see Sec.\,\ref{sec:candidates}) has been claimed as a potential survivor of this scenario, as argued by \citet{SBG2018}, but see \citet{MP2017}. 

\subsection{Super-Chandrasekhar mass explosions}

A straightforward scenario for the formation of a super-Chandrasekhar mass configuration that may lead to explosion is the merger of two sub-Chandrasekhar mass WDs with a combined mass larger than the Chandrasekhar mass.  In a single-degenerate scenario, explosions well above the canonical Chandrasekhar mass may be achieved if the WD is stabilised against ignition or collapse by rotation. Models of rotating WDs stable far above the canonical Chandrasekhar mass were provided by, for example, \citet{OB1968} and \citet{Hachisu1986}.  A distribution of the SN Ia explosion mass over a range of $0.08\Msun$, as a consequence of solid-body rotation in the accreted material, was investigated by \citet{UNH2003} as a potential source of diversity in observed SNe Ia (see also \citealt{YL2004b}).   However \cite{YL2005} found that, for plausible assumptions, the angular momentum gained by accreting WDs might allow differentially-rotating WDs which reach  $2.2\Msun$.  When accretion is no longer able to maintain the necessary differential rotation, the WD may lose or redistribute spin angular momentum \citep[e.g. under the influence of the r-mode instability or tidal interaction, see e.g.][]{YL2004b}, and so ignite and explode.  In the context of this paper, explosions of such highly super-Chandrasekhar WDs from single-degenerate SN Ia progenitors could increase the maximum ejection velocity of hypervelocity stars.  However, depending again on the microphysics of the WD \citep[see e.g.][]{NYL2017}, it is debated whether, and under what circumstances, differential rotation can be maintained  in accreting WDs \citep[see, e.g. ][]{YL2004b,Piro2008,HKSN2012}.  Nonetheless, even if differential rotation cannot be maintained in the accreting WD then, if ignition of the WD due to other factors (such as viscous heating) can be avoided, solid-body rotation may allow the WDs in some single-degenerate SN Ia progenitors to reach $\approx 1.5\Msun$ without explosion or instability \citep[e.g. ][]{Hachisu1986,UNH2003}. 

\citet{J2011} and \citet{DVC2011} argued that considering the increased WD mass at explosion as a consequence of accreted angular momentum plausibly also resolves some of the arguments which had been used against single-degenerate SN Ia progenitors, with the latter paper describing the scenario as the ``spin-up-spin-down'' model (see also, e.g.  \citealt{HKSN2012} for a clear investigation of potential advantages of this model).  A common simple picture of this scenario is of the subset of cases in which mass accretion totally ceases before the explosion, with the WD then spinning-down towards ignition.  The duration of that potential phase of spin-down, or internal angular-momentum redistribution, is one of the important remaining uncertainties for this scenario.   However, a long spin-down phase would be problematic for the ejection of high-velocity helium-burning donor stars. For producing a population of high-velocity He-sdO stars, a natural scenario within the spin-up-spin-down model may be if, in these cases, the rate of ongoing mass accretion becomes insufficient to sustain the rotation needed to avoid explosion (e.g.\ if the WD reaches a mass at which differential rotation is necessary to delay explosion further).

\section{Physical assumptions and methodology} \label{sec:assumptions}

\subsection{Numerical method}

This study employs the lightweight stellar kinematics code SHyRT \citep{N2020,NKGH2021} in order to model the movement of ejected SN survivors through the Galactic potential, which was assumed to be static and correspond to model 1 put forward by \cite{IWT2013} as a revision of \cite{AS1991} in the form 
\begin{equation} \label{eq:model1.1}
	\Phi_b(R)= -\frac{M_b}{\sqrt{R^2+b_b^2}}
\end{equation}
for the bulge, where $R$ is the distance from the Galactic centre (note that the gravitational constant $G=1$ here)
\begin{equation}
	\Phi_d(r,z)= -\frac{M_d}{\sqrt{r^2+(a_d+\sqrt{z^2+b_d^2})}}
\end{equation}
for the disc, where $r$ is the distance from the Galactic centre in the $x$-$y$-plane and $z$ is the distance from the $x$-$y$-plane and
\begin{equation} \label{eq:model1.4}
    \Phi_h(R) =
    \begin{cases}
        -\frac{M_h}{a_h}\left[ \frac{1}{\gamma -1} \ln\left( \frac{1+(R/a_h)^{\gamma-1}}{1+(\Lambda/a_h)^{\gamma-1}} \right) - \frac{(\Lambda/a_h)^{\gamma-1}}{(1 +\Lambda/a_h)^{\gamma-1}} \right] &\text{if}~R < \Lambda \\
        -\frac{M_h}{R} \frac{(\Lambda/a_h)^{\gamma-1}}{(1 +\Lambda/a_h)^{\gamma-1}} &\text{if}~R \geq \Lambda ,
    \end{cases}
\end{equation}
with $\gamma=2$ and the other parameters given in Table~\ref{tab:modelp}.
\begin{table} 
	\caption{Parameters used in Eqs.~\ref{eq:model1.1}-\ref{eq:model1.4}.} \label{tab:modelp}
	\centering
	\begin{tabular}{ccccc} 
		\hline 
		\hline
		& $M_\text{b/d/h}~[\text{M}_\text{Gal}]$ & $a_\text{d/h}~[\text{kpc}]$ & $b_\text{b/d}~[\text{kpc}]$ & $\Lambda~[\text{kpc}]$  \\  
		\hline
		Bulge$_\text{b}$& $409\pm63$  &  & $0.23\pm0.03$  &\\ 
		Disc$_\text{d}$ & $2856^{+376}_{-202}$  & $4.22^{+0.53}_{-0.99}$ & $0.292^{+0.020}_{-0.025}$ &\\ 
		Halo$_\text{h}$ & $1018^{+27933}_{-603}$ & $2.562^{+25.963}_{-1.419}$ &  &$200^{+0}_{-82}$\\ 
		\hline 
	\end{tabular} 
	\tablefoot{$\mathrm{M}_\mathrm{Gal} = 100 \times 1000^2 \mkpc\, [\mathrm{m}]/G\,[\mathrm{SI}] \Msun \sim 2.325\magten{7} \Msun$ \citep[as given by][]{IWT2013} is the Galactic mass unit with the gravitational constant $G=1$.}
\end{table}

This potential is axisymmetric and neglects Galactic substructure such as spiral arms and the bar in the disc as well as cold dark matter substructure in the halo. This choice may affect the ratio of bound to unbound objects on long timescales, but the effect is deemed unimportant for our simulation time limit of $300\Myr$.
Specifically, we analyse six different simulation runs, one each for one assumed terminal WD mass. Parameters are given in table\,\ref{tab:run_par}. Note that run M14 was not performed specifically for this study, but \citet{NKGH2021}, but we present a more detailed and extended analysis here. 
We draw attention to the fact that the upper boundary for the mass of the runaway differs between individual runs. This is a result of the ejection velocity spectra obtained by \citet{N2020} only covering the stated ranges. While this introduces some difficulty in comparing the different synthetic populations, we opt to remain within the bounds of the provided spectra, as opposed to presenting an extrapolation or constraining the samples overall. Instead we highlight the differences in sample size and construct comparable sub-samples if they would, quantitatively or qualitatively, affect our results.

\begin{table} 
	\caption{Parameters for each simulation run performed for this study} \label{tab:run_par}
	\centering
	\begin{tabular}{cccc} 
		\hline 
		\hline
		run & $\MWDf $ & $ \MWd $ range & $N_\mathrm{e}$  \\  
		\hline
		M10& $1.0\Msun$  & $0.20-1.00\Msun$  & $1\magten{6}$  \\ 
		M11& $1.1\Msun$  & $0.20-1.00\Msun$  & $1\magten{6}$  \\ 
		M12& $1.2\Msun$  & $0.20-0.95\Msun$  & $1\magten{6}$  \\ 
		M13& $1.3\Msun$  & $0.20-0.90\Msun$  & $1\magten{6}$  \\ 
		M14& $1.4\Msun$  & $0.20-0.80\Msun$  & $1\magten{6}$  \\ 
		M15& $1.5\Msun$  & $0.20-0.70\Msun$  & $1\magten{6}$  \\ 
		\hline 
	\end{tabular} 
	\tablefoot{ M14 is identical to the synthetic population used by \citet{NKGH2021}.} 
\end{table}

Each run contains a total number of $N_\mathrm{e}$ individual tracks, with initial positions randomly assigned in the Galactic $x$-$y$-plane, but weighted to correspond to the Galactic density profile as defined by model 1 provided by \citet{IWT2013} and see \citet{NKGH2021} for the initial distribution. The ejection direction is chosen randomly across $4\pi$. The ejection velocity is determined by the randomly chosen mass of the ejected star, randomly falling into the ranges (depending on mass) detailed in Table\,\ref{tab:vvalues}, as provided by \citet{N2020}.

\begin{table*} 
	\begin{center}
		 \caption{Ejection velocity spectra} \label{tab:vvalues}
		\begin{tabular}{cc|cccccccc} 
			\hline 
			\hline
			$\MWDf$ [$\Msun$]	&	velocity	&	$\MWd$ [$\Msun$]	&		&	&		&		&	&		&		\\
			& [$\mkms$]	&	0.20	&	0.25	&	0.30	&	0.35	&	0.40	&	0.45	&	0.50	&	0.55	\\
			\hline
			1.00	&	$v_\mathrm{max}$ 	&	968.1	&	960.1	&	860.7	&	765.2	&	665.6	&	650.1	&	606.5	&	571.6	\\
			&	$v_\mathrm{min}$ 	&	961.6	&	948.3	&	815.7	&	731.9	&	662.8	&	615.3	&	577.2	&	548.1	\\
			1.10	&	$v_\mathrm{max}$ 	&	1010.1	&	1000.1	&	903.3	&	798.9	&	735.0	&	683.8	&	638.6	&	603.8	\\
			&	$v_\mathrm{min}$ 	&	1000.1	&	986.7	&	845.3	&	756.1	&	686.9	&	639.3	&	601.8	&	572.0	\\
			1.20	&	$v_\mathrm{max}$ 	&	1054.9	&	1038.0	&	942.2	&	834.1	&	768.4	&	715.5	&	668.9	&	632.1	\\
			&	$v_\mathrm{min}$ 	&	1035.9	&	1027.6	&	870.8	&	776.2	&	708.0	&	661.7	&	625.0	&	594.3	\\
			1.30	&	$v_\mathrm{max}$ 	&	1098.2	&	1074.8	&	973.7	&	865.5	&	794.8	&	737.6	&	690.7	&	659.9	\\
			&	$v_\mathrm{min}$ 	&	1069.2	&	1052.5	&	893.3	&	800.5	&	735.2	&	687.6	&	652.7	&	620.2	\\
			1.40	&	$v_\mathrm{max}$ 	&	1140.0	&	1107.0	&	999.4	&	887.2	&	817.8	&	759.8	&	713.1	&	705.7	\\
			&	$v_\mathrm{min}$ 	&	1100.0	&	1088.0	&	918.2	&	830.1	&	762.0	&	712.3	&	677.8	&	644.8	\\
			1.50	&	$v_\mathrm{max}$ 	&	1194.3	&	1166.7	&	1069.9	&	916.2	&	824.4	&	758.5	&	710.8	&	671.5	\\
			&	$v_\mathrm{min}$ 	&	1128.8	&	1127.9	&	946.8	&	861.7	&	787.9	&	737.4	&	701.9	&	669.3	\\
			\hline 
		\end{tabular}  \\
	\vspace{0.3cm} 
	
	    \begin{tabular}{cc|ccccccccc} 
	    	\hline 
	    	\hline
	    	$\MWDf$ [$\Msun$]	&	velocity	&	$\MWd$ [$\Msun$]	&		&	&		&		&	&		&	&	\\
	    	& [$\mkms$] &	0.60	&	0.65	&	0.70	&	0.75	&	0.80	&	0.85	&	0.90	&	0.95	&	1.00	\\
	    	\hline
	    	1.00	&	$v_\mathrm{max}$ 	&	544.9	&	521.9	&	500.8	&	483.6	&	461.1	&	443.2	&	434.5	&	423.2	&	404.4	\\
	    	&	$v_\mathrm{min}$ 	&	524.6	&	502.3	&	483.9	&	469.8	&	454.8	&	442.6	&	429.2	&	416.2	&	404.4	\\
	    	1.10	&	$v_\mathrm{max}$ 	&	574.4	&	551.1	&	529.8	&	511.9	&	496.4	&	471.4	&	455.2	&	448.1	&	429.5	\\
	    	&	$v_\mathrm{min}$ 	&	547.8	&	525.4	&	506.5	&	492.8	&	484.5	&	469.0	&	455.2	&	441.7	&	429.5	\\
	    	1.20	&	$v_\mathrm{max}$ 	&	603.2	&	579.6	&	557.4	&	539.5	&	522.0	&	502.2	&	481.7	&	470.5	&		\\
	    	&	$v_\mathrm{min}$ 	&	569.0	&	548.1	&	528.9	&	514.6	&	502.1	&	492.6	&	479.8	&	465.9	&		\\
	    	1.30	&	$v_\mathrm{max}$ 	&	628.4	&	604.2	&	575.9	&	554.4	&	536.5	&	517.5	&	504.8	&		&		\\
	    	&	$v_\mathrm{min}$ 	&	597.6	&	576.1	&	553.3	&	538.3	&	525.4	&	514.5	&	504.8	&		&		\\
	    	1.40	&	$v_\mathrm{max}$ 	&	635.9	&	606.6	&	583.5	&	562.7	&	546.5	&		&		&		&		\\
	    	&	$v_\mathrm{min}$ 	&	620.8	&	598.9	&	578.4	&	558.5	&	546.5	&		&		&		&		\\
	    	1.50	&	$v_\mathrm{max}$ 	&	698.1	&	622.2	&	637.4	&		&		&		&		&		&		\\
	    	&	$v_\mathrm{min}$ 	&	643.9	&	622.2	&	637.4	&		&		&		&		&		&		\\
	    	\hline 
	    \end{tabular} 
    \tablefoot{Minimum $v_\mathrm{min}$ and maximum $v_\mathrm{max}$ ejection velocities in units of $\mkms$ for runaways of discrete terminal mass $\MWd$ and terminal accretor mass $\MWDf$ as provided by \citet{N2020} and utilised in this study.}
	\end{center}
\end{table*}
\subsection{Physical assumptions}
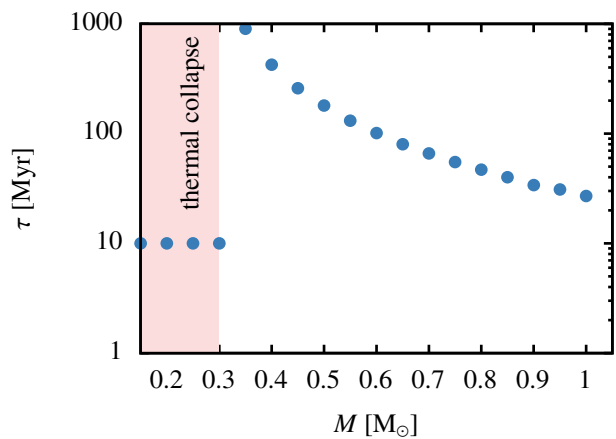
\begin{figure}
	\centering 
	\input{helife2}
	\caption{Assumed lifetimes $\tau$ of the ejected He-stars. Stars in the shaded region are below the helium-burning limit and are deemed to thermally collapse to form WDs within 10$\Myr$.} \label{fig:helife}
\end{figure}

Each trajectory is followed for a duration of 300$\Myr$, with one object assumed to be ejected every 300\,yr. This ejection rate corresponds to the inferred Galactic SN Ia rate \citep{CTT1997,SLP2006}. Unless explicitly stated otherwise, we characterise each of our synthetic populations as a snapshot of all ejected objects at the end of the simulated 300$\Myr$ time frame. Each object is assigned a predetermined lifetime corresponding to its nuclear timescale at the start of its helium-burning phase. The latter strategy is a result of our attempt to remain agnostic towards the explosion mechanism leading the ejection of the runaway. Different explosion mechanisms lead to a variety of requirements for the terminal accretor mass and the mass accretion rate it is subjected to. It follows that, by accepting the resulting agnostic ejection velocity spectra as input parameters of our study, we lose sensitivity to the remaining lifetime of the ejected star. Keeping this in mind, the nuclear timescale of a He-star of equal mass to the ejected star can serve as a reliable, though overestimated, upper limit. The mass-dependent lifetime assumed in this study is depicted in Fig.\,\ref{fig:helife}. The values shown in Fig.\,\ref{fig:helife} were derived from the nuclear timescales of MESA models of the appropriate mass \citep{MESA1,MESA2,MESA3,MESA4,MESA5} release 12778. As such, while all ejection trajectories are followed for the full duration mentioned above, stars exceeding their mass dependent lifetime are flagged appropriately.

Each terminal donor mass is chosen randomly in a range corresponding to the values given in Table\,\ref{tab:run_par}. The mass distribution is assumed to be flat. The latter assumption is again motivated by our desire to remain agnostic towards the explosion mechanism, as a non-flat initial mass distribution may be a result of either the explosion mechanism (i.e. certain terminal donor masses are favoured) or of the progenitor-binary's pre-evolution. Small number statistics in the observational record precludes reconstruction of an empirical initial mass distribution (i.e. the mass distribution of the observed population after ejection), while uncertainties in the theoretical modelling preclude a reliable construction of an initial mass distribution. We do note that population synthesis models predict that the initial mass distribution of the progenitor system (i.e. the mass distribution at the time of formation of the progenitor binary) is not flat \citep[see e.g.][]{WJH2013}. However, as seen in the aforementioned reference, the mass distribution of the donor star strongly depends on assumptions on the common envelope efficiency parameter $\alpha$, with lower end estimates producing an approximately flat distribution. As the observational record for these objects improves, reconstruction of an empirical initial mass distribution may become feasible, but for the moment we opt to assume a non-biased approach and allow the initial mass distribution to be flat as a baseline. However, we investigate some variation in initial conditions in Sec.\,\ref{sec:quantitative_properties}.
As seen in Table\,\ref{tab:vvalues}, each terminal donor mass is associated with a range of ejection velocities. Each individual ejection velocity is then randomly chosen from this range, again assuming a flat distribution. As shown by \citet{BWB2019} and, more recently \citet{LRYH2021}, interaction with the SN ejecta imparts a non-zero kick on the surviving donor star of the order of $200\mkms$. However, as this kick is (in circularised systems, as under consideration here) imparted perpendicular to the star's orbital motion, the pre-kick orbital motion dominates the post-kick velocity of the runaway (the difference is of the order of $20\mkms$). Since the magnitude of this kick depends on the pre-kick orbital configuration and the particulars of the SN (i.e. orbital separation at the time of the SN and energetics of the explosion), we neglect it in our calculations.
The SN progenitor system is deemed to be commoving with the surrounding stellar environment which, in turn, is assumed to orbit the Galactic centre. Therefore, the ejected star is given an additional initial velocity relative to its assumed rotation around the Galaxy as determined by model 1 of  \citet{IWT2013}.

\begin{figure*}
	\centering 
	\vspace{-2cm}
	\input{example-tracks-fig2}
	\vspace{-2cm}
	\caption{Illustration of 75 tracks, out of $1\magten{6}$ , chosen at random, from the $M\sub{WD,f} = 1.4\Msun$-population. Coordinates are Galactic Cartesian. Origins, all set at $z=0$, are indicated by red circles. Tracks dropping below the Galactic plane are indicated by dashed lines.} \label{fig:example-tracks}
\end{figure*}
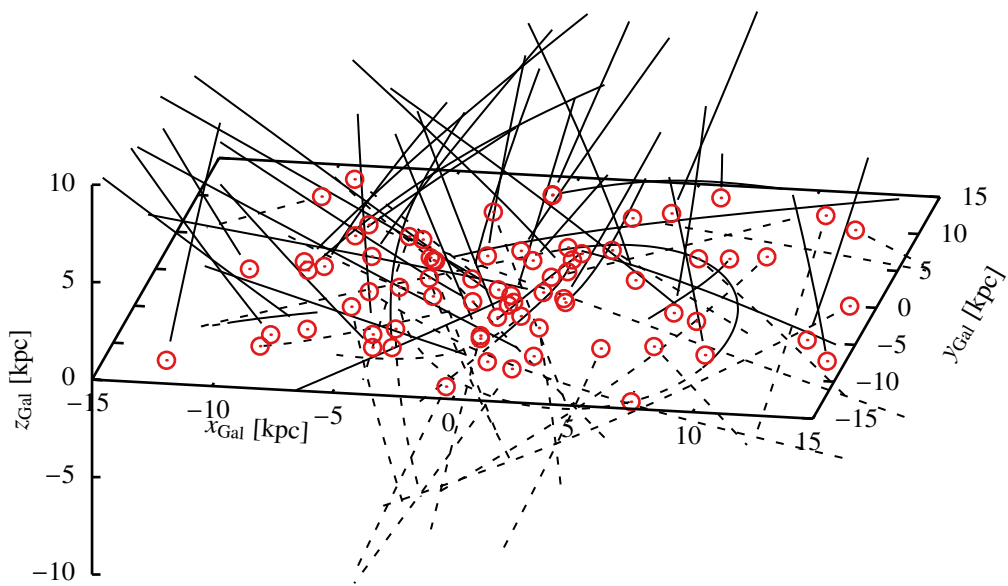
A sample of the resulting trajectories is shown in Fig.\,\ref{fig:example-tracks}. It should be noted that, while the number of experiments for each run is the same, the terminal mass of the exploding WD ($\MWDf$) is varied. However, since we opted to sample the ejection velocity spectra provided by \citet{N2020}, which are subject to a $\MWDf$-dependent upper limit for the terminal mass of the runaway sdB star, the number of runaway mass samples for each terminal accretor mass is different. These differences between our individual runs, as well as their designations, are summarised in Table\,\ref{tab:run_par}. 

In our simulations we neglect of the effects of dynamical friction for runaways crossing the Galactic stellar population. The timescale $T_\mathrm{DF}$ for DF to change the speed $v$ of the runaway (mass $M$) by order itself can be estimated as (\citealt{2008gady.book.....B}, S8.1) 

\begin{align}
	\label{eq:tDF}
	\nonumber T_\mathrm{DF} &\equiv \left ( \frac{1}{v} \frac{\mathrm{d} v}{\mathrm{d} t} \right )^{-1} \sim \frac{v^3}{4\pi G^2 M \rho \ln \Lambda} \left [ \mathrm{erf}(X) - \frac{2X}{\sqrt{\pi}} \exp \left (-X^2 \right ) \right ]^{-1} \\
	\nonumber &\approx \frac{v^3}{4\pi G^2 M \rho \ln \Lambda} \\
	\nonumber & \approx 8.4 \times 10^{17} \,\mathrm{yr} \, \left ( \frac{v}{600 \,\mathrm{km\,s^{-1}}} \right )^3 \left ( \frac{M}{0.4 \, \Msun} \right )^{-1} \left ( \frac{\rho}{0.1\,\Msun \,\mathrm{pc^{-3}}} \right )^{-1} \\
	&\quad \times \left ( \frac{\log \Lambda}{27} \right )^{-1},
\end{align}

with $X \equiv v/(\sqrt{2} \sigma)$ and $\sigma$ as the velocity dispersion, $\rho$ is the stellar density, and $\Lambda$ is the Coulomb factor, estimated as $\Lambda \sim b_\mathrm{max}/b_{90} \sim b_\mathrm{max} v^2/[G(M + m_\star)]$ with $b_\mathrm{max}$ being the maximum impact parameter and $m_\star$ the average stellar mass. Assuming $M = 0.4 \Msun$, $m_\star = 1 \, \Msun$, $v = 600\,\mathrm{km\,s^{-1}}$, which is a reasonable lower limit for stars ejected to unbound orbits,  and $b_\mathrm{max} = 10\,\mathrm{kpc}$, $\log \Lambda \sim 27$. The second line in Eq.~\ref{eq:tDF} applies since $X \sim 11$  assuming $\sigma = 40\,\mathrm{km\,s^{-1}}$ for the Solar neighbourhood  \citep[e.g. ][]{2008gady.book.....B}, implying that the factor in square brackets is extremely close to unity. For these assumed values and setting $\rho=0.1\,\Msun \,\mathrm{pc^{-3}}$ for the Solar neighbourhood \citep[e.g. ][]{2008gady.book.....B}, Eq.~\ref{eq:tDF} yields that $T_\mathrm{DF}$ exceeds the Hubble time by many orders of magnitude, thus showing that dynamical friction for unbound objects is unimportant. 
\section{Results - Kinematic overview} \label{sec:results}
We performed $6\magten{6}$ experiments as described above. However, in order to keep the length of this article manageable, we opt to focus our analysis on subsets of our data as required by the current topic of discussion. Specifically, we discuss general results (i.e. results easily transferred to the rest of the set) by using the M14 set as a baseline. Where necessary we expand to the M10, M12 and M15 sets as a comparison. The M11 and M13 sets are not discussed in detail and only mentioned where required.
For each synthetic population, we assume one ejection (i.e. one SN producing a runaway) every 300\,years, in line with estimates of the inferred Galactic SN Ia rate \citep{CTT1997,SLP2006}. Each trajectory is then randomly assigned an initial ejection time and its position at the end of a $300\Myr$ period is recorded. The final sample for each synthetic population is composed of the position and velocities, as well as the mass and state (pre-WD or WD) for each individual trajectory.

\subsection{Trajectories} \label{sec:trajectories}
The velocity of a star ejected in a SD-He SN is determined mainly by its terminal orbital velocity, to which is added the local orbital velocity of the commoving stellar environment. In the following, we discuss the correlation of general properties of population orbits as measured at a given location. 
\begin{figure}
	\centering 
	\input{local-space-earth-fig2}
	\caption{Trajectories of all stars crossing into Earth local space from the outside. Unbound trajectories are indicated as red lines, bound trajectories as blue lines. Panel (A) and (B) show Galactic Cartesian coordinates in the $x$-$y$-plane, with panel (A) showing the full trajectories and panel (B) the respective point of origin. Panel (C) shows full trajectories in the x-z-plane. The positions of Earth and the Galactic centre are as indicated. Note that, as seen from above the Galactic plane (with Earth located above the plane), the Galaxy rotates in a clockwise fashion.} \label{fig:local-space-earth}
\end{figure}
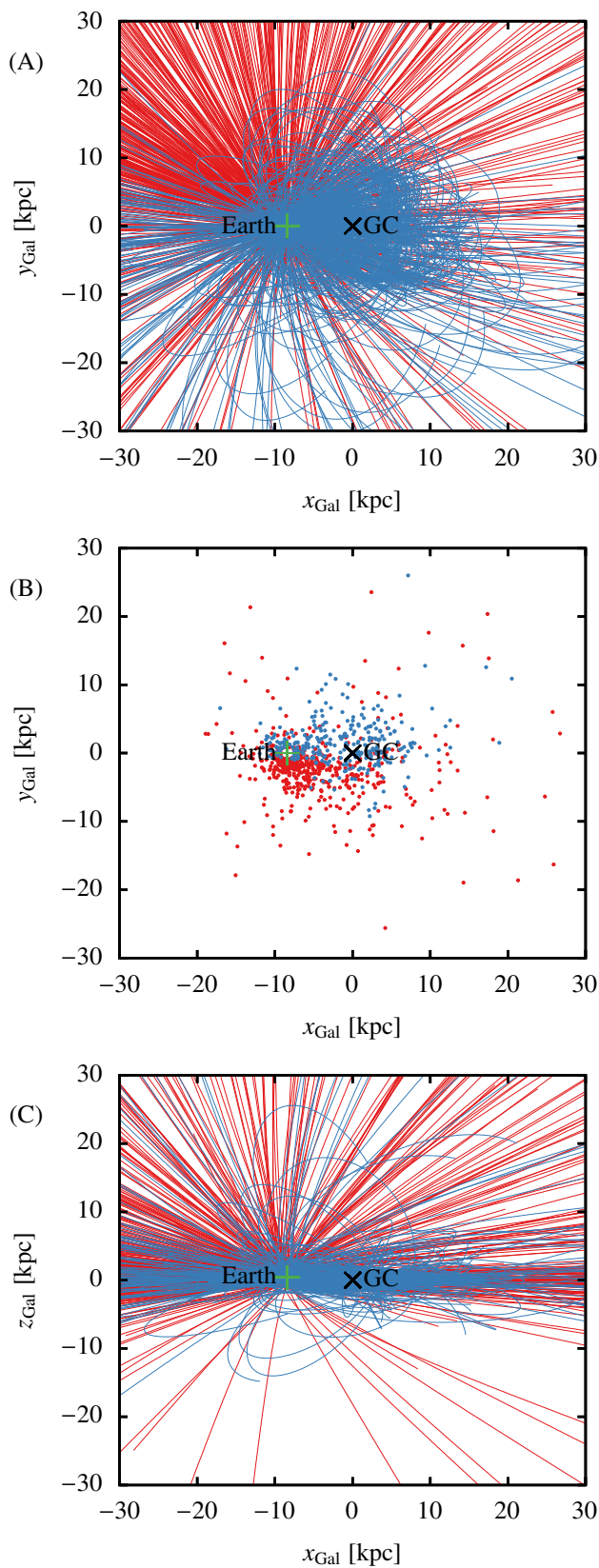
Here, as a practical choice, we measure the velocity field at two locations. Once at the position of Earth with respect to the Galactic disc, and once at the location of US\,708 in the same frame. A full discussion of the velocity field is possible given our data set, but is deemed beyond the intended scope of this study (and deemed more worthwhile if the observational record improves in the future). For this investigation we define a radius of 1\kpc around the locale of observation as "local space".  Fig.\,\ref{fig:local-space-earth} shows all trajectories (independent of ejection time) in the M14 data set which, at some point during the simulated time interval, cross into Earth's local space (though trajectories originating within this space are neglected) as well as the ejection locations of each one of these tracks, together with whether they are bound or not. During the following discussion it should be noted that the Galaxy rotates, as seen from above (with Earth located above the Galactic plane), in a clockwise direction \citep{BCK2014}. Fig.\,\ref{fig:local-space-earth} (A) shows a clear preference for unbound stars to leave Earth's local space (and then the Galaxy) in a prograde fashion with respect to Galactic rotation, while bound objects have a preference for retrograde motion.

 Fig.\,\ref{fig:local-space-earth} (B) shows the ejection locations for the same tracks. As can be seen, trajectories crossing into local space are dominated by stars ejected from the vicinity of Earth, with tracks originating from prograde locations dominated by bound stars and tracks originating from retrograde locations dominated by unbound stars. The Galactic centre, where, due to the higher assumed density of progenitor binaries, a higher number of stars are assumed to be ejected, plays a secondary role compared to Earth's immediate vicinity. There is further a preference for tracks originating from locations at shorter distance from the Galactic centre. This is in line with the expectation that higher density regions dominate. Fig.\,\ref{fig:local-space-earth} (C) shows the Galaxy in a the x-z-plane. As seen, the tracks leaving the Galaxy in a positive z-direction dominate. This is a consequence of or initial assumption of all ejection locations being uniformly located at $z_\mathrm{Gal}=0\mkpc$ but Earth being located at about $z_\mathrm{Gal}=0.4\mkpc$. If ejection locations were (more realistically) scattered throughout the volume of the disc, this preference would likely diminish, but not vanish.
\begin{figure}
	\centering 
	\input{chirality-M14-fig2}
	\caption{Chirality of the ejected stars crossing into Earth local space in the M14 data set with respect to Galactic rotation. On the x-axis (location), prograde and retrograde refer to the ejection location of the star being located either prograde or retrograde in the Galactic disc with respect to Earth. On the y-axis (direction), prograde and retrograde refer to the initial velocity vector of the ejected star being oriented either prograde or retrograde with respect to Galactic rotation. The position of Earth is as indicated. Blue dots imply bound stars, red unbound stars.} \label{fig:chirality-M14}
\end{figure}
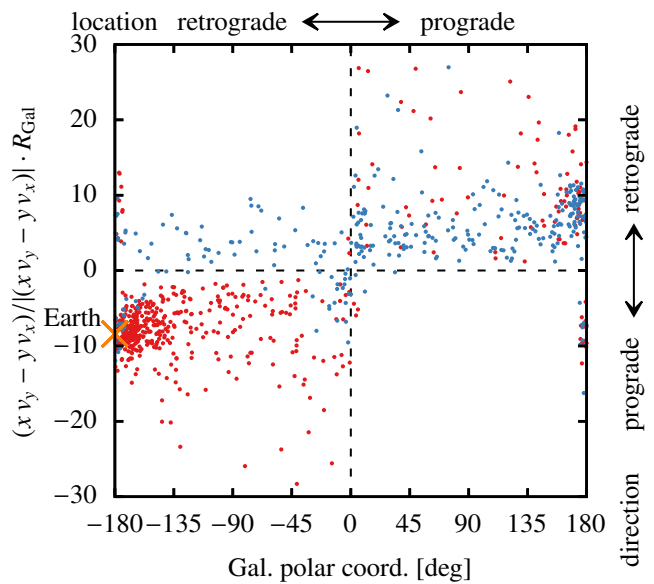

The correlation between kinematics and ejection location is further exemplified by Fig.\,\ref{fig:chirality-M14}. This figure shows the ejection location of each star in the Galactic plane, given in terms of angle centred on the Galactic centre with Earth located at 180 deg. This means that, the Galaxy seen from above, the left half of the plot shows the lower (retrograde) half of the Galactic disc while the right half of the plot indicates the upper (prograde) half. The upper half of the plot shows ejection direction aligned retrograde to Galactic rotation, the lower half prograde. The y-axis also provides a notion of the distance of the ejection location from the Galactic centre. This figure is thus divided into four quadrants. Unbound trajectories crossing into Earth's local space are clearly correlated with retrograde ejection locations and prograde ejection directions while bound trajectories tend (though not quite as strongly correlated) to originate from prograde ejection locations while having their ejection directions aligned in a retrograde fashion. The retrograde-retrograde quadrant is unsurprisingly dominated by bound trajectories, which orbit the Galaxy once (in a retrograde direction) before crossing into Earth's local space \footnote{We emphasise at this point that the position of Earth was merely chosen as a convenient point to sample the velocity field.}. As we do not take the movement of Earth into account in this investigation, the observer would have moved away from the sample location by the time stars on the retrograde-retrograde quadrant orbit the Galaxy. The lower right, prograde-prograde, quadrant is largely empty. The conclusion from this is that stars ejected prograde from a prograde location do not tend to cross into Earth's local space, except for a very small part of the parameter space where stars are ejected largely radially. The absence of any unbound trajectories in this part of the parameter space is not surprising, as these stars will simply be ejected. More surprising is the absence of any bound trajectories, which can be interpreted as Galactic rotation, all else being the same, being a decisive factor in whether an ejected star is bound or unbound. Further, stars on bound trajectories of this type will likely require more than the interval $300\Myr$ taken into account in our simulation to cross back into Earth local space. We further find that a star approaching Earth from a prograde position on a retrograde orbit is likely bound and vice versa.
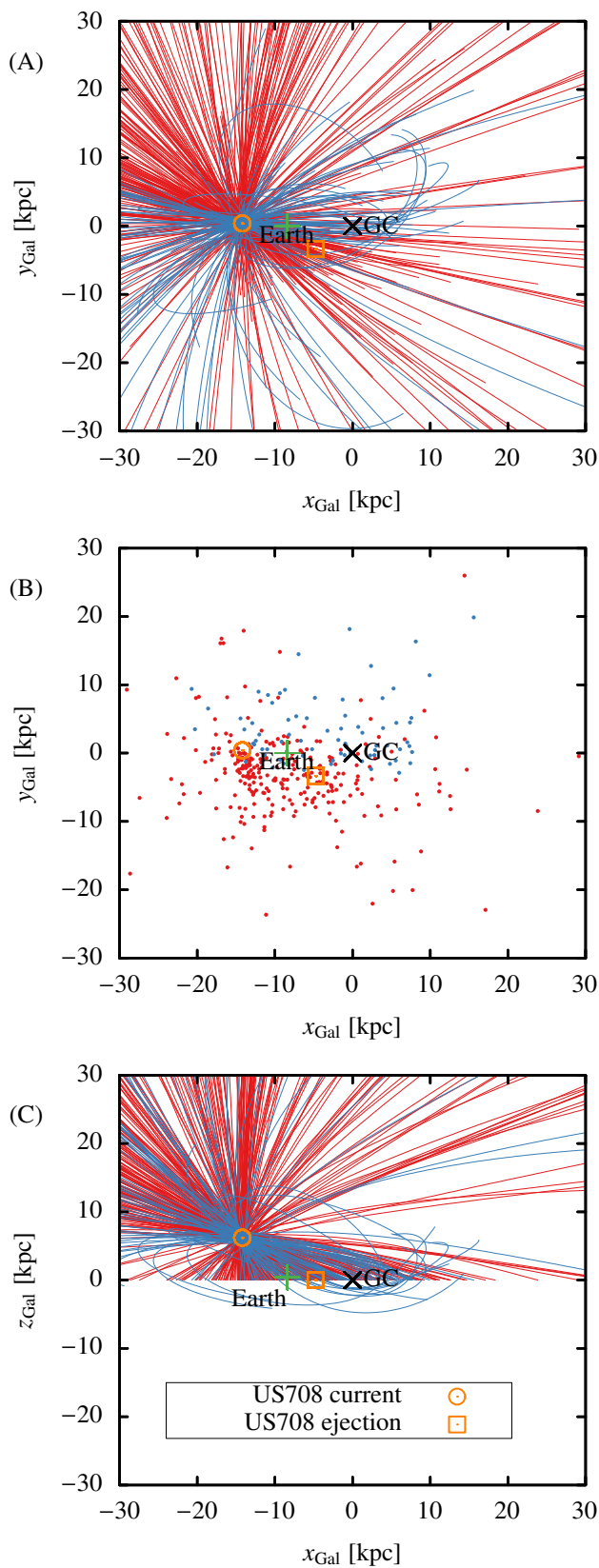
\begin{figure}[h!]
	\centering 
	\input{local-space-US708-fig2}
	\caption{Same as Fig.\,\ref{fig:local-space-earth} for the US\,708 local space. Here the inferred current positions of  US\,708 and its ejection location are indicated as well.} \label{fig:local-space-Us708}
\end{figure}

Similarly, Fig.\,\ref{fig:local-space-Us708} shows trajectories intersecting a 1\kpc volume surrounding the current position of US\,708 (i.e. US\,708 local space). As seen, there is a clear preference for trajectories intersecting US\,708 local space to be unbound rather than bound. This is equivalent to the statement that a star is more likely to reach a position such as that of US\,708 if it is unbound rather than bound. As in Earth local space, unbound trajectories tend to leave the Galaxy in a prograde direction. There is a clear preference for stars to reach the position of US\,708 if they are ejected perpendicular to the Galactic plane (note that this is not the case of US\,708 itself). As further seen, the ejection location of US\,708 is not clearly favoured over locations below (on the positive z-axis) its current position. As for Earth local space, there is a clear preference for ejections towards the positive z-axis to reach US\,708 local space.
\subsection{Number density} \label{ssec:numberdens}
The expected spacial distribution of the runaway population is modelled according to the precepts set out in Sec.\,\ref{sec:assumptions}. The population overview shown here represents a snapshot of the population at the end of the 300$\Myr$ simulation timeframe.
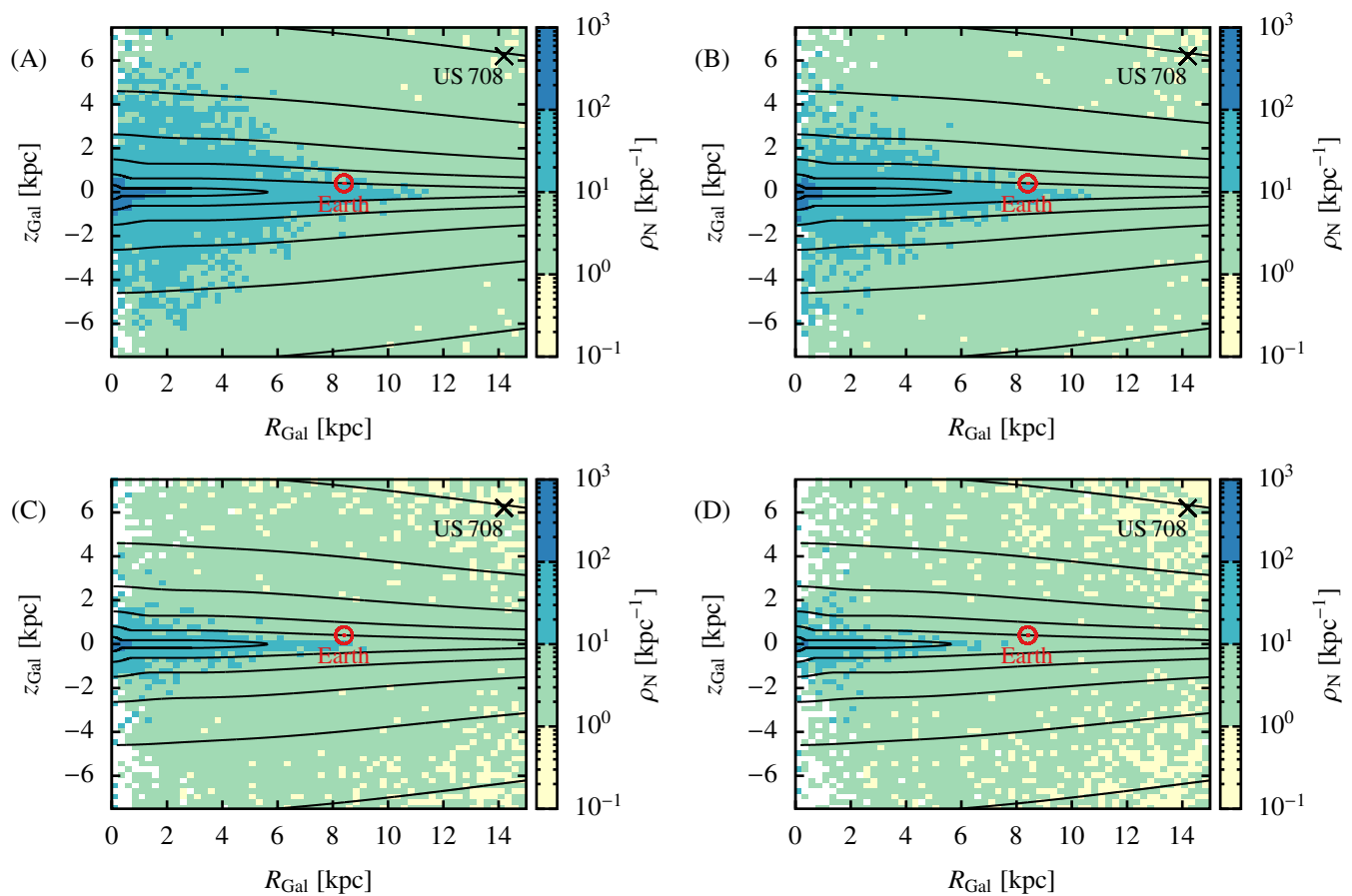
\begin{figure*}
	\centering 
	\input{r-z-fig2}
	\vspace{0.3cm}
	\caption{Expected number density distribution per \kkpc in cylindrical galactocentric coordinates for different terminal accretor masses with $M\sub{WD,f} = 1.0\Msun$, $M\sub{WD,f} = 1.2\Msun$, $M\sub{WD,f} = 1.4\Msun$ and $M\sub{WD,f} = 1.5\Msun$ depicted in panel (A), (B), (C) and (D) respectively. Here, the colour scale indicates the calculated number density. The current positions of Earth and US\,708 are indicated. Black lines indicate surfaces of constant baryonic matter density, logarithmic in base 10, as proposed by \cite{IWT2013}, model 1. White areas do not contain ejected stars in this realisation of our model.} \label{fig:r-z-fig2}
\end{figure*}
In Fig.\,\ref{fig:r-z-fig2} we construct a two-dimensional histogram along Galactic cylindrical coordinates for the M10, M12, M14 and M15 populations. This illustrates a  difference in distribution between the progenitor population which, (if initially distributed in the volume of the disc as opposed to only in the plane) would follow the black isopycnics. For the ejected population, the local space-number density is depicted in colour scale with each change in colour indicating a change in magnitude, and each black line indicating a drop in magnitude of baryonic matter density, starting at the Galactic centre. It is evident that the density gradient of the ejected population is shallower than for the progenitor population. At this point it is evident that the structure of the ejected population depends on the terminal accretor mass of the progenitor binary, with the lowest terminal accretor mass, in the M10 population, correlated with a more spherical space distribution, while higher terminal accretor masses would produce a flat, cylindrical space distribution. In general, all other parameters being the same, lower terminal accretor masses lead to a larger number of runaways remaining in the vicinity of the Galaxy. Distinctively, synthetic populations with higher terminal accretor masses (M12, M14 and M15) exhibit a roughly cone-shaped volume directly above and below the Galactic bulge largely devoid of stars. While also present, though less pronounced, in the M10 population, this feature is a result of Galactic rotation leading to radial ejection directions being favoured over ejection directly perpendicular to the Galactic plane.
\begin{table}
	\caption{Relative number densities}
	\label{table:relative-densities}      
	\centering
	\begin{tabular}{ c c c }
		\hline\hline
		$M_\mathrm{WD,f}$ & $\rho_\mathrm{Earth}$ $[\mathrm{kpc}^{-3}]$ & $\rho_\mathrm{US\,708}$ $[\mathrm{kpc}^{-3}]$  \\    
		\hline                        
		$1.0\Msun$ & $11.18\pm1.78$ & $1.67\pm0.21$  \\
		$1.1\Msun$ & $9.55\pm1.40$ & $1.50\pm0.15$  \\
		$1.2\Msun$ & $9.26\pm1.96$ & $1.26\pm0.22$ \\
		$1.3\Msun$ & $7.00\pm0.83$ & $1.20\pm0.07$ \\
		$1.4\Msun$ & $6.18\pm1.24$ & $1.02\pm0.18$\\
		$1.5\Msun$ & $4.40\pm0.81$ & $0.89\pm0.11$ \\
		\hline
	\end{tabular}
	\tablefoot{Calculated relative number densities of ejected hypervelocity sdB stars in Earth and US\,708 local space depending on the assumed terminal accretor mass. Errors indicate one $\sigma$.}
\end{table}
\begin{figure}
	\centering 
	\input{relative_n_dens-fig2}
	\caption{Comparative number densities in the vicinity of Earth and US\,708 as described in Table\,\ref{table:relative-densities}. Error bars indicate one $\sigma$.} \label{fig:relative_n_dens}
\end{figure}
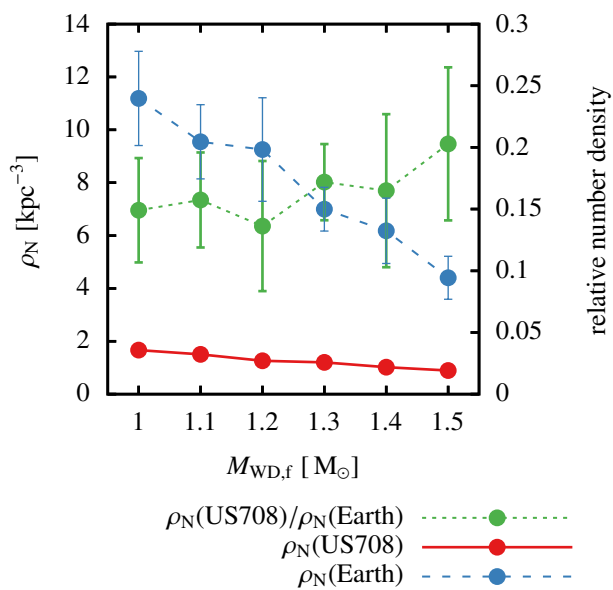
Comparison of the number density of objects in the vicinity of Earth and in the vicinity of US\,708, as summarised in Table\,\ref{table:relative-densities} and Fig.\,\ref{fig:relative_n_dens}, reveals, apart from a higher expected number density in the vicinity of Earth, an approximately constant relative density between the two locations, see green dashed line and error bars in Fig.\,\ref{fig:relative_n_dens}\footnote{Gaussian error bars were obtained by varying the sample region from 0.05\kpc to 5.00\kpc.}.  One of the possible reasons that US\,708 was discovered before any comparable objects closer to Earth may be (apart from the conclusion that they do not exist) that US\,708-like objects could be difficult to disentangle from a denser stellar environment. Another possible reason, namely that US\,708 (like all other candidates) is extreme in at least one observable, will be discussed in Sec.\,\ref{sec:astrometry}.
\section{Qualitative properties of the predicted population} \label{sec:qualitative properties}
\subsection{Defining criteria} \label{sec:criteria}
Taking into account the theoretical stellar properties of population members, observational evidence and kinematic properties as discussed here, a star strictly qualifies as a candidate member of this population by fulfilling the following criteria:
\begin{enumerate}
	\item One of:
	\begin{enumerate}
		\item Strong He-enrichment with $\log{n_\mathrm{H}/n_\mathrm{He}} \lesssim -2$, ideally spectroscopically similar to US\,708.
		\item Spectra compatible with a proto-WD.
		\item Currently a WD.
	\end{enumerate}
	\item Either of: 
	\begin{enumerate}
		\item Currently located in the disc.
		\item If part of the UI or BI subpopulations (see Sec.\,\ref{sec:subpopulations}): Time of flight to the most recent disc intersection compatible with typical sdB-lifetimes.
	\end{enumerate}
	\item Ejection velocities at the most recent disc crossing in the range $400-1150\mkms$, depending on mass (if determined).
	\item Likely mass $<1.0\Msun$.
	\item Time since ejection $<300\Myr$ \label{it:5}
\end{enumerate}
We note that item \ref{it:5} is only necessary as a consequence of the constraints adopted for our simulation. Bound WDs, i.e. descendants of bound non-degenerate runaways, may have been ejected earlier than in the given time frame.
Specifically, an inferred ejection velocity which includes the Galactic centre does not disqualify candidates. Distinction from HVS produced by interaction with the SMBH \citep[][and the like]{H1988} at the Galactic centre should chiefly rely on velocity and structure arguments, not the inferred ejection location, unless the trajectory can be traced back there directly \citep[as in][]{KBD2020}. Further, being unbound from the Galactic potential should also not be used as a criterion, as the question of boundness relies heavily on the intrinsically random parameter of the ejection direction. Bound and unbound objects instead form their own subpopulations.
\subsection{Subdivisions} \label{sec:subpopulations}
\subsubsection{Boundness}
As shown earlier in Sec.\,\ref{sec:results} and later in Fig.\ref{fig:bound-unbound-mass} (C), in the benchmark model, runaways of masses $\Mrun<0.3\Msun$ will not produce bound objects, while all other runaway masses may result in both bound and unbound objects. Since in our model further interactions with the remaining stellar population are neglected, an initially bound object will remain bound for the duration of our reference time. While, as stated, the question of an observed object being bound is, due to uncertainties in astrometry and insufficiently precise knowledge of the Galactic potential, not easily answered, boundness serves as a convenient division of the population. As such we designate objects bound to the Galactic potential (these would be called halo stars or extreme halo stars observationally) as "B-type". Conversely, unbound objects are designated "U". As will be shown later, terminal accretor masses $\MWDf < 1.4\Msun$ may produce B-type objects at all considered runaway masses. It should be mentioned that, if the Galaxy were to considered as a more realistic N-particle system instead of a static potential, as in this study, multi-body interaction post-ejection may lead to B-type objects becoming U-type and vice versa. However, since the size of the total population (as well as important parameters such as its mass spectrum) is currently not discernable observationally, this possibility is deemed unimportant for this study.
\subsubsection{Evolution}
As shown by \citet{N2020} the descendant of the donor star of the progenitor binary is unlikely to have the same appearance as a WD. Instead, if its mass at the time of ejection is larger than $0.3\Msun$, it is likely to appear observationally as a He-star (likely very similar to a He-sdO, like US\,708). While some introduction of additional entropy into the envelope of the ejected object cannot be ruled out, the star is expected to return to its previous state after approximately one Kelvin-Helmholtz timescale \citep{LSAW2015,BWB2019,LRYH2021}. The star, then a runaway, will continue its evolution as a He-star, eventually becoming a WD. If the mass of the ejected component prior to ejection decreases to below $0.3\Msun$ the star becomes unable to sustain He burning in its core, and its state at this point is more precisely described as a proto-WD. However thermal contraction during this stage proceeds slow enough that, under the influence of GWR emission, the system never detaches, which leads to the conclusion that, at ejection, the former donor star must still be in the state of a not fully degenerate proto-WD. Further, once the proto-WD reaches a mass $\sim0.2\Msun$, angular momentum transfer to the binary companion dominates of angular momentum loss through GWR, leading to an increase of the orbital period and a decrease of the ejection velocity \cite{TY1979,Yu2008,N2020} For these objects, as shown by \citet{BWB2019,LRYH2021}, interaction with the SN ejecta will lead to significant loss of material as well as an increase in entropy of the remnant, leading to an increase in radius as well as luminosity. The excess energy is then radiated away and the star contracts, analogously to the higher-mass case, thermally. However, unlike in the higher-mass case, the proto-WD does not return to its previous state. Instead it simply collapses to form a proper WD. As such, at ejection, all population members will be either core He-burning stars or proto-WDs, evolving to form degenerate objects independently (pre-WDs). Therefore the population naturally consists of pre-WDs (I-type stars) and WDs (II-type-stars), giving rise to the second subdivision. Population member therefore evolve naturally from being I-type objects to the state of II-type objects.
\subsubsection{Expected subpopulations}
The two criteria established above give rise to four distinct subpopulations: firstly, bound pre-WDs, designated BI; secondly, bound WDs, designated BII; thirdly, unbound pre-WDs, designated UI; and lastly, unbound WDs, designated UII.
As discussed, in our model, objects are assumed to evolve from subpopulation BI to BII and UI to UII, but not the reverse, but not from BI to UI or the reverse or BII to UII or the reverse. However, as stated, since stars of masses $<0.3\Msun$ are structurally different, and evolve differently, to those above this limit, we distinguish high mass ($>0.3\Msun$) UI and BI objects and low mass ($<0.3\Msun$) BI and UI objects.

\section{Results - quantitative properties} \label{sec:quantitative_properties}
This section is dedicated to a quantitative overview of the predicted population. We further discuss some observational properties, as well as some bulk properties of our sample. As in Sec.\,\ref{ssec:numberdens}, our analysis in this section is based on a snapshot of our population taken after 300$\Myr$.
\subsection{Fiducial volume} \label{sec:fiducial-vol}
Naked helium stars of even moderate masses can be quite luminous \citep{H2009,H2016} and are thus detectable out to considerable distances ($\sim 8.4\mkpc$ in the case of US 708). However, their luminosity is similar to intermediate mass main sequence stars, making helium stars in regions of significant reddening, such as the disc \citep{GRG2019}, liable to be misclassified as MS stars \citep{IGH2019}. Further, the part of Hertzsprung-Russell diagram populated by sdO-B stars is shared with other objects, such as hot WD stars. Thus, spectral analysis is the only way to reliably classify these objects. Accordingly of the $\sim 40000$ candidates reported by \citet{GRG2019}, only about 10\% have been spectrally classified. See also \citet{NKGH2021} for a discussion.
In order to facilitate comparison of our predictions with the observational record, we reject members of our synthetic population lying outside a pre-determined region deemed inaccessible to observation. Specifically, we assume that the likelihood of (eventually) detecting a member of our predicted population is unity inside our fiducial volume and zero everywhere else. We define this fiducial volume thus to consist of a 2\kpc sphere around the position of Earth and, additionally two cones, their axes of symmetry perpendicular to the Galactic plan and their vertices at Earth, one pointing up and the other down (with Earth above the plane). The angle between the surface of both the upward and downward pointing cone and the plane is $\alpha=10^\circ$, both cones extending out to a radial distance of 25\kpc from Earth. The limitation on the 2\kpc sphere is motivated by the approximate limit for obtaining reliable Gaia-parallaxes for very blue objects such as the ones under consideration here. The value for $\alpha$ is approximate for limits of negligible reddening as indicated by the results of \citet{GSF2018}. 
\begin{figure}[h]
	\centering
	\begin{tikzpicture}[scale=0.9]
	    \def\centerarc[#1](#2)(#3:#4:#5)
        { \draw[#1] ($(#2)+({#5*cos(#3)},{#5*sin(#3)})$) arc (#3:#4:#5); }
        \def\centersegment[#1](#2)(#3:#4:#5)
        { \draw[#1] ($(#2)$) -- ($(#2)+({#5*cos(#3)},{#5*sin(#3)})$) arc (#3:#4:#5) -- cycle; }
        \usetikzlibrary{calc}
		\definecolor{myblue}{HTML}{377eb8}
		\filldraw[fill=white, draw=gray, ,line width=0.5mm] (2.5,0) circle [radius=4.5];
		\filldraw[fill=gray, draw=gray, ,line width=0.5mm] (0,-0.1) to[out=180,in=180] (0,0.1) --  (2.3,0.1)  to[out=75,in=105] (2.7,0.1) -- (5,0.1) -- (5,0.1) to[out=0,in=0] (5,-0.1) -- (3,-0.1) -- (2.7,-0.1)  to[out=255,in=285] (2.3,-0.1)  -- cycle ;
		\draw[arrows =<->, ,line width=0.5mm,line cap=<type: round] (-0.9,-4) node[align=left,   below] {$x_\mathrm{Gal}$}  -- (-1.9,-4) -- (-1.9,-3)node[align=left,   left] {$z_\mathrm{Gal}$};
		\centersegment[fill=red, draw=red, opacity = 0.3](1.4,0)(20:160:2.5);
		\centersegment[fill=red, draw=red, opacity = 0.3](1.4,0)(-20:-160:2.5);
		\centersegment[fill=red, draw=red, opacity = 0.3](1.4,0)(20:-20:0.5);
		\centersegment[fill=red, draw=red, opacity = 0.3](1.4,0)(160:200:0.5);
		\centerarc[draw=black,line width=0.5mm, opacity = 0.3](1.4,0)(160:200:2.15);
		\draw[dashed, draw=black, arrows= |-|, ,line width=0.5mm] (1.4,0) -- (1.4,1.25)node[  left] {$25\mkpc$}-- (1.4,2.5);
		\draw[dashed, draw=black, arrows= |-|, ,line width=0.5mm] (2.3,0.4) -- (2.5,0.4)node[  above] {bulge}-- (2.7,0.4);
		\draw[dashed, draw=black, arrows= |-|, ,line width=0.5mm] (0,-0.9) -- (2.5,-0.9)node[  below] {disc}-- (5,-0.9);
		\draw[dashed, draw=black, arrows= |-|, ,line width=0.5mm] (0.9,-0.4) -- (1.4,-0.4)node[  below] {$4\mkpc$}-- (1.9,-0.4);
		\draw[ draw=black, ,line width=0.5mm, dashed] (-1.5,-0) -- (1.4,0);
		\draw (-0.55,0.3) node[right] {$\alpha$};
		\draw (-0.55,-0.3) node[right] {$\alpha$};
		\fill [fill=myblue] (1.4,0.03) circle [radius=0.1];
		\draw[myblue] (1.4,0.3) node[left] {Earth};
		\draw (5,-3.3)node[rotate = 38 ] {halo};
	\end{tikzpicture}
	\caption{Illustration of the fiducial volume chosen for the calculation of the expected detection and event rates. Only objects located in the red shaded area are taken into account. Specifically, all objects in a distance of $D<2\mkpc$ are assumed to be visible. Any objects at a distance $D>25\mkpc$ are assumed to be undetectable. Objects located in the disc are discounted by the assumption that objects with a Galactic latitude $\alpha < b < -\alpha$ are invisible, with $\alpha=10^\circ$. Earth's position is as indicated. Illustration not to scale.} \label{fig:fiducial-volume-fig2}
\end{figure}
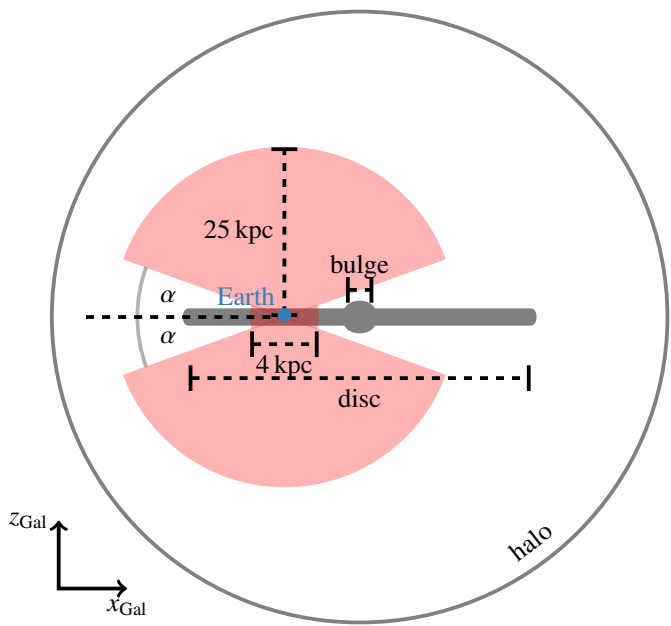
The shape of our fiducial volume is illustrated in Fig.\,\ref{fig:fiducial-volume-fig2}. We note that, as a consequence, any computation of the observationally supported production rate based on the arguments presented here should strictly reject any new observation outside of the fiducial volume as defined above. We further note that the shape of the fiducial volume is only an approximation. The geometry of Galactic extinction and reddening at high Galactic latitudes is non-trivial \citep{GSF2018,SJY2021}.

\subsection{Cumulative distributions}
\begin{figure*}
	\centering 
	\input{bound-unbound-cumulative-all-fig2}
	\vspace{-5cm}
	\caption{Cumulative distance distribution of runaway stars normalised to the total number of runaways in a non-degenerate state, $N_\mathrm{tot,n-d}$ for a ejection frequency of $f_\mathrm{ej} = 1/300\, \mathrm{yr}^{-1}$. The blue and orange curves add up to unity, with blue indicating the relative abundance of  the UI subpopulation and orange represents the BI subpopulation. Red indicates the BII subpopulation normalised to the UI+UII subpopulation. Dark orange and dark red indicate the admixture of objects in the mass range $0.7-0.8\Msun$. Panel A, B, C and D show the M10, M12, M14 and M15 populations respectively.} \label{fig:bound-unbound-cumulative}
\end{figure*}
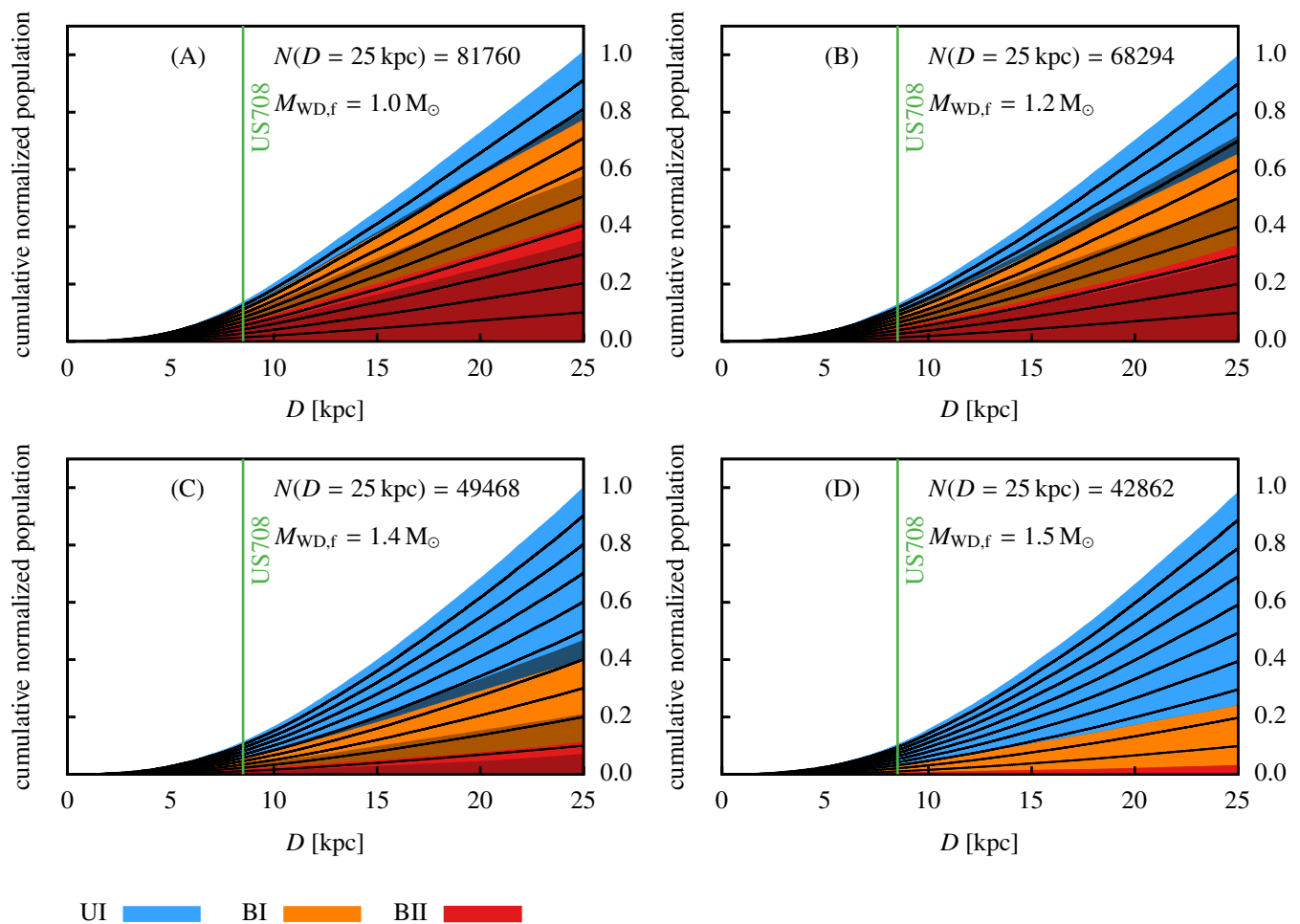
Fig.\,\ref{fig:bound-unbound-cumulative} (C) shows the sum of cumulative distributions of the UI, BI and BII subpopulations (the UII subpopulation is likely undetectable by virtue of being composed of fast, low mass WDs at large distance) in the M14 population with respect to distance from Earth. As can be seen, the BII subpopulation constitutes of the order of 10\% of the total population. The BI subpopulation constitutes of the order of 40\% at distances smaller than 15\kpc, decreasing to about 30\% at 25\kpc. Within a volume extending to the distance of US\,708 (itself a member of the UI subpopulation) about half of all objects are part of the UI subpopulation. 
The distribution resulting from the M10 population is shown in Fig.\,\ref{fig:bound-unbound-cumulative} (A). As can be seen, about 77\% (as opposed to about 40\% in the M14 sample) of the population is constituted by the B-subsample, with about 40\% being composed of the BII subpopulation and the remaining 37\% of the BI subpopulation. We note that this, to some extent, is a result of pollution by runaways in the $0.8-1.0\Msun$-range, which are not present in the M14 sample. These objects account for a pollution of about 10\% in the BI subpopulation and about 28\% in the BII subpopulation, illustrating the importance of the terminal donor mass for the structure of the resulting population. discounting the higher mass pollution, the BI subpopulation is about 50\% of the total population and the BII subpopulation about 20\%, the UI subpopulation accounting for the rest. It is clear, however, that the ratio of members of the different subpopulations, UI to BI+BII, is strongly influenced by the terminal accretor mass.
The distribution for the M12 population, as shown in Fig.\,\ref{fig:bound-unbound-cumulative} (B), falls between the M10 and the M14 set (as would naively be expected). Here, the UI-subpopulation constitutes about 66\% of the total population (including the $0.8-1.0\Msun$-admixture). This provides further evidence of the terminal accretor mass being the decisive factor in the determination of boundness to unboundness.
For the M15 population, shown in Fig.\,\ref{fig:bound-unbound-cumulative} (D), a clear preference for the UI population when compared to the other synthetic populations (only comparing the mass ranges present in all populations) becomes evident. As will be discussed in Sec.\,\ref{sec:BU-ratio}, the B-U ratio may constitute a viable criterion to distinguish terminal accretor masses.
A clear conclusion of the discussion in this section should be the realisation that the relative number of members of the UI-subpopulation compared to the B-subpopulation is strongly influenced by the terminal accretor mass of the generating SN. Further, if a significant number of runaways have a mass exceeding $0.7\Msun$, then a population of high-velocity WDs in the Galactic halo (i.e. the BII-subpopulation) is a necessary consequence. We mention in passing that this prediction is similar to that by  \citet{Hansen2003}, who considered a slightly different channel, requiring a donor in the mass range $1.3-3.3\Msun$, leading to lower ejection velocities incapable of producing HVS.
\subsection{The B-U-ratio as a measure of terminal accretor mass} \label{sec:BU-ratio}
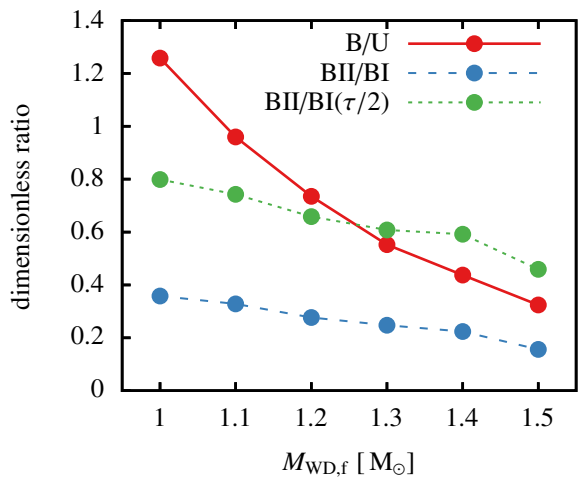
\begin{figure}
	\centering 
	\input{ratios-fig2}
	\vspace{-1.5cm}
	\caption{Relative occurrence of individual subpopulations in all synthetic populations in the fiducial volume depending on different terminal accretor masses. BII/BI($\tau/2$) was calculated under the assumption of a 50\% reduced nuclear burning lifetime of each ejected star.}  \label{fig:ratios}
\end{figure}
In Fig.\,\ref{fig:ratios} we show the relative occurrence rates of individual subpopulations as defined in Sec.\,\ref{sec:subpopulations} within the fiducial volume, comparing different synthetic populations, but excluding terminal donor masses of $\MWd >0.7\Msun$ in order to perform the comparison within a similar parameter space. Since the ratio of the BI to the BII subpopulation is affected by the assumed lifetime of each ejected star, we perform our analysis once under the assumption of our baseline lifetimes and once under the assumption of the baseline being reduced by half. As can be seen, the B/U-ratio is strongly affected by the choice of terminal accretor mass, reducing from about 1.25 at $\MWd=1.0\Msun$ to about 0.4 at $\MWd=1.5\Msun$. This indicates that, given a sufficiently large sample, a dominant accretor mass (assuming one dominates) should be discernable in the observational record with the B/U-ratio being a good indicator. Other than the BII/BI-ratio, this ratio is unaffected by any assumption on the runaway lifetime. In the same vein, while the BII/BI-ratio is a function (though a weaker one than the B/U-ratio) of the terminal accretor mass, this ratio is deemed to be a less viable indicator of the terminal accretor mass (though it may be a good indicator of the delay time, as it encodes the remaining lifetime of the donor star, but the necessary analysis is beyond the scope of the paper). 
\subsection{Dependence on the ejection rate} \label{ssec:ejection rate}
\begin{figure}
	\centering 
	\input{bound-unbound-cumulative-001-fig2}
	\caption{Same as Fig.\,\ref{fig:bound-unbound-cumulative} with $f_\mathrm{ej} = 1/3\cdot 10^{-4}\, \mathrm{yr}^{-1}$.} \label{fig:bound-unbound-cumulative-001}
\end{figure}
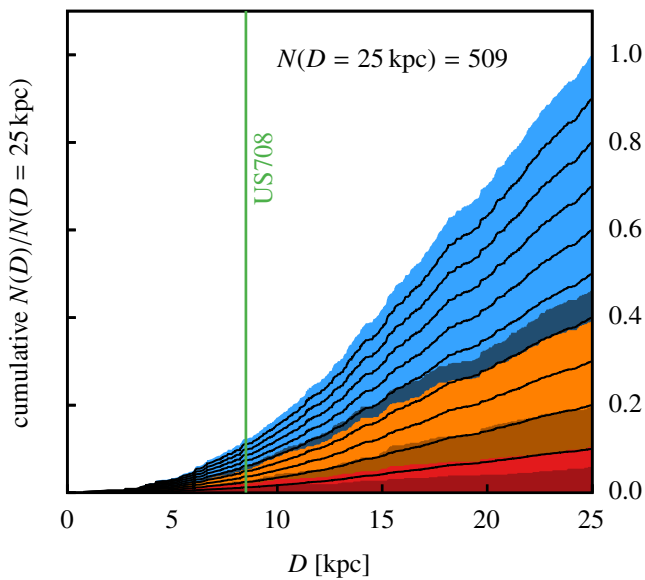

In the previous discussion, an ejection rate of  $f_\mathrm{ej} = 1/300\, \mathrm{yr}^{-1}$ was assumed as a baseline, roughly corresponding to the inferred Galactic SN Ia rate \citep{CTT1997,SLP2006}. In this section, we investigate whether changes in this assumption affect the qualitative conclusions drawn from our synthetic population. This investigation is effected by randomly choosing, from our initial synthetic population, a subset of tracks such that the resulting ejection rate is decreased either by a factor of ten, or a factor of one hundred from the baseline assumption. We perform this investigation explicitly for the M14 population.  
Fig.\,\ref{fig:bound-unbound-cumulative-001} shows the resulting cumulative distribution within the fiducial volume for an assumed ejection rate of $f_\mathrm{ej} = 1/3\cdot 10^{-4}\, \mathrm{yr}^{-1}$. This means that over the same 300$\Myr$ time frame, instead of $10^6$ objects being ejected, only $10^4$ objects are ejected. We note that, despite some statistical fluctuations due to the smaller set of objects, the relative numbers between the different subpopulations are only marginally affected.
We performed a comparison with a similar calculation assuming $f_\mathrm{ej} = 1/3\cdot 10^{-3}\, \mathrm{yr}^{-1}$. The number of objects contained in the fiducial volume, $N(D=25\mkpc)$, decreases by a factor of  $\sim10.02$ between $f_\mathrm{ej} = 1/3\cdot 10^{-2}\, \mathrm{yr}^{-1}$ and $f_\mathrm{ej} = 1/3\cdot 10^{-3}\, \mathrm{yr}^{-1}$, by a factor of $\sim9.61$ between $f_\mathrm{ej} = 1/3\cdot 10^{-3}\, \mathrm{yr}^{-1}$ and $f_\mathrm{ej} = 1/3\cdot 10^{-4}\, \mathrm{yr}^{-1}$ and, accordingly, by a factor of $\sim 96.27$ between $f_\mathrm{ej} = 1/3\cdot 10^{-2}\, \mathrm{yr}^{-1}$ and $f_\mathrm{ej} = 1/3\cdot 10^{-4}\, \mathrm{yr}^{-1}$. It is therefore reasonable to assume that the number of objects in the fiducial volume decreases linearly depending on the ejection rate, while the relative numbers in each subpopulation are, to first approximation, constant. We assume such a linear dependence, as well as constancy of the relative numbers, in the following discussion. 
At this point we note that SN Ia rates have been argued to be inversely correlated with the delay time via $\sim\frac{1}{t_\mathrm{delay}}$ \citep{RBF2009,MM2012}, which implies a time dependence of the ejection rate which is at odds with our assumption of a constant ejection rate. However, we note that, as $t_\mathrm{delay} = t_0+t_\mathrm{current}$, this effect would lead to a decrease of the ejection rate of the order of $2-3\%$ over the $300\Myr$ interval covered by our simulations. A decrease of this nature has a negligible impact on our results in every meaningful interpretation.

\subsection{Dependence on the terminal donor mass}
\begin{figure*}
	\centering 
	\input{bound-unbound-mass-fig2}
	\vspace{-4cm}
	\caption{Spectral distribution of subpopulation membership over terminal donor mass for the M10, M12, M14 and M15 populations for the total set (not merely the fiducial volume). Total numbers per mass at measured against the left hand y-axis. The black line indicates the BI/UI and is measured against the right hand y-axis.} \label{fig:bound-unbound-mass}
\end{figure*}
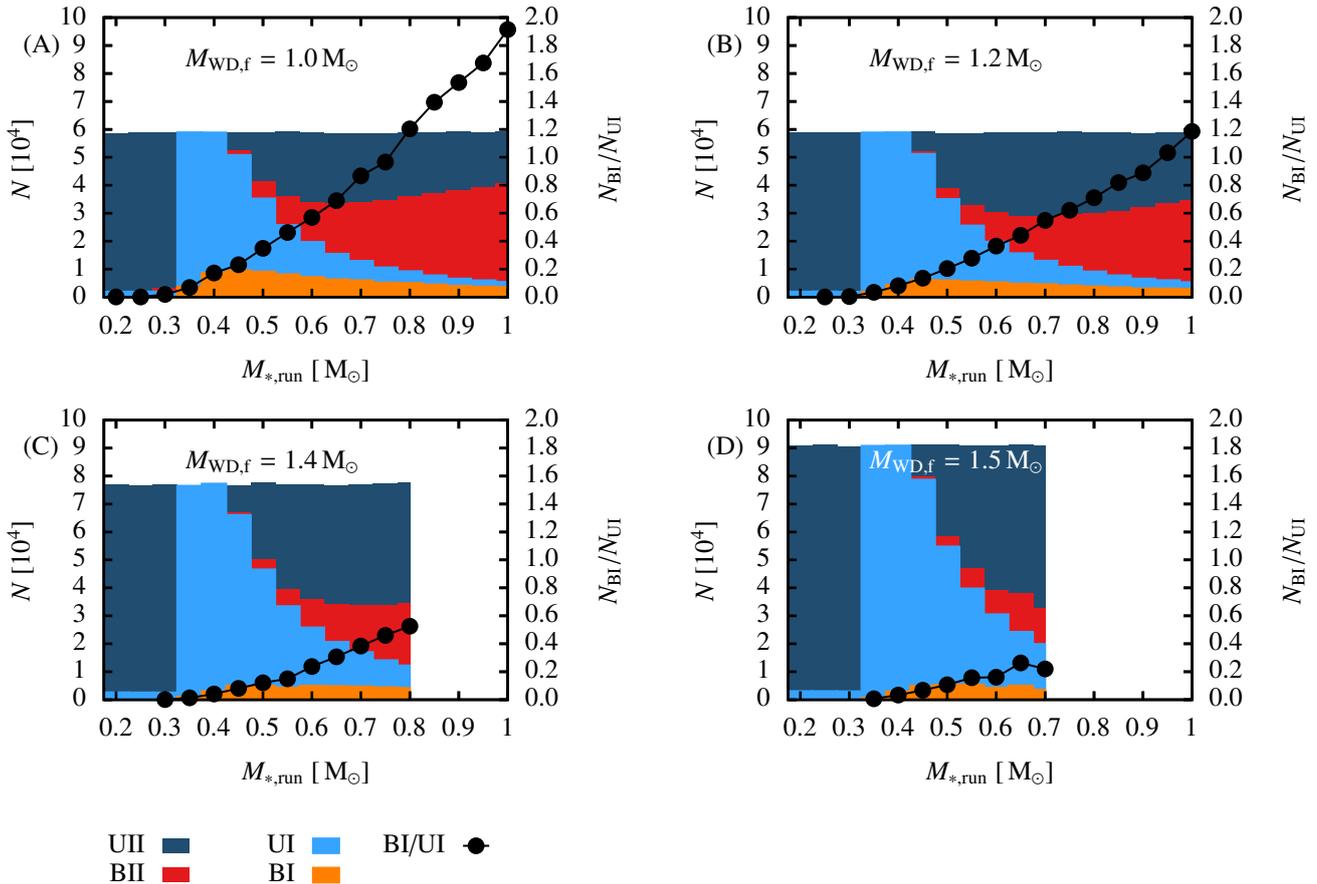

In this section we investigate the contributions of different assumed terminal donor masses to the total and subpopulations. Fig.\,\ref{fig:bound-unbound-mass} shows the contributions of each mass bin to the different subpopulations as defined in Sec.\,\ref{sec:subpopulations} in the M10, M12, M14 and M15 populations (subplots). We note that the flat initial mass distributions. Here, the entire population, independent of location in the vicinity of the Galaxy and ejection timing, is taken into account. The shape of the UI+BI subpopulations relative to the UII+BII subpopulations clearly reflects the assumed lifetimes (see Fig.\ref{fig:helife}).  
Consequently, as expected, the relative contributions of the BII subpopulation increases with the terminal donor mass, provided the terminal donor mass is greater than $0.35\Msun$. The expectation that there is no contribution of the BII subpopulation in the mass range $0.35-0.4\Msun$ is a consequence of the assumed lifetime of these objects exceeding the simulation time of $300\Myr$. However, this assumption represents an upper limit. In reality, a contribution of the BII subpopulation should be expected in this mass range (as supported by our $\tau/2$ test case). A striking result here is that we expect no contribution of the bound subpopulation (both BI and BII) in the $0.2-0.3\Msun$ mass range at any terminal accretor mass and only a contribution at terminal donor masses of $0.35\Msun$. This indicates that the Galactic potential should not be sufficient to bind runaways at this mass range if ejected via the SD-He channel. The only observed candidate object likely falling into this mass range (LP 398-9, see Sec.\,\ref{sec:candidates}), is very likely unbound as well. If bound objects at the lower end of the mass range ($<0.35\Msun$) are found, this would be an indication either that the production mechanism differs from the one under consideration here or that the Galactic potential has been underestimated. This conclusion is necessarily independent of the assumed lifetime as defined in Fig.\ref{fig:helife}, as these objects are proto-WDs and are thus observationally distinct from putative higher mass objects, first presenting as proto-WDs, then as ELM-WDs. Fig.\,\ref{fig:bound-unbound-mass} further shows the dependence of the ratio of the BI subpopulation to the UI subpopulation. This ratio decreases (at $\MWd=0.7\Msun$) from $\sim 0.9$ in the M10 population to $\sim0.2$ in the M15 population, thus showing a clear dependence on the terminal accretor mass. This ratio is also largely independent on the assumed runaway lifetimes, as these only affect the ratio of the UI+BI to UII+BII subpopulations. The dependence of the ratio of the UI+BI to UII+BII subpopulations on the lifetimes can be estimated the following way. The total number of objects in the population is simply the ejection rate times the simulation interval $t_\mathrm{sim}$
\begin{equation}
	N=f_\mathrm{ej}\cdot t_\mathrm{sim} .
\end{equation}
Then the number of objects in a certain terminal donor mass bin is
\begin{equation}
	N(\MWd)=f_\mathrm{ej}(\MWd)\cdot t_\mathrm{sim} ,
\end{equation}
where, due to the assumed flat mass distribution $f_\mathrm{ej}(\MWd) = f_\mathrm{ej}/n_\mathrm{bins}$. Then the number of objects in the UI+BI subpopulation is, with the assumed lifetime $\tau(\MWd)$, considering a constant $f_\mathrm{ej}$
\begin{equation}
	N(\MWd,I)=f_\mathrm{ej}(\MWd)\cdot \tau(\MWd)
\end{equation}
and the number of the UII+BII subpopulation given by $N(\MWd,II)=f_\mathrm{ej}(\MWd)\cdot (t_\mathrm{sim}-\tau(\MWd))$, the ratio is 
\begin{equation}
	\frac{N(\MWd,I)}{N(\MWd,II)} = \frac{\tau(\MWd)}{t_\mathrm{sim}-\tau(\MWd)} .
\end{equation}
As such, while inaccuracies in the assumed lifetimes are suppressed, the BI/UI ratio  of the total population is a yardstick for the determination of the terminal accretor mass. As will be shown later, the BI/UI ratio in the fiducial volume, on the other hand, does depend on the assumed lifetime.
\begin{figure*}
	\centering 
	\input{bound-unbound-mass-fiducial-fig2}
	\vspace{-4cm}
	\caption{Like Fig.\,\ref{fig:bound-unbound-mass} but restricted to the fiducial volume. In panel (A), the BI/UI ratio was cropped in order to preserve legibility. The missing value is $N_\mathrm{BI}/N_\mathrm{UI} = 23$ at $\MWd = 1.00\Msun$.} \label{fig:bound-unbound-mass-fiducial}
\end{figure*}

As the entirety of the population is spacially inaccessible to observation (as discussed in Sec.\ref{sec:fiducial-vol}), the population distribution needs to be studied only within the fiducial volume. We show the resulting population in Fig.\,\ref{fig:bound-unbound-mass-fiducial}. As seen, the clear dependence of the BI/UI ratio seen in the full spacial sample is replicated in the fiducial volume subsample. A number of features should be noted here: Firstly, the population size increases with increasing $\MWd$. This is a consequence of the slower ejection velocity of higher mass donor stars. Thus, higher mass donors tend to remain within the fiducial volume for a longer time. Secondly, the BI/UI ratio strongly increases with $\MWd$. This is, again, a result of the slower ejection velocity of higher mass donors. However, the ratios increase more strongly with decreasing $\MWDf$. For instance, not surprisingly, the slower the ejection velocity, the more likely an ejected star is bound. Thirdly, there is a maximum in the BI+UI distribution (not mirrored in the total population) which depends on the terminal accretor mass. While the increase of the population size is driven by the same effect as the continuous increase of the total population (as discussed in item 1.), the decrease at higher mass (leading to the maximum) is a result of the decreasing lifetime of the ejected stars as a result of their higher mass. The resulting maximum in the BI+UI subpopulation depends on the terminal donor mass and leads to a maximum at $\MWd=0.55\Msun$ at $\MWDf=1.0\Msun$, $\MWd=0.6\Msun$ at $\MWDf=1.2\Msun$, $\MWd=0.7\Msun$ at $\MWDf=1.4\Msun$; and $\MWd=0.7\Msun$ at $\MWDf=1.5\Msun$.

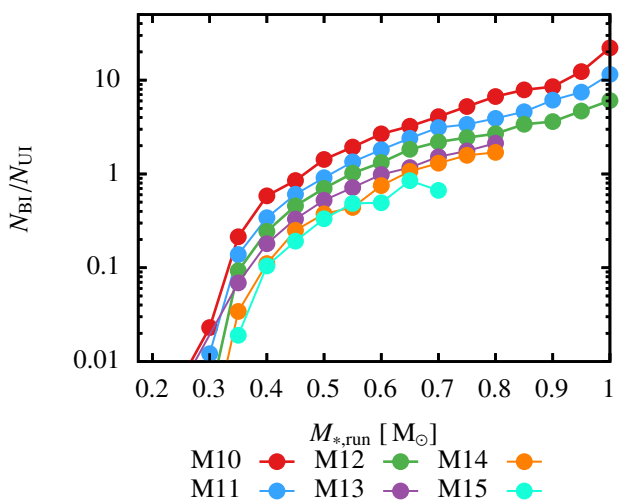
\begin{figure}
	\centering 
	\input{bound-unbound-mass-fiducial-ratios-fig2}
	\caption{Direct comparison of the BI/UI-ratio as presented in Fig.\,\ref{fig:bound-unbound-mass-fiducial}.} \label{fig:bound-unbound-mass-fiducial-ratios}
\end{figure}
For clarity, Fig.\,\ref{fig:bound-unbound-mass-fiducial-ratios} shows only the BI/UI ratios as presented in Fig.\,\ref{fig:bound-unbound-mass-fiducial}. As can be seen, the BI/UI-ratio is largely dependent on the terminal accretor mass (as designated by the synthetic population in this figure). The increase of the ratio with decreasing $\MWd$ should be noted.
\begin{figure*}
	
	\centering 
	\input{bound-unbound-mass-fiducial-tau-fig2}
	\vspace{-4cm}
	\caption{Like Fig.\,\ref{fig:bound-unbound-mass-fiducial} but with assumed lifetime $\tau = \tau/2$. In panel (A), the BI/UI ratio was cropped in order to preserve legibility. The missing value is $N_\mathrm{BI}/N_\mathrm{UI} = 55$ at $\MWd = 1.00\Msun$.} \label{fig:bound-unbound-mass-fiducial-tau}
\end{figure*}
In Fig.\,\ref{fig:bound-unbound-mass-fiducial-tau} we investigate the effect of a change of the assumed lifetime of the ejected star ($\tau$) by setting $\tau\rightarrow\tau/2$. As this affects the higher mass stars more than the lower mass ones, this change has the consequence of suppressing the UI+BI population vs the UII+BII population at higher masses, leading to a shift of the peak in the distribution compared to the baseline assumption towards lower masses. However, since all of our trajectories originate in the disc, a star necessarily has to be able to reach fiducial volume within its remaining lifetime in order for it to bee seen as a UI+BI subpopulation member,. As such, the BI/UI ratio is affected as well. While, naively, one would expect that this would lead to a decrease of the BI/UI-ratio, we instead, at lower terminal accretor masses ($\MWd<1.3\Msun$), find an increase. We attribute this to stars ejected in the vicinity of Earth having a higher likelihood of reaching the fiducial volume and stars ejected at larger distances from Earth not being sufficiently fast, even if unbound, to reach the fiducial volume.

\begin{figure}
	\centering 
	\input{bound-unbound-mass-fiducial-ratios-tau-fig2}
	\caption{Direct comparison of the BI/UI-ratio as presented in Fig.\,\ref{fig:bound-unbound-mass-fiducial-tau}. Grey lines indicate the data presented in \ref{fig:bound-unbound-mass-fiducial-ratios}.} \label{fig:bound-unbound-mass-fiducial-ratios-tau}
\end{figure}
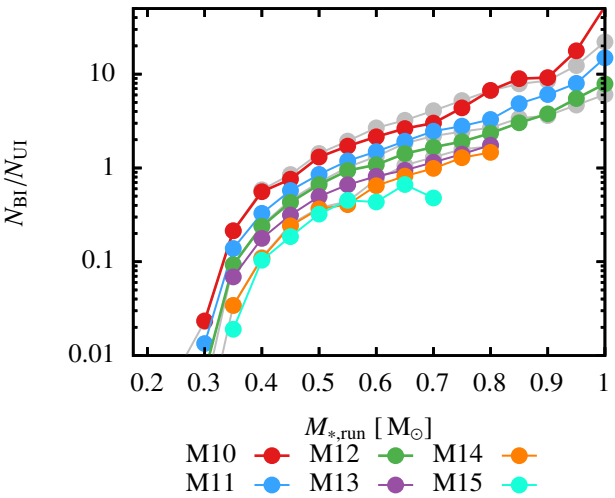

We compare the BI/UI-ratios for Fig.\,\ref{fig:bound-unbound-mass-fiducial-tau} with those presented in Fig.\,\ref{fig:bound-unbound-mass-fiducial} in Fig.\,\ref{fig:bound-unbound-mass-fiducial-ratios-tau}. As seen, the effect on the BI/UI-ratios is largest at lower $\MWDf$, becoming negligible at $\MWDf=1.4\Msun$ (i.e. in the M14 population).

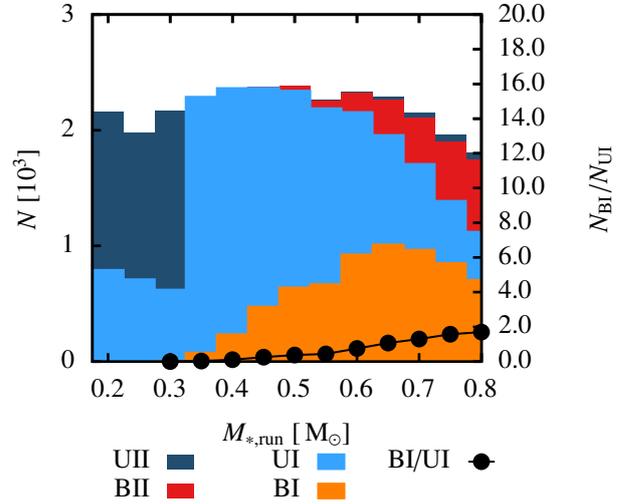
\begin{figure}
	\centering 
	\input{bound-unbound-mass-M14-fiducial-weighted-fig2}
	\caption{Like Fig.\,\ref{fig:bound-unbound-mass-fiducial} (C), but assuming a weighting function according to Eq.\,\ref{eq:weighting}.} \label{fig:bound-unbound-mass-fiducial-weighted-M14}
\end{figure} 
While we assumed a flat initial distribution for $\MWd$ for generality, we investigate the effect of a sloped distribution in Fig.\,\ref{fig:bound-unbound-mass-fiducial-weighted-M14} for the M14 population. Here, instead of a flat distribution, we arbitrarily impose a linearly decreasing weighting function (Eq.\,\ref{eq:weighting}), which decreases the relative abundance of objects relative to $\MWd=0.2\Msun$ to zero at $\MWd=1.05\Msun$ (we note, however, that the most massive runaway star in this synthetic population has a mass of $0.8\Msun$).
\begin{equation} \label{eq:weighting}
	N_\mathrm{ej}(\MWd) = N_\mathrm{ej}(\MWd=0.2\Msun)*\left(\frac{-\MWd}{0.85\Msun} +1+\frac{0.2}{0.85} \right) .
\end{equation}
As would be expected, the weighting has the effect of suppressing the population at higher values of $\MWd$, leading to a shift of the BI+UI subpopulation maximum to lower values. However, we note that the BI/UI ratio is unaffected by this change. The conclusion here is that, should the initial mass function differ from our assumptions (which is likely), and given that the population results from a single terminal accretor mass (which is currently unknown) the BI/UI-ratio would be independent of the relative terminal mass distribution of the donor stars.

\subsection{SNR association} \label{sec:SNR}
With each ejection of a HVS in the SD-He channel, a supernova remnant (SNR) is necessarily created. As such, the observation of a comparable number of SNRs may provide some insight into the viability of our predictions.  In Table\,\ref{table:SNR-assoc} we compare the number of runaways in each synthetic population with the number of associated SNRs. This is done by counting all runaways within the fiducial volume with a time of flight since ejection short enough for an associated SNR to not have dissipated. The lifetime of a SNR is subject to an number of uncertainties, such as the density of the local interstellar medium, ejecta mass and a number of other parameters. Estimates in literature \citep[e.g.][]{FGW1994} are of the order of $60\,\mathrm{kyr}$. A back of the envelope calculation reveals that, therefore, over a $300\Myr$ timespan, about 0.02\% of all stars should still have an associated non-dissipated SNR. As such, we assume an upper limit for the average SNR lifetime of $200\,\mathrm{kyr}$. We point out, however, that the expansion velocity of SNRs is generally faster than the highest ejection velocity attainable in this ejection scenario. Therefore, any HVS associated with an SNR should still be physically located inside its SNR except at late stages of SNR evolution. Surveys aimed at finding the ejected donor star in putative SN Ia remnants have generally failed to produce results \citep{KS2012,KC2014,KSS2018,KLW2018}, though \citet{RCM2004} claimed to have found a candidate for Tycho's SNR. A notable exception to this is one of the candidate objects for the population under consideration here, namely LP 398-9 (see more detailed discussion in Sec.\,\ref{sec:candidates}), which has been claimed to be associated with the SNR G70.0–21.5 \citep{SBG2018}, discovered by \citet{FNBK2015}. 
\begin{table}
	\caption{Population and number of associated supernova remnant}
	\label{table:SNR-assoc}      
	\centering
	\begin{tabular}{ c c c }
		\hline\hline
		run & $N_\mathrm{pop}$ & $N_\mathrm{SNR}$  \\    
		\hline                        
		M10 & 43000 & 2 \\
		M11 & 41313 & 3 \\
		M12 & 40136 & 5 \\
		M13 & 34994 & 3 \\
		M14 & 38949 & 6 \\
		M15 & 35748 & 4 \\
		\hline
	\end{tabular}
	\tablefoot{Total population count within the fiducial volume ($N_\mathrm{pop}$) in relation to the expected number of associated SNRs ($N_\mathrm{SNR}$) for all synthetic populations.}
\end{table}

As can be seen in Tab.\,\ref{table:SNR-assoc}, the number of objects in the fiducial volume dominates over the number of associated SNRs by at least a factor of $10^4$, which is on the same order of magnitude as the estimate presented earlier. Remembering that the assumed ejection frequency of $f_\mathrm{ej} = 1/3\cdot 10^{-2}\, \mathrm{yr}^{-1}$ is likely too high, we conclude that non-observation of associated SNRs is the expected outcome even for ideal circumstances. If SNR G70.0–21.5 is actually associated with LP 398-9, this observation should be viewed as coincidental.

\section{Candidate members of the population} \label{sec:candidates}

The observational record of objects fitting into this channel is, at the time of writing, extraordinarily small, with only a single object having been determined to fit with any reliable degree of certainty. However, using the predictions presented here and by \citet{NKGH2021}, it is possible to determine at least two additional candidate objects for this population. In this section, we briefly discuss the extant observational record. Keeping in mind the arguments presented in Sect.\,\ref{sec:assumptions}, we strive to be inclusive, rather than exclusive, in the construction of this sample. As such, we deem it useful to discuss the entirety of the current sample of candidate objects in detail. In the following, we follow up some astrometric properties of the objects under discussion using the Gaia DR2 and EDR3 data sets \citep{Gaia-mission2016,GAIA-DR2,Gaia-EDR32021}.

\subsection{US\,708}
First identified as a faint blue object showing an ultraviolet excess at high Galactic latitude by \citet{UMW1982}, then rediscovered as SDSS J093320.86+441705.4 and described by \citet{HHOB2005}, US\,708 was further determined it to possess an exceptionally high radial velocity. \citet{JWP2009} argued that a natural explanation for the properties of US\,708 is as a surviving helium donor star from a short-period single-degenerate SN Ia progenitor system and that this seemed more satisfying for US\,708 than the previously-suggested scenario of dynamical ejection from the Galactic centre (see also \citealt{WH2009} and \citealt{GMW2013}). \citet{GFZ2015} brought US\,708 to prominence as a compelling candidate survivor from a single-degenerate thermonuclear supernova, largely by demonstrating that the ejection location is traceable to the Galactic disc, with no intersection of the Galactic bulge. This largely excluded ejection from the Galactic centre via interaction with the SMBH located there. \citet{BAG2015} independently published similar conclusions for US\,708, although noted that the inferred space velocity for US\,708 was so high as to be in tension with published SN ejection models of the time, if US\,708 was ejected from the Galactic disc, and so speculated that the pre-SN system which ejected US\,708 might itself have been a member of the Galactic halo.  \citet{N2020} re-examined the past motion of US\,708 using Gaia DR2 data, which supported the conclusion that this star was not ejected from the Galactic centre. \citet{N2020} further found that the most likely inferred ejection velocity for US\,708 was lower by about 100\kms than in \citet{GFZ2015}.  Nonetheless, as shown in Fig.\,\ref{fig:RV-all-masses-fig2}, US\,708, if it indeed originates from the channel under consideration here, is an extreme member of its own population when its radial velocity is concerned (however, as stated, the extreme nature of US\,708 may well be the reason for its discovery). 

As shown by Fig.\,\ref{fig:RV-all-masses-pop-II-fig2}, the radial velocity exhibited by US\,708 is more compatible with low mass ($<0.3\Msun$) members of the UII subpopulation. However, taking into account US\,708's position in the HRD (generally associated with He-burning objects such as canonical sdB and sdO stars), as well as the time of flight to the most recent disc crossing (its assumed location of ejection) of $\sim 12-16\Myr$ suggest that it is currently He-burning, which suggests both a mass $>0.3\Msun$ as well as a pre-WD structure. As such, US\,708 is the prototype of this population as well as a member of the high-mass UI subpopulation.

If US\,708 does have a mass $>0.3\Msun$, we emphasise that the inferred ejection velocity is only marginally consistent with the systems from \citet{N2020} in which the WD explodes below the Chandrasekhar mass.  If the true mass of US\,708 can be determined, this could therefore provide evidence in favour of super-Chandrasekhar single-degenerate thermonuclear SNe. This would also be indirect evidence in favour of the ``spin-up/spin-down'' model, although the timescale for spin-down or angular-momentum redistribution in this case should be short compared to the helium-burning lifetime of US\,708.

\subsection{J2050}
 SDSS J205030.39-061957.8 (J2050) was first identified as part of the Hyper-MUCHFUSS catalogue \citep{GKH2015} and then characterised as as high-velocity bound spectroscopic twin of US\,708 by \citet{ZHG2017}, making it a candidate member of the BI subpopulation. With a time-of-flight of $>113\,\mathrm{Myr}$ to a disc-transit in the area $x_\mathrm{Gal}=10\mkpc < x_\mathrm{Gal,transit} <x_\mathrm{Gal}=20\mkpc$ and $y_\mathrm{Gal}= -2\mkpc < y_\mathrm{Gal,transit} < y_\mathrm{Gal}= 12\mkpc$, this object is consistent with a He-sdO of mass $<0.4\Msun$, which is an expected configuration for members of our population. As concluded by \citet{ZHG2017}, J2050 orbits the Galaxy in a retrograde direction, which makes it consistent with our prediction for members of the high mass BI subpopulation in Earth's local volume. However, while the aforementioned criteria make J2050 a promising candidate, one major caveat remains in that the inferred ejection velocity \citep[$\sim385\pm82\mkms$, as calculated by][]{ZHG2017} is low for products of the single degenerate He-donor scenario. This may be explained by the ejection also producing a bound remnant (as would be expected in certain production scenarios of Type Iax SNe) or some other unresolved mechanism. However, keeping this caveat in mind, J2050 remains one of the most promising candidates known so far.

\subsection{LP 398-9}

LP 398-9 was identified in the Luyten-Palomar catalogue and first appeared as a high proper motion star in the NLTT catalogue \citep[][as  NLTT 51732]{Lu1995} and again in the LPSM-North catalogue \citep[][as LSPM J2138+2522]{LSM2005}. Most notably, this object was identified by \citet{SBG2018} in the Gaia catalogue, as Gaia DR2 1798008584396457088, \cite{GAIA-DR2} as a candidate survivor of a double degenerate thermonuclear SN\footnote{The aforementioned paper nicknames this object D6-2.}. We note that the authors of that study base their conclusion not least on the argument that an inferred ejection velocity of $\sim 1200$\kms, by virtue of being too high, cannot be explained by the single degenerate channel. The contradiction arising by comparison with \citet{N2020} is explained by \citet{SBG2018} only taking into account terminal He-star of masses down to $0.5\Msun$ in the calculation of possible ejection velocities. We note that an ejection velocity of  $\sim 1200$\kms is too low to be easily explained by the double degenerate scenario, though terminal donor masses $<0.4\Msun$ and high effective temperatures $\sim 100\,\mathrm{kK}$ could marginally explain the observed value \citep[as shown very recently by][]{BCSH2021}. This problem is somewhat exacerbated by the ejection velocity derived from the Gaia EDR3 data set being on the order 100\kms lower than the velocity derived from the Gaia DR2 date set \citep[see ][]{BCSH2021}. However, the inferred ejection velocity is explained with the assumption of the donor star being a He-star or proto-WD with a terminal mass in the range of $0.2-0.3\Msun$ and terminal accretor masses approaching (or exceeding) the Chandrasekhar mass, which would make it a candidate member of the UI subpopulation. Further, with the appearance of  Gaia EDR3 \citep{Gaia-EDR32021}, the parallax of this object increased to $1.19\pm0.07\mas$ from $1.05\pm0.11\mas$, leading to a corresponding correction of the galactocentric space velocity and a decrease of the inferred ejection velocity. As such, LP 398-9 is a promising candidate member of the low mass UI subpopulation.
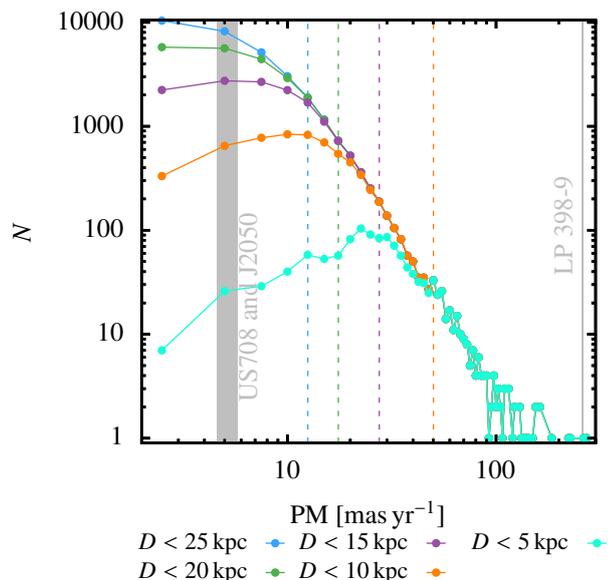
\begin{figure}
	\centering 
	\input{PM-14-fig2}
	\vspace{-0.3cm}
	\caption{Spectral diagram of the total proper motion of the predicted population for the M14 population. Lines of different colours indicate different distance-limited sub-samples. Dashed lines indicate where the distance-limited sub-sample of the same colour is the same as the next smaller limited sub-sample. Grey areas indicate the proper motion of the three population candidates as labelled. The PMs of US\,708 and J2050 relatively similar, so their corresponding lines were unified for legibility. Proper motion errors are of the order of $\sim \pm 0.2 \masyr$ for US\,708 and J2050 and $\sim \pm 0.1 \masyr$ for LP 398-9 \citep{Gaia-EDR32021}.} \label{fig:PM-14-fig2}
\end{figure}
We would like to draw attention at this point to the proper motion of LP 398-9 in relation to that of the predicted subpopulation as shown in Fig.\,\ref{fig:PM-14-fig2}. Comparison with the expected number of objects at the distance of LP 398-9 leads to the conclusion that, a priori, detection of an object at this distance with high proper motion such as this would be deemed very unlikely. Further, the reliability of the Gaia DR2 data, as utilised by \citet{SBG2018}, has been been called into question \citep{S2018}. 
\section{Astrometric properties} \label{sec:astrometry}
\begin{figure}
	\centering 
	\input{RV-all-masses-fig2}
	\vspace{-0.5cm}
	\caption{Spectral distribution of predicted radial velocities of the predicted UI+BI subpopulation for different terminal accretor masses within the fiducial volume. Lines of different colours indicate the sum over all runaways larger than the mass indicated in the key. Terminal accretor masses for each panel are $M_\mathrm{f,WD} = 1.0\Msun$ for panel (A), $M_\mathrm{f,WD} = 1.2\Msun$ for panel (B), $M_\mathrm{f,WD} = 1.4\Msun$ for panel (C) and $M_\mathrm{f,WD} = 1.5\Msun$ for panel (D).} \label{fig:RV-all-masses-fig2}
\end{figure}
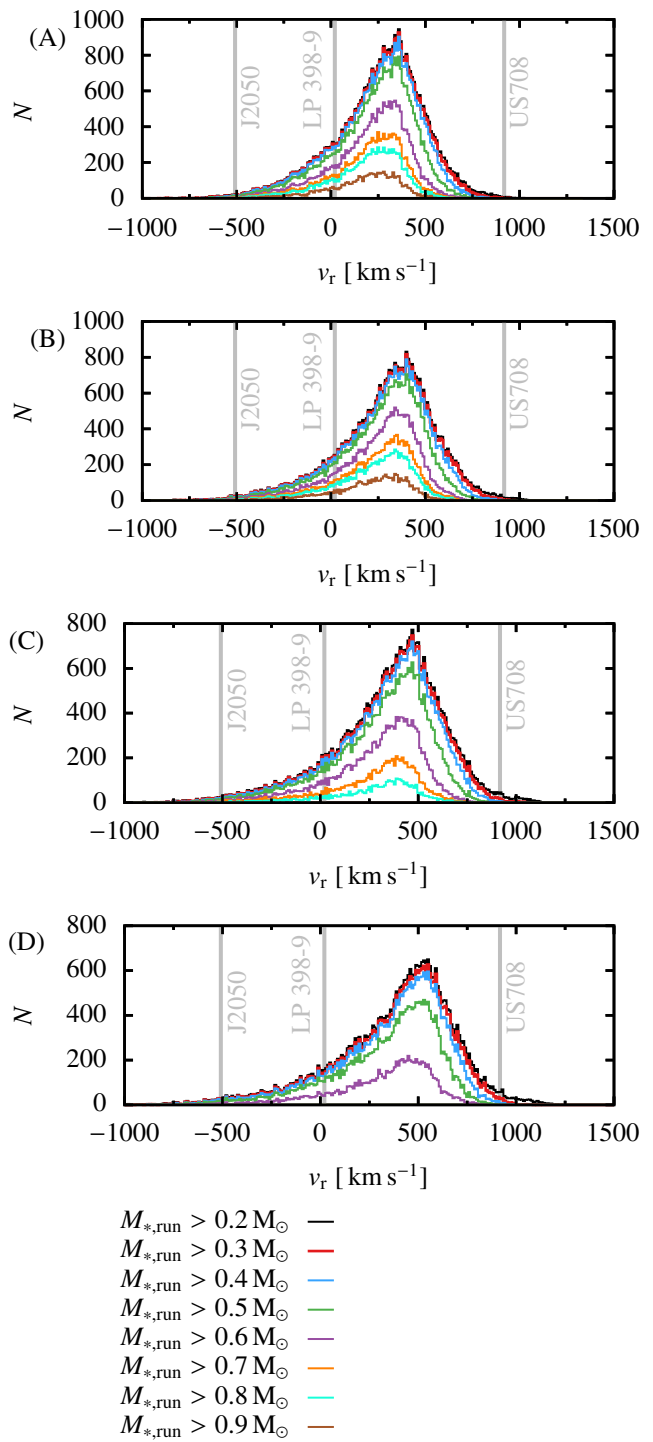
\begin{figure}
	\centering 
	\input{RV-all-masses-pop-II-fig2}
	\vspace{-0.5cm}
	\caption{Like Fig.\,\ref{fig:RV-all-masses-fig2}, but for the UII+BII subpopulation. Each line indicates the sum over all masses higher than indicated in the key; The secondary peak at about 750\kms indicates the peak of the low mass UII population.  The lifetimes of stars in the range $0.3 -0.45 \Msun$ are too long to contribute to this plot.} \label{fig:RV-all-masses-pop-II-fig2}
\end{figure}
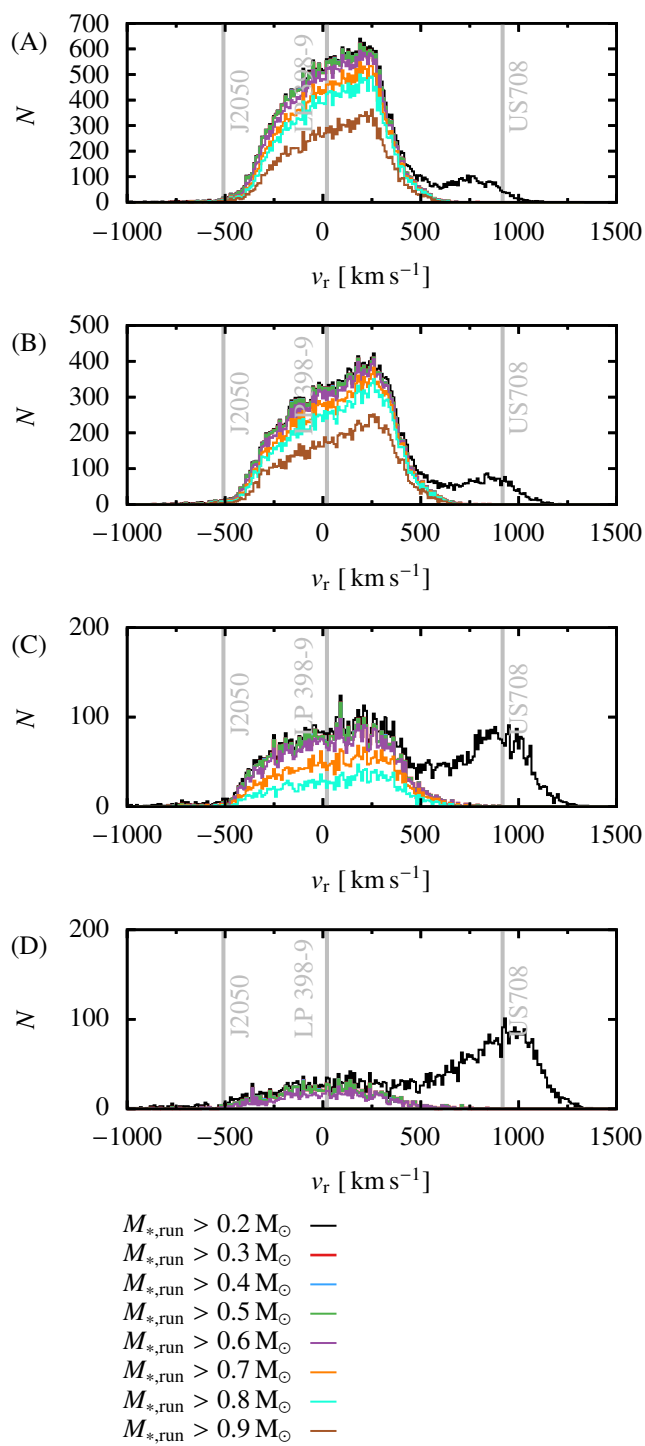
In this section we discuss the astrometric properties of the population, comparing them with the properties of the observed candidate objects. We again, like in sections Sec.\,\ref{ssec:numberdens} and Sec.\,\ref{sec:quantitative_properties}, focus on a snapshot of our populations taken at the end of the 300$\Myr$ simulation time frame.
\subsection{Proper motion}
In the era of Gaia, the most easily obtainable astrometric parameter is the proper motion. Thus, we investigate the proper motion distribution of our population in Fig.\,\ref{fig:PM-14-fig2} for the M14 population. As would be expected, the proper motion depends strongly on the distance of the object, even if high space velocities are considered. If the entire population in the fiducial volume is considered, the proper motion is expected to be scatter around zero, which is borne out by our sample. However, the maximum of the population moves towards higher values of the proper motion if the sample is restricted to smaller distances. Consequently, the maximum of the sample is located at $0\masyr$ in a volume extending to $25\mkpc$. Restricting the sample to $15\mkpc$ leads to a increase of the maximum of the distribution to $\sim 4 \masyr$, restriction to $10\mkpc$ to $\sim 10 \masyr$ and restriction to $5\mkpc$ to $\sim 20 \masyr$. The distribution approaches zero at $\sim 100 \masyr$. We note that the observed objects US\,708 and J2050, assuming they are part of this population, both exhibit lower proper motions than would be expected for objects of comparable distances (i.e. the sub-sample restricted to $10 \mkpc$). Most notably, the observed object LP 398-9 has been measured to exhibit a proper motion at the upper end of the distribution. In the case of LP 398-9, this high proper motion is a result of a number of factors. Objects at shorter distances (such as $\sim 1\mkpc$ for LP 398-9) from the observer will generally exhibit a higher proper motion. Further, high space velocity plays a role, with LP 398-9 being the fastest of these three objects. Together with LP 398-9's peculiar trajectory with a radial velocity not too different from zero, these factors combine to give LP 398-9 an exceptionally high proper motion. Observation of such an object should therefore, a priori, be judged as extremely unlikely. Detection of members of this population should therefore not rely on a lower proper motion cut-off. 
\subsection{Radial velocity}
The radial velocity is not as easily obtainable as the proper motion, but together with the proper motion, the space velocity, which is a primary observable for this population, is fully defined.  In Fig.\,\ref{fig:RV-all-masses-fig2} we show the radial velocity distribution for the entirety of the BI+UI subpopulation within the fiducial volume, depending on $\MWDf$ and decomposed into different values for $\MWd$. As can be seen, the maximum of the distribution is affected by the value of $\MWDf$, with the maximum increasing from $\sim450\mkms$ at $\MWDf=1.0\Msun$ to $\sim520\mkms$ at $\MWDf=1.5\Msun$. The distribution maxima for increasing $\MWd$ is, as would be expected, shifted to lower velocities.

We find that objects with the highest expected ejection velocities ($\MWd<0.3\Msun$) do, counter-intuitively, not significantly affect to the total distribution in RV, as their short lifetime means that they are removed from the sample before significantly contributing.
The importance of the lifetime in this discussion becomes evident when investigating the RV-distribution of the UII+BII subpopulation, as shown in Fig.\,\ref{fig:RV-all-masses-pop-II-fig2}. Here, as a larger fraction of ejected stars remain bound as $\MWDf$ decreases, the majority of the UII+BII subpopulation is clustered around zero, though exhibiting a peak at positive velocities. This peak is generally located at lower velocities than in the UI+BI subpopulation. As such, we expect this channel to result in a significant population of fast moving WDs in this velocity range. We highlight, however, the marked peak of low mass ($\MWd<0.3\Msun$) objects at around $1000\mkms$. This peak of fast moving, unbound and degenerate WDs is the final destination in phase space of the lower mass UI+BI objects in Fig.\,\ref{fig:RV-all-masses-fig2}. 

Comparing with the properties of the observed objects with the predicted populations, we find that US\,708, due to its high recession velocity, very likely does not originate from a system with terminal accretor mass $\MWDf<1.2\Msun$. However, as seen, both US\,708 and J2050 are notable as their distinguishing feature is their radial velocity. Both objects are located either in the negative or positive radial velocity tails of the population \citep[note that in the case of US\,708, this was already recognised by][]{NKGH2021}. We note that J2050's current radial velocity is an effect of its highly eccentric orbit with respect to the Galactic centre, not of an exceptionally high ejection velocity. As such, both of these objects are unusual in the sense of their radial motion being aligned with Earth in such a way as to provide either the maximum or minimum RV measurements. 

LP 398-9 is likely a member of the low mass ($\MWd<0.3\Msun$) UI population. As members of the low mass UI population have a very short lifetime, decaying into the UII population quickly, their velocity distribution can be assumed to be identical the UII of the same mass. Thus, LP 398-9 is extraordinary in its radial velocity ($\sim20\mkms$) when compared to the rest of this population, as can be concluded from Fig.\,\ref{fig:RV-all-masses-pop-II-fig2}. As seen, the low mass UII population (secondary peak in the black line in each panel) peaks at no less than $750\mkms$. We emphasise again that this radial velocity measurement was called into question in literature \citep{S2018}.
\subsection{Other population properties}
\begin{figure*}
	\centering 
	\includegraphics[width=\textwidth]{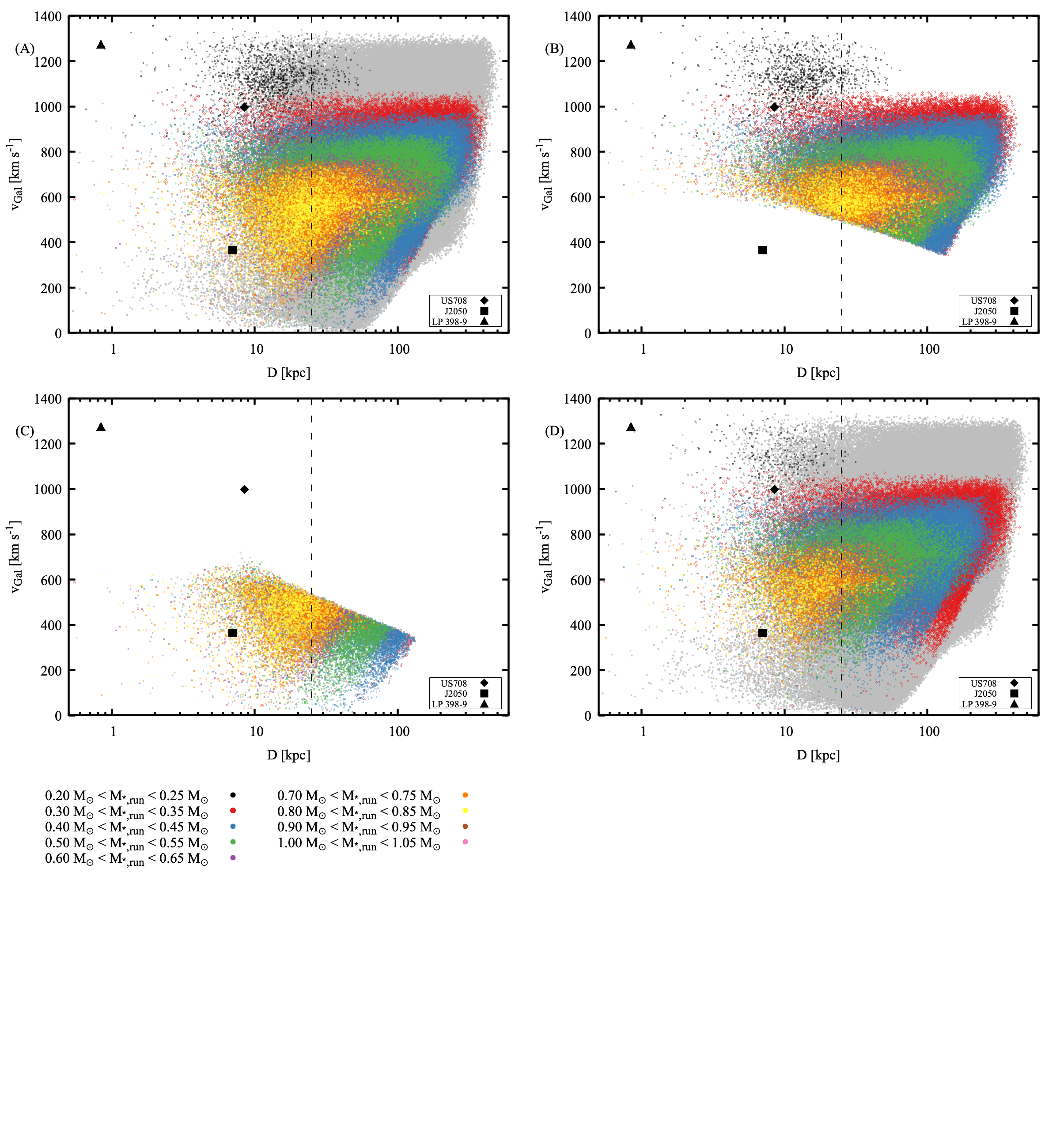}
	\vspace{-5cm}
	\caption{Galactocentric space velocity of I-type subpopulation in the M14 ($\MWDf=1.4\Msun$) population in the fiducial volume over distance $D$ from Earth. Colours indicate the mass of the ejected object. Panel (A) shows the entire population. Panel (B) and (C) only contain the UI and BI subpopulation respectively. Panel (D) is the same as panel (A), except assuming $\tau_\mathrm{lifetime} = 0.5\, \tau_\mathrm{lifetime,bm}$. Grey indicates (with no differentiation according to mass) the II-type subpopulation in panels (A) and (D). Positions of the observed candidates in the parameter space are as indicated. The dashed line indicates $D=25\mkpc$ (i.e. the parameter space left of the dashed line is equivalent to the fiducial volume as defined in Fig.\,\ref{fig:fiducial-volume-fig2}).} \label{fig:space-vel-distance-M14-fig2}
\end{figure*}
We show the expected distance-space velocity relation of the M14 population in Fig.\,\ref{fig:space-vel-distance-M14-fig2}. As can be seen, lower mass objects ($<0.3\Msun$), while higher velocity, are expected to occur less frequently at higher distances, owing to their short assumed lifetime. In panel (A), LP 398-9 can be distinguished as an extremely high space velocity object, but not unreasonable for a member of the low mass UI subpopulation. In general, low mass objects have a clear preference for being unbound over being bound. US\,708, with an assumed current mass of $\sim 0.3\Msun$ fits very well into the predicted parameters of its own mass range, straddling the line between low mass ($<0.3\Msun$) and higher mass ($>0.3\Msun$) objects. Further, the cut-off between bound and unbound objects is clearly distinguishable between panels (B) and (C), with J2050 clearly in the bound region occupied by higher mass objects. We emphasise again that this is not consistent with the expected lifetime of J2050 and a most recent disc crossing time of $\sim113 \Myr$. While the previous three panels contain the entire sample, panel (D) is the same as panel (A) but under the assumption of halved lifetimes. However, except for an, expected, reduction of the overall numbers and low and high mass objects being visible for a shorter time, the picture and drawn conclusions remain the same. Repeating this discussion for the other synthetic populations yields no significant changes other than for the M10 population, where US\,708 would instead be more likely a member of the low mass UI population \citep[which is in agreement with the conclusions drawn by][]{N2020} and LP 398-9 would be an even more extreme member of the low mass UI population. 
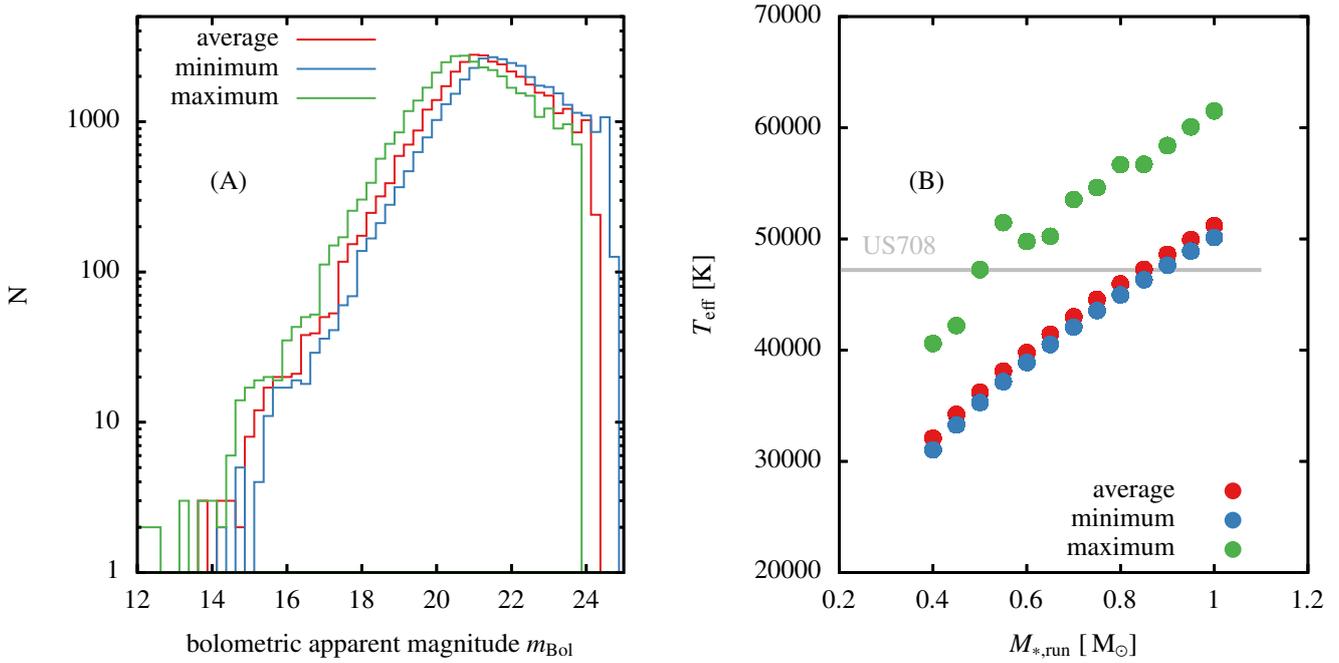
\begin{figure*}
	\centering 
	\input{M-T-diagram-M14-fig2}
	\vspace{-0.3cm}
	\caption{Expected number of objects per bolometric apparent magnitude within the fiducial volume (panel A), according to minimum, maximum and (time) averaged bolometric luminosity, and expected minimum and maximum effective temperature of He-stars in the mass range considered here (panel B). The data in panel (A) is binned along the x-axis into bins of 0.25.} \label{fig:M-T-diagram-M14-fig2}
\end{figure*}

In order to further study the observational properties of the population we calculate the average, as well as maximum and minimum luminosity and effective temperature of stars in our mass range. This approach is necessary because the exact state of the ejected star at the point of ejection, as well as its further evolutionary lifetime, is obscured to us through our methodology. For this, we use a set of models \citep[based on][]{N2020}, calculated with MESA \citep{MESA1,MESA2,MESA3,MESA4,MESA5} release 12778. The values given for the temperature, as a function of mass (as this parameter is, neglecting extinction, not affected by the distance of the object), are shown in Fig.\,\ref{fig:M-T-diagram-M14-fig2} (B). For the luminosity, we apply the recovered values to our M14 population (within the fiducial volume), in Fig.\,\ref{fig:M-T-diagram-M14-fig2} (A). The resulting histogram shows three peaks between bolometric apparent magnitudes of $\sim20.5$, $\sim21.0$ and $\sim21.5$, corresponding to assumptions of maximum, average and minimum luminosity respectively. Realistically, the expected peak in the distribution should be located somewhere between the peaks associated with these extreme assumptions. We note that US 708 is significantly brighter \citep[see e.g.][]{GFZ2015} than the majority of this population, again leading credence to the idea that, if this population exists in anything approaching significant numbers, a majority of it may simply have been unobservable in the past.
\section{Expected contribution to the Galactic SN Ia rate} \label{sec:rates}
Using our synthetic populations, we are in a position to use observations to independently calculate the birthrate of population member objects, and thus, the progenitor SN rate (which does not necessarily have to be of typical SN Ia, since the events may belong to the broader plethora of thermonuclear SNe). Specifically, as products of thermonuclear SNe, members of this putative population are, unlike the SNe themselves, or their SNRs, relatively long lived and, due to their location outside of the Galactic disc (if located within the fiducial volume), less affected by extinction, reddening and the stellar background than direct progenitor systems. Specifically, we draw attention to CD-30$^{\circ}$11223 \citep{VKO2012} and the very recently reported PTF1J2238+7430 \citep{KB2021}, both of which are expected to produce a runaway member of our low mass UI population with masses in the range $0.2 -0.3 \Msun$, that is, proto-WDs (though in these cases the terminal accretor mass, with $\sim 0.9\Msun$ is expected to be below the lowest terminal accretor mass considered here). However, we emphasise that these predictions are still, to some extent, model dependent. As such, observed products of SNe (assuming they can be claimed as such) present a model independent measure of the event rate.

A peculiarity of attempting to estimate an event rate from a population of high-velocity and hypervelocity stars is the expected inhomogeneous space density (see Fig.\,\ref{fig:r-z-fig2}) and observational properties of the objects (sub-luminous blue, only reliably classifiable with spectra) which makes them prone to misidentification as hot WDs, O or B type stars at shorter distances or, if reddened, main sequence stars. Selection criteria focussed on high proper motion may (as discussed in the previous section) miss a large number of unbound stars and determination of radial velocities requires a larger commitment of observational resources.
As the observational record may (or may not) be incomplete at the time of writing, we estimate the event rate as currently supported by observations. As, and if, additional population member are discovered, the supported event rate is likely to increase (or, less likely, to decrease, as will be explained later).

We base our calculation of the event rate on two assumptions. First, the abundance of each subpopulation (BI,BII,UI,UII) scales linearly with the ejection rate (i.e. the event rate). This assumption was tested in Sec.\,\ref{ssec:ejection rate}, and can be taken to be reasonably self-consistent. The second assumption is that the largest distance at which a population member can be detected is equal to the largest distance at which a population member has been detected. This assumption is to account for the unavailability of a reasonable estimate of the probability of detecting and correctly identifying a population member in the observational record. Further, a general assumption that individual stars, all else being equal, are more difficult to observe the more distant they are from the observer, is reasonable. We further assume the entirety of the population to be statistically distributed (i.e that the three candidate objects, as well as their distances, are no statistical aberration). As such, the farthest known population member serves as a marker of the volume used to calculate the average space density. We then use this volume to scale the space density of objects in our synthetic population to match the space density of observed objects in the same volume by linearly varying the ejection rate.
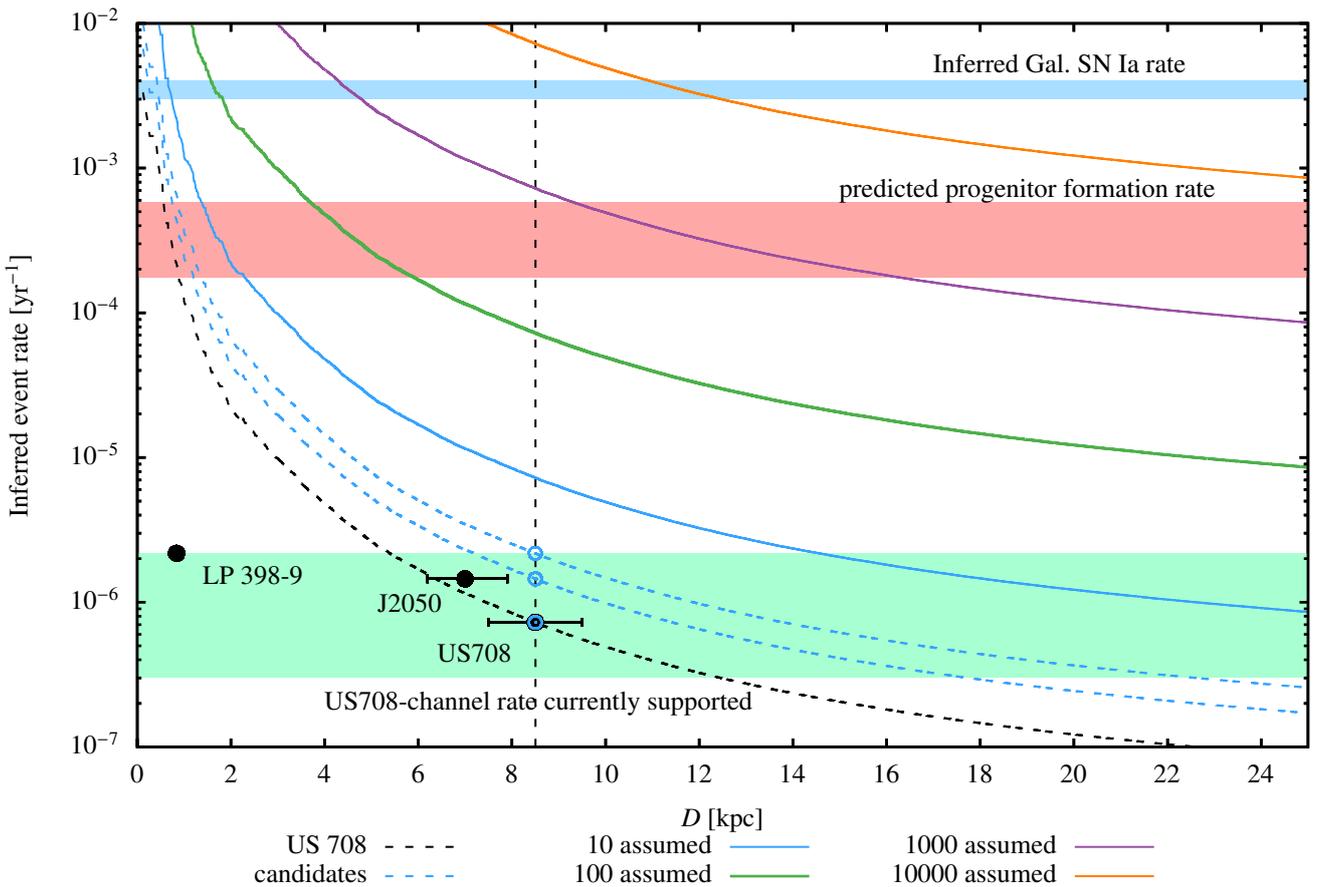
\begin{figure*}
	\centering 
	\input{rate-calculation-M14-fig2}
	\vspace{0.3cm}
	\caption{Calculated event rates for the production of population members. Dashed lines indicate the distance-dependent event rate based on the three observed candidates. Solid lines indicate event rates under the assumption of the indicated number of detections. Distances to the observed objects (including error bars) and associated event rates are indicated by filled and hollow circles. The error bars for LP 398-9 are smaller than the radius of the circles. Upper and lower limits for the observationally supported event rate, predicted formation rate of putative progenitor systems and the inferred Galactic SN Ia rate are as indicated. For details on calculation see text.} \label{fig:rate-calculation-fig2}
\end{figure*}
The supported ejection rate can thus be written
\begin{equation}
	f_\mathrm{ej,sup} = \frac{\rho_\mathrm{UI+BI,obs}(D_\mathrm{max})}{\rho_\mathrm{pop,sim}(D_\mathrm{max})}\cdot f_\mathrm{ej} =  \frac{N_\mathrm{UI+BI,obs}(D_\mathrm{max})}{N_\mathrm{pop,sim}(D_\mathrm{max})}\cdot f_\mathrm{ej} , 
\end{equation}
where $\rho_\mathrm{UI+BI,obs}(D_\mathrm{max}) = N_\mathrm{UI+BI,obs}(D_\mathrm{max})/V (D_\mathrm{max})$ and $\rho_\mathrm{pop,sim}(D_\mathrm{max}) = N_\mathrm{pop,sim}(D_\mathrm{max})/V (D_\mathrm{max})$ are the associated number densities within the volume defined by the most distant observed object. Then $N_\mathrm{UI+BI,obs}(D_\mathrm{max})$ is the number of observed objects (here three: US\,708, J2050, LP 398-9), $N_\mathrm{pop,sim}(D_\mathrm{max})$ is the number of expected objects, as predicted from our synthetic population, if the ejection rate is equal to $f_\mathrm{ej} = 1/300\,\mathrm{yr}^{-1}$. 
In principle, this methodology, namely leaving the calculation of the number density dependent on the observational record (as opposed to arbitrarily choosing a volume in which to calculate the number density), has the advantage of remaining as agnostic as possible to observational uncertainties. It has, however, the disadvantage of being susceptible to further observations decreasing the supported event rate, if the next object is detected at distances significantly greater than the previous most distant object. We take this into account be treating the space density within the volume defined by the currently most distant known object, US\,708, as an upper limit and the space density in the entirety of the fiducial volume (extending to $D=25\mkpc$) as a lower limit. In Fig.\,\ref{fig:rate-calculation-fig2} we show the resulting calculated rates for the M14 population. As can be seen, the currently supported event rate, based on three tentative detections, extends from $\sim 3\magten{-7}\pyr$ to $\sim 2\magten{-6}\pyr$. We note that this value is at least two orders of magnitude lower than the predicted formation rate of likely progenitor systems for this formation channel (see App.\,\ref{app:remarks-on-the-methodology}) and another order of magnitude lower than the inferred Galactic SN Ia rate \citep{CTT1997,SLP2006}. This indicates one of the following: The probability for He-star+WD binaries to form a SN, leading to the ejection of a high- or hypervelocity star is very small indeed. This would indicate that other outcomes, such as double degenerate mergers, with the He-star evolving into a WD prior to RLOF, are favoured for this type of system. In this case, comparison of the supported rate as calculated here, with the formation rate of other possible evolutionary products, should be performed. Another possibility is that observational uncertainties have precluded detection of further population members, in which case future observations may change this picture significantly. We note that, as seen in Fig.\,\ref{fig:rate-calculation-fig2}, if all possible progenitor systems result in the ejection of a high- or hypervelocity He-star, we would expect of the order of 200 objects within the volume defined by the current position of US\,708 (implying a detection bias of about 1/60) and of the order of 2000 objects within the $25\mkpc$ fiducial volume. The entirety of all Galactic SNe Ia resulting in the ejection of a high- or hypervelocity He-star would result in a population exceeding 2000 objects within the volume defined by the current position of US\,708 (implying a detection bias exceeding 1/600). We reasonably speculate that, even including future detections, the observational record is highly unlikely to support either the progenitor formation rate or the inferred Galactic SN Ia rate.

Another possibility arises from the result that the inferred ejection rate is lower by two orders of magnitude, than the calculated formation rate of putative progenitor binaries. Namely, only a small fraction of He-star+WD binaries produce a runaway He-star. Alternatively, the production rate of putative progenitor binaries may be overestimated. However, even if the ejection rate is indeed as low as we calculate here, and if the three putative objects are no statistical aberration, then we would expect at least of the order of 25 objects (22 of them currently undetected) within the fiducial volume.
\section{Conclusions} \label{sec:conclusions}
Using ejection velocity spectra for high- and hypervelocity runaway stars as presented by \citet{N2020}, we construct a synthetic population of SN products generated in the US\,708-channel. We calculate $1\magten{6}$ ejection trajectories each for one given terminal accretor mass, assuming ejection from the Galactic plane (i.e. $z_\mathrm{Gal} = 0$) with the ejection locations determined by the Galactic density distribution. Analysing the thus generated population, we constrain the defining characteristics of members of this putative population, as well as its structure. We find that this population is composed of stars both bound and unbound to the Galactic potential. Runaways ejected while identifiable as He-stars (He-sdOs) or proto-WDs will eventually evolve into unbound or high-velocity halo WDs. We distinguish four distinct subpopulations, separated according to two independent attributes: bound and unbound (B and U), pre-WD and WD (I and II), where the UI+BI subpopulation is split along low mass ($<0.3\Msun$) and high mass ($>0.3\Msun$) lines, where the low mass UI+BI subpopulation will present as a proto-WD, while the high mass UI+BI subpopulation will present as a core He burning star.
Thus, if the He-star+WD channel is responsible for any appreciable fraction of observable SNe, we predict the existence of a population composed of high- and hypervelocity He-stars and WDs.

In order to take account of observational constraints affecting hot subdwarf stars, we define a fiducial volume (Fig.\,\ref{fig:fiducial-volume-fig2}), excluding most of of the Galactic disc and bulge regions. Analysing our synthetic population at the end of a 300$\Myr$ time frame, we calculate the relative numbers of the expected subpopulations. We find that, within this fiducial volume, the profile of the BI/UI-ratio is directly related to the terminal mass of the accretor. Assuming that the channel responsible for the production of US\,708 favours a single terminal accretor mass but a different terminal donor masses, this represents a directly accessible observable. 
We further investigate the astrometric properties of our synthetic populations at the end of the 300$\Myr$ simulation time frame.
We find that three currently observed objects, US\,708/SDSS J093320.86+441705.4, as a member of the UI subpopulation, Gaia DR2 1798008584396457088 (LP 398-9), also as a member of the UI subpopulation, and SDSS J205030.39-061957.8 (J2050), as a member of the BI subpopulation, match the expected properties of the population, the former being the prototype of the population, the latter two matching sufficiently well to be designated as candidates. Of these, both US\,708 and LP 398-9 are generally compatible with ejection from either Chandrasekhar and super-Chandrasekhar mass events.

We further find that the relative numbers of the population are, to good approximation, independent of the assumed ejection rate. We use this property of our synthetic population to constrain the currently observationally supported ejection rate at $\sim 3\magten{-7}\pyr$ to $\sim 2\magten{-6}\pyr$, which is at least two orders of magnitude lower than the (theoretically predicted) formation rate of the putative progenitor binary and three orders of magnitude smaller than the inferred Galactic SN Ia rate. We conclude from this that the current observational record strongly disfavours the SN channel responsible for the creation of US\,708 being a major contributor to SNe Ia in general. 

We further conclude that the current observational records disfavours He-star+WD binaries generally producing a member of this population. However, we emphasise that this result cannot be called conclusive in the general sense, as our models further indicate that all of the currently known objects seem to be extreme members of their own population, US\,708 and J2050 exhibiting extreme radial velocities, LP 398-9 exhibiting extreme proper motion. We conclude from this that, while the extreme nature of these objects may have been a contributing factor in their original discovery over more ordinary population members, a larger population of more ordinary population members may as of yet be undetected. 

If the currently low ejection rate is taken at face value, a significant number of additional, possibly detectable, population members likely exist within the fiducial volume and outside, with a significantly smaller likelihood of detection. However, should the He-star+WD binary channel for thermonuclear SNe be responsible for any appreciable fraction of the observed thermonuclear SN rate, a much larger population of runaway He-stars such as He-sdBOs (e.g. US\,708 and J2050) and proto-WDs(e.g. LP 398-9), should exist, thus providing an immediately accessible observable of the associated rates. The study of high and hypervelocity stars ejected by channels not related to interaction with a SMBH should be considered a promising avenue to probe a number of unresolved astrophysical phenomena.\\
\\
The data underlying this study is available on \url{https://doi.org/10.5281/zenodo.6283669}

\begin{acknowledgements}
We thank the anonymous referee for insightful comments that improved the legibility of the manuscript. PN would like to thank Vedant Chandra for drawing attention an error in nomenclature.
MK is partly supported by Grant No 11521303, 11733008, 12090040, and 12090043 of the Natural Science Foundation of China. SJ acknowledges funding from the Netherlands Organisation for Scientific Research (NWO), as part of the Vidi research program BinWaves (project number 639.042.728, PI: de Mink).

This work has made use of data from the European Space Agency (ESA) mission
{\it Gaia} (\url{https://www.cosmos.esa.int/gaia}), processed by the {\it Gaia}
Data Processing and Analysis Consortium (DPAC,
\url{https://www.cosmos.esa.int/web/gaia/dpac/consortium}). Funding for the DPAC
has been provided by national institutions, in particular the institutions
participating in the {\it Gaia} Multilateral Agreement.  

This research made use of Astropy,\footnote{http://www.astropy.org} a community-developed core Python package for Astronomy \citep{Astropy2013, Astropy2018}.
\end{acknowledgements}

\bibliographystyle{aa}
\bibliography{paper}

\appendix

\section{Population synthesis} \label{app:remarks-on-the-methodology}
\begin{table}
    \centering
    \caption{\label{tab:WD+He-rates}Formation rates $[10^{-6}\,\mathrm{yr}^{-1}]$ of WD+He star binaries in a MW-like galaxy of three runs and using eight selection criteria. The uncertainties are pure Poissonian statistical uncertainties.}
    \label{tab:rates}
    \begin{tabular}{c c c c c}
        \hline
        \hline
        run & C1 & C2 & C3 & C4\\
        \hline
        R1 & $426.5\pm1.7$ & $419.1\pm1.7$ & $369.5\pm1.6$ & $362.1\pm1.6$\\
        R2 & $348.6\pm1.5$ & $331.3\pm1.5$ & $305.1\pm1.4$ & $287.8\pm1.4$\\
        R3 & $579.1\pm2.0$ & $466.1\pm1.8$ & $457.0\pm1.8$ & $344.1\pm1.5$\\
        \hline
    \end{tabular}\\[0.5\baselineskip]
    \begin{tabular}{c c c c c}
        \hline
        \hline
        run & C5 & C6 & C7 & C8\\
        \hline
        R1 & $340.4\pm1.5$ & $335.2\pm1.5$ & $293.3\pm1.4$ & $288.1\pm1.4$\\
        R2 & $249.5\pm1.3$ & $249.3\pm1.3$ & $225.1\pm1.2$ & $225.0\pm1.2$\\
        R3 & $289.2\pm1.4$ & $252.5\pm1.3$ & $211.2\pm1.2$ & $174.4\pm1.1$\\
        \hline
    \end{tabular}
    \tablefoot{
        runs:\\
        R1: Using the same parameters as given in Table 2 of \citet{KTL2018} but primary and secondary masses are both taken from the range 1 to $20\,\Msun$.\\
        R2: see R1 but primary and secondary masses between 1 and $10\,\Msun$, $\alpha_\mathrm{RLO}=0.01$, and $\beta_\mathrm{min}=0.0$.\\
        R3: see R2 but $\alpha_\mathrm{CE}=1.0$ and $\alpha_\mathrm{TH}=1.0$.\\
        selection criteria:\\
        C1: $0.4\,\Msun\leq M_\mathrm{He}\leq 1.1\,\Msun$, $M_\mathrm{WD}+M_\mathrm{He}>1.35\,\Msun$, $t_\mathrm{GW}<500\,\Myr$.\\
        C2: see C1 but $M_\mathrm{He}<1.0\,\Msun$.\\
        C3: see C1 but $M_\mathrm{WD}+M_\mathrm{He}>1.4\,\Msun$.\\
        C4: see C8 but $t_\mathrm{GW}<500\,\Myr$.\\
        C5: see C1 but $t_\mathrm{GW}<100\,\Myr$.\\
        C6: see C8 but $M_\mathrm{WD}+M_\mathrm{He}>1.35\,\Msun$.\\
        C7: see C8 but $M_\mathrm{He}\leq 1.1\,\Msun$.\\
        C8: $0.4\,\Msun\leq M_\mathrm{He}<1.0\,\Msun$, $M_\mathrm{WD}+M_\mathrm{He}>1.4\,\Msun$, $t_\mathrm{GW}<100\,\Myr$.
    }
\end{table}

We used the fast binary population synthesis code \textsc{ComBinE} \citep{KTL2018,K2018} in order to calculate the formation rate of putative progenitors of HVS as under consideration in this study. The same settings as given in Table 2 of \citet{KTL2018} are used, excepting a constraint of the initial stellar mass range to $[1:20]\,\Msun$ for the first run (R1). To form a binary consisting of a WD and a He-star, the binary had to undergo at least one, but usually several, episodes of mass transfer. In a second run (R2) we restrict the initial mass range further to $[1:10]\,\Msun$ and assume stable mass transfer ($\alpha_\mathrm{RLO}=0.01$, and $\beta_\mathrm{min}=0.0$) to be more conservative, which is more appropriate for low mass stars. Finally, we allow more efficient common envelope evolution ($\alpha_\mathrm{CE}=1.0$ and $\alpha_\mathrm{TH}=1.0$) in a third run (R3).

For each run, we calculate the expected formation rates in a Milky Way (MW)-like galaxy. We follow the same procedure as in \citet{KTL2018} by assuming a constant star formation rate of one binary per year with a primary star mass $>0.8\,\Msun$ \citep{HTP2002}. Systems of interest are selected by conditions on the mass of the He-star, $M_\mathrm{He}$, the total binary mass, $M_\mathrm{WD}+M_\mathrm{He}$, and the merger timescale by pure gravitational wave radiation, $t_\mathrm{GW}$. The last criterion, C8, is appropriate for the study presented in this paper. Table~\ref{tab:WD+He-rates} summarises the resulting formation rates of binaries consisting of a WD with a He-star companion. The range of $1.74$ to $5.79\times 10^{-4}\,\mathrm{yr}^{-1}$ shown in Fig.~\ref{fig:rate-calculation-fig2} arises from the lower and upper limits for the full set of population synthesis runs.

\end{document}

%% file: helife2.tex
\begingroup
  \makeatletter
  \providecommand\color[2][]{%
    \GenericError{(gnuplot) \space\space\space\@spaces}{%
      Package color not loaded in conjunction with
      terminal option `colourtext'%
    }{See the gnuplot documentation for explanation.%
    }{Either use 'blacktext' in gnuplot or load the package
      color.sty in LaTeX.}%
    \renewcommand\color[2][]{}%
  }%
  \providecommand\includegraphics[2][]{%
    \GenericError{(gnuplot) \space\space\space\@spaces}{%
      Package graphicx or graphics not loaded%
    }{See the gnuplot documentation for explanation.%
    }{The gnuplot epslatex terminal needs graphicx.sty or graphics.sty.}%
    \renewcommand\includegraphics[2][]{}%
  }%
  \providecommand\rotatebox[2]{#2}%
  \@ifundefined{ifGPcolor}{%
    \newif\ifGPcolor
    \GPcolortrue
  }{}%
  \@ifundefined{ifGPblacktext}{%
    \newif\ifGPblacktext
    \GPblacktextfalse
  }{}%
  \let\gplgaddtomacro\g@addto@macro
  \gdef\gplbacktext{}%
  \gdef\gplfronttext{}%
  \makeatother
  \ifGPblacktext
    \def\colorrgb#1{}%
    \def\colorgray#1{}%
  \else
    \ifGPcolor
      \def\colorrgb#1{\color[rgb]{#1}}%
      \def\colorgray#1{\color[gray]{#1}}%
      \expandafter\def\csname LTw\endcsname{\color{white}}%
      \expandafter\def\csname LTb\endcsname{\color{black}}%
      \expandafter\def\csname LTa\endcsname{\color{black}}%
      \expandafter\def\csname LT0\endcsname{\color[rgb]{1,0,0}}%
      \expandafter\def\csname LT1\endcsname{\color[rgb]{0,1,0}}%
      \expandafter\def\csname LT2\endcsname{\color[rgb]{0,0,1}}%
      \expandafter\def\csname LT3\endcsname{\color[rgb]{1,0,1}}%
      \expandafter\def\csname LT4\endcsname{\color[rgb]{0,1,1}}%
      \expandafter\def\csname LT5\endcsname{\color[rgb]{1,1,0}}%
      \expandafter\def\csname LT6\endcsname{\color[rgb]{0,0,0}}%
      \expandafter\def\csname LT7\endcsname{\color[rgb]{1,0.3,0}}%
      \expandafter\def\csname LT8\endcsname{\color[rgb]{0.5,0.5,0.5}}%
    \else
      \def\colorrgb#1{\color{black}}%
      \def\colorgray#1{\color[gray]{#1}}%
      \expandafter\def\csname LTw\endcsname{\color{white}}%
      \expandafter\def\csname LTb\endcsname{\color{black}}%
      \expandafter\def\csname LTa\endcsname{\color{black}}%
      \expandafter\def\csname LT0\endcsname{\color{black}}%
      \expandafter\def\csname LT1\endcsname{\color{black}}%
      \expandafter\def\csname LT2\endcsname{\color{black}}%
      \expandafter\def\csname LT3\endcsname{\color{black}}%
      \expandafter\def\csname LT4\endcsname{\color{black}}%
      \expandafter\def\csname LT5\endcsname{\color{black}}%
      \expandafter\def\csname LT6\endcsname{\color{black}}%
      \expandafter\def\csname LT7\endcsname{\color{black}}%
      \expandafter\def\csname LT8\endcsname{\color{black}}%
    \fi
  \fi
    \setlength{\unitlength}{0.0500bp}%
    \ifx\gptboxheight\undefined%
      \newlength{\gptboxheight}%
      \newlength{\gptboxwidth}%
      \newsavebox{\gptboxtext}%
    \fi%
    \setlength{\fboxrule}{0.5pt}%
    \setlength{\fboxsep}{1pt}%
    \definecolor{tbcol}{rgb}{1,1,1}%
\begin{picture}(4988.00,3400.00)%
    \gplgaddtomacro\gplbacktext{%
      \csname LTb\endcsname
      \put(946,704){\makebox(0,0)[r]{\strut{}$1$}}%
      \put(946,1529){\makebox(0,0)[r]{\strut{}$10$}}%
      \put(946,2354){\makebox(0,0)[r]{\strut{}$100$}}%
      \put(946,3179){\makebox(0,0)[r]{\strut{}$1000$}}%
      \put(1273,484){\makebox(0,0){\strut{}$0.2$}}%
      \put(1664,484){\makebox(0,0){\strut{}$0.3$}}%
      \put(2054,484){\makebox(0,0){\strut{}$0.4$}}%
      \put(2444,484){\makebox(0,0){\strut{}$0.5$}}%
      \put(2835,484){\makebox(0,0){\strut{}$0.6$}}%
      \put(3225,484){\makebox(0,0){\strut{}$0.7$}}%
      \put(3615,484){\makebox(0,0){\strut{}$0.8$}}%
      \put(4005,484){\makebox(0,0){\strut{}$0.9$}}%
      \put(4396,484){\makebox(0,0){\strut{}$1$}}%
    }%
    \gplgaddtomacro\gplfronttext{%
      \csname LTb\endcsname
      \put(209,1941){\rotatebox{-270}{\makebox(0,0){\strut{}$\tau$ [Myr]}}}%
      \put(2834,154){\makebox(0,0){\strut{}$M$ [$\text{M}_\odot$]}}%
      \csname LTb\endcsname
      \put(1468,1777){\rotatebox{90}{\makebox(0,0)[l]{\strut{}thermal collapse}}}%
    }%
    \gplbacktext
    \put(0,0){\includegraphics[width={249.40bp},height={170.00bp}]{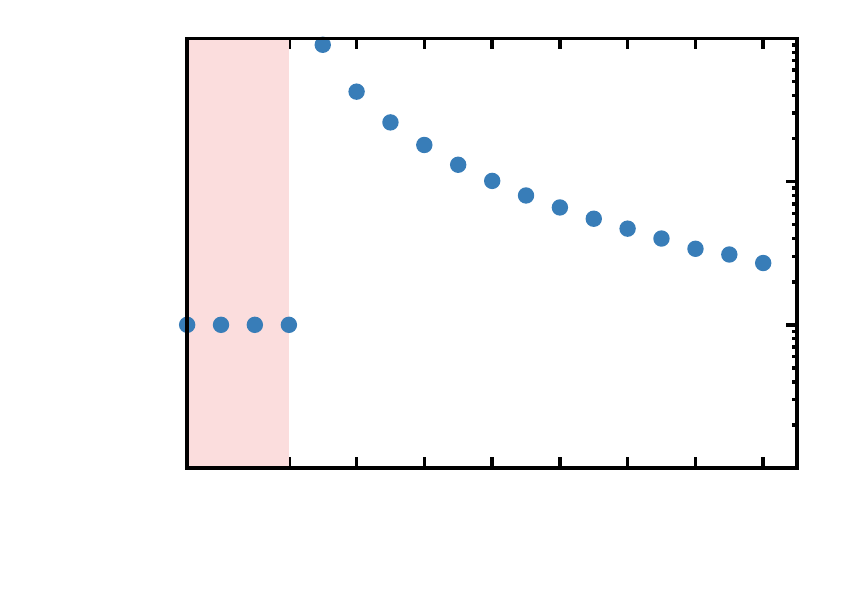}}%
    \gplfronttext
  \end{picture}%
\endgroup

%% file: example-tracks-fig2.tex
\begingroup
  \makeatletter
  \providecommand\color[2][]{%
    \GenericError{(gnuplot) \space\space\space\@spaces}{%
      Package color not loaded in conjunction with
      terminal option `colourtext'%
    }{See the gnuplot documentation for explanation.%
    }{Either use 'blacktext' in gnuplot or load the package
      color.sty in LaTeX.}%
    \renewcommand\color[2][]{}%
  }%
  \providecommand\includegraphics[2][]{%
    \GenericError{(gnuplot) \space\space\space\@spaces}{%
      Package graphicx or graphics not loaded%
    }{See the gnuplot documentation for explanation.%
    }{The gnuplot epslatex terminal needs graphicx.sty or graphics.sty.}%
    \renewcommand\includegraphics[2][]{}%
  }%
  \providecommand\rotatebox[2]{#2}%
  \@ifundefined{ifGPcolor}{%
    \newif\ifGPcolor
    \GPcolortrue
  }{}%
  \@ifundefined{ifGPblacktext}{%
    \newif\ifGPblacktext
    \GPblacktextfalse
  }{}%
  \let\gplgaddtomacro\g@addto@macro
  \gdef\gplbacktext{}%
  \gdef\gplfronttext{}%
  \makeatother
  \ifGPblacktext
    \def\colorrgb#1{}%
    \def\colorgray#1{}%
  \else
    \ifGPcolor
      \def\colorrgb#1{\color[rgb]{#1}}%
      \def\colorgray#1{\color[gray]{#1}}%
      \expandafter\def\csname LTw\endcsname{\color{white}}%
      \expandafter\def\csname LTb\endcsname{\color{black}}%
      \expandafter\def\csname LTa\endcsname{\color{black}}%
      \expandafter\def\csname LT0\endcsname{\color[rgb]{1,0,0}}%
      \expandafter\def\csname LT1\endcsname{\color[rgb]{0,1,0}}%
      \expandafter\def\csname LT2\endcsname{\color[rgb]{0,0,1}}%
      \expandafter\def\csname LT3\endcsname{\color[rgb]{1,0,1}}%
      \expandafter\def\csname LT4\endcsname{\color[rgb]{0,1,1}}%
      \expandafter\def\csname LT5\endcsname{\color[rgb]{1,1,0}}%
      \expandafter\def\csname LT6\endcsname{\color[rgb]{0,0,0}}%
      \expandafter\def\csname LT7\endcsname{\color[rgb]{1,0.3,0}}%
      \expandafter\def\csname LT8\endcsname{\color[rgb]{0.5,0.5,0.5}}%
    \else
      \def\colorrgb#1{\color{black}}%
      \def\colorgray#1{\color[gray]{#1}}%
      \expandafter\def\csname LTw\endcsname{\color{white}}%
      \expandafter\def\csname LTb\endcsname{\color{black}}%
      \expandafter\def\csname LTa\endcsname{\color{black}}%
      \expandafter\def\csname LT0\endcsname{\color{black}}%
      \expandafter\def\csname LT1\endcsname{\color{black}}%
      \expandafter\def\csname LT2\endcsname{\color{black}}%
      \expandafter\def\csname LT3\endcsname{\color{black}}%
      \expandafter\def\csname LT4\endcsname{\color{black}}%
      \expandafter\def\csname LT5\endcsname{\color{black}}%
      \expandafter\def\csname LT6\endcsname{\color{black}}%
      \expandafter\def\csname LT7\endcsname{\color{black}}%
      \expandafter\def\csname LT8\endcsname{\color{black}}%
    \fi
  \fi
    \setlength{\unitlength}{0.0500bp}%
    \ifx\gptboxheight\undefined%
      \newlength{\gptboxheight}%
      \newlength{\gptboxwidth}%
      \newsavebox{\gptboxtext}%
    \fi%
    \setlength{\fboxrule}{0.5pt}%
    \setlength{\fboxsep}{1pt}%
    \definecolor{tbcol}{rgb}{1,1,1}%
\begin{picture}(10204.00,6802.00)%
    \gplgaddtomacro\gplbacktext{%
      \csname LTb\endcsname
      \put(1913,2631){\makebox(0,0){\strut{}$-15$}}%
      \put(2808,2582){\makebox(0,0){\strut{}$-10$}}%
      \put(3704,2534){\makebox(0,0){\strut{}$-5$}}%
      \put(4600,2485){\makebox(0,0){\strut{}$0$}}%
      \put(5494,2436){\makebox(0,0){\strut{}$5$}}%
      \put(6390,2387){\makebox(0,0){\strut{}$10$}}%
      \put(7286,2338){\makebox(0,0){\strut{}$15$}}%
      \put(7480,2517){\makebox(0,0)[l]{\strut{}$-15$}}%
      \put(7638,2795){\makebox(0,0)[l]{\strut{}$-10$}}%
      \put(7796,3072){\makebox(0,0)[l]{\strut{}$-5$}}%
      \put(7954,3350){\makebox(0,0)[l]{\strut{}$0$}}%
      \put(8112,3626){\makebox(0,0)[l]{\strut{}$5$}}%
      \put(8270,3904){\makebox(0,0)[l]{\strut{}$10$}}%
      \put(8427,4181){\makebox(0,0)[l]{\strut{}$15$}}%
      \put(1816,1361){\makebox(0,0)[r]{\strut{}$-10$}}%
      \put(1816,2093){\makebox(0,0)[r]{\strut{}$-5$}}%
      \put(1816,2826){\makebox(0,0)[r]{\strut{}$0$}}%
      \put(1816,3557){\makebox(0,0)[r]{\strut{}$5$}}%
      \put(1816,4289){\makebox(0,0)[r]{\strut{}$10$}}%
    }%
    \gplgaddtomacro\gplfronttext{%
      \csname LTb\endcsname
      \put(3171,2446){\rotatebox{-4}{\makebox(0,0){\strut{}$x_\mathrm{Gal}$ [kpc]}}}%
      \put(8512,3322){\rotatebox{60}{\makebox(0,0){\strut{}$y_\mathrm{Gal}$ [kpc]}}}%
      \put(1414,2826){\rotatebox{90}{\makebox(0,0){\strut{}$z_\mathrm{Gal}$ [kpc]}}}%
    }%
    \gplbacktext
    \put(0,0){\includegraphics[width={510.20bp},height={340.10bp}]{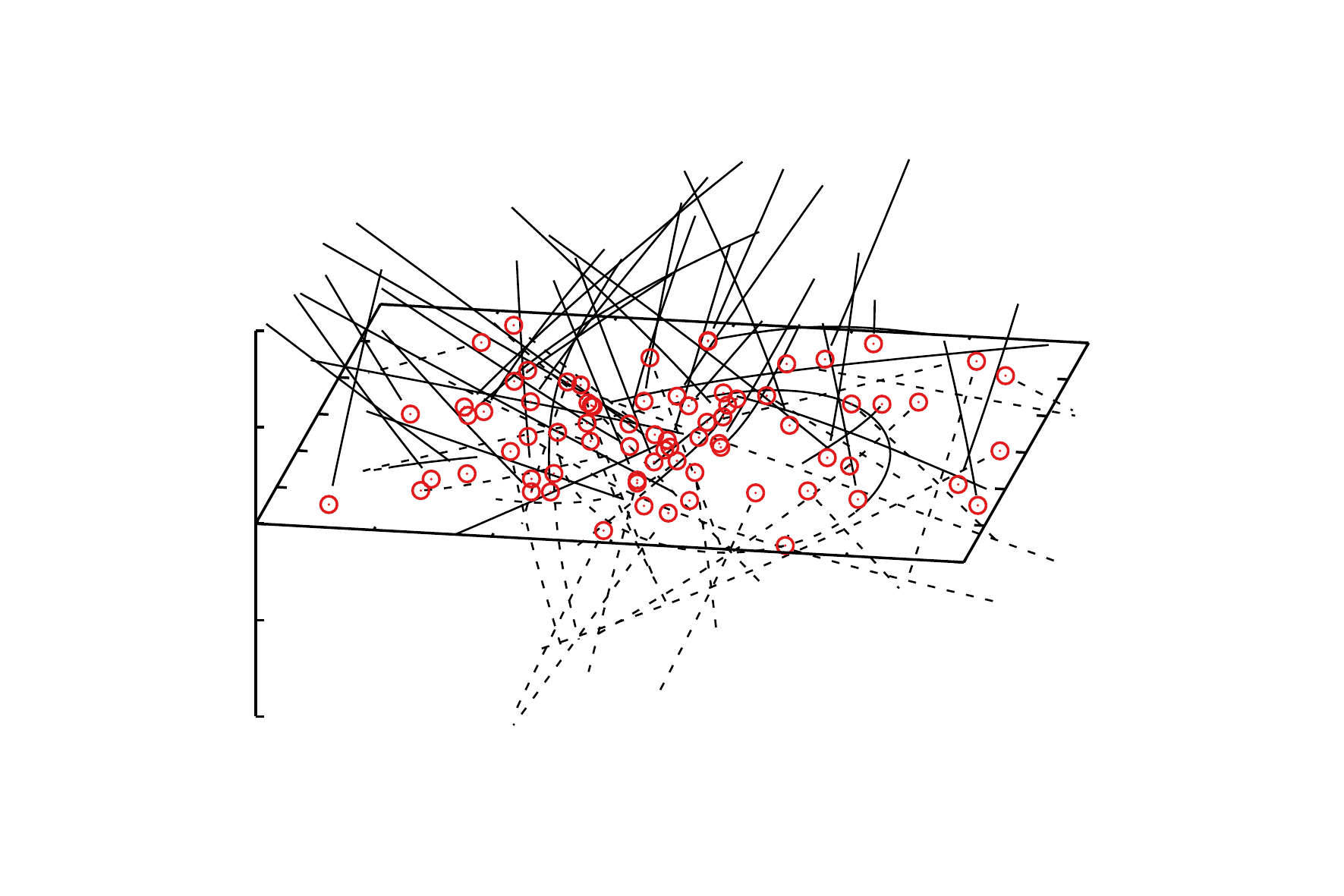}}%
    \gplfronttext
  \end{picture}%
\endgroup

%% file: local-space-earth-fig2.tex
\begingroup
  \makeatletter
  \providecommand\color[2][]{%
    \GenericError{(gnuplot) \space\space\space\@spaces}{%
      Package color not loaded in conjunction with
      terminal option `colourtext'%
    }{See the gnuplot documentation for explanation.%
    }{Either use 'blacktext' in gnuplot or load the package
      color.sty in LaTeX.}%
    \renewcommand\color[2][]{}%
  }%
  \providecommand\includegraphics[2][]{%
    \GenericError{(gnuplot) \space\space\space\@spaces}{%
      Package graphicx or graphics not loaded%
    }{See the gnuplot documentation for explanation.%
    }{The gnuplot epslatex terminal needs graphicx.sty or graphics.sty.}%
    \renewcommand\includegraphics[2][]{}%
  }%
  \providecommand\rotatebox[2]{#2}%
  \@ifundefined{ifGPcolor}{%
    \newif\ifGPcolor
    \GPcolortrue
  }{}%
  \@ifundefined{ifGPblacktext}{%
    \newif\ifGPblacktext
    \GPblacktextfalse
  }{}%
  \let\gplgaddtomacro\g@addto@macro
  \gdef\gplbacktext{}%
  \gdef\gplfronttext{}%
  \makeatother
  \ifGPblacktext
    \def\colorrgb#1{}%
    \def\colorgray#1{}%
  \else
    \ifGPcolor
      \def\colorrgb#1{\color[rgb]{#1}}%
      \def\colorgray#1{\color[gray]{#1}}%
      \expandafter\def\csname LTw\endcsname{\color{white}}%
      \expandafter\def\csname LTb\endcsname{\color{black}}%
      \expandafter\def\csname LTa\endcsname{\color{black}}%
      \expandafter\def\csname LT0\endcsname{\color[rgb]{1,0,0}}%
      \expandafter\def\csname LT1\endcsname{\color[rgb]{0,1,0}}%
      \expandafter\def\csname LT2\endcsname{\color[rgb]{0,0,1}}%
      \expandafter\def\csname LT3\endcsname{\color[rgb]{1,0,1}}%
      \expandafter\def\csname LT4\endcsname{\color[rgb]{0,1,1}}%
      \expandafter\def\csname LT5\endcsname{\color[rgb]{1,1,0}}%
      \expandafter\def\csname LT6\endcsname{\color[rgb]{0,0,0}}%
      \expandafter\def\csname LT7\endcsname{\color[rgb]{1,0.3,0}}%
      \expandafter\def\csname LT8\endcsname{\color[rgb]{0.5,0.5,0.5}}%
    \else
      \def\colorrgb#1{\color{black}}%
      \def\colorgray#1{\color[gray]{#1}}%
      \expandafter\def\csname LTw\endcsname{\color{white}}%
      \expandafter\def\csname LTb\endcsname{\color{black}}%
      \expandafter\def\csname LTa\endcsname{\color{black}}%
      \expandafter\def\csname LT0\endcsname{\color{black}}%
      \expandafter\def\csname LT1\endcsname{\color{black}}%
      \expandafter\def\csname LT2\endcsname{\color{black}}%
      \expandafter\def\csname LT3\endcsname{\color{black}}%
      \expandafter\def\csname LT4\endcsname{\color{black}}%
      \expandafter\def\csname LT5\endcsname{\color{black}}%
      \expandafter\def\csname LT6\endcsname{\color{black}}%
      \expandafter\def\csname LT7\endcsname{\color{black}}%
      \expandafter\def\csname LT8\endcsname{\color{black}}%
    \fi
  \fi
    \setlength{\unitlength}{0.0500bp}%
    \ifx\gptboxheight\undefined%
      \newlength{\gptboxheight}%
      \newlength{\gptboxwidth}%
      \newsavebox{\gptboxtext}%
    \fi%
    \setlength{\fboxrule}{0.5pt}%
    \setlength{\fboxsep}{1pt}%
    \definecolor{tbcol}{rgb}{1,1,1}%
\begin{picture}(4988.00,12472.00)%
    \gplgaddtomacro\gplbacktext{%
      \csname LTb\endcsname
      \put(217,11928){\makebox(0,0){\strut{}(A)}}%
    }%
    \gplgaddtomacro\gplfronttext{%
      \csname LTb\endcsname
      \put(209,10634){\rotatebox{-270}{\makebox(0,0){\strut{}$y_\mathrm{Gal}$ [kpc]}}}%
      \put(2768,8468){\makebox(0,0){\strut{}$x_\mathrm{Gal}$ [kpc]}}%
      \csname LTb\endcsname
      \put(814,9018){\makebox(0,0)[r]{\strut{}$-30$}}%
      \put(814,9557){\makebox(0,0)[r]{\strut{}$-20$}}%
      \put(814,10096){\makebox(0,0)[r]{\strut{}$-10$}}%
      \put(814,10635){\makebox(0,0)[r]{\strut{}$0$}}%
      \put(814,11173){\makebox(0,0)[r]{\strut{}$10$}}%
      \put(814,11712){\makebox(0,0)[r]{\strut{}$20$}}%
      \put(814,12251){\makebox(0,0)[r]{\strut{}$30$}}%
      \put(946,8798){\makebox(0,0){\strut{}$-30$}}%
      \put(1554,8798){\makebox(0,0){\strut{}$-20$}}%
      \put(2161,8798){\makebox(0,0){\strut{}$-10$}}%
      \put(2769,8798){\makebox(0,0){\strut{}$0$}}%
      \put(3376,8798){\makebox(0,0){\strut{}$10$}}%
      \put(3984,8798){\makebox(0,0){\strut{}$20$}}%
      \put(4591,8798){\makebox(0,0){\strut{}$30$}}%
      \put(2179,10635){\makebox(0,0)[r]{\strut{}Earth}}%
      \put(2860,10635){\makebox(0,0)[l]{\strut{}GC}}%
    }%
    \gplgaddtomacro\gplbacktext{%
      \csname LTb\endcsname
      \put(217,7771){\makebox(0,0){\strut{}(B)}}%
    }%
    \gplgaddtomacro\gplfronttext{%
      \csname LTb\endcsname
      \put(209,6477){\rotatebox{-270}{\makebox(0,0){\strut{}$y_\mathrm{Gal}$ [kpc]}}}%
      \put(2768,4311){\makebox(0,0){\strut{}$x_\mathrm{Gal}$ [kpc]}}%
      \csname LTb\endcsname
      \put(814,4861){\makebox(0,0)[r]{\strut{}$-30$}}%
      \put(814,5400){\makebox(0,0)[r]{\strut{}$-20$}}%
      \put(814,5939){\makebox(0,0)[r]{\strut{}$-10$}}%
      \put(814,6478){\makebox(0,0)[r]{\strut{}$0$}}%
      \put(814,7016){\makebox(0,0)[r]{\strut{}$10$}}%
      \put(814,7555){\makebox(0,0)[r]{\strut{}$20$}}%
      \put(814,8094){\makebox(0,0)[r]{\strut{}$30$}}%
      \put(946,4641){\makebox(0,0){\strut{}$-30$}}%
      \put(1554,4641){\makebox(0,0){\strut{}$-20$}}%
      \put(2161,4641){\makebox(0,0){\strut{}$-10$}}%
      \put(2769,4641){\makebox(0,0){\strut{}$0$}}%
      \put(3376,4641){\makebox(0,0){\strut{}$10$}}%
      \put(3984,4641){\makebox(0,0){\strut{}$20$}}%
      \put(4591,4641){\makebox(0,0){\strut{}$30$}}%
      \put(2179,6478){\makebox(0,0)[r]{\strut{}Earth}}%
      \put(2860,6478){\makebox(0,0)[l]{\strut{}GC}}%
    }%
    \gplgaddtomacro\gplbacktext{%
      \csname LTb\endcsname
      \put(217,3614){\makebox(0,0){\strut{}(C)}}%
    }%
    \gplgaddtomacro\gplfronttext{%
      \csname LTb\endcsname
      \put(209,2320){\rotatebox{-270}{\makebox(0,0){\strut{}$z_\mathrm{Gal}$ [kpc]}}}%
      \put(2768,154){\makebox(0,0){\strut{}$x_\mathrm{Gal}$ [kpc]}}%
      \csname LTb\endcsname
      \put(814,704){\makebox(0,0)[r]{\strut{}$-30$}}%
      \put(814,1243){\makebox(0,0)[r]{\strut{}$-20$}}%
      \put(814,1782){\makebox(0,0)[r]{\strut{}$-10$}}%
      \put(814,2321){\makebox(0,0)[r]{\strut{}$0$}}%
      \put(814,2859){\makebox(0,0)[r]{\strut{}$10$}}%
      \put(814,3398){\makebox(0,0)[r]{\strut{}$20$}}%
      \put(814,3937){\makebox(0,0)[r]{\strut{}$30$}}%
      \put(946,484){\makebox(0,0){\strut{}$-30$}}%
      \put(1554,484){\makebox(0,0){\strut{}$-20$}}%
      \put(2161,484){\makebox(0,0){\strut{}$-10$}}%
      \put(2769,484){\makebox(0,0){\strut{}$0$}}%
      \put(3376,484){\makebox(0,0){\strut{}$10$}}%
      \put(3984,484){\makebox(0,0){\strut{}$20$}}%
      \put(4591,484){\makebox(0,0){\strut{}$30$}}%
      \put(2179,2342){\makebox(0,0)[r]{\strut{}Earth}}%
      \put(2860,2321){\makebox(0,0)[l]{\strut{}GC}}%
    }%
    \gplbacktext
    \put(0,0){\includegraphics[width={249.40bp},height={623.60bp}]{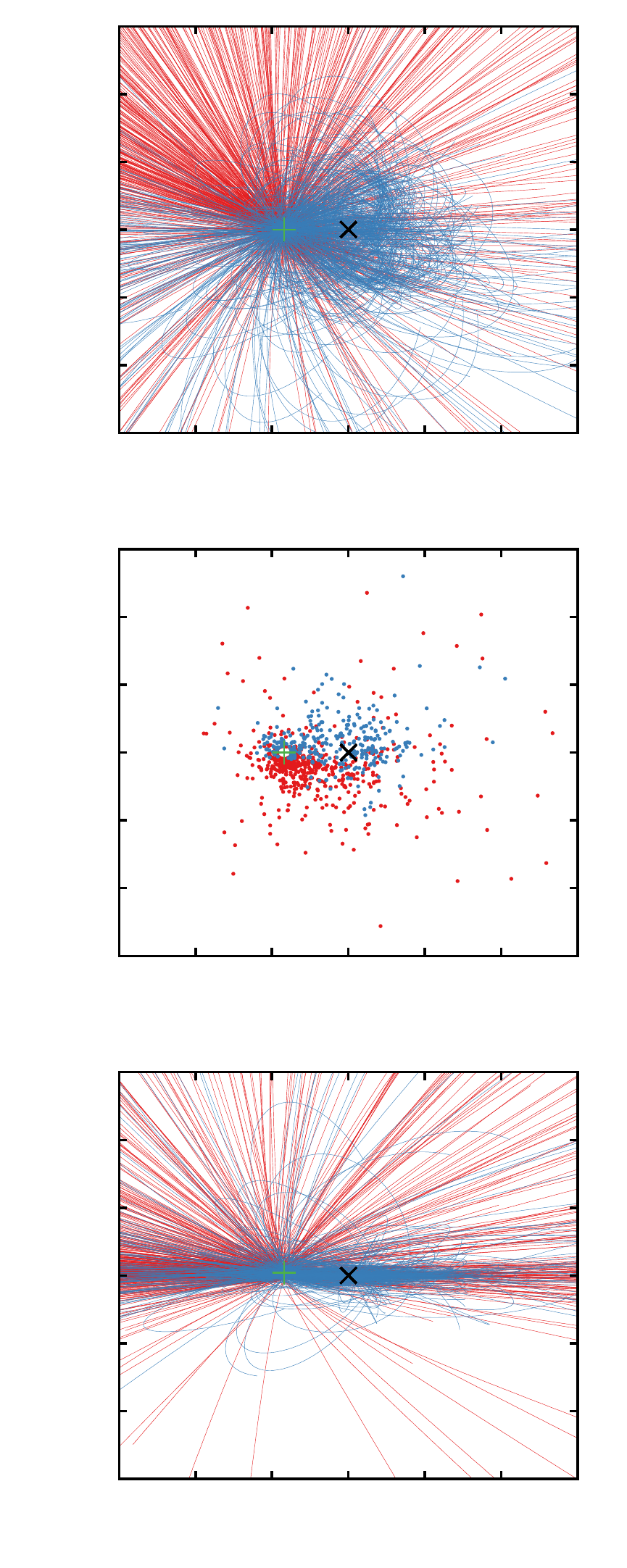}}%
    \gplfronttext
  \end{picture}%
\endgroup

%% file: chirality-M14-fig2.tex
\begingroup
  \makeatletter
  \providecommand\color[2][]{%
    \GenericError{(gnuplot) \space\space\space\@spaces}{%
      Package color not loaded in conjunction with
      terminal option `colourtext'%
    }{See the gnuplot documentation for explanation.%
    }{Either use 'blacktext' in gnuplot or load the package
      color.sty in LaTeX.}%
    \renewcommand\color[2][]{}%
  }%
  \providecommand\includegraphics[2][]{%
    \GenericError{(gnuplot) \space\space\space\@spaces}{%
      Package graphicx or graphics not loaded%
    }{See the gnuplot documentation for explanation.%
    }{The gnuplot epslatex terminal needs graphicx.sty or graphics.sty.}%
    \renewcommand\includegraphics[2][]{}%
  }%
  \providecommand\rotatebox[2]{#2}%
  \@ifundefined{ifGPcolor}{%
    \newif\ifGPcolor
    \GPcolortrue
  }{}%
  \@ifundefined{ifGPblacktext}{%
    \newif\ifGPblacktext
    \GPblacktextfalse
  }{}%
  \let\gplgaddtomacro\g@addto@macro
  \gdef\gplbacktext{}%
  \gdef\gplfronttext{}%
  \makeatother
  \ifGPblacktext
    \def\colorrgb#1{}%
    \def\colorgray#1{}%
  \else
    \ifGPcolor
      \def\colorrgb#1{\color[rgb]{#1}}%
      \def\colorgray#1{\color[gray]{#1}}%
      \expandafter\def\csname LTw\endcsname{\color{white}}%
      \expandafter\def\csname LTb\endcsname{\color{black}}%
      \expandafter\def\csname LTa\endcsname{\color{black}}%
      \expandafter\def\csname LT0\endcsname{\color[rgb]{1,0,0}}%
      \expandafter\def\csname LT1\endcsname{\color[rgb]{0,1,0}}%
      \expandafter\def\csname LT2\endcsname{\color[rgb]{0,0,1}}%
      \expandafter\def\csname LT3\endcsname{\color[rgb]{1,0,1}}%
      \expandafter\def\csname LT4\endcsname{\color[rgb]{0,1,1}}%
      \expandafter\def\csname LT5\endcsname{\color[rgb]{1,1,0}}%
      \expandafter\def\csname LT6\endcsname{\color[rgb]{0,0,0}}%
      \expandafter\def\csname LT7\endcsname{\color[rgb]{1,0.3,0}}%
      \expandafter\def\csname LT8\endcsname{\color[rgb]{0.5,0.5,0.5}}%
    \else
      \def\colorrgb#1{\color{black}}%
      \def\colorgray#1{\color[gray]{#1}}%
      \expandafter\def\csname LTw\endcsname{\color{white}}%
      \expandafter\def\csname LTb\endcsname{\color{black}}%
      \expandafter\def\csname LTa\endcsname{\color{black}}%
      \expandafter\def\csname LT0\endcsname{\color{black}}%
      \expandafter\def\csname LT1\endcsname{\color{black}}%
      \expandafter\def\csname LT2\endcsname{\color{black}}%
      \expandafter\def\csname LT3\endcsname{\color{black}}%
      \expandafter\def\csname LT4\endcsname{\color{black}}%
      \expandafter\def\csname LT5\endcsname{\color{black}}%
      \expandafter\def\csname LT6\endcsname{\color{black}}%
      \expandafter\def\csname LT7\endcsname{\color{black}}%
      \expandafter\def\csname LT8\endcsname{\color{black}}%
    \fi
  \fi
    \setlength{\unitlength}{0.0500bp}%
    \ifx\gptboxheight\undefined%
      \newlength{\gptboxheight}%
      \newlength{\gptboxwidth}%
      \newsavebox{\gptboxtext}%
    \fi%
    \setlength{\fboxrule}{0.5pt}%
    \setlength{\fboxsep}{1pt}%
    \definecolor{tbcol}{rgb}{1,1,1}%
\begin{picture}(4988.00,4534.00)%
    \gplgaddtomacro\gplbacktext{%
      \csname LTb\endcsname
      \put(1824,4262){\makebox(0,0){\strut{}retrograde}}%
      \put(3581,4262){\makebox(0,0){\strut{}prograde}}%
      \put(946,4262){\makebox(0,0){\strut{}location}}%
      \put(4810,1551){\rotatebox{90}{\makebox(0,0){\strut{}prograde}}}%
      \put(4810,3246){\rotatebox{90}{\makebox(0,0){\strut{}retrograde}}}%
      \put(4810,535){\rotatebox{90}{\makebox(0,0){\strut{}direction}}}%
    }%
    \gplgaddtomacro\gplfronttext{%
      \csname LTb\endcsname
      \put(275,2398){\rotatebox{-270}{\makebox(0,0){\strut{}$(x\,v_y-y\,v_x)/|(x\,v_y-y\,v_x)| \cdot R_\mathrm{Gal}$}}}%
      \put(2702,154){\makebox(0,0){\strut{}Gal. polar coord. [deg]}}%
      \csname LTb\endcsname
      \put(814,704){\makebox(0,0)[r]{\strut{}$-30$}}%
      \put(814,1269){\makebox(0,0)[r]{\strut{}$-20$}}%
      \put(814,1834){\makebox(0,0)[r]{\strut{}$-10$}}%
      \put(814,2399){\makebox(0,0)[r]{\strut{}$0$}}%
      \put(814,2963){\makebox(0,0)[r]{\strut{}$10$}}%
      \put(814,3528){\makebox(0,0)[r]{\strut{}$20$}}%
      \put(814,4093){\makebox(0,0)[r]{\strut{}$30$}}%
      \put(946,484){\makebox(0,0){\strut{}$-180$}}%
      \put(1385,484){\makebox(0,0){\strut{}$-135$}}%
      \put(1824,484){\makebox(0,0){\strut{}$-90$}}%
      \put(2263,484){\makebox(0,0){\strut{}$-45$}}%
      \put(2703,484){\makebox(0,0){\strut{}$0$}}%
      \put(3142,484){\makebox(0,0){\strut{}$45$}}%
      \put(3581,484){\makebox(0,0){\strut{}$90$}}%
      \put(4020,484){\makebox(0,0){\strut{}$135$}}%
      \put(4459,484){\makebox(0,0){\strut{}$180$}}%
      \put(848,2031){\makebox(0,0)[r]{\strut{}Earth}}%
    }%
    \gplbacktext
    \put(0,0){\includegraphics[width={249.40bp},height={226.70bp}]{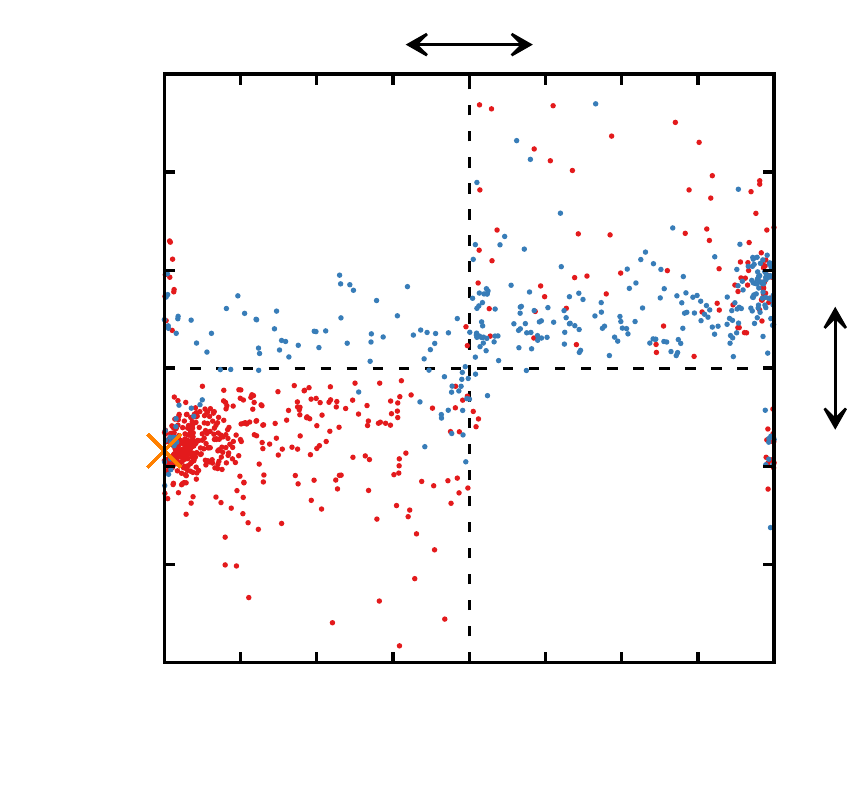}}%
    \gplfronttext
  \end{picture}%
\endgroup

%% file: local-space-US708-fig2.tex
\begingroup
  \makeatletter
  \providecommand\color[2][]{%
    \GenericError{(gnuplot) \space\space\space\@spaces}{%
      Package color not loaded in conjunction with
      terminal option `colourtext'%
    }{See the gnuplot documentation for explanation.%
    }{Either use 'blacktext' in gnuplot or load the package
      color.sty in LaTeX.}%
    \renewcommand\color[2][]{}%
  }%
  \providecommand\includegraphics[2][]{%
    \GenericError{(gnuplot) \space\space\space\@spaces}{%
      Package graphicx or graphics not loaded%
    }{See the gnuplot documentation for explanation.%
    }{The gnuplot epslatex terminal needs graphicx.sty or graphics.sty.}%
    \renewcommand\includegraphics[2][]{}%
  }%
  \providecommand\rotatebox[2]{#2}%
  \@ifundefined{ifGPcolor}{%
    \newif\ifGPcolor
    \GPcolortrue
  }{}%
  \@ifundefined{ifGPblacktext}{%
    \newif\ifGPblacktext
    \GPblacktextfalse
  }{}%
  \let\gplgaddtomacro\g@addto@macro
  \gdef\gplbacktext{}%
  \gdef\gplfronttext{}%
  \makeatother
  \ifGPblacktext
    \def\colorrgb#1{}%
    \def\colorgray#1{}%
  \else
    \ifGPcolor
      \def\colorrgb#1{\color[rgb]{#1}}%
      \def\colorgray#1{\color[gray]{#1}}%
      \expandafter\def\csname LTw\endcsname{\color{white}}%
      \expandafter\def\csname LTb\endcsname{\color{black}}%
      \expandafter\def\csname LTa\endcsname{\color{black}}%
      \expandafter\def\csname LT0\endcsname{\color[rgb]{1,0,0}}%
      \expandafter\def\csname LT1\endcsname{\color[rgb]{0,1,0}}%
      \expandafter\def\csname LT2\endcsname{\color[rgb]{0,0,1}}%
      \expandafter\def\csname LT3\endcsname{\color[rgb]{1,0,1}}%
      \expandafter\def\csname LT4\endcsname{\color[rgb]{0,1,1}}%
      \expandafter\def\csname LT5\endcsname{\color[rgb]{1,1,0}}%
      \expandafter\def\csname LT6\endcsname{\color[rgb]{0,0,0}}%
      \expandafter\def\csname LT7\endcsname{\color[rgb]{1,0.3,0}}%
      \expandafter\def\csname LT8\endcsname{\color[rgb]{0.5,0.5,0.5}}%
    \else
      \def\colorrgb#1{\color{black}}%
      \def\colorgray#1{\color[gray]{#1}}%
      \expandafter\def\csname LTw\endcsname{\color{white}}%
      \expandafter\def\csname LTb\endcsname{\color{black}}%
      \expandafter\def\csname LTa\endcsname{\color{black}}%
      \expandafter\def\csname LT0\endcsname{\color{black}}%
      \expandafter\def\csname LT1\endcsname{\color{black}}%
      \expandafter\def\csname LT2\endcsname{\color{black}}%
      \expandafter\def\csname LT3\endcsname{\color{black}}%
      \expandafter\def\csname LT4\endcsname{\color{black}}%
      \expandafter\def\csname LT5\endcsname{\color{black}}%
      \expandafter\def\csname LT6\endcsname{\color{black}}%
      \expandafter\def\csname LT7\endcsname{\color{black}}%
      \expandafter\def\csname LT8\endcsname{\color{black}}%
    \fi
  \fi
    \setlength{\unitlength}{0.0500bp}%
    \ifx\gptboxheight\undefined%
      \newlength{\gptboxheight}%
      \newlength{\gptboxwidth}%
      \newsavebox{\gptboxtext}%
    \fi%
    \setlength{\fboxrule}{0.5pt}%
    \setlength{\fboxsep}{1pt}%
    \definecolor{tbcol}{rgb}{1,1,1}%
\begin{picture}(4988.00,12472.00)%
    \gplgaddtomacro\gplbacktext{%
      \csname LTb\endcsname
      \put(217,11928){\makebox(0,0){\strut{}(A)}}%
    }%
    \gplgaddtomacro\gplfronttext{%
      \csname LTb\endcsname
      \put(209,10634){\rotatebox{-270}{\makebox(0,0){\strut{}$y_\mathrm{Gal}$ [kpc]}}}%
      \put(2768,8468){\makebox(0,0){\strut{}$x_\mathrm{Gal}$ [kpc]}}%
      \csname LTb\endcsname
      \put(814,9018){\makebox(0,0)[r]{\strut{}$-30$}}%
      \put(814,9557){\makebox(0,0)[r]{\strut{}$-20$}}%
      \put(814,10096){\makebox(0,0)[r]{\strut{}$-10$}}%
      \put(814,10635){\makebox(0,0)[r]{\strut{}$0$}}%
      \put(814,11173){\makebox(0,0)[r]{\strut{}$10$}}%
      \put(814,11712){\makebox(0,0)[r]{\strut{}$20$}}%
      \put(814,12251){\makebox(0,0)[r]{\strut{}$30$}}%
      \put(946,8798){\makebox(0,0){\strut{}$-30$}}%
      \put(1554,8798){\makebox(0,0){\strut{}$-20$}}%
      \put(2161,8798){\makebox(0,0){\strut{}$-10$}}%
      \put(2769,8798){\makebox(0,0){\strut{}$0$}}%
      \put(3376,8798){\makebox(0,0){\strut{}$10$}}%
      \put(3984,8798){\makebox(0,0){\strut{}$20$}}%
      \put(4591,8798){\makebox(0,0){\strut{}$30$}}%
      \put(2258,10554){\makebox(0,0){\strut{}Earth}}%
      \put(2860,10635){\makebox(0,0)[l]{\strut{}GC}}%
    }%
    \gplgaddtomacro\gplbacktext{%
      \csname LTb\endcsname
      \put(217,7771){\makebox(0,0){\strut{}(B)}}%
    }%
    \gplgaddtomacro\gplfronttext{%
      \csname LTb\endcsname
      \put(209,6477){\rotatebox{-270}{\makebox(0,0){\strut{}$y_\mathrm{Gal}$ [kpc]}}}%
      \put(2768,4311){\makebox(0,0){\strut{}$x_\mathrm{Gal}$ [kpc]}}%
      \csname LTb\endcsname
      \put(814,4861){\makebox(0,0)[r]{\strut{}$-30$}}%
      \put(814,5400){\makebox(0,0)[r]{\strut{}$-20$}}%
      \put(814,5939){\makebox(0,0)[r]{\strut{}$-10$}}%
      \put(814,6478){\makebox(0,0)[r]{\strut{}$0$}}%
      \put(814,7016){\makebox(0,0)[r]{\strut{}$10$}}%
      \put(814,7555){\makebox(0,0)[r]{\strut{}$20$}}%
      \put(814,8094){\makebox(0,0)[r]{\strut{}$30$}}%
      \put(946,4641){\makebox(0,0){\strut{}$-30$}}%
      \put(1554,4641){\makebox(0,0){\strut{}$-20$}}%
      \put(2161,4641){\makebox(0,0){\strut{}$-10$}}%
      \put(2769,4641){\makebox(0,0){\strut{}$0$}}%
      \put(3376,4641){\makebox(0,0){\strut{}$10$}}%
      \put(3984,4641){\makebox(0,0){\strut{}$20$}}%
      \put(4591,4641){\makebox(0,0){\strut{}$30$}}%
      \put(2258,6397){\makebox(0,0){\strut{}Earth}}%
      \put(2860,6478){\makebox(0,0)[l]{\strut{}GC}}%
    }%
    \gplgaddtomacro\gplbacktext{%
      \csname LTb\endcsname
      \put(217,3614){\makebox(0,0){\strut{}(C)}}%
    }%
    \gplgaddtomacro\gplfronttext{%
      \csname LTb\endcsname
      \put(209,2320){\rotatebox{-270}{\makebox(0,0){\strut{}$z_\mathrm{Gal}$ [kpc]}}}%
      \put(2768,154){\makebox(0,0){\strut{}$x_\mathrm{Gal}$ [kpc]}}%
      \csname LTb\endcsname
      \put(3159,1402){\makebox(0,0)[r]{\strut{}US708 current}}%
      \csname LTb\endcsname
      \put(3159,1182){\makebox(0,0)[r]{\strut{}US708 ejection}}%
      \csname LTb\endcsname
      \put(814,704){\makebox(0,0)[r]{\strut{}$-30$}}%
      \put(814,1243){\makebox(0,0)[r]{\strut{}$-20$}}%
      \put(814,1782){\makebox(0,0)[r]{\strut{}$-10$}}%
      \put(814,2321){\makebox(0,0)[r]{\strut{}$0$}}%
      \put(814,2859){\makebox(0,0)[r]{\strut{}$10$}}%
      \put(814,3398){\makebox(0,0)[r]{\strut{}$20$}}%
      \put(814,3937){\makebox(0,0)[r]{\strut{}$30$}}%
      \put(946,484){\makebox(0,0){\strut{}$-30$}}%
      \put(1554,484){\makebox(0,0){\strut{}$-20$}}%
      \put(2161,484){\makebox(0,0){\strut{}$-10$}}%
      \put(2769,484){\makebox(0,0){\strut{}$0$}}%
      \put(3376,484){\makebox(0,0){\strut{}$10$}}%
      \put(3984,484){\makebox(0,0){\strut{}$20$}}%
      \put(4591,484){\makebox(0,0){\strut{}$30$}}%
      \put(2258,2159){\makebox(0,0)[r]{\strut{}Earth}}%
      \put(2860,2321){\makebox(0,0)[l]{\strut{}GC}}%
    }%
    \gplbacktext
    \put(0,0){\includegraphics[width={249.40bp},height={623.60bp}]{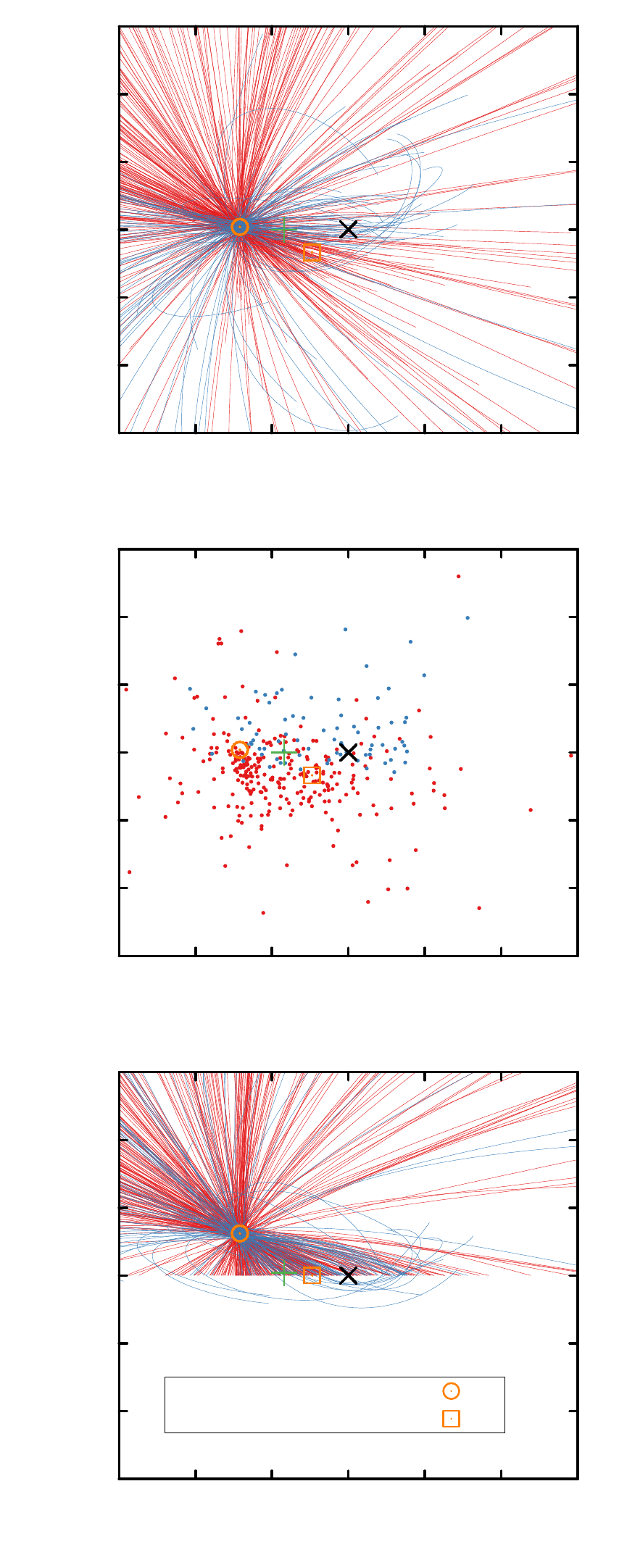}}%
    \gplfronttext
  \end{picture}%
\endgroup

%% file: r-z-fig2.tex
\begingroup
  \makeatletter
  \providecommand\color[2][]{%
    \GenericError{(gnuplot) \space\space\space\@spaces}{%
      Package color not loaded in conjunction with
      terminal option `colourtext'%
    }{See the gnuplot documentation for explanation.%
    }{Either use 'blacktext' in gnuplot or load the package
      color.sty in LaTeX.}%
    \renewcommand\color[2][]{}%
  }%
  \providecommand\includegraphics[2][]{%
    \GenericError{(gnuplot) \space\space\space\@spaces}{%
      Package graphicx or graphics not loaded%
    }{See the gnuplot documentation for explanation.%
    }{The gnuplot epslatex terminal needs graphicx.sty or graphics.sty.}%
    \renewcommand\includegraphics[2][]{}%
  }%
  \providecommand\rotatebox[2]{#2}%
  \@ifundefined{ifGPcolor}{%
    \newif\ifGPcolor
    \GPcolortrue
  }{}%
  \@ifundefined{ifGPblacktext}{%
    \newif\ifGPblacktext
    \GPblacktextfalse
  }{}%
  \let\gplgaddtomacro\g@addto@macro
  \gdef\gplbacktext{}%
  \gdef\gplfronttext{}%
  \makeatother
  \ifGPblacktext
    \def\colorrgb#1{}%
    \def\colorgray#1{}%
  \else
    \ifGPcolor
      \def\colorrgb#1{\color[rgb]{#1}}%
      \def\colorgray#1{\color[gray]{#1}}%
      \expandafter\def\csname LTw\endcsname{\color{white}}%
      \expandafter\def\csname LTb\endcsname{\color{black}}%
      \expandafter\def\csname LTa\endcsname{\color{black}}%
      \expandafter\def\csname LT0\endcsname{\color[rgb]{1,0,0}}%
      \expandafter\def\csname LT1\endcsname{\color[rgb]{0,1,0}}%
      \expandafter\def\csname LT2\endcsname{\color[rgb]{0,0,1}}%
      \expandafter\def\csname LT3\endcsname{\color[rgb]{1,0,1}}%
      \expandafter\def\csname LT4\endcsname{\color[rgb]{0,1,1}}%
      \expandafter\def\csname LT5\endcsname{\color[rgb]{1,1,0}}%
      \expandafter\def\csname LT6\endcsname{\color[rgb]{0,0,0}}%
      \expandafter\def\csname LT7\endcsname{\color[rgb]{1,0.3,0}}%
      \expandafter\def\csname LT8\endcsname{\color[rgb]{0.5,0.5,0.5}}%
    \else
      \def\colorrgb#1{\color{black}}%
      \def\colorgray#1{\color[gray]{#1}}%
      \expandafter\def\csname LTw\endcsname{\color{white}}%
      \expandafter\def\csname LTb\endcsname{\color{black}}%
      \expandafter\def\csname LTa\endcsname{\color{black}}%
      \expandafter\def\csname LT0\endcsname{\color{black}}%
      \expandafter\def\csname LT1\endcsname{\color{black}}%
      \expandafter\def\csname LT2\endcsname{\color{black}}%
      \expandafter\def\csname LT3\endcsname{\color{black}}%
      \expandafter\def\csname LT4\endcsname{\color{black}}%
      \expandafter\def\csname LT5\endcsname{\color{black}}%
      \expandafter\def\csname LT6\endcsname{\color{black}}%
      \expandafter\def\csname LT7\endcsname{\color{black}}%
      \expandafter\def\csname LT8\endcsname{\color{black}}%
    \fi
  \fi
    \setlength{\unitlength}{0.0500bp}%
    \ifx\gptboxheight\undefined%
      \newlength{\gptboxheight}%
      \newlength{\gptboxwidth}%
      \newsavebox{\gptboxtext}%
    \fi%
    \setlength{\fboxrule}{0.5pt}%
    \setlength{\fboxsep}{1pt}%
    \definecolor{tbcol}{rgb}{1,1,1}%
\begin{picture}(10204.00,6802.00)%
    \gplgaddtomacro\gplbacktext{%
      \csname LTb\endcsname
      \put(196,6333){\makebox(0,0){\strut{}(A)}}%
    }%
    \gplgaddtomacro\gplfronttext{%
      \csname LTb\endcsname
      \put(209,5343){\rotatebox{-270}{\makebox(0,0){\strut{}$z_\mathrm{Gal}$ [kpc]}}}%
      \put(2360,3555){\makebox(0,0){\strut{}$R_\mathrm{Gal}$ [kpc]}}%
      \csname LTb\endcsname
      \put(682,4353){\makebox(0,0)[r]{\strut{}$-6$}}%
      \put(682,4683){\makebox(0,0)[r]{\strut{}$-4$}}%
      \put(682,5013){\makebox(0,0)[r]{\strut{}$-2$}}%
      \put(682,5343){\makebox(0,0)[r]{\strut{}$0$}}%
      \put(682,5673){\makebox(0,0)[r]{\strut{}$2$}}%
      \put(682,6003){\makebox(0,0)[r]{\strut{}$4$}}%
      \put(682,6333){\makebox(0,0)[r]{\strut{}$6$}}%
      \put(814,3885){\makebox(0,0){\strut{}$0$}}%
      \put(1226,3885){\makebox(0,0){\strut{}$2$}}%
      \put(1639,3885){\makebox(0,0){\strut{}$4$}}%
      \put(2051,3885){\makebox(0,0){\strut{}$6$}}%
      \put(2463,3885){\makebox(0,0){\strut{}$8$}}%
      \put(2875,3885){\makebox(0,0){\strut{}$10$}}%
      \put(3288,3885){\makebox(0,0){\strut{}$12$}}%
      \put(3700,3885){\makebox(0,0){\strut{}$14$}}%
      \put(4269,4105){\makebox(0,0)[l]{\strut{}$10^{ -1}$}}%
      \put(4269,4724){\makebox(0,0)[l]{\strut{}$10^{  0}$}}%
      \put(4269,5343){\makebox(0,0)[l]{\strut{}$10^{  1}$}}%
      \put(4269,5962){\makebox(0,0)[l]{\strut{}$10^{  2}$}}%
      \put(4269,6581){\makebox(0,0)[l]{\strut{}$10^{  3}$}}%
      \put(4797,5343){\rotatebox{-270}{\makebox(0,0){\strut{}$\rho_\mathrm{N}$ [$\mathrm{kpc}^{-1}$]}}}%
      \colorrgb{0.89,0.10,0.11}
      \put(2546,5244){\makebox(0,0){\strut{}\small{Earth}}}%
      \colorrgb{0.00,0.00,0.00}
      \put(3741,6201){\makebox(0,0)[r]{\strut{}\small{US\,708}}}%
    }%
    \gplgaddtomacro\gplbacktext{%
      \csname LTb\endcsname
      \put(5298,6333){\makebox(0,0){\strut{}(B)}}%
    }%
    \gplgaddtomacro\gplfronttext{%
      \csname LTb\endcsname
      \put(5311,5343){\rotatebox{-270}{\makebox(0,0){\strut{}$z_\mathrm{Gal}$ [kpc]}}}%
      \put(7462,3555){\makebox(0,0){\strut{}$R_\mathrm{Gal}$ [kpc]}}%
      \csname LTb\endcsname
      \put(5784,4353){\makebox(0,0)[r]{\strut{}$-6$}}%
      \put(5784,4683){\makebox(0,0)[r]{\strut{}$-4$}}%
      \put(5784,5013){\makebox(0,0)[r]{\strut{}$-2$}}%
      \put(5784,5343){\makebox(0,0)[r]{\strut{}$0$}}%
      \put(5784,5673){\makebox(0,0)[r]{\strut{}$2$}}%
      \put(5784,6003){\makebox(0,0)[r]{\strut{}$4$}}%
      \put(5784,6333){\makebox(0,0)[r]{\strut{}$6$}}%
      \put(5916,3885){\makebox(0,0){\strut{}$0$}}%
      \put(6328,3885){\makebox(0,0){\strut{}$2$}}%
      \put(6741,3885){\makebox(0,0){\strut{}$4$}}%
      \put(7153,3885){\makebox(0,0){\strut{}$6$}}%
      \put(7565,3885){\makebox(0,0){\strut{}$8$}}%
      \put(7977,3885){\makebox(0,0){\strut{}$10$}}%
      \put(8390,3885){\makebox(0,0){\strut{}$12$}}%
      \put(8802,3885){\makebox(0,0){\strut{}$14$}}%
      \put(9371,4105){\makebox(0,0)[l]{\strut{}$10^{ -1}$}}%
      \put(9371,4724){\makebox(0,0)[l]{\strut{}$10^{  0}$}}%
      \put(9371,5343){\makebox(0,0)[l]{\strut{}$10^{  1}$}}%
      \put(9371,5962){\makebox(0,0)[l]{\strut{}$10^{  2}$}}%
      \put(9371,6581){\makebox(0,0)[l]{\strut{}$10^{  3}$}}%
      \put(9899,5343){\rotatebox{-270}{\makebox(0,0){\strut{}$\rho_\mathrm{N}$ [$\mathrm{kpc}^{-1}$]}}}%
      \colorrgb{0.89,0.10,0.11}
      \put(7648,5244){\makebox(0,0){\strut{}\small{Earth}}}%
      \colorrgb{0.00,0.00,0.00}
      \put(8843,6201){\makebox(0,0)[r]{\strut{}\small{US\,708}}}%
    }%
    \gplgaddtomacro\gplbacktext{%
      \csname LTb\endcsname
      \put(196,2933){\makebox(0,0){\strut{}(C)}}%
    }%
    \gplgaddtomacro\gplfronttext{%
      \csname LTb\endcsname
      \put(209,1942){\rotatebox{-270}{\makebox(0,0){\strut{}$z_\mathrm{Gal}$ [kpc]}}}%
      \put(2360,154){\makebox(0,0){\strut{}$R_\mathrm{Gal}$ [kpc]}}%
      \csname LTb\endcsname
      \put(682,952){\makebox(0,0)[r]{\strut{}$-6$}}%
      \put(682,1282){\makebox(0,0)[r]{\strut{}$-4$}}%
      \put(682,1612){\makebox(0,0)[r]{\strut{}$-2$}}%
      \put(682,1943){\makebox(0,0)[r]{\strut{}$0$}}%
      \put(682,2273){\makebox(0,0)[r]{\strut{}$2$}}%
      \put(682,2603){\makebox(0,0)[r]{\strut{}$4$}}%
      \put(682,2933){\makebox(0,0)[r]{\strut{}$6$}}%
      \put(814,484){\makebox(0,0){\strut{}$0$}}%
      \put(1226,484){\makebox(0,0){\strut{}$2$}}%
      \put(1639,484){\makebox(0,0){\strut{}$4$}}%
      \put(2051,484){\makebox(0,0){\strut{}$6$}}%
      \put(2463,484){\makebox(0,0){\strut{}$8$}}%
      \put(2875,484){\makebox(0,0){\strut{}$10$}}%
      \put(3288,484){\makebox(0,0){\strut{}$12$}}%
      \put(3700,484){\makebox(0,0){\strut{}$14$}}%
      \put(4269,703){\makebox(0,0)[l]{\strut{}$10^{ -1}$}}%
      \put(4269,1323){\makebox(0,0)[l]{\strut{}$10^{  0}$}}%
      \put(4269,1942){\makebox(0,0)[l]{\strut{}$10^{  1}$}}%
      \put(4269,2561){\makebox(0,0)[l]{\strut{}$10^{  2}$}}%
      \put(4269,3181){\makebox(0,0)[l]{\strut{}$10^{  3}$}}%
      \put(4797,1942){\rotatebox{-270}{\makebox(0,0){\strut{}$\rho_\mathrm{N}$ [$\mathrm{kpc}^{-1}$]}}}%
      \colorrgb{0.89,0.10,0.11}
      \put(2546,1843){\makebox(0,0){\strut{}\small{Earth}}}%
      \colorrgb{0.00,0.00,0.00}
      \put(3741,2801){\makebox(0,0)[r]{\strut{}\small{US\,708}}}%
    }%
    \gplgaddtomacro\gplbacktext{%
      \csname LTb\endcsname
      \put(5298,2933){\makebox(0,0){\strut{}(D)}}%
    }%
    \gplgaddtomacro\gplfronttext{%
      \csname LTb\endcsname
      \put(5311,1942){\rotatebox{-270}{\makebox(0,0){\strut{}$z_\mathrm{Gal}$ [kpc]}}}%
      \put(7462,154){\makebox(0,0){\strut{}$R_\mathrm{Gal}$ [kpc]}}%
      \csname LTb\endcsname
      \put(5784,952){\makebox(0,0)[r]{\strut{}$-6$}}%
      \put(5784,1282){\makebox(0,0)[r]{\strut{}$-4$}}%
      \put(5784,1612){\makebox(0,0)[r]{\strut{}$-2$}}%
      \put(5784,1943){\makebox(0,0)[r]{\strut{}$0$}}%
      \put(5784,2273){\makebox(0,0)[r]{\strut{}$2$}}%
      \put(5784,2603){\makebox(0,0)[r]{\strut{}$4$}}%
      \put(5784,2933){\makebox(0,0)[r]{\strut{}$6$}}%
      \put(5916,484){\makebox(0,0){\strut{}$0$}}%
      \put(6328,484){\makebox(0,0){\strut{}$2$}}%
      \put(6741,484){\makebox(0,0){\strut{}$4$}}%
      \put(7153,484){\makebox(0,0){\strut{}$6$}}%
      \put(7565,484){\makebox(0,0){\strut{}$8$}}%
      \put(7977,484){\makebox(0,0){\strut{}$10$}}%
      \put(8390,484){\makebox(0,0){\strut{}$12$}}%
      \put(8802,484){\makebox(0,0){\strut{}$14$}}%
      \put(9371,703){\makebox(0,0)[l]{\strut{}$10^{ -1}$}}%
      \put(9371,1323){\makebox(0,0)[l]{\strut{}$10^{  0}$}}%
      \put(9371,1942){\makebox(0,0)[l]{\strut{}$10^{  1}$}}%
      \put(9371,2561){\makebox(0,0)[l]{\strut{}$10^{  2}$}}%
      \put(9371,3181){\makebox(0,0)[l]{\strut{}$10^{  3}$}}%
      \put(9899,1942){\rotatebox{-270}{\makebox(0,0){\strut{}$\rho_\mathrm{N}$ [$\mathrm{kpc}^{-1}$]}}}%
      \colorrgb{0.89,0.10,0.11}
      \put(7648,1843){\makebox(0,0){\strut{}\small{Earth}}}%
      \colorrgb{0.00,0.00,0.00}
      \put(8843,2801){\makebox(0,0)[r]{\strut{}\small{US\,708}}}%
    }%
    \gplbacktext
    \put(0,0){\includegraphics[width={510.20bp},height={340.10bp}]{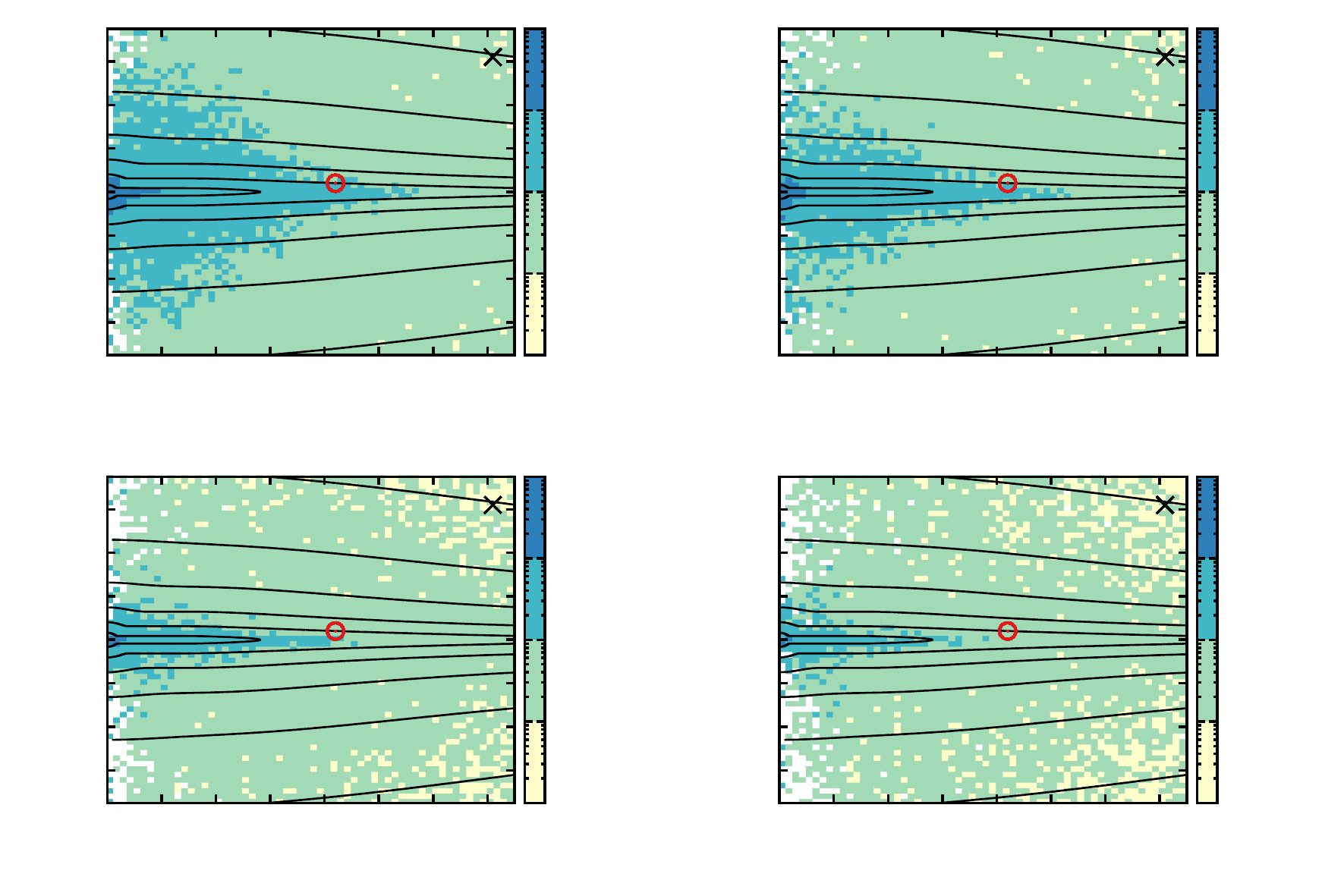}}%
    \gplfronttext
  \end{picture}%
\endgroup

%% file: relative_n_dens-fig2.tex
\begingroup
  \makeatletter
  \providecommand\color[2][]{%
    \GenericError{(gnuplot) \space\space\space\@spaces}{%
      Package color not loaded in conjunction with
      terminal option `colourtext'%
    }{See the gnuplot documentation for explanation.%
    }{Either use 'blacktext' in gnuplot or load the package
      color.sty in LaTeX.}%
    \renewcommand\color[2][]{}%
  }%
  \providecommand\includegraphics[2][]{%
    \GenericError{(gnuplot) \space\space\space\@spaces}{%
      Package graphicx or graphics not loaded%
    }{See the gnuplot documentation for explanation.%
    }{The gnuplot epslatex terminal needs graphicx.sty or graphics.sty.}%
    \renewcommand\includegraphics[2][]{}%
  }%
  \providecommand\rotatebox[2]{#2}%
  \@ifundefined{ifGPcolor}{%
    \newif\ifGPcolor
    \GPcolortrue
  }{}%
  \@ifundefined{ifGPblacktext}{%
    \newif\ifGPblacktext
    \GPblacktextfalse
  }{}%
  \let\gplgaddtomacro\g@addto@macro
  \gdef\gplbacktext{}%
  \gdef\gplfronttext{}%
  \makeatother
  \ifGPblacktext
    \def\colorrgb#1{}%
    \def\colorgray#1{}%
  \else
    \ifGPcolor
      \def\colorrgb#1{\color[rgb]{#1}}%
      \def\colorgray#1{\color[gray]{#1}}%
      \expandafter\def\csname LTw\endcsname{\color{white}}%
      \expandafter\def\csname LTb\endcsname{\color{black}}%
      \expandafter\def\csname LTa\endcsname{\color{black}}%
      \expandafter\def\csname LT0\endcsname{\color[rgb]{1,0,0}}%
      \expandafter\def\csname LT1\endcsname{\color[rgb]{0,1,0}}%
      \expandafter\def\csname LT2\endcsname{\color[rgb]{0,0,1}}%
      \expandafter\def\csname LT3\endcsname{\color[rgb]{1,0,1}}%
      \expandafter\def\csname LT4\endcsname{\color[rgb]{0,1,1}}%
      \expandafter\def\csname LT5\endcsname{\color[rgb]{1,1,0}}%
      \expandafter\def\csname LT6\endcsname{\color[rgb]{0,0,0}}%
      \expandafter\def\csname LT7\endcsname{\color[rgb]{1,0.3,0}}%
      \expandafter\def\csname LT8\endcsname{\color[rgb]{0.5,0.5,0.5}}%
    \else
      \def\colorrgb#1{\color{black}}%
      \def\colorgray#1{\color[gray]{#1}}%
      \expandafter\def\csname LTw\endcsname{\color{white}}%
      \expandafter\def\csname LTb\endcsname{\color{black}}%
      \expandafter\def\csname LTa\endcsname{\color{black}}%
      \expandafter\def\csname LT0\endcsname{\color{black}}%
      \expandafter\def\csname LT1\endcsname{\color{black}}%
      \expandafter\def\csname LT2\endcsname{\color{black}}%
      \expandafter\def\csname LT3\endcsname{\color{black}}%
      \expandafter\def\csname LT4\endcsname{\color{black}}%
      \expandafter\def\csname LT5\endcsname{\color{black}}%
      \expandafter\def\csname LT6\endcsname{\color{black}}%
      \expandafter\def\csname LT7\endcsname{\color{black}}%
      \expandafter\def\csname LT8\endcsname{\color{black}}%
    \fi
  \fi
    \setlength{\unitlength}{0.0500bp}%
    \ifx\gptboxheight\undefined%
      \newlength{\gptboxheight}%
      \newlength{\gptboxwidth}%
      \newsavebox{\gptboxtext}%
    \fi%
    \setlength{\fboxrule}{0.5pt}%
    \setlength{\fboxsep}{1pt}%
    \definecolor{tbcol}{rgb}{1,1,1}%
\begin{picture}(4988.00,4534.00)%
    \gplgaddtomacro\gplbacktext{%
    }%
    \gplgaddtomacro\gplfronttext{%
      \csname LTb\endcsname
      \put(209,2926){\rotatebox{-270}{\makebox(0,0){\strut{}$\rho_\mathrm{N}$ [\kkpc]}}}%
      \put(4481,2926){\rotatebox{-270}{\makebox(0,0){\strut{}relative number density}}}%
      \put(2196,990){\makebox(0,0){\strut{}$M_\mathrm{WD,f}$ [$\Msun$]}}%
      \csname LTb\endcsname
      \put(3023,613){\makebox(0,0)[r]{\strut{}$\rho_\mathrm{N}(\mathrm{US708})/\rho_\mathrm{N}(\mathrm{Earth})$}}%
      \csname LTb\endcsname
      \put(3023,393){\makebox(0,0)[r]{\strut{}$\rho_\mathrm{N}(\mathrm{US708})$}}%
      \csname LTb\endcsname
      \put(3023,173){\makebox(0,0)[r]{\strut{}$\rho_\mathrm{N}(\mathrm{Earth})$}}%
      \csname LTb\endcsname
      \put(682,1540){\makebox(0,0)[r]{\strut{}$0$}}%
      \put(682,1936){\makebox(0,0)[r]{\strut{}$2$}}%
      \put(682,2332){\makebox(0,0)[r]{\strut{}$4$}}%
      \put(682,2728){\makebox(0,0)[r]{\strut{}$6$}}%
      \put(682,3125){\makebox(0,0)[r]{\strut{}$8$}}%
      \put(682,3521){\makebox(0,0)[r]{\strut{}$10$}}%
      \put(682,3917){\makebox(0,0)[r]{\strut{}$12$}}%
      \put(682,4313){\makebox(0,0)[r]{\strut{}$14$}}%
      \put(1044,1320){\makebox(0,0){\strut{}$1$}}%
      \put(1505,1320){\makebox(0,0){\strut{}$1.1$}}%
      \put(1966,1320){\makebox(0,0){\strut{}$1.2$}}%
      \put(2427,1320){\makebox(0,0){\strut{}$1.3$}}%
      \put(2888,1320){\makebox(0,0){\strut{}$1.4$}}%
      \put(3349,1320){\makebox(0,0){\strut{}$1.5$}}%
      \put(3711,1540){\makebox(0,0)[l]{\strut{}$0$}}%
      \put(3711,2002){\makebox(0,0)[l]{\strut{}$0.05$}}%
      \put(3711,2464){\makebox(0,0)[l]{\strut{}$0.1$}}%
      \put(3711,2927){\makebox(0,0)[l]{\strut{}$0.15$}}%
      \put(3711,3389){\makebox(0,0)[l]{\strut{}$0.2$}}%
      \put(3711,3851){\makebox(0,0)[l]{\strut{}$0.25$}}%
      \put(3711,4313){\makebox(0,0)[l]{\strut{}$0.3$}}%
    }%
    \gplbacktext
    \put(0,0){\includegraphics[width={249.40bp},height={226.70bp}]{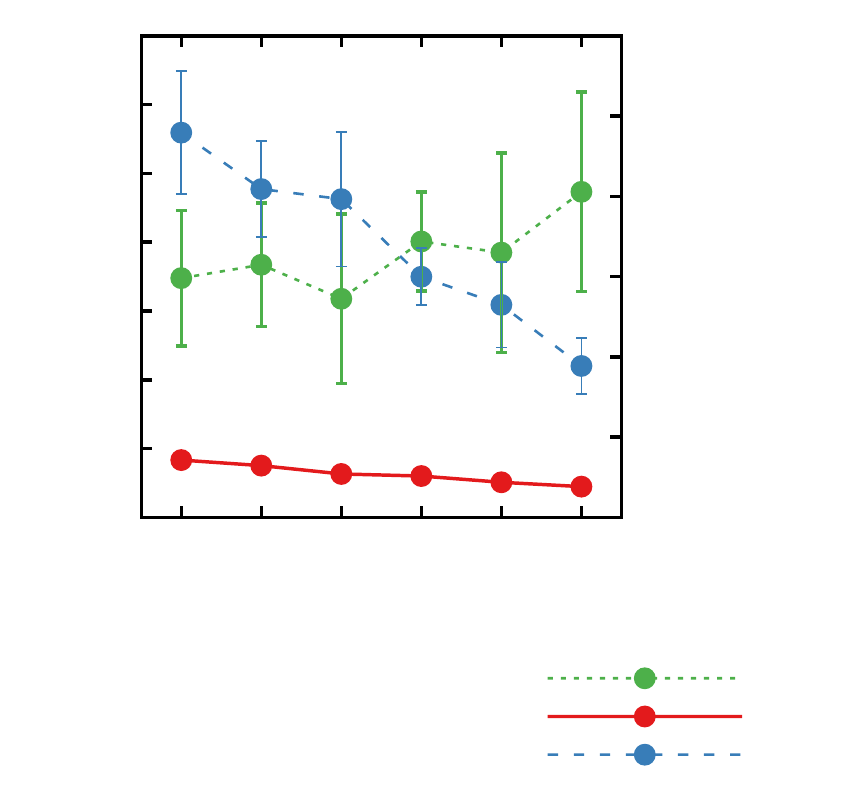}}%
    \gplfronttext
  \end{picture}%
\endgroup

%% file: bound-unbound-cumulative-all-fig2.tex
\begingroup
  \makeatletter
  \providecommand\color[2][]{%
    \GenericError{(gnuplot) \space\space\space\@spaces}{%
      Package color not loaded in conjunction with
      terminal option `colourtext'%
    }{See the gnuplot documentation for explanation.%
    }{Either use 'blacktext' in gnuplot or load the package
      color.sty in LaTeX.}%
    \renewcommand\color[2][]{}%
  }%
  \providecommand\includegraphics[2][]{%
    \GenericError{(gnuplot) \space\space\space\@spaces}{%
      Package graphicx or graphics not loaded%
    }{See the gnuplot documentation for explanation.%
    }{The gnuplot epslatex terminal needs graphicx.sty or graphics.sty.}%
    \renewcommand\includegraphics[2][]{}%
  }%
  \providecommand\rotatebox[2]{#2}%
  \@ifundefined{ifGPcolor}{%
    \newif\ifGPcolor
    \GPcolortrue
  }{}%
  \@ifundefined{ifGPblacktext}{%
    \newif\ifGPblacktext
    \GPblacktextfalse
  }{}%
  \let\gplgaddtomacro\g@addto@macro
  \gdef\gplbacktext{}%
  \gdef\gplfronttext{}%
  \makeatother
  \ifGPblacktext
    \def\colorrgb#1{}%
    \def\colorgray#1{}%
  \else
    \ifGPcolor
      \def\colorrgb#1{\color[rgb]{#1}}%
      \def\colorgray#1{\color[gray]{#1}}%
      \expandafter\def\csname LTw\endcsname{\color{white}}%
      \expandafter\def\csname LTb\endcsname{\color{black}}%
      \expandafter\def\csname LTa\endcsname{\color{black}}%
      \expandafter\def\csname LT0\endcsname{\color[rgb]{1,0,0}}%
      \expandafter\def\csname LT1\endcsname{\color[rgb]{0,1,0}}%
      \expandafter\def\csname LT2\endcsname{\color[rgb]{0,0,1}}%
      \expandafter\def\csname LT3\endcsname{\color[rgb]{1,0,1}}%
      \expandafter\def\csname LT4\endcsname{\color[rgb]{0,1,1}}%
      \expandafter\def\csname LT5\endcsname{\color[rgb]{1,1,0}}%
      \expandafter\def\csname LT6\endcsname{\color[rgb]{0,0,0}}%
      \expandafter\def\csname LT7\endcsname{\color[rgb]{1,0.3,0}}%
      \expandafter\def\csname LT8\endcsname{\color[rgb]{0.5,0.5,0.5}}%
    \else
      \def\colorrgb#1{\color{black}}%
      \def\colorgray#1{\color[gray]{#1}}%
      \expandafter\def\csname LTw\endcsname{\color{white}}%
      \expandafter\def\csname LTb\endcsname{\color{black}}%
      \expandafter\def\csname LTa\endcsname{\color{black}}%
      \expandafter\def\csname LT0\endcsname{\color{black}}%
      \expandafter\def\csname LT1\endcsname{\color{black}}%
      \expandafter\def\csname LT2\endcsname{\color{black}}%
      \expandafter\def\csname LT3\endcsname{\color{black}}%
      \expandafter\def\csname LT4\endcsname{\color{black}}%
      \expandafter\def\csname LT5\endcsname{\color{black}}%
      \expandafter\def\csname LT6\endcsname{\color{black}}%
      \expandafter\def\csname LT7\endcsname{\color{black}}%
      \expandafter\def\csname LT8\endcsname{\color{black}}%
    \fi
  \fi
    \setlength{\unitlength}{0.0500bp}%
    \ifx\gptboxheight\undefined%
      \newlength{\gptboxheight}%
      \newlength{\gptboxwidth}%
      \newsavebox{\gptboxtext}%
    \fi%
    \setlength{\fboxrule}{0.5pt}%
    \setlength{\fboxsep}{1pt}%
    \definecolor{tbcol}{rgb}{1,1,1}%
\begin{picture}(10204.00,10204.00)%
    \gplgaddtomacro\gplbacktext{%
      \csname LTb\endcsname
      \put(1355,9735){\makebox(0,0)[l]{\strut{}(A)}}%
    }%
    \gplgaddtomacro\gplfronttext{%
      \csname LTb\endcsname
      \put(209,8744){\rotatebox{-270}{\makebox(0,0){\strut{}cumulative normalized population}}}%
      \put(2561,6956){\makebox(0,0){\strut{}$D$ [kpc]}}%
      \csname LTb\endcsname
      \put(418,7506){\makebox(0,0)[r]{\strut{}}}%
      \put(418,7956){\makebox(0,0)[r]{\strut{}}}%
      \put(418,8407){\makebox(0,0)[r]{\strut{}}}%
      \put(418,8857){\makebox(0,0)[r]{\strut{}}}%
      \put(418,9307){\makebox(0,0)[r]{\strut{}}}%
      \put(418,9758){\makebox(0,0)[r]{\strut{}}}%
      \put(550,7286){\makebox(0,0){\strut{}$0$}}%
      \put(1355,7286){\makebox(0,0){\strut{}$5$}}%
      \put(2159,7286){\makebox(0,0){\strut{}$10$}}%
      \put(2964,7286){\makebox(0,0){\strut{}$15$}}%
      \put(3768,7286){\makebox(0,0){\strut{}$20$}}%
      \put(4573,7286){\makebox(0,0){\strut{}$25$}}%
      \put(4705,7506){\makebox(0,0)[l]{\strut{}$0.0$}}%
      \put(4705,7956){\makebox(0,0)[l]{\strut{}$0.2$}}%
      \put(4705,8407){\makebox(0,0)[l]{\strut{}$0.4$}}%
      \put(4705,8857){\makebox(0,0)[l]{\strut{}$0.6$}}%
      \put(4705,9307){\makebox(0,0)[l]{\strut{}$0.8$}}%
      \put(4705,9758){\makebox(0,0)[l]{\strut{}$1.0$}}%
      \put(2159,9364){\makebox(0,0)[l]{\strut{}$\MWDf=1.0\Msun$}}%
      \colorrgb{0.30,0.69,0.29}
      \put(2079,8992){\rotatebox{90}{\makebox(0,0)[l]{\strut{}US708}}}%
      \colorrgb{0.00,0.00,0.00}
      \put(2159,9735){\makebox(0,0)[l]{\strut{}$N(D=25\,\mathrm{kpc})= 81760$}}%
    }%
    \gplgaddtomacro\gplbacktext{%
      \csname LTb\endcsname
      \put(6457,9735){\makebox(0,0)[l]{\strut{}(B)}}%
    }%
    \gplgaddtomacro\gplfronttext{%
      \csname LTb\endcsname
      \put(5311,8744){\rotatebox{-270}{\makebox(0,0){\strut{}cumulative normalized population}}}%
      \put(7663,6956){\makebox(0,0){\strut{}$D$ [kpc]}}%
      \csname LTb\endcsname
      \put(5520,7506){\makebox(0,0)[r]{\strut{}}}%
      \put(5520,7956){\makebox(0,0)[r]{\strut{}}}%
      \put(5520,8407){\makebox(0,0)[r]{\strut{}}}%
      \put(5520,8857){\makebox(0,0)[r]{\strut{}}}%
      \put(5520,9307){\makebox(0,0)[r]{\strut{}}}%
      \put(5520,9758){\makebox(0,0)[r]{\strut{}}}%
      \put(5652,7286){\makebox(0,0){\strut{}$0$}}%
      \put(6457,7286){\makebox(0,0){\strut{}$5$}}%
      \put(7261,7286){\makebox(0,0){\strut{}$10$}}%
      \put(8066,7286){\makebox(0,0){\strut{}$15$}}%
      \put(8870,7286){\makebox(0,0){\strut{}$20$}}%
      \put(9675,7286){\makebox(0,0){\strut{}$25$}}%
      \put(9807,7506){\makebox(0,0)[l]{\strut{}$0.0$}}%
      \put(9807,7956){\makebox(0,0)[l]{\strut{}$0.2$}}%
      \put(9807,8407){\makebox(0,0)[l]{\strut{}$0.4$}}%
      \put(9807,8857){\makebox(0,0)[l]{\strut{}$0.6$}}%
      \put(9807,9307){\makebox(0,0)[l]{\strut{}$0.8$}}%
      \put(9807,9758){\makebox(0,0)[l]{\strut{}$1.0$}}%
      \put(7261,9364){\makebox(0,0)[l]{\strut{}$\MWDf=1.2\Msun$}}%
      \colorrgb{0.30,0.69,0.29}
      \put(7181,8992){\rotatebox{90}{\makebox(0,0)[l]{\strut{}US708}}}%
      \colorrgb{0.00,0.00,0.00}
      \put(7261,9735){\makebox(0,0)[l]{\strut{}$N(D=25\,\mathrm{kpc})= 68294$}}%
    }%
    \gplgaddtomacro\gplbacktext{%
      \csname LTb\endcsname
      \put(1355,6334){\makebox(0,0)[l]{\strut{}(C)}}%
    }%
    \gplgaddtomacro\gplfronttext{%
      \csname LTb\endcsname
      \put(209,5343){\rotatebox{-270}{\makebox(0,0){\strut{}cumulative normalized population}}}%
      \put(2561,3555){\makebox(0,0){\strut{}$D$ [kpc]}}%
      \csname LTb\endcsname
      \put(418,4105){\makebox(0,0)[r]{\strut{}}}%
      \put(418,4555){\makebox(0,0)[r]{\strut{}}}%
      \put(418,5006){\makebox(0,0)[r]{\strut{}}}%
      \put(418,5456){\makebox(0,0)[r]{\strut{}}}%
      \put(418,5906){\makebox(0,0)[r]{\strut{}}}%
      \put(418,6357){\makebox(0,0)[r]{\strut{}}}%
      \put(550,3885){\makebox(0,0){\strut{}$0$}}%
      \put(1355,3885){\makebox(0,0){\strut{}$5$}}%
      \put(2159,3885){\makebox(0,0){\strut{}$10$}}%
      \put(2964,3885){\makebox(0,0){\strut{}$15$}}%
      \put(3768,3885){\makebox(0,0){\strut{}$20$}}%
      \put(4573,3885){\makebox(0,0){\strut{}$25$}}%
      \put(4705,4105){\makebox(0,0)[l]{\strut{}$0.0$}}%
      \put(4705,4555){\makebox(0,0)[l]{\strut{}$0.2$}}%
      \put(4705,5006){\makebox(0,0)[l]{\strut{}$0.4$}}%
      \put(4705,5456){\makebox(0,0)[l]{\strut{}$0.6$}}%
      \put(4705,5906){\makebox(0,0)[l]{\strut{}$0.8$}}%
      \put(4705,6357){\makebox(0,0)[l]{\strut{}$1.0$}}%
      \put(2159,5963){\makebox(0,0)[l]{\strut{}$\MWDf=1.4\Msun$}}%
      \colorrgb{0.30,0.69,0.29}
      \put(2079,5591){\rotatebox{90}{\makebox(0,0)[l]{\strut{}US708}}}%
      \colorrgb{0.00,0.00,0.00}
      \put(2159,6334){\makebox(0,0)[l]{\strut{}$N(D=25\,\mathrm{kpc})= 49468$}}%
    }%
    \gplgaddtomacro\gplbacktext{%
      \csname LTb\endcsname
      \put(6457,6334){\makebox(0,0)[l]{\strut{}(D)}}%
    }%
    \gplgaddtomacro\gplfronttext{%
      \csname LTb\endcsname
      \put(5311,5343){\rotatebox{-270}{\makebox(0,0){\strut{}cumulative normalized population}}}%
      \put(7663,3555){\makebox(0,0){\strut{}$D$ [kpc]}}%
      \csname LTb\endcsname
      \put(5520,4105){\makebox(0,0)[r]{\strut{}}}%
      \put(5520,4555){\makebox(0,0)[r]{\strut{}}}%
      \put(5520,5006){\makebox(0,0)[r]{\strut{}}}%
      \put(5520,5456){\makebox(0,0)[r]{\strut{}}}%
      \put(5520,5906){\makebox(0,0)[r]{\strut{}}}%
      \put(5520,6357){\makebox(0,0)[r]{\strut{}}}%
      \put(5652,3885){\makebox(0,0){\strut{}$0$}}%
      \put(6457,3885){\makebox(0,0){\strut{}$5$}}%
      \put(7261,3885){\makebox(0,0){\strut{}$10$}}%
      \put(8066,3885){\makebox(0,0){\strut{}$15$}}%
      \put(8870,3885){\makebox(0,0){\strut{}$20$}}%
      \put(9675,3885){\makebox(0,0){\strut{}$25$}}%
      \put(9807,4105){\makebox(0,0)[l]{\strut{}$0.0$}}%
      \put(9807,4555){\makebox(0,0)[l]{\strut{}$0.2$}}%
      \put(9807,5006){\makebox(0,0)[l]{\strut{}$0.4$}}%
      \put(9807,5456){\makebox(0,0)[l]{\strut{}$0.6$}}%
      \put(9807,5906){\makebox(0,0)[l]{\strut{}$0.8$}}%
      \put(9807,6357){\makebox(0,0)[l]{\strut{}$1.0$}}%
      \put(7261,5963){\makebox(0,0)[l]{\strut{}$\MWDf=1.5\Msun$}}%
      \colorrgb{0.30,0.69,0.29}
      \put(7181,5591){\rotatebox{90}{\makebox(0,0)[l]{\strut{}US708}}}%
      \colorrgb{0.00,0.00,0.00}
      \put(7261,6334){\makebox(0,0)[l]{\strut{}$N(D=25\,\mathrm{kpc})= 42862$}}%
    }%
    \gplgaddtomacro\gplbacktext{%
    }%
    \gplgaddtomacro\gplfronttext{%
      \csname LTb\endcsname
      \put(858,3008){\makebox(0,0)[r]{\strut{}UI}}%
      \csname LTb\endcsname
      \put(2109,3008){\makebox(0,0)[r]{\strut{}BI}}%
      \csname LTb\endcsname
      \put(3360,3008){\makebox(0,0)[r]{\strut{}BII}}%
    }%
    \gplgaddtomacro\gplbacktext{%
    }%
    \gplgaddtomacro\gplfronttext{%
    }%
    \gplbacktext
    \put(0,0){\includegraphics[width={510.20bp},height={510.20bp}]{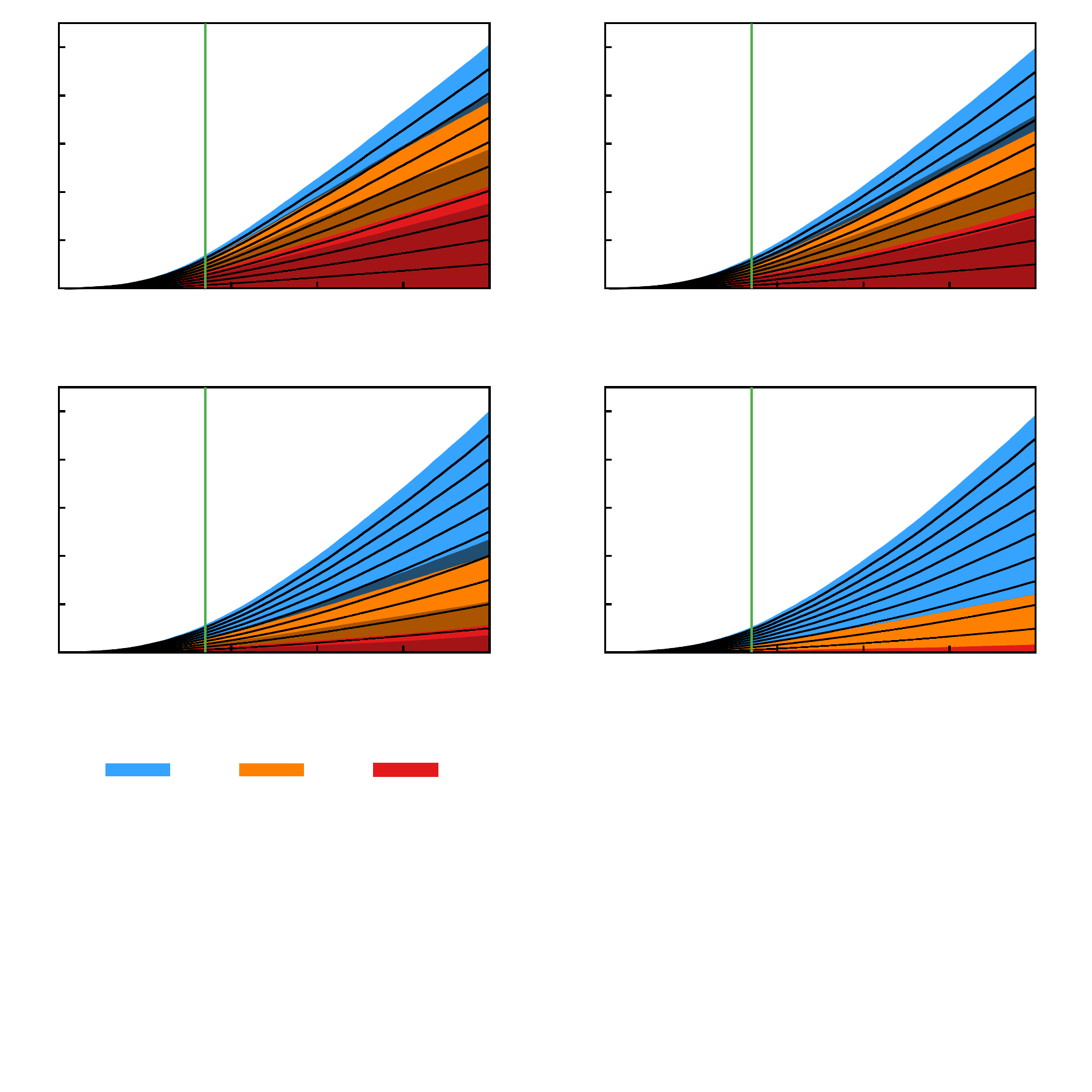}}%
    \gplfronttext
  \end{picture}%
\endgroup

%% file: ratios-fig2.tex
\begingroup
  \makeatletter
  \providecommand\color[2][]{%
    \GenericError{(gnuplot) \space\space\space\@spaces}{%
      Package color not loaded in conjunction with
      terminal option `colourtext'%
    }{See the gnuplot documentation for explanation.%
    }{Either use 'blacktext' in gnuplot or load the package
      color.sty in LaTeX.}%
    \renewcommand\color[2][]{}%
  }%
  \providecommand\includegraphics[2][]{%
    \GenericError{(gnuplot) \space\space\space\@spaces}{%
      Package graphicx or graphics not loaded%
    }{See the gnuplot documentation for explanation.%
    }{The gnuplot epslatex terminal needs graphicx.sty or graphics.sty.}%
    \renewcommand\includegraphics[2][]{}%
  }%
  \providecommand\rotatebox[2]{#2}%
  \@ifundefined{ifGPcolor}{%
    \newif\ifGPcolor
    \GPcolortrue
  }{}%
  \@ifundefined{ifGPblacktext}{%
    \newif\ifGPblacktext
    \GPblacktextfalse
  }{}%
  \let\gplgaddtomacro\g@addto@macro
  \gdef\gplbacktext{}%
  \gdef\gplfronttext{}%
  \makeatother
  \ifGPblacktext
    \def\colorrgb#1{}%
    \def\colorgray#1{}%
  \else
    \ifGPcolor
      \def\colorrgb#1{\color[rgb]{#1}}%
      \def\colorgray#1{\color[gray]{#1}}%
      \expandafter\def\csname LTw\endcsname{\color{white}}%
      \expandafter\def\csname LTb\endcsname{\color{black}}%
      \expandafter\def\csname LTa\endcsname{\color{black}}%
      \expandafter\def\csname LT0\endcsname{\color[rgb]{1,0,0}}%
      \expandafter\def\csname LT1\endcsname{\color[rgb]{0,1,0}}%
      \expandafter\def\csname LT2\endcsname{\color[rgb]{0,0,1}}%
      \expandafter\def\csname LT3\endcsname{\color[rgb]{1,0,1}}%
      \expandafter\def\csname LT4\endcsname{\color[rgb]{0,1,1}}%
      \expandafter\def\csname LT5\endcsname{\color[rgb]{1,1,0}}%
      \expandafter\def\csname LT6\endcsname{\color[rgb]{0,0,0}}%
      \expandafter\def\csname LT7\endcsname{\color[rgb]{1,0.3,0}}%
      \expandafter\def\csname LT8\endcsname{\color[rgb]{0.5,0.5,0.5}}%
    \else
      \def\colorrgb#1{\color{black}}%
      \def\colorgray#1{\color[gray]{#1}}%
      \expandafter\def\csname LTw\endcsname{\color{white}}%
      \expandafter\def\csname LTb\endcsname{\color{black}}%
      \expandafter\def\csname LTa\endcsname{\color{black}}%
      \expandafter\def\csname LT0\endcsname{\color{black}}%
      \expandafter\def\csname LT1\endcsname{\color{black}}%
      \expandafter\def\csname LT2\endcsname{\color{black}}%
      \expandafter\def\csname LT3\endcsname{\color{black}}%
      \expandafter\def\csname LT4\endcsname{\color{black}}%
      \expandafter\def\csname LT5\endcsname{\color{black}}%
      \expandafter\def\csname LT6\endcsname{\color{black}}%
      \expandafter\def\csname LT7\endcsname{\color{black}}%
      \expandafter\def\csname LT8\endcsname{\color{black}}%
    \fi
  \fi
    \setlength{\unitlength}{0.0500bp}%
    \ifx\gptboxheight\undefined%
      \newlength{\gptboxheight}%
      \newlength{\gptboxwidth}%
      \newsavebox{\gptboxtext}%
    \fi%
    \setlength{\fboxrule}{0.5pt}%
    \setlength{\fboxsep}{1pt}%
    \definecolor{tbcol}{rgb}{1,1,1}%
\begin{picture}(4988.00,4534.00)%
    \gplgaddtomacro\gplbacktext{%
    }%
    \gplgaddtomacro\gplfronttext{%
      \csname LTb\endcsname
      \put(209,2926){\rotatebox{-270}{\makebox(0,0){\strut{}dimensionless ratio}}}%
      \put(2636,990){\makebox(0,0){\strut{}$M_\mathrm{WD,f}$ [$\Msun$]}}%
      \csname LTb\endcsname
      \put(2944,4140){\makebox(0,0)[r]{\strut{}B/U}}%
      \csname LTb\endcsname
      \put(2944,3920){\makebox(0,0)[r]{\strut{}BII/BI}}%
      \csname LTb\endcsname
      \put(2944,3700){\makebox(0,0)[r]{\strut{}BII/BI($\tau/2$)}}%
      \csname LTb\endcsname
      \put(814,1540){\makebox(0,0)[r]{\strut{}$0$}}%
      \put(814,1936){\makebox(0,0)[r]{\strut{}$0.2$}}%
      \put(814,2332){\makebox(0,0)[r]{\strut{}$0.4$}}%
      \put(814,2728){\makebox(0,0)[r]{\strut{}$0.6$}}%
      \put(814,3125){\makebox(0,0)[r]{\strut{}$0.8$}}%
      \put(814,3521){\makebox(0,0)[r]{\strut{}$1$}}%
      \put(814,3917){\makebox(0,0)[r]{\strut{}$1.2$}}%
      \put(814,4313){\makebox(0,0)[r]{\strut{}$1.4$}}%
      \put(1228,1320){\makebox(0,0){\strut{}$1$}}%
      \put(1791,1320){\makebox(0,0){\strut{}$1.1$}}%
      \put(2355,1320){\makebox(0,0){\strut{}$1.2$}}%
      \put(2918,1320){\makebox(0,0){\strut{}$1.3$}}%
      \put(3482,1320){\makebox(0,0){\strut{}$1.4$}}%
      \put(4045,1320){\makebox(0,0){\strut{}$1.5$}}%
    }%
    \gplbacktext
    \put(0,0){\includegraphics[width={249.40bp},height={226.70bp}]{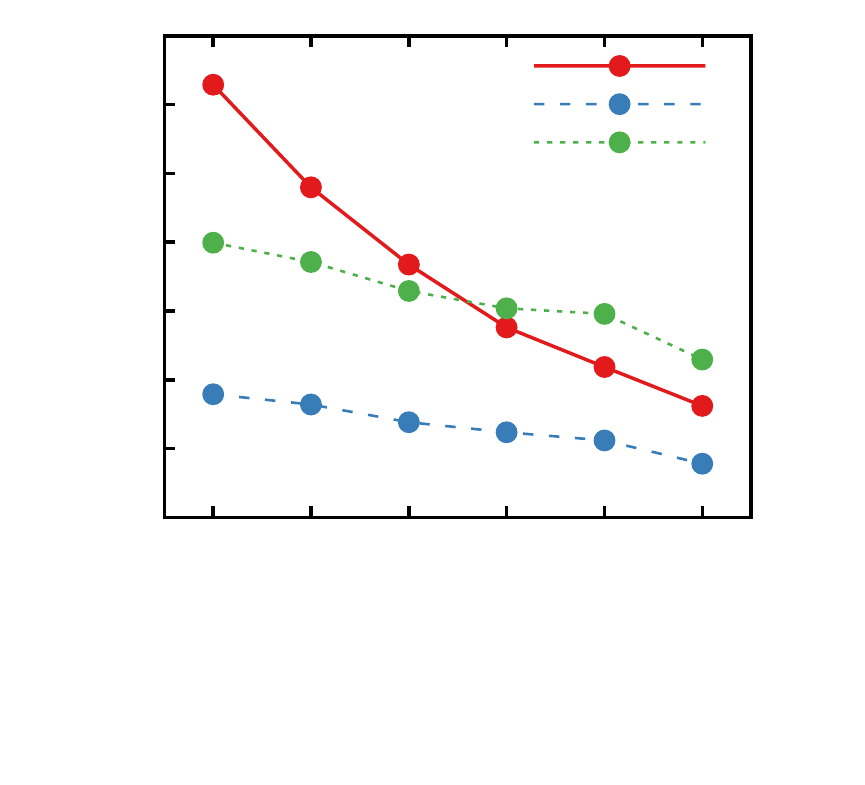}}%
    \gplfronttext
  \end{picture}%
\endgroup

%% file: bound-unbound-cumulative-001-fig2.tex
\begingroup
  \makeatletter
  \providecommand\color[2][]{%
    \GenericError{(gnuplot) \space\space\space\@spaces}{%
      Package color not loaded in conjunction with
      terminal option `colourtext'%
    }{See the gnuplot documentation for explanation.%
    }{Either use 'blacktext' in gnuplot or load the package
      color.sty in LaTeX.}%
    \renewcommand\color[2][]{}%
  }%
  \providecommand\includegraphics[2][]{%
    \GenericError{(gnuplot) \space\space\space\@spaces}{%
      Package graphicx or graphics not loaded%
    }{See the gnuplot documentation for explanation.%
    }{The gnuplot epslatex terminal needs graphicx.sty or graphics.sty.}%
    \renewcommand\includegraphics[2][]{}%
  }%
  \providecommand\rotatebox[2]{#2}%
  \@ifundefined{ifGPcolor}{%
    \newif\ifGPcolor
    \GPcolortrue
  }{}%
  \@ifundefined{ifGPblacktext}{%
    \newif\ifGPblacktext
    \GPblacktextfalse
  }{}%
  \let\gplgaddtomacro\g@addto@macro
  \gdef\gplbacktext{}%
  \gdef\gplfronttext{}%
  \makeatother
  \ifGPblacktext
    \def\colorrgb#1{}%
    \def\colorgray#1{}%
  \else
    \ifGPcolor
      \def\colorrgb#1{\color[rgb]{#1}}%
      \def\colorgray#1{\color[gray]{#1}}%
      \expandafter\def\csname LTw\endcsname{\color{white}}%
      \expandafter\def\csname LTb\endcsname{\color{black}}%
      \expandafter\def\csname LTa\endcsname{\color{black}}%
      \expandafter\def\csname LT0\endcsname{\color[rgb]{1,0,0}}%
      \expandafter\def\csname LT1\endcsname{\color[rgb]{0,1,0}}%
      \expandafter\def\csname LT2\endcsname{\color[rgb]{0,0,1}}%
      \expandafter\def\csname LT3\endcsname{\color[rgb]{1,0,1}}%
      \expandafter\def\csname LT4\endcsname{\color[rgb]{0,1,1}}%
      \expandafter\def\csname LT5\endcsname{\color[rgb]{1,1,0}}%
      \expandafter\def\csname LT6\endcsname{\color[rgb]{0,0,0}}%
      \expandafter\def\csname LT7\endcsname{\color[rgb]{1,0.3,0}}%
      \expandafter\def\csname LT8\endcsname{\color[rgb]{0.5,0.5,0.5}}%
    \else
      \def\colorrgb#1{\color{black}}%
      \def\colorgray#1{\color[gray]{#1}}%
      \expandafter\def\csname LTw\endcsname{\color{white}}%
      \expandafter\def\csname LTb\endcsname{\color{black}}%
      \expandafter\def\csname LTa\endcsname{\color{black}}%
      \expandafter\def\csname LT0\endcsname{\color{black}}%
      \expandafter\def\csname LT1\endcsname{\color{black}}%
      \expandafter\def\csname LT2\endcsname{\color{black}}%
      \expandafter\def\csname LT3\endcsname{\color{black}}%
      \expandafter\def\csname LT4\endcsname{\color{black}}%
      \expandafter\def\csname LT5\endcsname{\color{black}}%
      \expandafter\def\csname LT6\endcsname{\color{black}}%
      \expandafter\def\csname LT7\endcsname{\color{black}}%
      \expandafter\def\csname LT8\endcsname{\color{black}}%
    \fi
  \fi
    \setlength{\unitlength}{0.0500bp}%
    \ifx\gptboxheight\undefined%
      \newlength{\gptboxheight}%
      \newlength{\gptboxwidth}%
      \newsavebox{\gptboxtext}%
    \fi%
    \setlength{\fboxrule}{0.5pt}%
    \setlength{\fboxsep}{1pt}%
    \definecolor{tbcol}{rgb}{1,1,1}%
\begin{picture}(4988.00,4534.00)%
    \gplgaddtomacro\gplbacktext{%
    }%
    \gplgaddtomacro\gplfronttext{%
      \csname LTb\endcsname
      \put(209,2508){\rotatebox{-270}{\makebox(0,0){\strut{}cumulative $N(D)/N(D=25\,\mathrm{kpc})$}}}%
      \put(2504,154){\makebox(0,0){\strut{}$D$ [kpc]}}%
      \csname LTb\endcsname
      \put(418,704){\makebox(0,0)[r]{\strut{}}}%
      \put(418,1360){\makebox(0,0)[r]{\strut{}}}%
      \put(418,2016){\makebox(0,0)[r]{\strut{}}}%
      \put(418,2673){\makebox(0,0)[r]{\strut{}}}%
      \put(418,3329){\makebox(0,0)[r]{\strut{}}}%
      \put(418,3985){\makebox(0,0)[r]{\strut{}}}%
      \put(550,484){\makebox(0,0){\strut{}$0$}}%
      \put(1332,484){\makebox(0,0){\strut{}$5$}}%
      \put(2114,484){\makebox(0,0){\strut{}$10$}}%
      \put(2895,484){\makebox(0,0){\strut{}$15$}}%
      \put(3677,484){\makebox(0,0){\strut{}$20$}}%
      \put(4459,484){\makebox(0,0){\strut{}$25$}}%
      \put(4591,704){\makebox(0,0)[l]{\strut{}$0.0$}}%
      \put(4591,1360){\makebox(0,0)[l]{\strut{}$0.2$}}%
      \put(4591,2016){\makebox(0,0)[l]{\strut{}$0.4$}}%
      \put(4591,2673){\makebox(0,0)[l]{\strut{}$0.6$}}%
      \put(4591,3329){\makebox(0,0)[l]{\strut{}$0.8$}}%
      \put(4591,3985){\makebox(0,0)[l]{\strut{}$1.0$}}%
      \colorrgb{0.30,0.69,0.29}
      \put(2035,2869){\rotatebox{90}{\makebox(0,0)[l]{\strut{}US708}}}%
      \colorrgb{0.00,0.00,0.00}
      \put(2114,3952){\makebox(0,0)[l]{\strut{}$N(D=25\,\mathrm{kpc})= 509$}}%
    }%
    \gplbacktext
    \put(0,0){\includegraphics[width={249.40bp},height={226.70bp}]{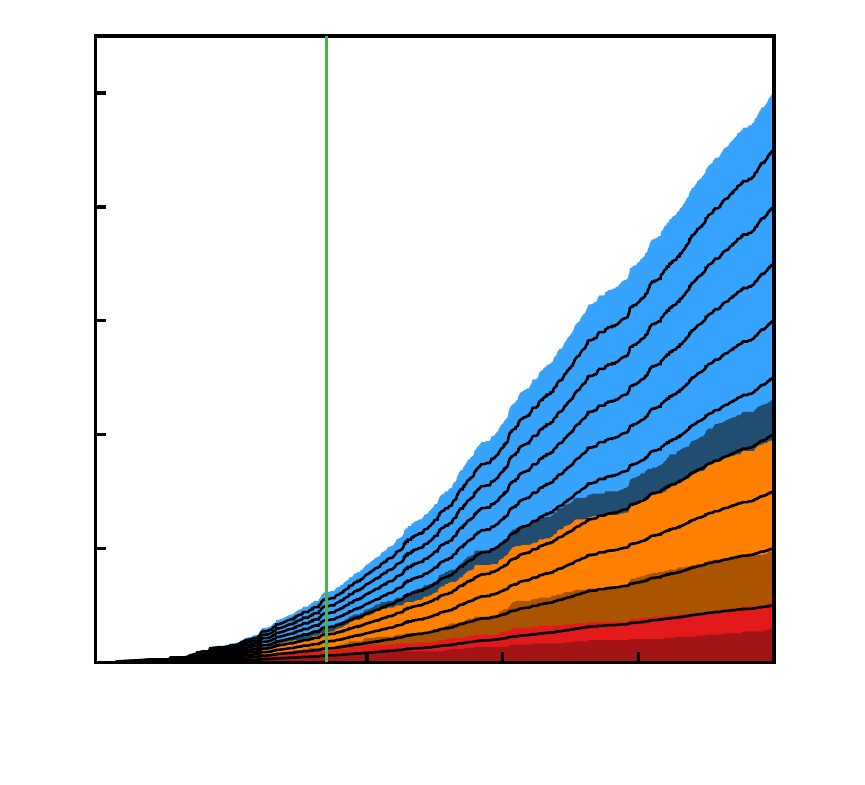}}%
    \gplfronttext
  \end{picture}%
\endgroup

%% file: bound-unbound-mass-fig2.tex
\begingroup
  \makeatletter
  \providecommand\color[2][]{%
    \GenericError{(gnuplot) \space\space\space\@spaces}{%
      Package color not loaded in conjunction with
      terminal option `colourtext'%
    }{See the gnuplot documentation for explanation.%
    }{Either use 'blacktext' in gnuplot or load the package
      color.sty in LaTeX.}%
    \renewcommand\color[2][]{}%
  }%
  \providecommand\includegraphics[2][]{%
    \GenericError{(gnuplot) \space\space\space\@spaces}{%
      Package graphicx or graphics not loaded%
    }{See the gnuplot documentation for explanation.%
    }{The gnuplot epslatex terminal needs graphicx.sty or graphics.sty.}%
    \renewcommand\includegraphics[2][]{}%
  }%
  \providecommand\rotatebox[2]{#2}%
  \@ifundefined{ifGPcolor}{%
    \newif\ifGPcolor
    \GPcolortrue
  }{}%
  \@ifundefined{ifGPblacktext}{%
    \newif\ifGPblacktext
    \GPblacktextfalse
  }{}%
  \let\gplgaddtomacro\g@addto@macro
  \gdef\gplbacktext{}%
  \gdef\gplfronttext{}%
  \makeatother
  \ifGPblacktext
    \def\colorrgb#1{}%
    \def\colorgray#1{}%
  \else
    \ifGPcolor
      \def\colorrgb#1{\color[rgb]{#1}}%
      \def\colorgray#1{\color[gray]{#1}}%
      \expandafter\def\csname LTw\endcsname{\color{white}}%
      \expandafter\def\csname LTb\endcsname{\color{black}}%
      \expandafter\def\csname LTa\endcsname{\color{black}}%
      \expandafter\def\csname LT0\endcsname{\color[rgb]{1,0,0}}%
      \expandafter\def\csname LT1\endcsname{\color[rgb]{0,1,0}}%
      \expandafter\def\csname LT2\endcsname{\color[rgb]{0,0,1}}%
      \expandafter\def\csname LT3\endcsname{\color[rgb]{1,0,1}}%
      \expandafter\def\csname LT4\endcsname{\color[rgb]{0,1,1}}%
      \expandafter\def\csname LT5\endcsname{\color[rgb]{1,1,0}}%
      \expandafter\def\csname LT6\endcsname{\color[rgb]{0,0,0}}%
      \expandafter\def\csname LT7\endcsname{\color[rgb]{1,0.3,0}}%
      \expandafter\def\csname LT8\endcsname{\color[rgb]{0.5,0.5,0.5}}%
    \else
      \def\colorrgb#1{\color{black}}%
      \def\colorgray#1{\color[gray]{#1}}%
      \expandafter\def\csname LTw\endcsname{\color{white}}%
      \expandafter\def\csname LTb\endcsname{\color{black}}%
      \expandafter\def\csname LTa\endcsname{\color{black}}%
      \expandafter\def\csname LT0\endcsname{\color{black}}%
      \expandafter\def\csname LT1\endcsname{\color{black}}%
      \expandafter\def\csname LT2\endcsname{\color{black}}%
      \expandafter\def\csname LT3\endcsname{\color{black}}%
      \expandafter\def\csname LT4\endcsname{\color{black}}%
      \expandafter\def\csname LT5\endcsname{\color{black}}%
      \expandafter\def\csname LT6\endcsname{\color{black}}%
      \expandafter\def\csname LT7\endcsname{\color{black}}%
      \expandafter\def\csname LT8\endcsname{\color{black}}%
    \fi
  \fi
    \setlength{\unitlength}{0.0500bp}%
    \ifx\gptboxheight\undefined%
      \newlength{\gptboxheight}%
      \newlength{\gptboxwidth}%
      \newsavebox{\gptboxtext}%
    \fi%
    \setlength{\fboxrule}{0.5pt}%
    \setlength{\fboxsep}{1pt}%
    \definecolor{tbcol}{rgb}{1,1,1}%
\begin{picture}(10204.00,9070.00)%
    \gplgaddtomacro\gplbacktext{%
      \csname LTb\endcsname
      \put(212,8639){\makebox(0,0)[l]{\strut{}(A)}}%
    }%
    \gplgaddtomacro\gplfronttext{%
      \csname LTb\endcsname
      \put(209,7799){\rotatebox{-270}{\makebox(0,0){\strut{}$N$ [$10^4$]}}}%
      \put(4595,7799){\rotatebox{-270}{\makebox(0,0){\strut{}$N_\mathrm{BI}/N_\mathrm{UI}$}}}%
      \put(2319,6200){\makebox(0,0){\strut{}$M_{\ast,\mathrm{run}}$ [$\Msun$]}}%
      \csname LTb\endcsname
      \put(682,6750){\makebox(0,0)[r]{\strut{}$0$}}%
      \put(682,6960){\makebox(0,0)[r]{\strut{}$1$}}%
      \put(682,7170){\makebox(0,0)[r]{\strut{}$2$}}%
      \put(682,7380){\makebox(0,0)[r]{\strut{}$3$}}%
      \put(682,7590){\makebox(0,0)[r]{\strut{}$4$}}%
      \put(682,7800){\makebox(0,0)[r]{\strut{}$5$}}%
      \put(682,8009){\makebox(0,0)[r]{\strut{}$6$}}%
      \put(682,8219){\makebox(0,0)[r]{\strut{}$7$}}%
      \put(682,8429){\makebox(0,0)[r]{\strut{}$8$}}%
      \put(682,8639){\makebox(0,0)[r]{\strut{}$9$}}%
      \put(682,8849){\makebox(0,0)[r]{\strut{}$10$}}%
      \put(905,6530){\makebox(0,0){\strut{}$0.2$}}%
      \put(1270,6530){\makebox(0,0){\strut{}$0.3$}}%
      \put(1635,6530){\makebox(0,0){\strut{}$0.4$}}%
      \put(2000,6530){\makebox(0,0){\strut{}$0.5$}}%
      \put(2365,6530){\makebox(0,0){\strut{}$0.6$}}%
      \put(2730,6530){\makebox(0,0){\strut{}$0.7$}}%
      \put(3095,6530){\makebox(0,0){\strut{}$0.8$}}%
      \put(3460,6530){\makebox(0,0){\strut{}$0.9$}}%
      \put(3825,6530){\makebox(0,0){\strut{}$1$}}%
      \put(3957,6750){\makebox(0,0)[l]{\strut{}$0.0$}}%
      \put(3957,6960){\makebox(0,0)[l]{\strut{}$0.2$}}%
      \put(3957,7170){\makebox(0,0)[l]{\strut{}$0.4$}}%
      \put(3957,7380){\makebox(0,0)[l]{\strut{}$0.6$}}%
      \put(3957,7590){\makebox(0,0)[l]{\strut{}$0.8$}}%
      \put(3957,7800){\makebox(0,0)[l]{\strut{}$1.0$}}%
      \put(3957,8009){\makebox(0,0)[l]{\strut{}$1.2$}}%
      \put(3957,8219){\makebox(0,0)[l]{\strut{}$1.4$}}%
      \put(3957,8429){\makebox(0,0)[l]{\strut{}$1.6$}}%
      \put(3957,8639){\makebox(0,0)[l]{\strut{}$1.8$}}%
      \put(3957,8849){\makebox(0,0)[l]{\strut{}$2.0$}}%
      \put(1416,8534){\makebox(0,0)[l]{\strut{}$\MWDf=1.0\Msun$}}%
    }%
    \gplgaddtomacro\gplbacktext{%
      \csname LTb\endcsname
      \put(5314,8639){\makebox(0,0)[l]{\strut{}(B)}}%
    }%
    \gplgaddtomacro\gplfronttext{%
      \csname LTb\endcsname
      \put(5311,7799){\rotatebox{-270}{\makebox(0,0){\strut{}$N$ [$10^4$]}}}%
      \put(9697,7799){\rotatebox{-270}{\makebox(0,0){\strut{}$N_\mathrm{BI}/N_\mathrm{UI}$}}}%
      \put(7421,6200){\makebox(0,0){\strut{}$M_{\ast,\mathrm{run}}$ [$\Msun$]}}%
      \csname LTb\endcsname
      \put(5784,6750){\makebox(0,0)[r]{\strut{}$0$}}%
      \put(5784,6960){\makebox(0,0)[r]{\strut{}$1$}}%
      \put(5784,7170){\makebox(0,0)[r]{\strut{}$2$}}%
      \put(5784,7380){\makebox(0,0)[r]{\strut{}$3$}}%
      \put(5784,7590){\makebox(0,0)[r]{\strut{}$4$}}%
      \put(5784,7800){\makebox(0,0)[r]{\strut{}$5$}}%
      \put(5784,8009){\makebox(0,0)[r]{\strut{}$6$}}%
      \put(5784,8219){\makebox(0,0)[r]{\strut{}$7$}}%
      \put(5784,8429){\makebox(0,0)[r]{\strut{}$8$}}%
      \put(5784,8639){\makebox(0,0)[r]{\strut{}$9$}}%
      \put(5784,8849){\makebox(0,0)[r]{\strut{}$10$}}%
      \put(6007,6530){\makebox(0,0){\strut{}$0.2$}}%
      \put(6372,6530){\makebox(0,0){\strut{}$0.3$}}%
      \put(6737,6530){\makebox(0,0){\strut{}$0.4$}}%
      \put(7102,6530){\makebox(0,0){\strut{}$0.5$}}%
      \put(7467,6530){\makebox(0,0){\strut{}$0.6$}}%
      \put(7832,6530){\makebox(0,0){\strut{}$0.7$}}%
      \put(8197,6530){\makebox(0,0){\strut{}$0.8$}}%
      \put(8562,6530){\makebox(0,0){\strut{}$0.9$}}%
      \put(8927,6530){\makebox(0,0){\strut{}$1$}}%
      \put(9059,6750){\makebox(0,0)[l]{\strut{}$0.0$}}%
      \put(9059,6960){\makebox(0,0)[l]{\strut{}$0.2$}}%
      \put(9059,7170){\makebox(0,0)[l]{\strut{}$0.4$}}%
      \put(9059,7380){\makebox(0,0)[l]{\strut{}$0.6$}}%
      \put(9059,7590){\makebox(0,0)[l]{\strut{}$0.8$}}%
      \put(9059,7800){\makebox(0,0)[l]{\strut{}$1.0$}}%
      \put(9059,8009){\makebox(0,0)[l]{\strut{}$1.2$}}%
      \put(9059,8219){\makebox(0,0)[l]{\strut{}$1.4$}}%
      \put(9059,8429){\makebox(0,0)[l]{\strut{}$1.6$}}%
      \put(9059,8639){\makebox(0,0)[l]{\strut{}$1.8$}}%
      \put(9059,8849){\makebox(0,0)[l]{\strut{}$2.0$}}%
      \put(6518,8534){\makebox(0,0)[l]{\strut{}$\MWDf=1.2\Msun$}}%
    }%
    \gplgaddtomacro\gplbacktext{%
      \csname LTb\endcsname
      \put(212,5616){\makebox(0,0)[l]{\strut{}(C)}}%
    }%
    \gplgaddtomacro\gplfronttext{%
      \csname LTb\endcsname
      \put(209,4776){\rotatebox{-270}{\makebox(0,0){\strut{}$N$ [$10^4$]}}}%
      \put(4595,4776){\rotatebox{-270}{\makebox(0,0){\strut{}$N_\mathrm{BI}/N_\mathrm{UI}$}}}%
      \put(2319,3177){\makebox(0,0){\strut{}$M_{\ast,\mathrm{run}}$ [$\Msun$]}}%
      \csname LTb\endcsname
      \put(682,3727){\makebox(0,0)[r]{\strut{}$0$}}%
      \put(682,3937){\makebox(0,0)[r]{\strut{}$1$}}%
      \put(682,4147){\makebox(0,0)[r]{\strut{}$2$}}%
      \put(682,4357){\makebox(0,0)[r]{\strut{}$3$}}%
      \put(682,4567){\makebox(0,0)[r]{\strut{}$4$}}%
      \put(682,4777){\makebox(0,0)[r]{\strut{}$5$}}%
      \put(682,4986){\makebox(0,0)[r]{\strut{}$6$}}%
      \put(682,5196){\makebox(0,0)[r]{\strut{}$7$}}%
      \put(682,5406){\makebox(0,0)[r]{\strut{}$8$}}%
      \put(682,5616){\makebox(0,0)[r]{\strut{}$9$}}%
      \put(682,5826){\makebox(0,0)[r]{\strut{}$10$}}%
      \put(905,3507){\makebox(0,0){\strut{}$0.2$}}%
      \put(1270,3507){\makebox(0,0){\strut{}$0.3$}}%
      \put(1635,3507){\makebox(0,0){\strut{}$0.4$}}%
      \put(2000,3507){\makebox(0,0){\strut{}$0.5$}}%
      \put(2365,3507){\makebox(0,0){\strut{}$0.6$}}%
      \put(2730,3507){\makebox(0,0){\strut{}$0.7$}}%
      \put(3095,3507){\makebox(0,0){\strut{}$0.8$}}%
      \put(3460,3507){\makebox(0,0){\strut{}$0.9$}}%
      \put(3825,3507){\makebox(0,0){\strut{}$1$}}%
      \put(3957,3727){\makebox(0,0)[l]{\strut{}$0.0$}}%
      \put(3957,3937){\makebox(0,0)[l]{\strut{}$0.2$}}%
      \put(3957,4147){\makebox(0,0)[l]{\strut{}$0.4$}}%
      \put(3957,4357){\makebox(0,0)[l]{\strut{}$0.6$}}%
      \put(3957,4567){\makebox(0,0)[l]{\strut{}$0.8$}}%
      \put(3957,4777){\makebox(0,0)[l]{\strut{}$1.0$}}%
      \put(3957,4986){\makebox(0,0)[l]{\strut{}$1.2$}}%
      \put(3957,5196){\makebox(0,0)[l]{\strut{}$1.4$}}%
      \put(3957,5406){\makebox(0,0)[l]{\strut{}$1.6$}}%
      \put(3957,5616){\makebox(0,0)[l]{\strut{}$1.8$}}%
      \put(3957,5826){\makebox(0,0)[l]{\strut{}$2.0$}}%
      \put(1416,5511){\makebox(0,0)[l]{\strut{}$\MWDf=1.4\Msun$}}%
    }%
    \gplgaddtomacro\gplbacktext{%
      \csname LTb\endcsname
      \put(5314,5616){\makebox(0,0)[l]{\strut{}(D)}}%
    }%
    \gplgaddtomacro\gplfronttext{%
      \csname LTb\endcsname
      \put(5311,4776){\rotatebox{-270}{\makebox(0,0){\strut{}$N$ [$10^4$]}}}%
      \put(9697,4776){\rotatebox{-270}{\makebox(0,0){\strut{}$N_\mathrm{BI}/N_\mathrm{UI}$}}}%
      \put(7421,3177){\makebox(0,0){\strut{}$M_{\ast,\mathrm{run}}$ [$\Msun$]}}%
      \csname LTb\endcsname
      \put(5784,3727){\makebox(0,0)[r]{\strut{}$0$}}%
      \put(5784,3937){\makebox(0,0)[r]{\strut{}$1$}}%
      \put(5784,4147){\makebox(0,0)[r]{\strut{}$2$}}%
      \put(5784,4357){\makebox(0,0)[r]{\strut{}$3$}}%
      \put(5784,4567){\makebox(0,0)[r]{\strut{}$4$}}%
      \put(5784,4777){\makebox(0,0)[r]{\strut{}$5$}}%
      \put(5784,4986){\makebox(0,0)[r]{\strut{}$6$}}%
      \put(5784,5196){\makebox(0,0)[r]{\strut{}$7$}}%
      \put(5784,5406){\makebox(0,0)[r]{\strut{}$8$}}%
      \put(5784,5616){\makebox(0,0)[r]{\strut{}$9$}}%
      \put(5784,5826){\makebox(0,0)[r]{\strut{}$10$}}%
      \put(6007,3507){\makebox(0,0){\strut{}$0.2$}}%
      \put(6372,3507){\makebox(0,0){\strut{}$0.3$}}%
      \put(6737,3507){\makebox(0,0){\strut{}$0.4$}}%
      \put(7102,3507){\makebox(0,0){\strut{}$0.5$}}%
      \put(7467,3507){\makebox(0,0){\strut{}$0.6$}}%
      \put(7832,3507){\makebox(0,0){\strut{}$0.7$}}%
      \put(8197,3507){\makebox(0,0){\strut{}$0.8$}}%
      \put(8562,3507){\makebox(0,0){\strut{}$0.9$}}%
      \put(8927,3507){\makebox(0,0){\strut{}$1$}}%
      \put(9059,3727){\makebox(0,0)[l]{\strut{}$0.0$}}%
      \put(9059,3937){\makebox(0,0)[l]{\strut{}$0.2$}}%
      \put(9059,4147){\makebox(0,0)[l]{\strut{}$0.4$}}%
      \put(9059,4357){\makebox(0,0)[l]{\strut{}$0.6$}}%
      \put(9059,4567){\makebox(0,0)[l]{\strut{}$0.8$}}%
      \put(9059,4777){\makebox(0,0)[l]{\strut{}$1.0$}}%
      \put(9059,4986){\makebox(0,0)[l]{\strut{}$1.2$}}%
      \put(9059,5196){\makebox(0,0)[l]{\strut{}$1.4$}}%
      \put(9059,5406){\makebox(0,0)[l]{\strut{}$1.6$}}%
      \put(9059,5616){\makebox(0,0)[l]{\strut{}$1.8$}}%
      \put(9059,5826){\makebox(0,0)[l]{\strut{}$2.0$}}%
      \colorrgb{1.00,1.00,1.00}
      \put(6518,5511){\makebox(0,0)[l]{\strut{}$\MWDf=1.5\Msun$}}%
    }%
    \gplgaddtomacro\gplbacktext{%
    }%
    \gplgaddtomacro\gplfronttext{%
      \csname LTb\endcsname
      \put(1122,2630){\makebox(0,0)[r]{\strut{}UII}}%
      \csname LTb\endcsname
      \put(1122,2410){\makebox(0,0)[r]{\strut{}BII}}%
      \csname LTb\endcsname
      \put(2241,2630){\makebox(0,0)[r]{\strut{}UI}}%
      \csname LTb\endcsname
      \put(2241,2410){\makebox(0,0)[r]{\strut{}BI}}%
      \csname LTb\endcsname
      \put(3360,2630){\makebox(0,0)[r]{\strut{}BI/UI}}%
    }%
    \gplgaddtomacro\gplbacktext{%
    }%
    \gplgaddtomacro\gplfronttext{%
    }%
    \gplbacktext
    \put(0,0){\includegraphics[width={510.20bp},height={453.50bp}]{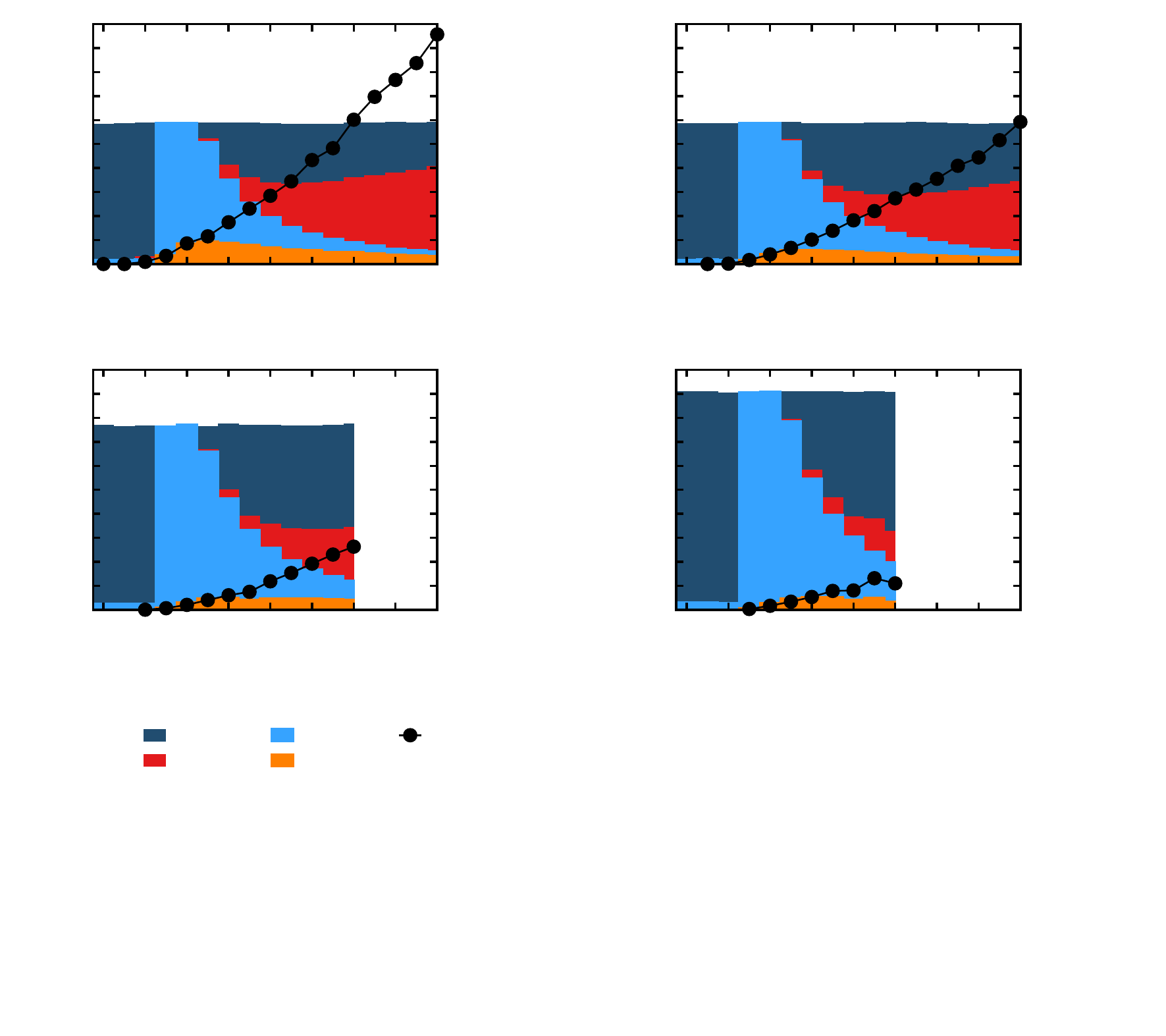}}%
    \gplfronttext
  \end{picture}%
\endgroup

%% file: bound-unbound-mass-fiducial-fig2.tex
\begingroup
  \makeatletter
  \providecommand\color[2][]{%
    \GenericError{(gnuplot) \space\space\space\@spaces}{%
      Package color not loaded in conjunction with
      terminal option `colourtext'%
    }{See the gnuplot documentation for explanation.%
    }{Either use 'blacktext' in gnuplot or load the package
      color.sty in LaTeX.}%
    \renewcommand\color[2][]{}%
  }%
  \providecommand\includegraphics[2][]{%
    \GenericError{(gnuplot) \space\space\space\@spaces}{%
      Package graphicx or graphics not loaded%
    }{See the gnuplot documentation for explanation.%
    }{The gnuplot epslatex terminal needs graphicx.sty or graphics.sty.}%
    \renewcommand\includegraphics[2][]{}%
  }%
  \providecommand\rotatebox[2]{#2}%
  \@ifundefined{ifGPcolor}{%
    \newif\ifGPcolor
    \GPcolortrue
  }{}%
  \@ifundefined{ifGPblacktext}{%
    \newif\ifGPblacktext
    \GPblacktextfalse
  }{}%
  \let\gplgaddtomacro\g@addto@macro
  \gdef\gplbacktext{}%
  \gdef\gplfronttext{}%
  \makeatother
  \ifGPblacktext
    \def\colorrgb#1{}%
    \def\colorgray#1{}%
  \else
    \ifGPcolor
      \def\colorrgb#1{\color[rgb]{#1}}%
      \def\colorgray#1{\color[gray]{#1}}%
      \expandafter\def\csname LTw\endcsname{\color{white}}%
      \expandafter\def\csname LTb\endcsname{\color{black}}%
      \expandafter\def\csname LTa\endcsname{\color{black}}%
      \expandafter\def\csname LT0\endcsname{\color[rgb]{1,0,0}}%
      \expandafter\def\csname LT1\endcsname{\color[rgb]{0,1,0}}%
      \expandafter\def\csname LT2\endcsname{\color[rgb]{0,0,1}}%
      \expandafter\def\csname LT3\endcsname{\color[rgb]{1,0,1}}%
      \expandafter\def\csname LT4\endcsname{\color[rgb]{0,1,1}}%
      \expandafter\def\csname LT5\endcsname{\color[rgb]{1,1,0}}%
      \expandafter\def\csname LT6\endcsname{\color[rgb]{0,0,0}}%
      \expandafter\def\csname LT7\endcsname{\color[rgb]{1,0.3,0}}%
      \expandafter\def\csname LT8\endcsname{\color[rgb]{0.5,0.5,0.5}}%
    \else
      \def\colorrgb#1{\color{black}}%
      \def\colorgray#1{\color[gray]{#1}}%
      \expandafter\def\csname LTw\endcsname{\color{white}}%
      \expandafter\def\csname LTb\endcsname{\color{black}}%
      \expandafter\def\csname LTa\endcsname{\color{black}}%
      \expandafter\def\csname LT0\endcsname{\color{black}}%
      \expandafter\def\csname LT1\endcsname{\color{black}}%
      \expandafter\def\csname LT2\endcsname{\color{black}}%
      \expandafter\def\csname LT3\endcsname{\color{black}}%
      \expandafter\def\csname LT4\endcsname{\color{black}}%
      \expandafter\def\csname LT5\endcsname{\color{black}}%
      \expandafter\def\csname LT6\endcsname{\color{black}}%
      \expandafter\def\csname LT7\endcsname{\color{black}}%
      \expandafter\def\csname LT8\endcsname{\color{black}}%
    \fi
  \fi
    \setlength{\unitlength}{0.0500bp}%
    \ifx\gptboxheight\undefined%
      \newlength{\gptboxheight}%
      \newlength{\gptboxwidth}%
      \newsavebox{\gptboxtext}%
    \fi%
    \setlength{\fboxrule}{0.5pt}%
    \setlength{\fboxsep}{1pt}%
    \definecolor{tbcol}{rgb}{1,1,1}%
\begin{picture}(10204.00,9070.00)%
    \gplgaddtomacro\gplbacktext{%
      \csname LTb\endcsname
      \put(185,8639){\makebox(0,0)[l]{\strut{}(A)}}%
      \put(1443,8429){\makebox(0,0)[l]{\strut{}$\MWDf=1.0\Msun$}}%
    }%
    \gplgaddtomacro\gplfronttext{%
      \csname LTb\endcsname
      \put(209,7799){\rotatebox{-270}{\makebox(0,0){\strut{}$N$ [$10^3$]}}}%
      \put(4595,7799){\rotatebox{-270}{\makebox(0,0){\strut{}$N_\mathrm{BI}/N_\mathrm{UI}$}}}%
      \put(2385,6200){\makebox(0,0){\strut{}$M_{\ast,\mathrm{run}}$ [$\Msun$]}}%
      \csname LTb\endcsname
      \put(682,6750){\makebox(0,0)[r]{\strut{}$0$}}%
      \put(682,6960){\makebox(0,0)[r]{\strut{}$1$}}%
      \put(682,7170){\makebox(0,0)[r]{\strut{}$2$}}%
      \put(682,7380){\makebox(0,0)[r]{\strut{}$3$}}%
      \put(682,7590){\makebox(0,0)[r]{\strut{}$4$}}%
      \put(682,7800){\makebox(0,0)[r]{\strut{}$5$}}%
      \put(682,8009){\makebox(0,0)[r]{\strut{}$6$}}%
      \put(682,8219){\makebox(0,0)[r]{\strut{}$7$}}%
      \put(682,8429){\makebox(0,0)[r]{\strut{}$8$}}%
      \put(682,8639){\makebox(0,0)[r]{\strut{}$9$}}%
      \put(682,8849){\makebox(0,0)[r]{\strut{}$10$}}%
      \put(909,6530){\makebox(0,0){\strut{}$0.2$}}%
      \put(1290,6530){\makebox(0,0){\strut{}$0.3$}}%
      \put(1671,6530){\makebox(0,0){\strut{}$0.4$}}%
      \put(2052,6530){\makebox(0,0){\strut{}$0.5$}}%
      \put(2433,6530){\makebox(0,0){\strut{}$0.6$}}%
      \put(2814,6530){\makebox(0,0){\strut{}$0.7$}}%
      \put(3195,6530){\makebox(0,0){\strut{}$0.8$}}%
      \put(3576,6530){\makebox(0,0){\strut{}$0.9$}}%
      \put(3957,6530){\makebox(0,0){\strut{}$1$}}%
      \put(4089,6750){\makebox(0,0)[l]{\strut{}$0$}}%
      \put(4089,6960){\makebox(0,0)[l]{\strut{}$2$}}%
      \put(4089,7170){\makebox(0,0)[l]{\strut{}$4$}}%
      \put(4089,7380){\makebox(0,0)[l]{\strut{}$6$}}%
      \put(4089,7590){\makebox(0,0)[l]{\strut{}$8$}}%
      \put(4089,7800){\makebox(0,0)[l]{\strut{}$10$}}%
      \put(4089,8009){\makebox(0,0)[l]{\strut{}$12$}}%
      \put(4089,8219){\makebox(0,0)[l]{\strut{}$14$}}%
      \put(4089,8429){\makebox(0,0)[l]{\strut{}$16$}}%
      \put(4089,8639){\makebox(0,0)[l]{\strut{}$18$}}%
      \put(4089,8849){\makebox(0,0)[l]{\strut{}$20$}}%
    }%
    \gplgaddtomacro\gplbacktext{%
      \csname LTb\endcsname
      \put(5287,8639){\makebox(0,0)[l]{\strut{}(B)}}%
      \put(6545,8429){\makebox(0,0)[l]{\strut{}$\MWDf=1.2\Msun$}}%
    }%
    \gplgaddtomacro\gplfronttext{%
      \csname LTb\endcsname
      \put(5311,7799){\rotatebox{-270}{\makebox(0,0){\strut{}$N$ [$10^3$]}}}%
      \put(9697,7799){\rotatebox{-270}{\makebox(0,0){\strut{}$N_\mathrm{BI}/N_\mathrm{UI}$}}}%
      \put(7487,6200){\makebox(0,0){\strut{}$M_{\ast,\mathrm{run}}$ [$\Msun$]}}%
      \csname LTb\endcsname
      \put(5784,6750){\makebox(0,0)[r]{\strut{}$0$}}%
      \put(5784,6960){\makebox(0,0)[r]{\strut{}$1$}}%
      \put(5784,7170){\makebox(0,0)[r]{\strut{}$2$}}%
      \put(5784,7380){\makebox(0,0)[r]{\strut{}$3$}}%
      \put(5784,7590){\makebox(0,0)[r]{\strut{}$4$}}%
      \put(5784,7800){\makebox(0,0)[r]{\strut{}$5$}}%
      \put(5784,8009){\makebox(0,0)[r]{\strut{}$6$}}%
      \put(5784,8219){\makebox(0,0)[r]{\strut{}$7$}}%
      \put(5784,8429){\makebox(0,0)[r]{\strut{}$8$}}%
      \put(5784,8639){\makebox(0,0)[r]{\strut{}$9$}}%
      \put(5784,8849){\makebox(0,0)[r]{\strut{}$10$}}%
      \put(6011,6530){\makebox(0,0){\strut{}$0.2$}}%
      \put(6392,6530){\makebox(0,0){\strut{}$0.3$}}%
      \put(6773,6530){\makebox(0,0){\strut{}$0.4$}}%
      \put(7154,6530){\makebox(0,0){\strut{}$0.5$}}%
      \put(7535,6530){\makebox(0,0){\strut{}$0.6$}}%
      \put(7916,6530){\makebox(0,0){\strut{}$0.7$}}%
      \put(8297,6530){\makebox(0,0){\strut{}$0.8$}}%
      \put(8678,6530){\makebox(0,0){\strut{}$0.9$}}%
      \put(9059,6530){\makebox(0,0){\strut{}$1$}}%
      \put(9191,6750){\makebox(0,0)[l]{\strut{}$0$}}%
      \put(9191,6960){\makebox(0,0)[l]{\strut{}$2$}}%
      \put(9191,7170){\makebox(0,0)[l]{\strut{}$4$}}%
      \put(9191,7380){\makebox(0,0)[l]{\strut{}$6$}}%
      \put(9191,7590){\makebox(0,0)[l]{\strut{}$8$}}%
      \put(9191,7800){\makebox(0,0)[l]{\strut{}$10$}}%
      \put(9191,8009){\makebox(0,0)[l]{\strut{}$12$}}%
      \put(9191,8219){\makebox(0,0)[l]{\strut{}$14$}}%
      \put(9191,8429){\makebox(0,0)[l]{\strut{}$16$}}%
      \put(9191,8639){\makebox(0,0)[l]{\strut{}$18$}}%
      \put(9191,8849){\makebox(0,0)[l]{\strut{}$20$}}%
    }%
    \gplgaddtomacro\gplbacktext{%
      \csname LTb\endcsname
      \put(185,5616){\makebox(0,0)[l]{\strut{}(C)}}%
      \put(1443,5406){\makebox(0,0)[l]{\strut{}$\MWDf=1.4\Msun$}}%
    }%
    \gplgaddtomacro\gplfronttext{%
      \csname LTb\endcsname
      \put(209,4776){\rotatebox{-270}{\makebox(0,0){\strut{}$N$ [$10^3$]}}}%
      \put(4595,4776){\rotatebox{-270}{\makebox(0,0){\strut{}$N_\mathrm{BI}/N_\mathrm{UI}$}}}%
      \put(2385,3177){\makebox(0,0){\strut{}$M_{\ast,\mathrm{run}}$ [$\Msun$]}}%
      \csname LTb\endcsname
      \put(682,3727){\makebox(0,0)[r]{\strut{}$0$}}%
      \put(682,3937){\makebox(0,0)[r]{\strut{}$1$}}%
      \put(682,4147){\makebox(0,0)[r]{\strut{}$2$}}%
      \put(682,4357){\makebox(0,0)[r]{\strut{}$3$}}%
      \put(682,4567){\makebox(0,0)[r]{\strut{}$4$}}%
      \put(682,4777){\makebox(0,0)[r]{\strut{}$5$}}%
      \put(682,4986){\makebox(0,0)[r]{\strut{}$6$}}%
      \put(682,5196){\makebox(0,0)[r]{\strut{}$7$}}%
      \put(682,5406){\makebox(0,0)[r]{\strut{}$8$}}%
      \put(682,5616){\makebox(0,0)[r]{\strut{}$9$}}%
      \put(682,5826){\makebox(0,0)[r]{\strut{}$10$}}%
      \put(909,3507){\makebox(0,0){\strut{}$0.2$}}%
      \put(1290,3507){\makebox(0,0){\strut{}$0.3$}}%
      \put(1671,3507){\makebox(0,0){\strut{}$0.4$}}%
      \put(2052,3507){\makebox(0,0){\strut{}$0.5$}}%
      \put(2433,3507){\makebox(0,0){\strut{}$0.6$}}%
      \put(2814,3507){\makebox(0,0){\strut{}$0.7$}}%
      \put(3195,3507){\makebox(0,0){\strut{}$0.8$}}%
      \put(3576,3507){\makebox(0,0){\strut{}$0.9$}}%
      \put(3957,3507){\makebox(0,0){\strut{}$1$}}%
      \put(4089,3727){\makebox(0,0)[l]{\strut{}$0$}}%
      \put(4089,3937){\makebox(0,0)[l]{\strut{}$2$}}%
      \put(4089,4147){\makebox(0,0)[l]{\strut{}$4$}}%
      \put(4089,4357){\makebox(0,0)[l]{\strut{}$6$}}%
      \put(4089,4567){\makebox(0,0)[l]{\strut{}$8$}}%
      \put(4089,4777){\makebox(0,0)[l]{\strut{}$10$}}%
      \put(4089,4986){\makebox(0,0)[l]{\strut{}$12$}}%
      \put(4089,5196){\makebox(0,0)[l]{\strut{}$14$}}%
      \put(4089,5406){\makebox(0,0)[l]{\strut{}$16$}}%
      \put(4089,5616){\makebox(0,0)[l]{\strut{}$18$}}%
      \put(4089,5826){\makebox(0,0)[l]{\strut{}$20$}}%
    }%
    \gplgaddtomacro\gplbacktext{%
      \csname LTb\endcsname
      \put(5287,5616){\makebox(0,0)[l]{\strut{}(D)}}%
      \put(6545,5406){\makebox(0,0)[l]{\strut{}$\MWDf=1.5\Msun$}}%
    }%
    \gplgaddtomacro\gplfronttext{%
      \csname LTb\endcsname
      \put(5311,4776){\rotatebox{-270}{\makebox(0,0){\strut{}$N$ [$10^3$]}}}%
      \put(9697,4776){\rotatebox{-270}{\makebox(0,0){\strut{}$N_\mathrm{BI}/N_\mathrm{UI}$}}}%
      \put(7487,3177){\makebox(0,0){\strut{}$M_{\ast,\mathrm{run}}$ [$\Msun$]}}%
      \csname LTb\endcsname
      \put(5784,3727){\makebox(0,0)[r]{\strut{}$0$}}%
      \put(5784,3937){\makebox(0,0)[r]{\strut{}$1$}}%
      \put(5784,4147){\makebox(0,0)[r]{\strut{}$2$}}%
      \put(5784,4357){\makebox(0,0)[r]{\strut{}$3$}}%
      \put(5784,4567){\makebox(0,0)[r]{\strut{}$4$}}%
      \put(5784,4777){\makebox(0,0)[r]{\strut{}$5$}}%
      \put(5784,4986){\makebox(0,0)[r]{\strut{}$6$}}%
      \put(5784,5196){\makebox(0,0)[r]{\strut{}$7$}}%
      \put(5784,5406){\makebox(0,0)[r]{\strut{}$8$}}%
      \put(5784,5616){\makebox(0,0)[r]{\strut{}$9$}}%
      \put(5784,5826){\makebox(0,0)[r]{\strut{}$10$}}%
      \put(6011,3507){\makebox(0,0){\strut{}$0.2$}}%
      \put(6392,3507){\makebox(0,0){\strut{}$0.3$}}%
      \put(6773,3507){\makebox(0,0){\strut{}$0.4$}}%
      \put(7154,3507){\makebox(0,0){\strut{}$0.5$}}%
      \put(7535,3507){\makebox(0,0){\strut{}$0.6$}}%
      \put(7916,3507){\makebox(0,0){\strut{}$0.7$}}%
      \put(8297,3507){\makebox(0,0){\strut{}$0.8$}}%
      \put(8678,3507){\makebox(0,0){\strut{}$0.9$}}%
      \put(9059,3507){\makebox(0,0){\strut{}$1$}}%
      \put(9191,3727){\makebox(0,0)[l]{\strut{}$0$}}%
      \put(9191,3937){\makebox(0,0)[l]{\strut{}$2$}}%
      \put(9191,4147){\makebox(0,0)[l]{\strut{}$4$}}%
      \put(9191,4357){\makebox(0,0)[l]{\strut{}$6$}}%
      \put(9191,4567){\makebox(0,0)[l]{\strut{}$8$}}%
      \put(9191,4777){\makebox(0,0)[l]{\strut{}$10$}}%
      \put(9191,4986){\makebox(0,0)[l]{\strut{}$12$}}%
      \put(9191,5196){\makebox(0,0)[l]{\strut{}$14$}}%
      \put(9191,5406){\makebox(0,0)[l]{\strut{}$16$}}%
      \put(9191,5616){\makebox(0,0)[l]{\strut{}$18$}}%
      \put(9191,5826){\makebox(0,0)[l]{\strut{}$20$}}%
    }%
    \gplgaddtomacro\gplbacktext{%
    }%
    \gplgaddtomacro\gplfronttext{%
      \csname LTb\endcsname
      \put(1122,2630){\makebox(0,0)[r]{\strut{}UII}}%
      \csname LTb\endcsname
      \put(1122,2410){\makebox(0,0)[r]{\strut{}BII}}%
      \csname LTb\endcsname
      \put(2241,2630){\makebox(0,0)[r]{\strut{}UI}}%
      \csname LTb\endcsname
      \put(2241,2410){\makebox(0,0)[r]{\strut{}BI}}%
      \csname LTb\endcsname
      \put(3360,2630){\makebox(0,0)[r]{\strut{}BI/UI}}%
    }%
    \gplgaddtomacro\gplbacktext{%
    }%
    \gplgaddtomacro\gplfronttext{%
    }%
    \gplbacktext
    \put(0,0){\includegraphics[width={510.20bp},height={453.50bp}]{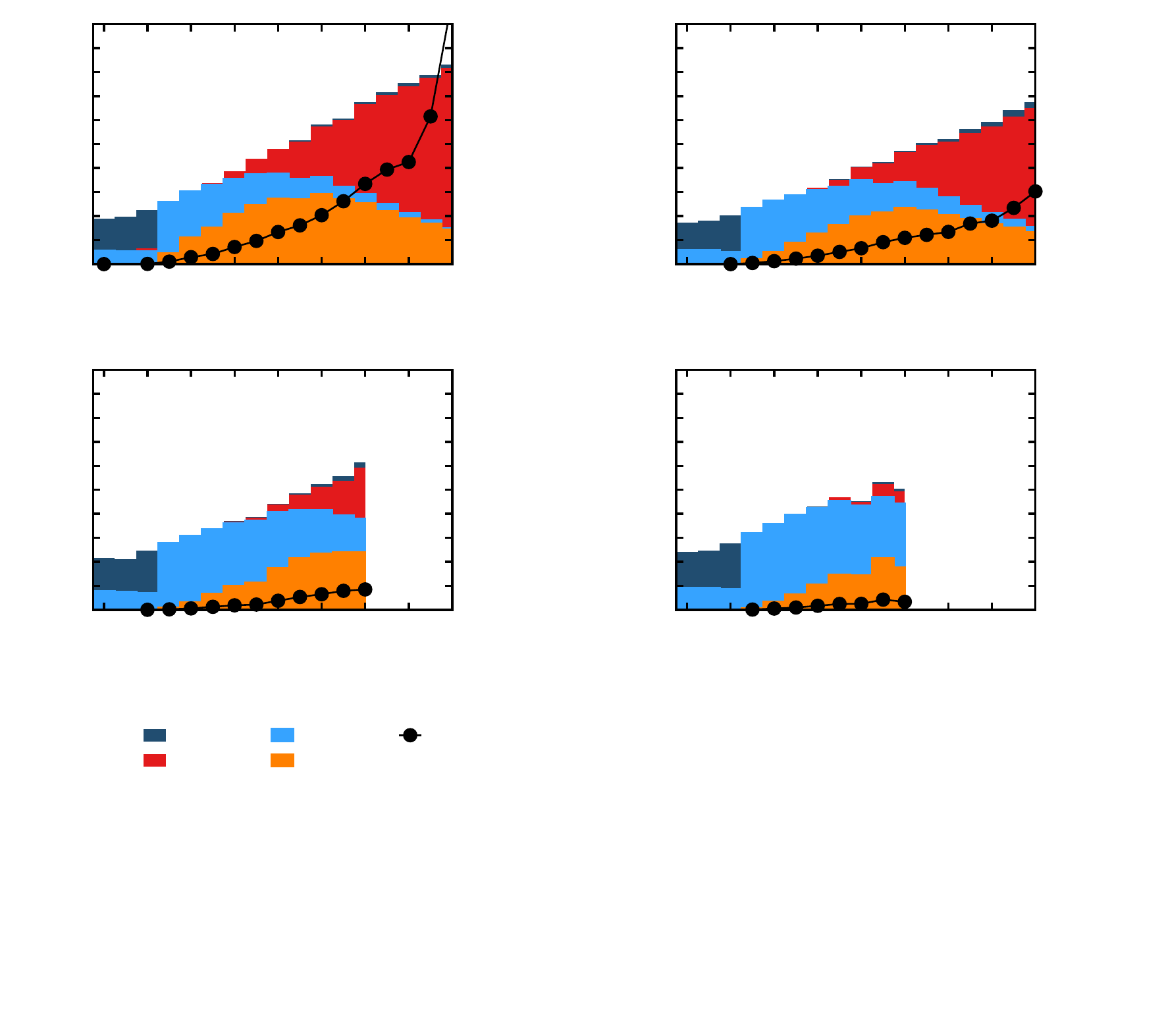}}%
    \gplfronttext
  \end{picture}%
\endgroup

%% file: bound-unbound-mass-fiducial-ratios-fig2.tex
\begingroup
  \makeatletter
  \providecommand\color[2][]{%
    \GenericError{(gnuplot) \space\space\space\@spaces}{%
      Package color not loaded in conjunction with
      terminal option `colourtext'%
    }{See the gnuplot documentation for explanation.%
    }{Either use 'blacktext' in gnuplot or load the package
      color.sty in LaTeX.}%
    \renewcommand\color[2][]{}%
  }%
  \providecommand\includegraphics[2][]{%
    \GenericError{(gnuplot) \space\space\space\@spaces}{%
      Package graphicx or graphics not loaded%
    }{See the gnuplot documentation for explanation.%
    }{The gnuplot epslatex terminal needs graphicx.sty or graphics.sty.}%
    \renewcommand\includegraphics[2][]{}%
  }%
  \providecommand\rotatebox[2]{#2}%
  \@ifundefined{ifGPcolor}{%
    \newif\ifGPcolor
    \GPcolortrue
  }{}%
  \@ifundefined{ifGPblacktext}{%
    \newif\ifGPblacktext
    \GPblacktextfalse
  }{}%
  \let\gplgaddtomacro\g@addto@macro
  \gdef\gplbacktext{}%
  \gdef\gplfronttext{}%
  \makeatother
  \ifGPblacktext
    \def\colorrgb#1{}%
    \def\colorgray#1{}%
  \else
    \ifGPcolor
      \def\colorrgb#1{\color[rgb]{#1}}%
      \def\colorgray#1{\color[gray]{#1}}%
      \expandafter\def\csname LTw\endcsname{\color{white}}%
      \expandafter\def\csname LTb\endcsname{\color{black}}%
      \expandafter\def\csname LTa\endcsname{\color{black}}%
      \expandafter\def\csname LT0\endcsname{\color[rgb]{1,0,0}}%
      \expandafter\def\csname LT1\endcsname{\color[rgb]{0,1,0}}%
      \expandafter\def\csname LT2\endcsname{\color[rgb]{0,0,1}}%
      \expandafter\def\csname LT3\endcsname{\color[rgb]{1,0,1}}%
      \expandafter\def\csname LT4\endcsname{\color[rgb]{0,1,1}}%
      \expandafter\def\csname LT5\endcsname{\color[rgb]{1,1,0}}%
      \expandafter\def\csname LT6\endcsname{\color[rgb]{0,0,0}}%
      \expandafter\def\csname LT7\endcsname{\color[rgb]{1,0.3,0}}%
      \expandafter\def\csname LT8\endcsname{\color[rgb]{0.5,0.5,0.5}}%
    \else
      \def\colorrgb#1{\color{black}}%
      \def\colorgray#1{\color[gray]{#1}}%
      \expandafter\def\csname LTw\endcsname{\color{white}}%
      \expandafter\def\csname LTb\endcsname{\color{black}}%
      \expandafter\def\csname LTa\endcsname{\color{black}}%
      \expandafter\def\csname LT0\endcsname{\color{black}}%
      \expandafter\def\csname LT1\endcsname{\color{black}}%
      \expandafter\def\csname LT2\endcsname{\color{black}}%
      \expandafter\def\csname LT3\endcsname{\color{black}}%
      \expandafter\def\csname LT4\endcsname{\color{black}}%
      \expandafter\def\csname LT5\endcsname{\color{black}}%
      \expandafter\def\csname LT6\endcsname{\color{black}}%
      \expandafter\def\csname LT7\endcsname{\color{black}}%
      \expandafter\def\csname LT8\endcsname{\color{black}}%
    \fi
  \fi
    \setlength{\unitlength}{0.0500bp}%
    \ifx\gptboxheight\undefined%
      \newlength{\gptboxheight}%
      \newlength{\gptboxwidth}%
      \newsavebox{\gptboxtext}%
    \fi%
    \setlength{\fboxrule}{0.5pt}%
    \setlength{\fboxsep}{1pt}%
    \definecolor{tbcol}{rgb}{1,1,1}%
\begin{picture}(4988.00,3968.00)%
    \gplgaddtomacro\gplbacktext{%
    }%
    \gplgaddtomacro\gplfronttext{%
      \csname LTb\endcsname
      \put(209,2445){\rotatebox{-270}{\makebox(0,0){\strut{}$N_\mathrm{BI}/N_\mathrm{UI}$}}}%
      \put(2834,594){\makebox(0,0){\strut{}$M_{\ast,\mathrm{run}}$ [$\Msun$]}}%
      \csname LTb\endcsname
      \put(1849,393){\makebox(0,0)[r]{\strut{}M10}}%
      \csname LTb\endcsname
      \put(1849,173){\makebox(0,0)[r]{\strut{}M11}}%
      \csname LTb\endcsname
      \put(2770,393){\makebox(0,0)[r]{\strut{}M12}}%
      \csname LTb\endcsname
      \put(2770,173){\makebox(0,0)[r]{\strut{}M13}}%
      \csname LTb\endcsname
      \put(3691,393){\makebox(0,0)[r]{\strut{}M14}}%
      \csname LTb\endcsname
      \put(3691,173){\makebox(0,0)[r]{\strut{}M15}}%
      \csname LTb\endcsname
      \put(946,1144){\makebox(0,0)[r]{\strut{}$0.01$}}%
      \put(946,1848){\makebox(0,0)[r]{\strut{}$0.1$}}%
      \put(946,2551){\makebox(0,0)[r]{\strut{}$1$}}%
      \put(946,3255){\makebox(0,0)[r]{\strut{}$10$}}%
      \put(1184,924){\makebox(0,0){\strut{}$0.2$}}%
      \put(1610,924){\makebox(0,0){\strut{}$0.3$}}%
      \put(2036,924){\makebox(0,0){\strut{}$0.4$}}%
      \put(2462,924){\makebox(0,0){\strut{}$0.5$}}%
      \put(2888,924){\makebox(0,0){\strut{}$0.6$}}%
      \put(3314,924){\makebox(0,0){\strut{}$0.7$}}%
      \put(3739,924){\makebox(0,0){\strut{}$0.8$}}%
      \put(4165,924){\makebox(0,0){\strut{}$0.9$}}%
      \put(4591,924){\makebox(0,0){\strut{}$1$}}%
    }%
    \gplbacktext
    \put(0,0){\includegraphics[width={249.40bp},height={198.40bp}]{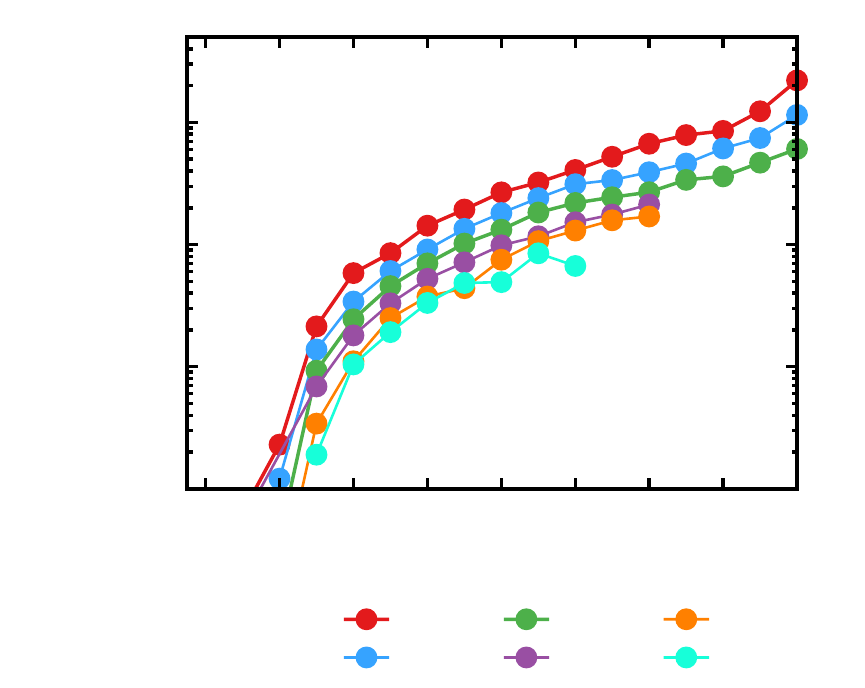}}%
    \gplfronttext
  \end{picture}%
\endgroup

%% file: bound-unbound-mass-fiducial-tau-fig2.tex
\begingroup
  \makeatletter
  \providecommand\color[2][]{%
    \GenericError{(gnuplot) \space\space\space\@spaces}{%
      Package color not loaded in conjunction with
      terminal option `colourtext'%
    }{See the gnuplot documentation for explanation.%
    }{Either use 'blacktext' in gnuplot or load the package
      color.sty in LaTeX.}%
    \renewcommand\color[2][]{}%
  }%
  \providecommand\includegraphics[2][]{%
    \GenericError{(gnuplot) \space\space\space\@spaces}{%
      Package graphicx or graphics not loaded%
    }{See the gnuplot documentation for explanation.%
    }{The gnuplot epslatex terminal needs graphicx.sty or graphics.sty.}%
    \renewcommand\includegraphics[2][]{}%
  }%
  \providecommand\rotatebox[2]{#2}%
  \@ifundefined{ifGPcolor}{%
    \newif\ifGPcolor
    \GPcolortrue
  }{}%
  \@ifundefined{ifGPblacktext}{%
    \newif\ifGPblacktext
    \GPblacktextfalse
  }{}%
  \let\gplgaddtomacro\g@addto@macro
  \gdef\gplbacktext{}%
  \gdef\gplfronttext{}%
  \makeatother
  \ifGPblacktext
    \def\colorrgb#1{}%
    \def\colorgray#1{}%
  \else
    \ifGPcolor
      \def\colorrgb#1{\color[rgb]{#1}}%
      \def\colorgray#1{\color[gray]{#1}}%
      \expandafter\def\csname LTw\endcsname{\color{white}}%
      \expandafter\def\csname LTb\endcsname{\color{black}}%
      \expandafter\def\csname LTa\endcsname{\color{black}}%
      \expandafter\def\csname LT0\endcsname{\color[rgb]{1,0,0}}%
      \expandafter\def\csname LT1\endcsname{\color[rgb]{0,1,0}}%
      \expandafter\def\csname LT2\endcsname{\color[rgb]{0,0,1}}%
      \expandafter\def\csname LT3\endcsname{\color[rgb]{1,0,1}}%
      \expandafter\def\csname LT4\endcsname{\color[rgb]{0,1,1}}%
      \expandafter\def\csname LT5\endcsname{\color[rgb]{1,1,0}}%
      \expandafter\def\csname LT6\endcsname{\color[rgb]{0,0,0}}%
      \expandafter\def\csname LT7\endcsname{\color[rgb]{1,0.3,0}}%
      \expandafter\def\csname LT8\endcsname{\color[rgb]{0.5,0.5,0.5}}%
    \else
      \def\colorrgb#1{\color{black}}%
      \def\colorgray#1{\color[gray]{#1}}%
      \expandafter\def\csname LTw\endcsname{\color{white}}%
      \expandafter\def\csname LTb\endcsname{\color{black}}%
      \expandafter\def\csname LTa\endcsname{\color{black}}%
      \expandafter\def\csname LT0\endcsname{\color{black}}%
      \expandafter\def\csname LT1\endcsname{\color{black}}%
      \expandafter\def\csname LT2\endcsname{\color{black}}%
      \expandafter\def\csname LT3\endcsname{\color{black}}%
      \expandafter\def\csname LT4\endcsname{\color{black}}%
      \expandafter\def\csname LT5\endcsname{\color{black}}%
      \expandafter\def\csname LT6\endcsname{\color{black}}%
      \expandafter\def\csname LT7\endcsname{\color{black}}%
      \expandafter\def\csname LT8\endcsname{\color{black}}%
    \fi
  \fi
    \setlength{\unitlength}{0.0500bp}%
    \ifx\gptboxheight\undefined%
      \newlength{\gptboxheight}%
      \newlength{\gptboxwidth}%
      \newsavebox{\gptboxtext}%
    \fi%
    \setlength{\fboxrule}{0.5pt}%
    \setlength{\fboxsep}{1pt}%
    \definecolor{tbcol}{rgb}{1,1,1}%
\begin{picture}(10204.00,9070.00)%
    \gplgaddtomacro\gplbacktext{%
      \csname LTb\endcsname
      \put(185,8639){\makebox(0,0)[l]{\strut{}(A)}}%
      \put(1443,8429){\makebox(0,0)[l]{\strut{}$\MWDf=1.0\Msun$}}%
    }%
    \gplgaddtomacro\gplfronttext{%
      \csname LTb\endcsname
      \put(209,7799){\rotatebox{-270}{\makebox(0,0){\strut{}$N$ [$10^3$]}}}%
      \put(4595,7799){\rotatebox{-270}{\makebox(0,0){\strut{}$N_\mathrm{BI}/N_\mathrm{UI}$}}}%
      \put(2385,6200){\makebox(0,0){\strut{}$M_{\ast,\mathrm{run}}$ [$\Msun$]}}%
      \csname LTb\endcsname
      \put(682,6750){\makebox(0,0)[r]{\strut{}$0$}}%
      \put(682,6960){\makebox(0,0)[r]{\strut{}$1$}}%
      \put(682,7170){\makebox(0,0)[r]{\strut{}$2$}}%
      \put(682,7380){\makebox(0,0)[r]{\strut{}$3$}}%
      \put(682,7590){\makebox(0,0)[r]{\strut{}$4$}}%
      \put(682,7800){\makebox(0,0)[r]{\strut{}$5$}}%
      \put(682,8009){\makebox(0,0)[r]{\strut{}$6$}}%
      \put(682,8219){\makebox(0,0)[r]{\strut{}$7$}}%
      \put(682,8429){\makebox(0,0)[r]{\strut{}$8$}}%
      \put(682,8639){\makebox(0,0)[r]{\strut{}$9$}}%
      \put(682,8849){\makebox(0,0)[r]{\strut{}$10$}}%
      \put(909,6530){\makebox(0,0){\strut{}$0.2$}}%
      \put(1290,6530){\makebox(0,0){\strut{}$0.3$}}%
      \put(1671,6530){\makebox(0,0){\strut{}$0.4$}}%
      \put(2052,6530){\makebox(0,0){\strut{}$0.5$}}%
      \put(2433,6530){\makebox(0,0){\strut{}$0.6$}}%
      \put(2814,6530){\makebox(0,0){\strut{}$0.7$}}%
      \put(3195,6530){\makebox(0,0){\strut{}$0.8$}}%
      \put(3576,6530){\makebox(0,0){\strut{}$0.9$}}%
      \put(3957,6530){\makebox(0,0){\strut{}$1$}}%
      \put(4089,6750){\makebox(0,0)[l]{\strut{}$0$}}%
      \put(4089,6960){\makebox(0,0)[l]{\strut{}$2$}}%
      \put(4089,7170){\makebox(0,0)[l]{\strut{}$4$}}%
      \put(4089,7380){\makebox(0,0)[l]{\strut{}$6$}}%
      \put(4089,7590){\makebox(0,0)[l]{\strut{}$8$}}%
      \put(4089,7800){\makebox(0,0)[l]{\strut{}$10$}}%
      \put(4089,8009){\makebox(0,0)[l]{\strut{}$12$}}%
      \put(4089,8219){\makebox(0,0)[l]{\strut{}$14$}}%
      \put(4089,8429){\makebox(0,0)[l]{\strut{}$16$}}%
      \put(4089,8639){\makebox(0,0)[l]{\strut{}$18$}}%
      \put(4089,8849){\makebox(0,0)[l]{\strut{}$20$}}%
    }%
    \gplgaddtomacro\gplbacktext{%
      \csname LTb\endcsname
      \put(5287,8639){\makebox(0,0)[l]{\strut{}(B)}}%
      \put(6545,8429){\makebox(0,0)[l]{\strut{}$\MWDf=1.2\Msun$}}%
    }%
    \gplgaddtomacro\gplfronttext{%
      \csname LTb\endcsname
      \put(5311,7799){\rotatebox{-270}{\makebox(0,0){\strut{}$N$ [$10^3$]}}}%
      \put(9697,7799){\rotatebox{-270}{\makebox(0,0){\strut{}$N_\mathrm{BI}/N_\mathrm{UI}$}}}%
      \put(7487,6200){\makebox(0,0){\strut{}$M_{\ast,\mathrm{run}}$ [$\Msun$]}}%
      \csname LTb\endcsname
      \put(5784,6750){\makebox(0,0)[r]{\strut{}$0$}}%
      \put(5784,6960){\makebox(0,0)[r]{\strut{}$1$}}%
      \put(5784,7170){\makebox(0,0)[r]{\strut{}$2$}}%
      \put(5784,7380){\makebox(0,0)[r]{\strut{}$3$}}%
      \put(5784,7590){\makebox(0,0)[r]{\strut{}$4$}}%
      \put(5784,7800){\makebox(0,0)[r]{\strut{}$5$}}%
      \put(5784,8009){\makebox(0,0)[r]{\strut{}$6$}}%
      \put(5784,8219){\makebox(0,0)[r]{\strut{}$7$}}%
      \put(5784,8429){\makebox(0,0)[r]{\strut{}$8$}}%
      \put(5784,8639){\makebox(0,0)[r]{\strut{}$9$}}%
      \put(5784,8849){\makebox(0,0)[r]{\strut{}$10$}}%
      \put(6011,6530){\makebox(0,0){\strut{}$0.2$}}%
      \put(6392,6530){\makebox(0,0){\strut{}$0.3$}}%
      \put(6773,6530){\makebox(0,0){\strut{}$0.4$}}%
      \put(7154,6530){\makebox(0,0){\strut{}$0.5$}}%
      \put(7535,6530){\makebox(0,0){\strut{}$0.6$}}%
      \put(7916,6530){\makebox(0,0){\strut{}$0.7$}}%
      \put(8297,6530){\makebox(0,0){\strut{}$0.8$}}%
      \put(8678,6530){\makebox(0,0){\strut{}$0.9$}}%
      \put(9059,6530){\makebox(0,0){\strut{}$1$}}%
      \put(9191,6750){\makebox(0,0)[l]{\strut{}$0$}}%
      \put(9191,6960){\makebox(0,0)[l]{\strut{}$2$}}%
      \put(9191,7170){\makebox(0,0)[l]{\strut{}$4$}}%
      \put(9191,7380){\makebox(0,0)[l]{\strut{}$6$}}%
      \put(9191,7590){\makebox(0,0)[l]{\strut{}$8$}}%
      \put(9191,7800){\makebox(0,0)[l]{\strut{}$10$}}%
      \put(9191,8009){\makebox(0,0)[l]{\strut{}$12$}}%
      \put(9191,8219){\makebox(0,0)[l]{\strut{}$14$}}%
      \put(9191,8429){\makebox(0,0)[l]{\strut{}$16$}}%
      \put(9191,8639){\makebox(0,0)[l]{\strut{}$18$}}%
      \put(9191,8849){\makebox(0,0)[l]{\strut{}$20$}}%
    }%
    \gplgaddtomacro\gplbacktext{%
      \csname LTb\endcsname
      \put(185,5616){\makebox(0,0)[l]{\strut{}(C)}}%
      \put(1443,5406){\makebox(0,0)[l]{\strut{}$\MWDf=1.4\Msun$}}%
    }%
    \gplgaddtomacro\gplfronttext{%
      \csname LTb\endcsname
      \put(209,4776){\rotatebox{-270}{\makebox(0,0){\strut{}$N$ [$10^3$]}}}%
      \put(4595,4776){\rotatebox{-270}{\makebox(0,0){\strut{}$N_\mathrm{BI}/N_\mathrm{UI}$}}}%
      \put(2385,3177){\makebox(0,0){\strut{}$M_{\ast,\mathrm{run}}$ [$\Msun$]}}%
      \csname LTb\endcsname
      \put(682,3727){\makebox(0,0)[r]{\strut{}$0$}}%
      \put(682,3937){\makebox(0,0)[r]{\strut{}$1$}}%
      \put(682,4147){\makebox(0,0)[r]{\strut{}$2$}}%
      \put(682,4357){\makebox(0,0)[r]{\strut{}$3$}}%
      \put(682,4567){\makebox(0,0)[r]{\strut{}$4$}}%
      \put(682,4777){\makebox(0,0)[r]{\strut{}$5$}}%
      \put(682,4986){\makebox(0,0)[r]{\strut{}$6$}}%
      \put(682,5196){\makebox(0,0)[r]{\strut{}$7$}}%
      \put(682,5406){\makebox(0,0)[r]{\strut{}$8$}}%
      \put(682,5616){\makebox(0,0)[r]{\strut{}$9$}}%
      \put(682,5826){\makebox(0,0)[r]{\strut{}$10$}}%
      \put(909,3507){\makebox(0,0){\strut{}$0.2$}}%
      \put(1290,3507){\makebox(0,0){\strut{}$0.3$}}%
      \put(1671,3507){\makebox(0,0){\strut{}$0.4$}}%
      \put(2052,3507){\makebox(0,0){\strut{}$0.5$}}%
      \put(2433,3507){\makebox(0,0){\strut{}$0.6$}}%
      \put(2814,3507){\makebox(0,0){\strut{}$0.7$}}%
      \put(3195,3507){\makebox(0,0){\strut{}$0.8$}}%
      \put(3576,3507){\makebox(0,0){\strut{}$0.9$}}%
      \put(3957,3507){\makebox(0,0){\strut{}$1$}}%
      \put(4089,3727){\makebox(0,0)[l]{\strut{}$0$}}%
      \put(4089,3937){\makebox(0,0)[l]{\strut{}$2$}}%
      \put(4089,4147){\makebox(0,0)[l]{\strut{}$4$}}%
      \put(4089,4357){\makebox(0,0)[l]{\strut{}$6$}}%
      \put(4089,4567){\makebox(0,0)[l]{\strut{}$8$}}%
      \put(4089,4777){\makebox(0,0)[l]{\strut{}$10$}}%
      \put(4089,4986){\makebox(0,0)[l]{\strut{}$12$}}%
      \put(4089,5196){\makebox(0,0)[l]{\strut{}$14$}}%
      \put(4089,5406){\makebox(0,0)[l]{\strut{}$16$}}%
      \put(4089,5616){\makebox(0,0)[l]{\strut{}$18$}}%
      \put(4089,5826){\makebox(0,0)[l]{\strut{}$20$}}%
    }%
    \gplgaddtomacro\gplbacktext{%
      \csname LTb\endcsname
      \put(5287,5616){\makebox(0,0)[l]{\strut{}(D)}}%
      \put(6545,5406){\makebox(0,0)[l]{\strut{}$\MWDf=1.5\Msun$}}%
    }%
    \gplgaddtomacro\gplfronttext{%
      \csname LTb\endcsname
      \put(5311,4776){\rotatebox{-270}{\makebox(0,0){\strut{}$N$ [$10^3$]}}}%
      \put(9697,4776){\rotatebox{-270}{\makebox(0,0){\strut{}$N_\mathrm{BI}/N_\mathrm{UI}$}}}%
      \put(7487,3177){\makebox(0,0){\strut{}$M_{\ast,\mathrm{run}}$ [$\Msun$]}}%
      \csname LTb\endcsname
      \put(5784,3727){\makebox(0,0)[r]{\strut{}$0$}}%
      \put(5784,3937){\makebox(0,0)[r]{\strut{}$1$}}%
      \put(5784,4147){\makebox(0,0)[r]{\strut{}$2$}}%
      \put(5784,4357){\makebox(0,0)[r]{\strut{}$3$}}%
      \put(5784,4567){\makebox(0,0)[r]{\strut{}$4$}}%
      \put(5784,4777){\makebox(0,0)[r]{\strut{}$5$}}%
      \put(5784,4986){\makebox(0,0)[r]{\strut{}$6$}}%
      \put(5784,5196){\makebox(0,0)[r]{\strut{}$7$}}%
      \put(5784,5406){\makebox(0,0)[r]{\strut{}$8$}}%
      \put(5784,5616){\makebox(0,0)[r]{\strut{}$9$}}%
      \put(5784,5826){\makebox(0,0)[r]{\strut{}$10$}}%
      \put(6011,3507){\makebox(0,0){\strut{}$0.2$}}%
      \put(6392,3507){\makebox(0,0){\strut{}$0.3$}}%
      \put(6773,3507){\makebox(0,0){\strut{}$0.4$}}%
      \put(7154,3507){\makebox(0,0){\strut{}$0.5$}}%
      \put(7535,3507){\makebox(0,0){\strut{}$0.6$}}%
      \put(7916,3507){\makebox(0,0){\strut{}$0.7$}}%
      \put(8297,3507){\makebox(0,0){\strut{}$0.8$}}%
      \put(8678,3507){\makebox(0,0){\strut{}$0.9$}}%
      \put(9059,3507){\makebox(0,0){\strut{}$1$}}%
      \put(9191,3727){\makebox(0,0)[l]{\strut{}$0$}}%
      \put(9191,3937){\makebox(0,0)[l]{\strut{}$2$}}%
      \put(9191,4147){\makebox(0,0)[l]{\strut{}$4$}}%
      \put(9191,4357){\makebox(0,0)[l]{\strut{}$6$}}%
      \put(9191,4567){\makebox(0,0)[l]{\strut{}$8$}}%
      \put(9191,4777){\makebox(0,0)[l]{\strut{}$10$}}%
      \put(9191,4986){\makebox(0,0)[l]{\strut{}$12$}}%
      \put(9191,5196){\makebox(0,0)[l]{\strut{}$14$}}%
      \put(9191,5406){\makebox(0,0)[l]{\strut{}$16$}}%
      \put(9191,5616){\makebox(0,0)[l]{\strut{}$18$}}%
      \put(9191,5826){\makebox(0,0)[l]{\strut{}$20$}}%
    }%
    \gplgaddtomacro\gplbacktext{%
    }%
    \gplgaddtomacro\gplfronttext{%
      \csname LTb\endcsname
      \put(1122,2630){\makebox(0,0)[r]{\strut{}UII}}%
      \csname LTb\endcsname
      \put(1122,2410){\makebox(0,0)[r]{\strut{}BII}}%
      \csname LTb\endcsname
      \put(2241,2630){\makebox(0,0)[r]{\strut{}UI}}%
      \csname LTb\endcsname
      \put(2241,2410){\makebox(0,0)[r]{\strut{}BI}}%
      \csname LTb\endcsname
      \put(3360,2630){\makebox(0,0)[r]{\strut{}BI/UI}}%
    }%
    \gplgaddtomacro\gplbacktext{%
    }%
    \gplgaddtomacro\gplfronttext{%
    }%
    \gplbacktext
    \put(0,0){\includegraphics[width={510.20bp},height={453.50bp}]{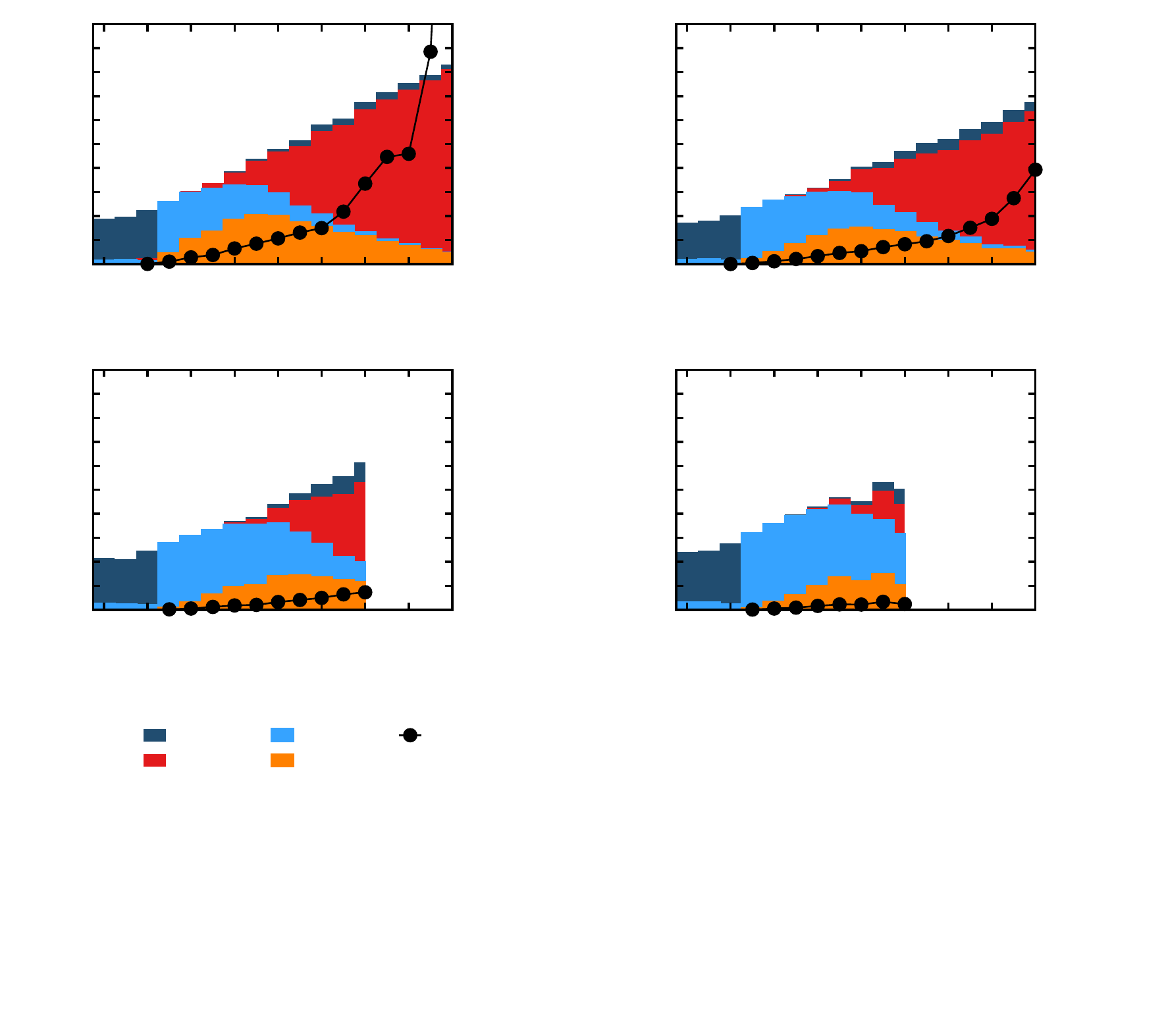}}%
    \gplfronttext
  \end{picture}%
\endgroup

%% file: bound-unbound-mass-fiducial-ratios-tau-fig2.tex
\begingroup
  \makeatletter
  \providecommand\color[2][]{%
    \GenericError{(gnuplot) \space\space\space\@spaces}{%
      Package color not loaded in conjunction with
      terminal option `colourtext'%
    }{See the gnuplot documentation for explanation.%
    }{Either use 'blacktext' in gnuplot or load the package
      color.sty in LaTeX.}%
    \renewcommand\color[2][]{}%
  }%
  \providecommand\includegraphics[2][]{%
    \GenericError{(gnuplot) \space\space\space\@spaces}{%
      Package graphicx or graphics not loaded%
    }{See the gnuplot documentation for explanation.%
    }{The gnuplot epslatex terminal needs graphicx.sty or graphics.sty.}%
    \renewcommand\includegraphics[2][]{}%
  }%
  \providecommand\rotatebox[2]{#2}%
  \@ifundefined{ifGPcolor}{%
    \newif\ifGPcolor
    \GPcolortrue
  }{}%
  \@ifundefined{ifGPblacktext}{%
    \newif\ifGPblacktext
    \GPblacktextfalse
  }{}%
  \let\gplgaddtomacro\g@addto@macro
  \gdef\gplbacktext{}%
  \gdef\gplfronttext{}%
  \makeatother
  \ifGPblacktext
    \def\colorrgb#1{}%
    \def\colorgray#1{}%
  \else
    \ifGPcolor
      \def\colorrgb#1{\color[rgb]{#1}}%
      \def\colorgray#1{\color[gray]{#1}}%
      \expandafter\def\csname LTw\endcsname{\color{white}}%
      \expandafter\def\csname LTb\endcsname{\color{black}}%
      \expandafter\def\csname LTa\endcsname{\color{black}}%
      \expandafter\def\csname LT0\endcsname{\color[rgb]{1,0,0}}%
      \expandafter\def\csname LT1\endcsname{\color[rgb]{0,1,0}}%
      \expandafter\def\csname LT2\endcsname{\color[rgb]{0,0,1}}%
      \expandafter\def\csname LT3\endcsname{\color[rgb]{1,0,1}}%
      \expandafter\def\csname LT4\endcsname{\color[rgb]{0,1,1}}%
      \expandafter\def\csname LT5\endcsname{\color[rgb]{1,1,0}}%
      \expandafter\def\csname LT6\endcsname{\color[rgb]{0,0,0}}%
      \expandafter\def\csname LT7\endcsname{\color[rgb]{1,0.3,0}}%
      \expandafter\def\csname LT8\endcsname{\color[rgb]{0.5,0.5,0.5}}%
    \else
      \def\colorrgb#1{\color{black}}%
      \def\colorgray#1{\color[gray]{#1}}%
      \expandafter\def\csname LTw\endcsname{\color{white}}%
      \expandafter\def\csname LTb\endcsname{\color{black}}%
      \expandafter\def\csname LTa\endcsname{\color{black}}%
      \expandafter\def\csname LT0\endcsname{\color{black}}%
      \expandafter\def\csname LT1\endcsname{\color{black}}%
      \expandafter\def\csname LT2\endcsname{\color{black}}%
      \expandafter\def\csname LT3\endcsname{\color{black}}%
      \expandafter\def\csname LT4\endcsname{\color{black}}%
      \expandafter\def\csname LT5\endcsname{\color{black}}%
      \expandafter\def\csname LT6\endcsname{\color{black}}%
      \expandafter\def\csname LT7\endcsname{\color{black}}%
      \expandafter\def\csname LT8\endcsname{\color{black}}%
    \fi
  \fi
    \setlength{\unitlength}{0.0500bp}%
    \ifx\gptboxheight\undefined%
      \newlength{\gptboxheight}%
      \newlength{\gptboxwidth}%
      \newsavebox{\gptboxtext}%
    \fi%
    \setlength{\fboxrule}{0.5pt}%
    \setlength{\fboxsep}{1pt}%
    \definecolor{tbcol}{rgb}{1,1,1}%
\begin{picture}(4988.00,3968.00)%
    \gplgaddtomacro\gplbacktext{%
    }%
    \gplgaddtomacro\gplfronttext{%
      \csname LTb\endcsname
      \put(209,2445){\rotatebox{-270}{\makebox(0,0){\strut{}$N_\mathrm{BI}/N_\mathrm{UI}$}}}%
      \put(2834,594){\makebox(0,0){\strut{}$M_{\ast,\mathrm{run}}$ [$\Msun$]}}%
      \csname LTb\endcsname
      \put(1849,393){\makebox(0,0)[r]{\strut{}M10}}%
      \csname LTb\endcsname
      \put(1849,173){\makebox(0,0)[r]{\strut{}M11}}%
      \csname LTb\endcsname
      \put(2770,393){\makebox(0,0)[r]{\strut{}M12}}%
      \csname LTb\endcsname
      \put(2770,173){\makebox(0,0)[r]{\strut{}M13}}%
      \csname LTb\endcsname
      \put(3691,393){\makebox(0,0)[r]{\strut{}M14}}%
      \csname LTb\endcsname
      \put(3691,173){\makebox(0,0)[r]{\strut{}M15}}%
      \csname LTb\endcsname
      \put(946,1144){\makebox(0,0)[r]{\strut{}$0.01$}}%
      \put(946,1848){\makebox(0,0)[r]{\strut{}$0.1$}}%
      \put(946,2551){\makebox(0,0)[r]{\strut{}$1$}}%
      \put(946,3255){\makebox(0,0)[r]{\strut{}$10$}}%
      \put(1184,924){\makebox(0,0){\strut{}$0.2$}}%
      \put(1610,924){\makebox(0,0){\strut{}$0.3$}}%
      \put(2036,924){\makebox(0,0){\strut{}$0.4$}}%
      \put(2462,924){\makebox(0,0){\strut{}$0.5$}}%
      \put(2888,924){\makebox(0,0){\strut{}$0.6$}}%
      \put(3314,924){\makebox(0,0){\strut{}$0.7$}}%
      \put(3739,924){\makebox(0,0){\strut{}$0.8$}}%
      \put(4165,924){\makebox(0,0){\strut{}$0.9$}}%
      \put(4591,924){\makebox(0,0){\strut{}$1$}}%
    }%
    \gplbacktext
    \put(0,0){\includegraphics[width={249.40bp},height={198.40bp}]{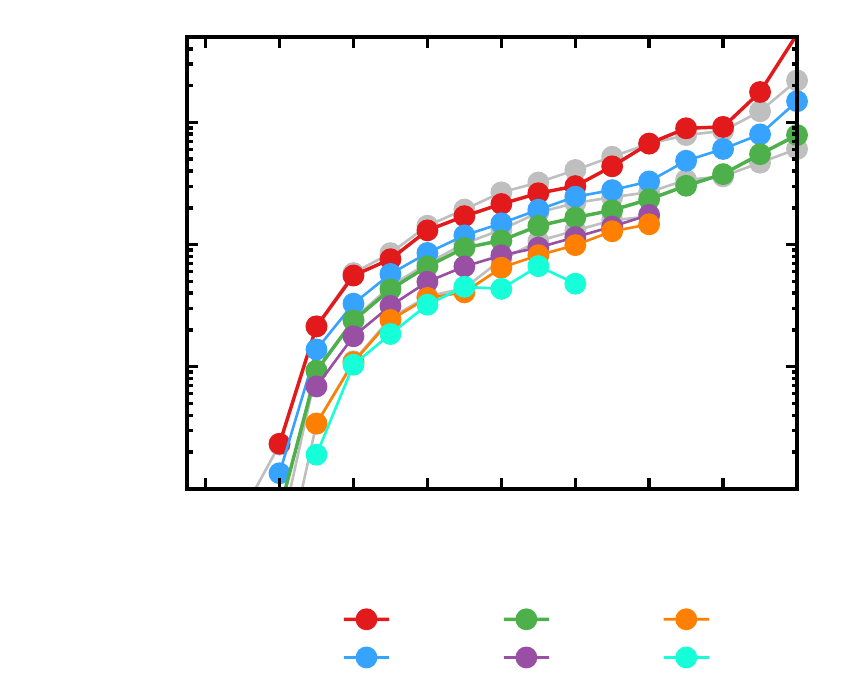}}%
    \gplfronttext
  \end{picture}%
\endgroup

%% file: bound-unbound-mass-M14-fiducial-weighted-fig2.tex
\begingroup
  \makeatletter
  \providecommand\color[2][]{%
    \GenericError{(gnuplot) \space\space\space\@spaces}{%
      Package color not loaded in conjunction with
      terminal option `colourtext'%
    }{See the gnuplot documentation for explanation.%
    }{Either use 'blacktext' in gnuplot or load the package
      color.sty in LaTeX.}%
    \renewcommand\color[2][]{}%
  }%
  \providecommand\includegraphics[2][]{%
    \GenericError{(gnuplot) \space\space\space\@spaces}{%
      Package graphicx or graphics not loaded%
    }{See the gnuplot documentation for explanation.%
    }{The gnuplot epslatex terminal needs graphicx.sty or graphics.sty.}%
    \renewcommand\includegraphics[2][]{}%
  }%
  \providecommand\rotatebox[2]{#2}%
  \@ifundefined{ifGPcolor}{%
    \newif\ifGPcolor
    \GPcolortrue
  }{}%
  \@ifundefined{ifGPblacktext}{%
    \newif\ifGPblacktext
    \GPblacktextfalse
  }{}%
  \let\gplgaddtomacro\g@addto@macro
  \gdef\gplbacktext{}%
  \gdef\gplfronttext{}%
  \makeatother
  \ifGPblacktext
    \def\colorrgb#1{}%
    \def\colorgray#1{}%
  \else
    \ifGPcolor
      \def\colorrgb#1{\color[rgb]{#1}}%
      \def\colorgray#1{\color[gray]{#1}}%
      \expandafter\def\csname LTw\endcsname{\color{white}}%
      \expandafter\def\csname LTb\endcsname{\color{black}}%
      \expandafter\def\csname LTa\endcsname{\color{black}}%
      \expandafter\def\csname LT0\endcsname{\color[rgb]{1,0,0}}%
      \expandafter\def\csname LT1\endcsname{\color[rgb]{0,1,0}}%
      \expandafter\def\csname LT2\endcsname{\color[rgb]{0,0,1}}%
      \expandafter\def\csname LT3\endcsname{\color[rgb]{1,0,1}}%
      \expandafter\def\csname LT4\endcsname{\color[rgb]{0,1,1}}%
      \expandafter\def\csname LT5\endcsname{\color[rgb]{1,1,0}}%
      \expandafter\def\csname LT6\endcsname{\color[rgb]{0,0,0}}%
      \expandafter\def\csname LT7\endcsname{\color[rgb]{1,0.3,0}}%
      \expandafter\def\csname LT8\endcsname{\color[rgb]{0.5,0.5,0.5}}%
    \else
      \def\colorrgb#1{\color{black}}%
      \def\colorgray#1{\color[gray]{#1}}%
      \expandafter\def\csname LTw\endcsname{\color{white}}%
      \expandafter\def\csname LTb\endcsname{\color{black}}%
      \expandafter\def\csname LTa\endcsname{\color{black}}%
      \expandafter\def\csname LT0\endcsname{\color{black}}%
      \expandafter\def\csname LT1\endcsname{\color{black}}%
      \expandafter\def\csname LT2\endcsname{\color{black}}%
      \expandafter\def\csname LT3\endcsname{\color{black}}%
      \expandafter\def\csname LT4\endcsname{\color{black}}%
      \expandafter\def\csname LT5\endcsname{\color{black}}%
      \expandafter\def\csname LT6\endcsname{\color{black}}%
      \expandafter\def\csname LT7\endcsname{\color{black}}%
      \expandafter\def\csname LT8\endcsname{\color{black}}%
    \fi
  \fi
    \setlength{\unitlength}{0.0500bp}%
    \ifx\gptboxheight\undefined%
      \newlength{\gptboxheight}%
      \newlength{\gptboxwidth}%
      \newsavebox{\gptboxtext}%
    \fi%
    \setlength{\fboxrule}{0.5pt}%
    \setlength{\fboxsep}{1pt}%
    \definecolor{tbcol}{rgb}{1,1,1}%
\begin{picture}(4988.00,3968.00)%
    \gplgaddtomacro\gplbacktext{%
    }%
    \gplgaddtomacro\gplfronttext{%
      \csname LTb\endcsname
      \put(209,2445){\rotatebox{-270}{\makebox(0,0){\strut{}$N$ [$10^3$]}}}%
      \put(4481,2445){\rotatebox{-270}{\makebox(0,0){\strut{}$N_\mathrm{BI}/N_\mathrm{UI}$}}}%
      \put(2130,594){\makebox(0,0){\strut{}$M_{\ast,\mathrm{run}}$ [$\Msun$]}}%
      \csname LTb\endcsname
      \put(1112,393){\makebox(0,0)[r]{\strut{}UII}}%
      \csname LTb\endcsname
      \put(1112,173){\makebox(0,0)[r]{\strut{}BII}}%
      \csname LTb\endcsname
      \put(2231,393){\makebox(0,0)[r]{\strut{}UI}}%
      \csname LTb\endcsname
      \put(2231,173){\makebox(0,0)[r]{\strut{}BI}}%
      \csname LTb\endcsname
      \put(3350,393){\makebox(0,0)[r]{\strut{}BI/UI}}%
      \csname LTb\endcsname
      \put(550,1144){\makebox(0,0)[r]{\strut{}$0$}}%
      \put(550,2012){\makebox(0,0)[r]{\strut{}$1$}}%
      \put(550,2879){\makebox(0,0)[r]{\strut{}$2$}}%
      \put(550,3747){\makebox(0,0)[r]{\strut{}$3$}}%
      \put(798,924){\makebox(0,0){\strut{}$0.2$}}%
      \put(1261,924){\makebox(0,0){\strut{}$0.3$}}%
      \put(1725,924){\makebox(0,0){\strut{}$0.4$}}%
      \put(2188,924){\makebox(0,0){\strut{}$0.5$}}%
      \put(2652,924){\makebox(0,0){\strut{}$0.6$}}%
      \put(3115,924){\makebox(0,0){\strut{}$0.7$}}%
      \put(3579,924){\makebox(0,0){\strut{}$0.8$}}%
      \put(3711,1144){\makebox(0,0)[l]{\strut{}$0.0$}}%
      \put(3711,1404){\makebox(0,0)[l]{\strut{}$2.0$}}%
      \put(3711,1665){\makebox(0,0)[l]{\strut{}$4.0$}}%
      \put(3711,1925){\makebox(0,0)[l]{\strut{}$6.0$}}%
      \put(3711,2185){\makebox(0,0)[l]{\strut{}$8.0$}}%
      \put(3711,2446){\makebox(0,0)[l]{\strut{}$10.0$}}%
      \put(3711,2706){\makebox(0,0)[l]{\strut{}$12.0$}}%
      \put(3711,2966){\makebox(0,0)[l]{\strut{}$14.0$}}%
      \put(3711,3226){\makebox(0,0)[l]{\strut{}$16.0$}}%
      \put(3711,3487){\makebox(0,0)[l]{\strut{}$18.0$}}%
      \put(3711,3747){\makebox(0,0)[l]{\strut{}$20.0$}}%
    }%
    \gplbacktext
    \put(0,0){\includegraphics[width={249.40bp},height={198.40bp}]{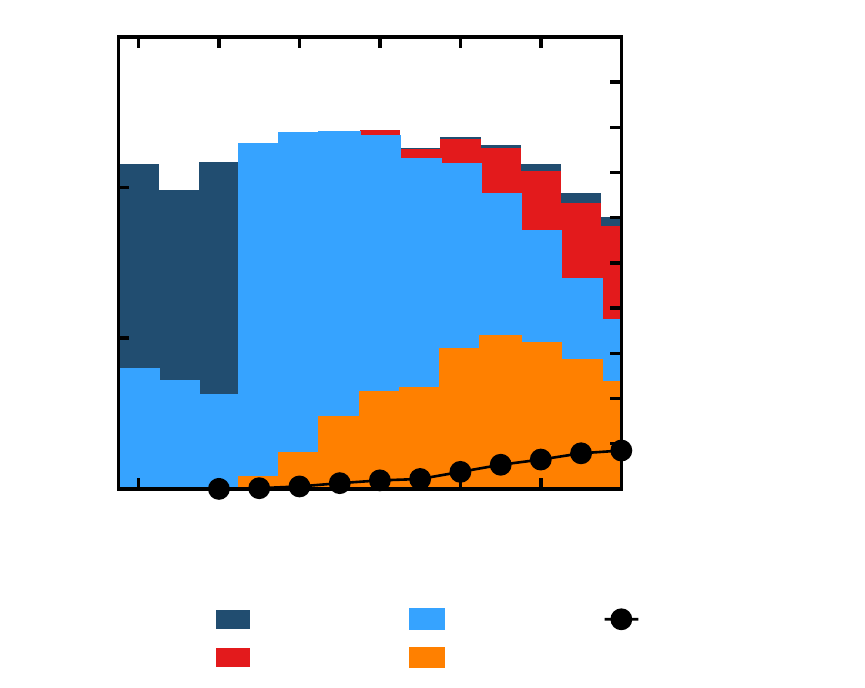}}%
    \gplfronttext
  \end{picture}%
\endgroup

%% file: PM-14-fig2.tex
\begingroup
  \makeatletter
  \providecommand\color[2][]{%
    \GenericError{(gnuplot) \space\space\space\@spaces}{%
      Package color not loaded in conjunction with
      terminal option `colourtext'%
    }{See the gnuplot documentation for explanation.%
    }{Either use 'blacktext' in gnuplot or load the package
      color.sty in LaTeX.}%
    \renewcommand\color[2][]{}%
  }%
  \providecommand\includegraphics[2][]{%
    \GenericError{(gnuplot) \space\space\space\@spaces}{%
      Package graphicx or graphics not loaded%
    }{See the gnuplot documentation for explanation.%
    }{The gnuplot epslatex terminal needs graphicx.sty or graphics.sty.}%
    \renewcommand\includegraphics[2][]{}%
  }%
  \providecommand\rotatebox[2]{#2}%
  \@ifundefined{ifGPcolor}{%
    \newif\ifGPcolor
    \GPcolortrue
  }{}%
  \@ifundefined{ifGPblacktext}{%
    \newif\ifGPblacktext
    \GPblacktextfalse
  }{}%
  \let\gplgaddtomacro\g@addto@macro
  \gdef\gplbacktext{}%
  \gdef\gplfronttext{}%
  \makeatother
  \ifGPblacktext
    \def\colorrgb#1{}%
    \def\colorgray#1{}%
  \else
    \ifGPcolor
      \def\colorrgb#1{\color[rgb]{#1}}%
      \def\colorgray#1{\color[gray]{#1}}%
      \expandafter\def\csname LTw\endcsname{\color{white}}%
      \expandafter\def\csname LTb\endcsname{\color{black}}%
      \expandafter\def\csname LTa\endcsname{\color{black}}%
      \expandafter\def\csname LT0\endcsname{\color[rgb]{1,0,0}}%
      \expandafter\def\csname LT1\endcsname{\color[rgb]{0,1,0}}%
      \expandafter\def\csname LT2\endcsname{\color[rgb]{0,0,1}}%
      \expandafter\def\csname LT3\endcsname{\color[rgb]{1,0,1}}%
      \expandafter\def\csname LT4\endcsname{\color[rgb]{0,1,1}}%
      \expandafter\def\csname LT5\endcsname{\color[rgb]{1,1,0}}%
      \expandafter\def\csname LT6\endcsname{\color[rgb]{0,0,0}}%
      \expandafter\def\csname LT7\endcsname{\color[rgb]{1,0.3,0}}%
      \expandafter\def\csname LT8\endcsname{\color[rgb]{0.5,0.5,0.5}}%
    \else
      \def\colorrgb#1{\color{black}}%
      \def\colorgray#1{\color[gray]{#1}}%
      \expandafter\def\csname LTw\endcsname{\color{white}}%
      \expandafter\def\csname LTb\endcsname{\color{black}}%
      \expandafter\def\csname LTa\endcsname{\color{black}}%
      \expandafter\def\csname LT0\endcsname{\color{black}}%
      \expandafter\def\csname LT1\endcsname{\color{black}}%
      \expandafter\def\csname LT2\endcsname{\color{black}}%
      \expandafter\def\csname LT3\endcsname{\color{black}}%
      \expandafter\def\csname LT4\endcsname{\color{black}}%
      \expandafter\def\csname LT5\endcsname{\color{black}}%
      \expandafter\def\csname LT6\endcsname{\color{black}}%
      \expandafter\def\csname LT7\endcsname{\color{black}}%
      \expandafter\def\csname LT8\endcsname{\color{black}}%
    \fi
  \fi
    \setlength{\unitlength}{0.0500bp}%
    \ifx\gptboxheight\undefined%
      \newlength{\gptboxheight}%
      \newlength{\gptboxwidth}%
      \newsavebox{\gptboxtext}%
    \fi%
    \setlength{\fboxrule}{0.5pt}%
    \setlength{\fboxsep}{1pt}%
    \definecolor{tbcol}{rgb}{1,1,1}%
\begin{picture}(4988.00,4534.00)%
    \gplgaddtomacro\gplbacktext{%
    }%
    \gplgaddtomacro\gplfronttext{%
      \csname LTb\endcsname
      \put(341,2728){\rotatebox{-270}{\makebox(0,0){\strut{}$N$}}}%
      \put(2900,594){\makebox(0,0){\strut{}PM [$\mathrm{mas}\,\mathrm{yr}^{-1}$]}}%
      \csname LTb\endcsname
      \put(2002,393){\makebox(0,0)[r]{\strut{}\small{$D<25\mkpc$}}}%
      \csname LTb\endcsname
      \put(2002,173){\makebox(0,0)[r]{\strut{}\small{$D<20\mkpc$}}}%
      \csname LTb\endcsname
      \put(3200,393){\makebox(0,0)[r]{\strut{}\small{$D<15\mkpc$}}}%
      \csname LTb\endcsname
      \put(3200,173){\makebox(0,0)[r]{\strut{}\small{$D<10\mkpc$}}}%
      \csname LTb\endcsname
      \put(4398,393){\makebox(0,0)[r]{\strut{}\small{$D<5\mkpc$}}}%
      \csname LTb\endcsname
      \put(1078,1180){\makebox(0,0)[r]{\strut{}$1$}}%
      \put(1078,1959){\makebox(0,0)[r]{\strut{}$10$}}%
      \put(1078,2738){\makebox(0,0)[r]{\strut{}$100$}}%
      \put(1078,3517){\makebox(0,0)[r]{\strut{}$1000$}}%
      \put(1078,4296){\makebox(0,0)[r]{\strut{}$10000$}}%
      \put(2296,924){\makebox(0,0){\strut{}$10$}}%
      \put(3850,924){\makebox(0,0){\strut{}$100$}}%
      \colorrgb{0.75,0.75,0.75}
      \put(2029,1461){\rotatebox{90}{\makebox(0,0)[l]{\strut{}US708 and J2050}}}%
      \put(4380,2412){\rotatebox{90}{\makebox(0,0)[l]{\strut{}LP 398-9}}}%
    }%
    \gplbacktext
    \put(0,0){\includegraphics[width={249.40bp},height={226.70bp}]{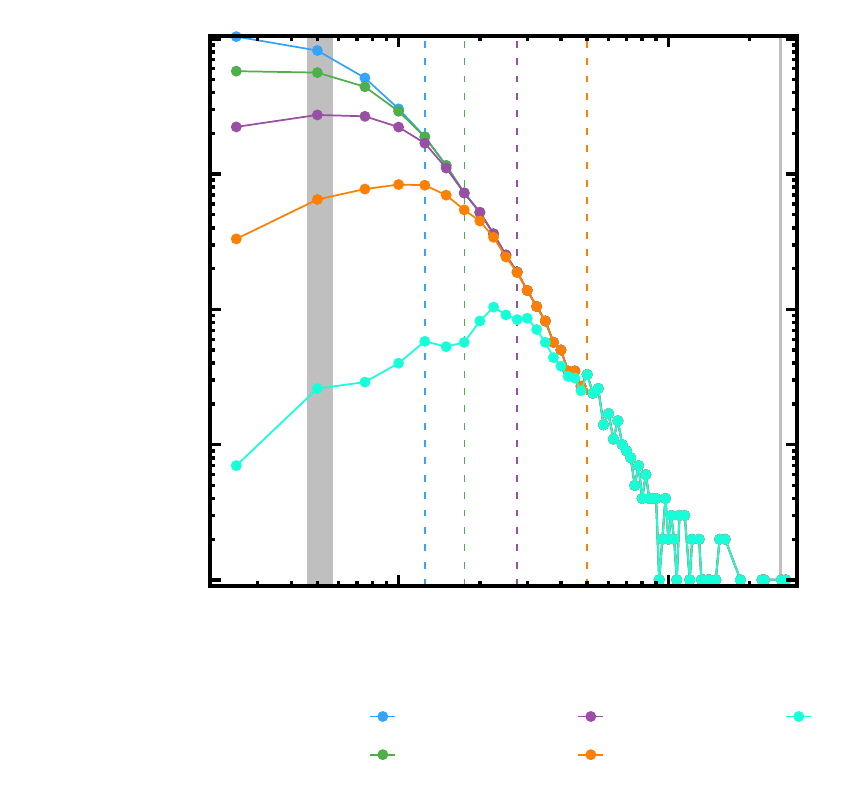}}%
    \gplfronttext
  \end{picture}%
\endgroup

%% file: RV-all-masses-fig2.tex
\begingroup
  \makeatletter
  \providecommand\color[2][]{%
    \GenericError{(gnuplot) \space\space\space\@spaces}{%
      Package color not loaded in conjunction with
      terminal option `colourtext'%
    }{See the gnuplot documentation for explanation.%
    }{Either use 'blacktext' in gnuplot or load the package
      color.sty in LaTeX.}%
    \renewcommand\color[2][]{}%
  }%
  \providecommand\includegraphics[2][]{%
    \GenericError{(gnuplot) \space\space\space\@spaces}{%
      Package graphicx or graphics not loaded%
    }{See the gnuplot documentation for explanation.%
    }{The gnuplot epslatex terminal needs graphicx.sty or graphics.sty.}%
    \renewcommand\includegraphics[2][]{}%
  }%
  \providecommand\rotatebox[2]{#2}%
  \@ifundefined{ifGPcolor}{%
    \newif\ifGPcolor
    \GPcolortrue
  }{}%
  \@ifundefined{ifGPblacktext}{%
    \newif\ifGPblacktext
    \GPblacktextfalse
  }{}%
  \let\gplgaddtomacro\g@addto@macro
  \gdef\gplbacktext{}%
  \gdef\gplfronttext{}%
  \makeatother
  \ifGPblacktext
    \def\colorrgb#1{}%
    \def\colorgray#1{}%
  \else
    \ifGPcolor
      \def\colorrgb#1{\color[rgb]{#1}}%
      \def\colorgray#1{\color[gray]{#1}}%
      \expandafter\def\csname LTw\endcsname{\color{white}}%
      \expandafter\def\csname LTb\endcsname{\color{black}}%
      \expandafter\def\csname LTa\endcsname{\color{black}}%
      \expandafter\def\csname LT0\endcsname{\color[rgb]{1,0,0}}%
      \expandafter\def\csname LT1\endcsname{\color[rgb]{0,1,0}}%
      \expandafter\def\csname LT2\endcsname{\color[rgb]{0,0,1}}%
      \expandafter\def\csname LT3\endcsname{\color[rgb]{1,0,1}}%
      \expandafter\def\csname LT4\endcsname{\color[rgb]{0,1,1}}%
      \expandafter\def\csname LT5\endcsname{\color[rgb]{1,1,0}}%
      \expandafter\def\csname LT6\endcsname{\color[rgb]{0,0,0}}%
      \expandafter\def\csname LT7\endcsname{\color[rgb]{1,0.3,0}}%
      \expandafter\def\csname LT8\endcsname{\color[rgb]{0.5,0.5,0.5}}%
    \else
      \def\colorrgb#1{\color{black}}%
      \def\colorgray#1{\color[gray]{#1}}%
      \expandafter\def\csname LTw\endcsname{\color{white}}%
      \expandafter\def\csname LTb\endcsname{\color{black}}%
      \expandafter\def\csname LTa\endcsname{\color{black}}%
      \expandafter\def\csname LT0\endcsname{\color{black}}%
      \expandafter\def\csname LT1\endcsname{\color{black}}%
      \expandafter\def\csname LT2\endcsname{\color{black}}%
      \expandafter\def\csname LT3\endcsname{\color{black}}%
      \expandafter\def\csname LT4\endcsname{\color{black}}%
      \expandafter\def\csname LT5\endcsname{\color{black}}%
      \expandafter\def\csname LT6\endcsname{\color{black}}%
      \expandafter\def\csname LT7\endcsname{\color{black}}%
      \expandafter\def\csname LT8\endcsname{\color{black}}%
    \fi
  \fi
    \setlength{\unitlength}{0.0500bp}%
    \ifx\gptboxheight\undefined%
      \newlength{\gptboxheight}%
      \newlength{\gptboxwidth}%
      \newsavebox{\gptboxtext}%
    \fi%
    \setlength{\fboxrule}{0.5pt}%
    \setlength{\fboxsep}{1pt}%
    \definecolor{tbcol}{rgb}{1,1,1}%
\begin{picture}(4988.00,11338.00)%
    \gplgaddtomacro\gplbacktext{%
      \csname LTb\endcsname
      \put(375,10983){\makebox(0,0){\strut{}(A)}}%
    }%
    \gplgaddtomacro\gplfronttext{%
      \csname LTb\endcsname
      \put(209,10445){\rotatebox{-270}{\makebox(0,0){\strut{}$N$}}}%
      \put(2834,9224){\makebox(0,0){\strut{}$v_\mathrm{r}$ [\kms]}}%
      \csname LTb\endcsname
      \put(946,9774){\makebox(0,0)[r]{\strut{}$0$}}%
      \put(946,10043){\makebox(0,0)[r]{\strut{}$200$}}%
      \put(946,10311){\makebox(0,0)[r]{\strut{}$400$}}%
      \put(946,10580){\makebox(0,0)[r]{\strut{}$600$}}%
      \put(946,10848){\makebox(0,0)[r]{\strut{}$800$}}%
      \put(946,11117){\makebox(0,0)[r]{\strut{}$1000$}}%
      \put(1078,9554){\makebox(0,0){\strut{}$-1000$}}%
      \put(1781,9554){\makebox(0,0){\strut{}$-500$}}%
      \put(2483,9554){\makebox(0,0){\strut{}$0$}}%
      \put(3186,9554){\makebox(0,0){\strut{}$500$}}%
      \put(3888,9554){\makebox(0,0){\strut{}$1000$}}%
      \put(4591,9554){\makebox(0,0){\strut{}$1500$}}%
      \colorrgb{0.75,0.75,0.75}
      \put(3912,10311){\rotatebox{90}{\makebox(0,0)[l]{\strut{}US708}}}%
      \put(2371,10311){\rotatebox{90}{\makebox(0,0)[l]{\strut{}LP 398-9}}}%
      \put(1908,10311){\rotatebox{90}{\makebox(0,0)[l]{\strut{}J2050}}}%
    }%
    \gplgaddtomacro\gplbacktext{%
      \csname LTb\endcsname
      \put(375,8716){\makebox(0,0){\strut{}(B)}}%
    }%
    \gplgaddtomacro\gplfronttext{%
      \csname LTb\endcsname
      \put(209,8178){\rotatebox{-270}{\makebox(0,0){\strut{}$N$}}}%
      \put(2834,6956){\makebox(0,0){\strut{}$v_\mathrm{r}$ [\kms]}}%
      \csname LTb\endcsname
      \put(946,7506){\makebox(0,0)[r]{\strut{}$0$}}%
      \put(946,7775){\makebox(0,0)[r]{\strut{}$200$}}%
      \put(946,8044){\makebox(0,0)[r]{\strut{}$400$}}%
      \put(946,8312){\makebox(0,0)[r]{\strut{}$600$}}%
      \put(946,8581){\makebox(0,0)[r]{\strut{}$800$}}%
      \put(946,8850){\makebox(0,0)[r]{\strut{}$1000$}}%
      \put(1078,7286){\makebox(0,0){\strut{}$-1000$}}%
      \put(1781,7286){\makebox(0,0){\strut{}$-500$}}%
      \put(2483,7286){\makebox(0,0){\strut{}$0$}}%
      \put(3186,7286){\makebox(0,0){\strut{}$500$}}%
      \put(3888,7286){\makebox(0,0){\strut{}$1000$}}%
      \put(4591,7286){\makebox(0,0){\strut{}$1500$}}%
      \colorrgb{0.75,0.75,0.75}
      \put(3912,8044){\rotatebox{90}{\makebox(0,0)[l]{\strut{}US708}}}%
      \put(2371,8044){\rotatebox{90}{\makebox(0,0)[l]{\strut{}LP 398-9}}}%
      \put(1908,8044){\rotatebox{90}{\makebox(0,0)[l]{\strut{}J2050}}}%
    }%
    \gplgaddtomacro\gplbacktext{%
      \csname LTb\endcsname
      \put(217,6448){\makebox(0,0){\strut{}(C)}}%
    }%
    \gplgaddtomacro\gplfronttext{%
      \csname LTb\endcsname
      \put(209,5910){\rotatebox{-270}{\makebox(0,0){\strut{}$N$}}}%
      \put(2768,4689){\makebox(0,0){\strut{}$v_\mathrm{r}$ [\kms]}}%
      \csname LTb\endcsname
      \put(814,5239){\makebox(0,0)[r]{\strut{}$0$}}%
      \put(814,5575){\makebox(0,0)[r]{\strut{}$200$}}%
      \put(814,5911){\makebox(0,0)[r]{\strut{}$400$}}%
      \put(814,6246){\makebox(0,0)[r]{\strut{}$600$}}%
      \put(814,6582){\makebox(0,0)[r]{\strut{}$800$}}%
      \put(946,5019){\makebox(0,0){\strut{}$-1000$}}%
      \put(1675,5019){\makebox(0,0){\strut{}$-500$}}%
      \put(2404,5019){\makebox(0,0){\strut{}$0$}}%
      \put(3133,5019){\makebox(0,0){\strut{}$500$}}%
      \put(3862,5019){\makebox(0,0){\strut{}$1000$}}%
      \put(4591,5019){\makebox(0,0){\strut{}$1500$}}%
      \colorrgb{0.75,0.75,0.75}
      \put(3887,5776){\rotatebox{90}{\makebox(0,0)[l]{\strut{}US708}}}%
      \put(2287,5776){\rotatebox{90}{\makebox(0,0)[l]{\strut{}LP 398-9}}}%
      \put(1808,5776){\rotatebox{90}{\makebox(0,0)[l]{\strut{}J2050}}}%
    }%
    \gplgaddtomacro\gplbacktext{%
      \csname LTb\endcsname
      \put(217,4181){\makebox(0,0){\strut{}(D)}}%
    }%
    \gplgaddtomacro\gplfronttext{%
      \csname LTb\endcsname
      \put(209,3643){\rotatebox{-270}{\makebox(0,0){\strut{}$N$}}}%
      \put(2768,2421){\makebox(0,0){\strut{}$v_\mathrm{r}$ [\kms]}}%
      \csname LTb\endcsname
      \put(814,2971){\makebox(0,0)[r]{\strut{}$0$}}%
      \put(814,3307){\makebox(0,0)[r]{\strut{}$200$}}%
      \put(814,3643){\makebox(0,0)[r]{\strut{}$400$}}%
      \put(814,3979){\makebox(0,0)[r]{\strut{}$600$}}%
      \put(814,4315){\makebox(0,0)[r]{\strut{}$800$}}%
      \put(946,2751){\makebox(0,0){\strut{}$-1000$}}%
      \put(1675,2751){\makebox(0,0){\strut{}$-500$}}%
      \put(2404,2751){\makebox(0,0){\strut{}$0$}}%
      \put(3133,2751){\makebox(0,0){\strut{}$500$}}%
      \put(3862,2751){\makebox(0,0){\strut{}$1000$}}%
      \put(4591,2751){\makebox(0,0){\strut{}$1500$}}%
      \colorrgb{0.75,0.75,0.75}
      \put(3887,3509){\rotatebox{90}{\makebox(0,0)[l]{\strut{}US708}}}%
      \put(2287,3509){\rotatebox{90}{\makebox(0,0)[l]{\strut{}LP 398-9}}}%
      \put(1808,3509){\rotatebox{90}{\makebox(0,0)[l]{\strut{}J2050}}}%
    }%
    \gplgaddtomacro\gplbacktext{%
    }%
    \gplgaddtomacro\gplfronttext{%
      \csname LTb\endcsname
      \put(2178,2094){\makebox(0,0)[r]{\strut{}$M_\mathrm{\ast,run} > 0.2\Msun$}}%
      \csname LTb\endcsname
      \put(2178,1874){\makebox(0,0)[r]{\strut{}$M_\mathrm{\ast,run} > 0.3\Msun$}}%
      \csname LTb\endcsname
      \put(2178,1654){\makebox(0,0)[r]{\strut{}$M_\mathrm{\ast,run} > 0.4\Msun$}}%
      \csname LTb\endcsname
      \put(2178,1434){\makebox(0,0)[r]{\strut{}$M_\mathrm{\ast,run} > 0.5\Msun$}}%
      \csname LTb\endcsname
      \put(2178,1214){\makebox(0,0)[r]{\strut{}$M_\mathrm{\ast,run} > 0.6\Msun$}}%
      \csname LTb\endcsname
      \put(2178,994){\makebox(0,0)[r]{\strut{}$M_\mathrm{\ast,run} > 0.7\Msun$}}%
      \csname LTb\endcsname
      \put(2178,774){\makebox(0,0)[r]{\strut{}$M_\mathrm{\ast,run} > 0.8\Msun$}}%
      \csname LTb\endcsname
      \put(2178,554){\makebox(0,0)[r]{\strut{}$M_\mathrm{\ast,run} > 0.9\Msun$}}%
    }%
    \gplbacktext
    \put(0,0){\includegraphics[width={249.40bp},height={566.90bp}]{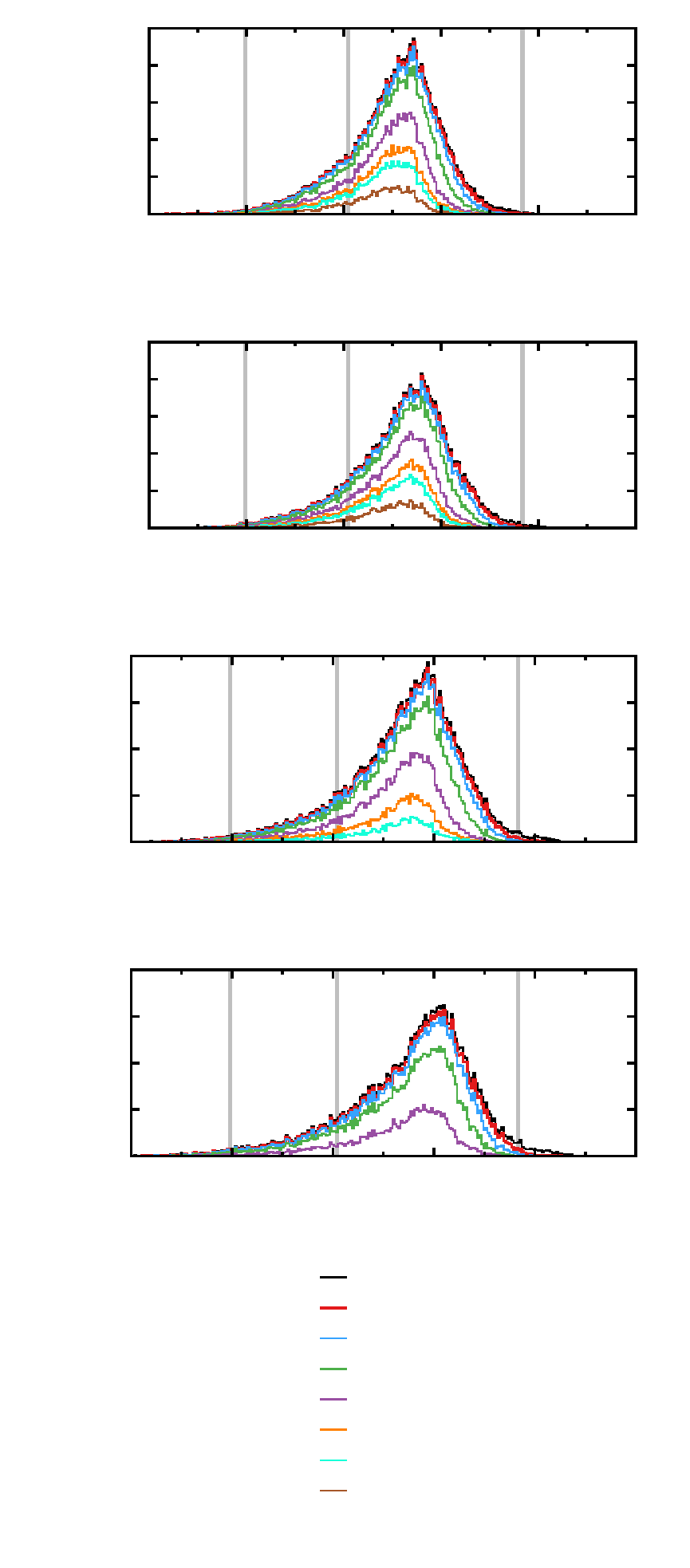}}%
    \gplfronttext
  \end{picture}%
\endgroup

%% file: RV-all-masses-pop-II-fig2.tex
\begingroup
  \makeatletter
  \providecommand\color[2][]{%
    \GenericError{(gnuplot) \space\space\space\@spaces}{%
      Package color not loaded in conjunction with
      terminal option `colourtext'%
    }{See the gnuplot documentation for explanation.%
    }{Either use 'blacktext' in gnuplot or load the package
      color.sty in LaTeX.}%
    \renewcommand\color[2][]{}%
  }%
  \providecommand\includegraphics[2][]{%
    \GenericError{(gnuplot) \space\space\space\@spaces}{%
      Package graphicx or graphics not loaded%
    }{See the gnuplot documentation for explanation.%
    }{The gnuplot epslatex terminal needs graphicx.sty or graphics.sty.}%
    \renewcommand\includegraphics[2][]{}%
  }%
  \providecommand\rotatebox[2]{#2}%
  \@ifundefined{ifGPcolor}{%
    \newif\ifGPcolor
    \GPcolortrue
  }{}%
  \@ifundefined{ifGPblacktext}{%
    \newif\ifGPblacktext
    \GPblacktextfalse
  }{}%
  \let\gplgaddtomacro\g@addto@macro
  \gdef\gplbacktext{}%
  \gdef\gplfronttext{}%
  \makeatother
  \ifGPblacktext
    \def\colorrgb#1{}%
    \def\colorgray#1{}%
  \else
    \ifGPcolor
      \def\colorrgb#1{\color[rgb]{#1}}%
      \def\colorgray#1{\color[gray]{#1}}%
      \expandafter\def\csname LTw\endcsname{\color{white}}%
      \expandafter\def\csname LTb\endcsname{\color{black}}%
      \expandafter\def\csname LTa\endcsname{\color{black}}%
      \expandafter\def\csname LT0\endcsname{\color[rgb]{1,0,0}}%
      \expandafter\def\csname LT1\endcsname{\color[rgb]{0,1,0}}%
      \expandafter\def\csname LT2\endcsname{\color[rgb]{0,0,1}}%
      \expandafter\def\csname LT3\endcsname{\color[rgb]{1,0,1}}%
      \expandafter\def\csname LT4\endcsname{\color[rgb]{0,1,1}}%
      \expandafter\def\csname LT5\endcsname{\color[rgb]{1,1,0}}%
      \expandafter\def\csname LT6\endcsname{\color[rgb]{0,0,0}}%
      \expandafter\def\csname LT7\endcsname{\color[rgb]{1,0.3,0}}%
      \expandafter\def\csname LT8\endcsname{\color[rgb]{0.5,0.5,0.5}}%
    \else
      \def\colorrgb#1{\color{black}}%
      \def\colorgray#1{\color[gray]{#1}}%
      \expandafter\def\csname LTw\endcsname{\color{white}}%
      \expandafter\def\csname LTb\endcsname{\color{black}}%
      \expandafter\def\csname LTa\endcsname{\color{black}}%
      \expandafter\def\csname LT0\endcsname{\color{black}}%
      \expandafter\def\csname LT1\endcsname{\color{black}}%
      \expandafter\def\csname LT2\endcsname{\color{black}}%
      \expandafter\def\csname LT3\endcsname{\color{black}}%
      \expandafter\def\csname LT4\endcsname{\color{black}}%
      \expandafter\def\csname LT5\endcsname{\color{black}}%
      \expandafter\def\csname LT6\endcsname{\color{black}}%
      \expandafter\def\csname LT7\endcsname{\color{black}}%
      \expandafter\def\csname LT8\endcsname{\color{black}}%
    \fi
  \fi
    \setlength{\unitlength}{0.0500bp}%
    \ifx\gptboxheight\undefined%
      \newlength{\gptboxheight}%
      \newlength{\gptboxwidth}%
      \newsavebox{\gptboxtext}%
    \fi%
    \setlength{\fboxrule}{0.5pt}%
    \setlength{\fboxsep}{1pt}%
    \definecolor{tbcol}{rgb}{1,1,1}%
\begin{picture}(4988.00,11338.00)%
    \gplgaddtomacro\gplbacktext{%
      \csname LTb\endcsname
      \put(217,10983){\makebox(0,0){\strut{}(A)}}%
    }%
    \gplgaddtomacro\gplfronttext{%
      \csname LTb\endcsname
      \put(209,10445){\rotatebox{-270}{\makebox(0,0){\strut{}$N$}}}%
      \put(2768,9224){\makebox(0,0){\strut{}$v_\mathrm{r}$ [\kms]}}%
      \csname LTb\endcsname
      \put(814,9774){\makebox(0,0)[r]{\strut{}$0$}}%
      \put(814,9966){\makebox(0,0)[r]{\strut{}$100$}}%
      \put(814,10158){\makebox(0,0)[r]{\strut{}$200$}}%
      \put(814,10350){\makebox(0,0)[r]{\strut{}$300$}}%
      \put(814,10541){\makebox(0,0)[r]{\strut{}$400$}}%
      \put(814,10733){\makebox(0,0)[r]{\strut{}$500$}}%
      \put(814,10925){\makebox(0,0)[r]{\strut{}$600$}}%
      \put(814,11117){\makebox(0,0)[r]{\strut{}$700$}}%
      \put(946,9554){\makebox(0,0){\strut{}$-1000$}}%
      \put(1675,9554){\makebox(0,0){\strut{}$-500$}}%
      \put(2404,9554){\makebox(0,0){\strut{}$0$}}%
      \put(3133,9554){\makebox(0,0){\strut{}$500$}}%
      \put(3862,9554){\makebox(0,0){\strut{}$1000$}}%
      \put(4591,9554){\makebox(0,0){\strut{}$1500$}}%
      \colorrgb{0.75,0.75,0.75}
      \put(3887,10311){\rotatebox{90}{\makebox(0,0)[l]{\strut{}US708}}}%
      \put(2287,10311){\rotatebox{90}{\makebox(0,0)[l]{\strut{}LP 398-9}}}%
      \put(1808,10311){\rotatebox{90}{\makebox(0,0)[l]{\strut{}J2050}}}%
    }%
    \gplgaddtomacro\gplbacktext{%
      \csname LTb\endcsname
      \put(217,8716){\makebox(0,0){\strut{}(B)}}%
    }%
    \gplgaddtomacro\gplfronttext{%
      \csname LTb\endcsname
      \put(209,8178){\rotatebox{-270}{\makebox(0,0){\strut{}$N$}}}%
      \put(2768,6956){\makebox(0,0){\strut{}$v_\mathrm{r}$ [\kms]}}%
      \csname LTb\endcsname
      \put(814,7506){\makebox(0,0)[r]{\strut{}$0$}}%
      \put(814,7775){\makebox(0,0)[r]{\strut{}$100$}}%
      \put(814,8044){\makebox(0,0)[r]{\strut{}$200$}}%
      \put(814,8312){\makebox(0,0)[r]{\strut{}$300$}}%
      \put(814,8581){\makebox(0,0)[r]{\strut{}$400$}}%
      \put(814,8850){\makebox(0,0)[r]{\strut{}$500$}}%
      \put(946,7286){\makebox(0,0){\strut{}$-1000$}}%
      \put(1675,7286){\makebox(0,0){\strut{}$-500$}}%
      \put(2404,7286){\makebox(0,0){\strut{}$0$}}%
      \put(3133,7286){\makebox(0,0){\strut{}$500$}}%
      \put(3862,7286){\makebox(0,0){\strut{}$1000$}}%
      \put(4591,7286){\makebox(0,0){\strut{}$1500$}}%
      \colorrgb{0.75,0.75,0.75}
      \put(3887,8044){\rotatebox{90}{\makebox(0,0)[l]{\strut{}US708}}}%
      \put(2287,8044){\rotatebox{90}{\makebox(0,0)[l]{\strut{}LP 398-9}}}%
      \put(1808,8044){\rotatebox{90}{\makebox(0,0)[l]{\strut{}J2050}}}%
    }%
    \gplgaddtomacro\gplbacktext{%
      \csname LTb\endcsname
      \put(217,6448){\makebox(0,0){\strut{}(C)}}%
    }%
    \gplgaddtomacro\gplfronttext{%
      \csname LTb\endcsname
      \put(209,5910){\rotatebox{-270}{\makebox(0,0){\strut{}$N$}}}%
      \put(2768,4689){\makebox(0,0){\strut{}$v_\mathrm{r}$ [\kms]}}%
      \csname LTb\endcsname
      \put(814,5239){\makebox(0,0)[r]{\strut{}$0$}}%
      \put(814,5911){\makebox(0,0)[r]{\strut{}$100$}}%
      \put(814,6582){\makebox(0,0)[r]{\strut{}$200$}}%
      \put(946,5019){\makebox(0,0){\strut{}$-1000$}}%
      \put(1675,5019){\makebox(0,0){\strut{}$-500$}}%
      \put(2404,5019){\makebox(0,0){\strut{}$0$}}%
      \put(3133,5019){\makebox(0,0){\strut{}$500$}}%
      \put(3862,5019){\makebox(0,0){\strut{}$1000$}}%
      \put(4591,5019){\makebox(0,0){\strut{}$1500$}}%
      \colorrgb{0.75,0.75,0.75}
      \put(3887,5776){\rotatebox{90}{\makebox(0,0)[l]{\strut{}US708}}}%
      \put(2287,5776){\rotatebox{90}{\makebox(0,0)[l]{\strut{}LP 398-9}}}%
      \put(1808,5776){\rotatebox{90}{\makebox(0,0)[l]{\strut{}J2050}}}%
    }%
    \gplgaddtomacro\gplbacktext{%
      \csname LTb\endcsname
      \put(217,4181){\makebox(0,0){\strut{}(D)}}%
    }%
    \gplgaddtomacro\gplfronttext{%
      \csname LTb\endcsname
      \put(209,3643){\rotatebox{-270}{\makebox(0,0){\strut{}$N$}}}%
      \put(2768,2421){\makebox(0,0){\strut{}$v_\mathrm{r}$ [\kms]}}%
      \csname LTb\endcsname
      \put(814,2971){\makebox(0,0)[r]{\strut{}$0$}}%
      \put(814,3643){\makebox(0,0)[r]{\strut{}$100$}}%
      \put(814,4315){\makebox(0,0)[r]{\strut{}$200$}}%
      \put(946,2751){\makebox(0,0){\strut{}$-1000$}}%
      \put(1675,2751){\makebox(0,0){\strut{}$-500$}}%
      \put(2404,2751){\makebox(0,0){\strut{}$0$}}%
      \put(3133,2751){\makebox(0,0){\strut{}$500$}}%
      \put(3862,2751){\makebox(0,0){\strut{}$1000$}}%
      \put(4591,2751){\makebox(0,0){\strut{}$1500$}}%
      \colorrgb{0.75,0.75,0.75}
      \put(3887,3509){\rotatebox{90}{\makebox(0,0)[l]{\strut{}US708}}}%
      \put(2287,3509){\rotatebox{90}{\makebox(0,0)[l]{\strut{}LP 398-9}}}%
      \put(1808,3509){\rotatebox{90}{\makebox(0,0)[l]{\strut{}J2050}}}%
    }%
    \gplgaddtomacro\gplbacktext{%
    }%
    \gplgaddtomacro\gplfronttext{%
      \csname LTb\endcsname
      \put(2178,2094){\makebox(0,0)[r]{\strut{}$M_\mathrm{\ast,run} > 0.2\Msun$}}%
      \csname LTb\endcsname
      \put(2178,1874){\makebox(0,0)[r]{\strut{}$M_\mathrm{\ast,run} > 0.3\Msun$}}%
      \csname LTb\endcsname
      \put(2178,1654){\makebox(0,0)[r]{\strut{}$M_\mathrm{\ast,run} > 0.4\Msun$}}%
      \csname LTb\endcsname
      \put(2178,1434){\makebox(0,0)[r]{\strut{}$M_\mathrm{\ast,run} > 0.5\Msun$}}%
      \csname LTb\endcsname
      \put(2178,1214){\makebox(0,0)[r]{\strut{}$M_\mathrm{\ast,run} > 0.6\Msun$}}%
      \csname LTb\endcsname
      \put(2178,994){\makebox(0,0)[r]{\strut{}$M_\mathrm{\ast,run} > 0.7\Msun$}}%
      \csname LTb\endcsname
      \put(2178,774){\makebox(0,0)[r]{\strut{}$M_\mathrm{\ast,run} > 0.8\Msun$}}%
      \csname LTb\endcsname
      \put(2178,554){\makebox(0,0)[r]{\strut{}$M_\mathrm{\ast,run} > 0.9\Msun$}}%
    }%
    \gplbacktext
    \put(0,0){\includegraphics[width={249.40bp},height={566.90bp}]{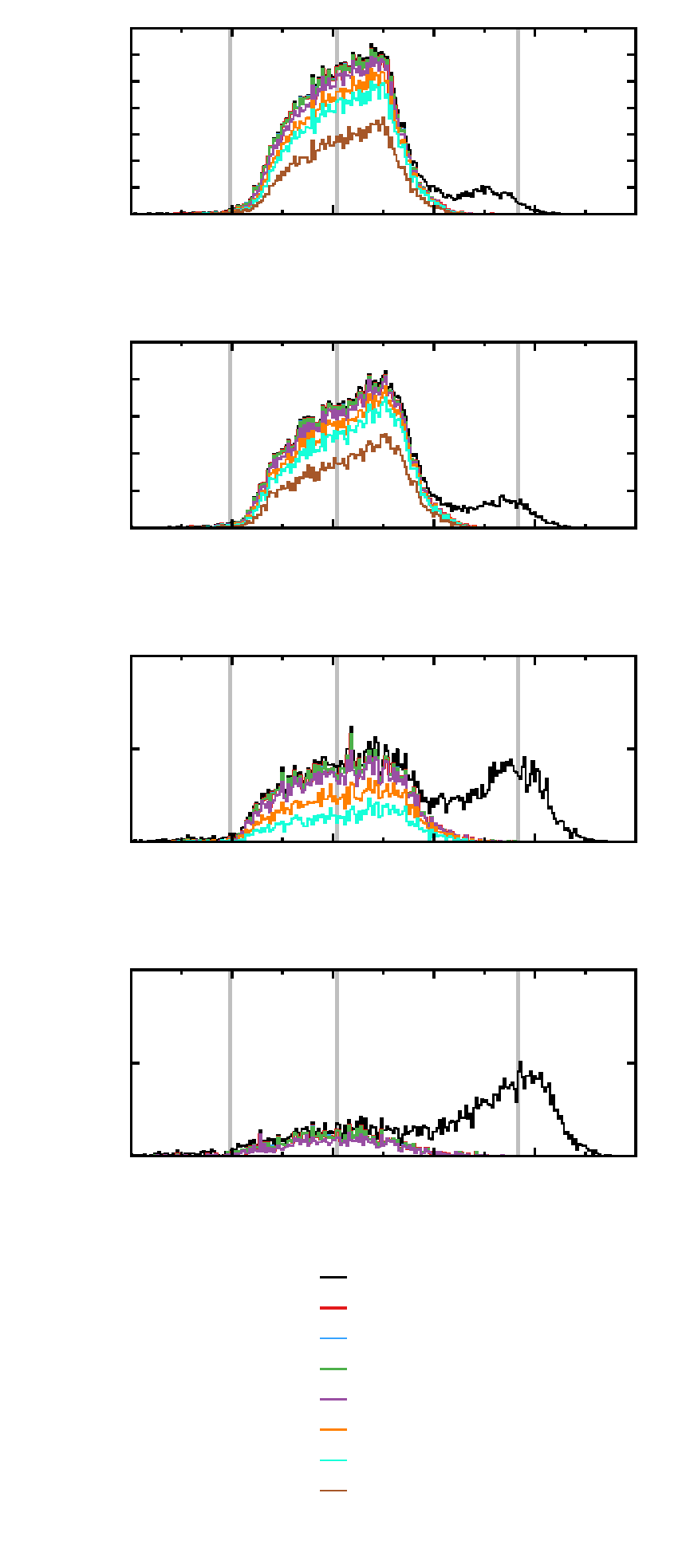}}%
    \gplfronttext
  \end{picture}%
\endgroup

%% file: M-T-diagram-M14-fig2.tex
\begingroup
  \makeatletter
  \providecommand\color[2][]{%
    \GenericError{(gnuplot) \space\space\space\@spaces}{%
      Package color not loaded in conjunction with
      terminal option `colourtext'%
    }{See the gnuplot documentation for explanation.%
    }{Either use 'blacktext' in gnuplot or load the package
      color.sty in LaTeX.}%
    \renewcommand\color[2][]{}%
  }%
  \providecommand\includegraphics[2][]{%
    \GenericError{(gnuplot) \space\space\space\@spaces}{%
      Package graphicx or graphics not loaded%
    }{See the gnuplot documentation for explanation.%
    }{The gnuplot epslatex terminal needs graphicx.sty or graphics.sty.}%
    \renewcommand\includegraphics[2][]{}%
  }%
  \providecommand\rotatebox[2]{#2}%
  \@ifundefined{ifGPcolor}{%
    \newif\ifGPcolor
    \GPcolortrue
  }{}%
  \@ifundefined{ifGPblacktext}{%
    \newif\ifGPblacktext
    \GPblacktextfalse
  }{}%
  \let\gplgaddtomacro\g@addto@macro
  \gdef\gplbacktext{}%
  \gdef\gplfronttext{}%
  \makeatother
  \ifGPblacktext
    \def\colorrgb#1{}%
    \def\colorgray#1{}%
  \else
    \ifGPcolor
      \def\colorrgb#1{\color[rgb]{#1}}%
      \def\colorgray#1{\color[gray]{#1}}%
      \expandafter\def\csname LTw\endcsname{\color{white}}%
      \expandafter\def\csname LTb\endcsname{\color{black}}%
      \expandafter\def\csname LTa\endcsname{\color{black}}%
      \expandafter\def\csname LT0\endcsname{\color[rgb]{1,0,0}}%
      \expandafter\def\csname LT1\endcsname{\color[rgb]{0,1,0}}%
      \expandafter\def\csname LT2\endcsname{\color[rgb]{0,0,1}}%
      \expandafter\def\csname LT3\endcsname{\color[rgb]{1,0,1}}%
      \expandafter\def\csname LT4\endcsname{\color[rgb]{0,1,1}}%
      \expandafter\def\csname LT5\endcsname{\color[rgb]{1,1,0}}%
      \expandafter\def\csname LT6\endcsname{\color[rgb]{0,0,0}}%
      \expandafter\def\csname LT7\endcsname{\color[rgb]{1,0.3,0}}%
      \expandafter\def\csname LT8\endcsname{\color[rgb]{0.5,0.5,0.5}}%
    \else
      \def\colorrgb#1{\color{black}}%
      \def\colorgray#1{\color[gray]{#1}}%
      \expandafter\def\csname LTw\endcsname{\color{white}}%
      \expandafter\def\csname LTb\endcsname{\color{black}}%
      \expandafter\def\csname LTa\endcsname{\color{black}}%
      \expandafter\def\csname LT0\endcsname{\color{black}}%
      \expandafter\def\csname LT1\endcsname{\color{black}}%
      \expandafter\def\csname LT2\endcsname{\color{black}}%
      \expandafter\def\csname LT3\endcsname{\color{black}}%
      \expandafter\def\csname LT4\endcsname{\color{black}}%
      \expandafter\def\csname LT5\endcsname{\color{black}}%
      \expandafter\def\csname LT6\endcsname{\color{black}}%
      \expandafter\def\csname LT7\endcsname{\color{black}}%
      \expandafter\def\csname LT8\endcsname{\color{black}}%
    \fi
  \fi
    \setlength{\unitlength}{0.0500bp}%
    \ifx\gptboxheight\undefined%
      \newlength{\gptboxheight}%
      \newlength{\gptboxwidth}%
      \newsavebox{\gptboxtext}%
    \fi%
    \setlength{\fboxrule}{0.5pt}%
    \setlength{\fboxsep}{1pt}%
    \definecolor{tbcol}{rgb}{1,1,1}%
\begin{picture}(10204.00,5102.00)%
    \gplgaddtomacro\gplbacktext{%
      \csname LTb\endcsname
      \put(1622,3628){\makebox(0,0)[l]{\strut{}(A)}}%
    }%
    \gplgaddtomacro\gplfronttext{%
      \csname LTb\endcsname
      \put(209,2792){\rotatebox{-270}{\makebox(0,0){\strut{}N}}}%
      \put(2891,154){\makebox(0,0){\strut{}bolometric apparent magnitude $m_\mathrm{Bol}$}}%
      \csname LTb\endcsname
      \put(2134,4708){\makebox(0,0)[r]{\strut{}average}}%
      \csname LTb\endcsname
      \put(2134,4488){\makebox(0,0)[r]{\strut{}minimum}}%
      \csname LTb\endcsname
      \put(2134,4268){\makebox(0,0)[r]{\strut{}maximum}}%
      \csname LTb\endcsname
      \put(946,704){\makebox(0,0)[r]{\strut{}$1$}}%
      \put(946,1833){\makebox(0,0)[r]{\strut{}$10$}}%
      \put(946,2962){\makebox(0,0)[r]{\strut{}$100$}}%
      \put(946,4092){\makebox(0,0)[r]{\strut{}$1000$}}%
      \put(1078,484){\makebox(0,0){\strut{}$12$}}%
      \put(1636,484){\makebox(0,0){\strut{}$14$}}%
      \put(2194,484){\makebox(0,0){\strut{}$16$}}%
      \put(2752,484){\makebox(0,0){\strut{}$18$}}%
      \put(3310,484){\makebox(0,0){\strut{}$20$}}%
      \put(3868,484){\makebox(0,0){\strut{}$22$}}%
      \put(4426,484){\makebox(0,0){\strut{}$24$}}%
    }%
    \gplgaddtomacro\gplbacktext{%
      \csname LTb\endcsname
      \put(6836,3628){\makebox(0,0)[l]{\strut{}(B)}}%
    }%
    \gplgaddtomacro\gplfronttext{%
      \csname LTb\endcsname
      \put(5311,2792){\rotatebox{-270}{\makebox(0,0){\strut{}$T_\mathrm{eff}$ [K]}}}%
      \put(8059,154){\makebox(0,0){\strut{}$M_\mathrm{\ast,run}$ [$\Msun$]}}%
      \csname LTb\endcsname
      \put(8820,1317){\makebox(0,0)[r]{\strut{}average}}%
      \csname LTb\endcsname
      \put(8820,1097){\makebox(0,0)[r]{\strut{}minimum}}%
      \csname LTb\endcsname
      \put(8820,877){\makebox(0,0)[r]{\strut{}maximum}}%
      \csname LTb\endcsname
      \put(6180,704){\makebox(0,0)[r]{\strut{}$20000$}}%
      \put(6180,1539){\makebox(0,0)[r]{\strut{}$30000$}}%
      \put(6180,2375){\makebox(0,0)[r]{\strut{}$40000$}}%
      \put(6180,3210){\makebox(0,0)[r]{\strut{}$50000$}}%
      \put(6180,4046){\makebox(0,0)[r]{\strut{}$60000$}}%
      \put(6180,4881){\makebox(0,0)[r]{\strut{}$70000$}}%
      \put(6312,484){\makebox(0,0){\strut{}$0.2$}}%
      \put(7011,484){\makebox(0,0){\strut{}$0.4$}}%
      \put(7710,484){\makebox(0,0){\strut{}$0.6$}}%
      \put(8409,484){\makebox(0,0){\strut{}$0.8$}}%
      \put(9108,484){\makebox(0,0){\strut{}$1$}}%
      \put(9807,484){\makebox(0,0){\strut{}$1.2$}}%
      \colorrgb{0.75,0.75,0.75}
      \put(6487,3143){\makebox(0,0)[l]{\strut{}US708}}%
    }%
    \gplbacktext
    \put(0,0){\includegraphics[width={510.20bp},height={255.10bp}]{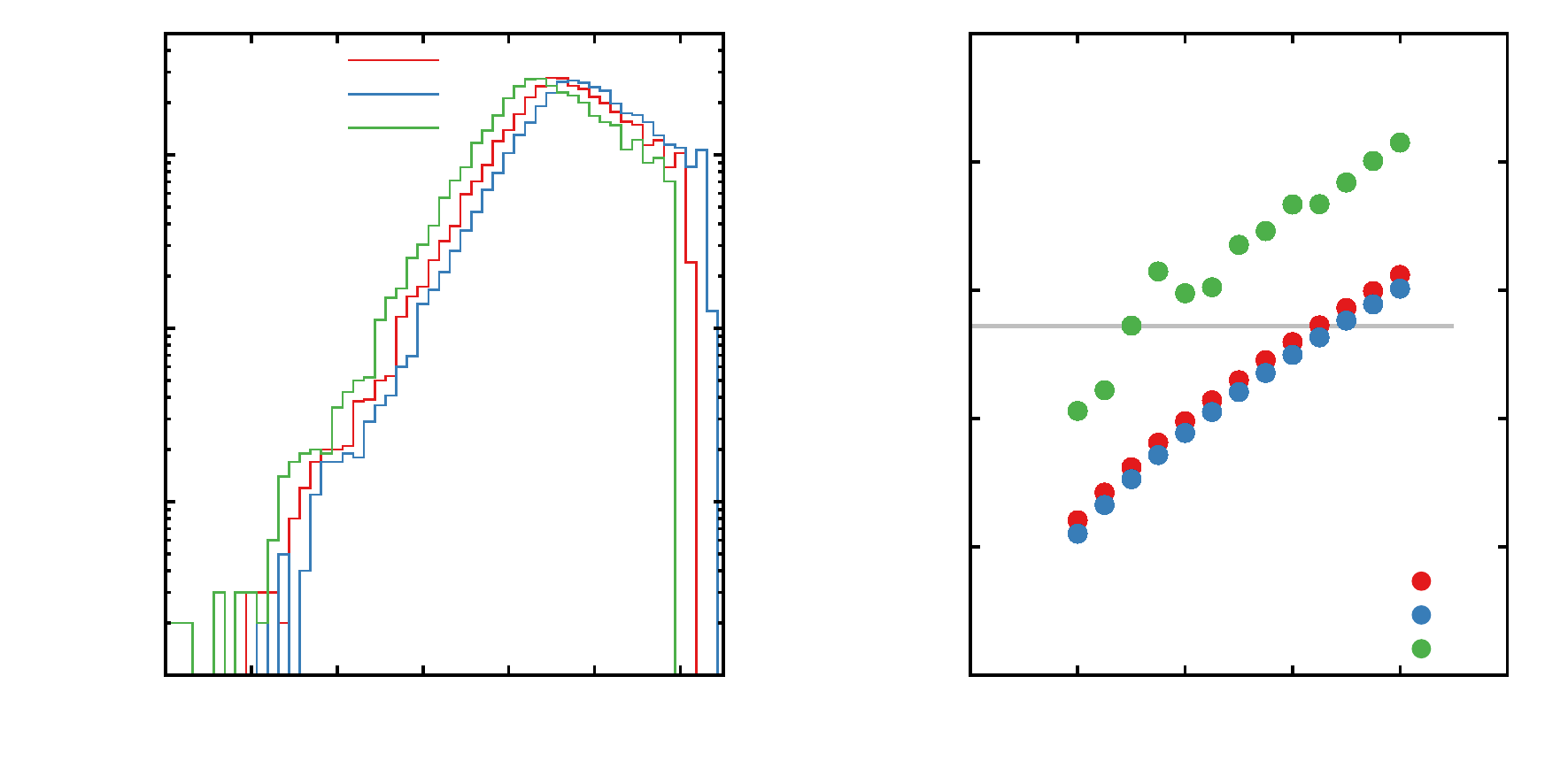}}%
    \gplfronttext
  \end{picture}%
\endgroup

%% file: rate-calculation-M14-fig2.tex
\begingroup
  \makeatletter
  \providecommand\color[2][]{%
    \GenericError{(gnuplot) \space\space\space\@spaces}{%
      Package color not loaded in conjunction with
      terminal option `colourtext'%
    }{See the gnuplot documentation for explanation.%
    }{Either use 'blacktext' in gnuplot or load the package
      color.sty in LaTeX.}%
    \renewcommand\color[2][]{}%
  }%
  \providecommand\includegraphics[2][]{%
    \GenericError{(gnuplot) \space\space\space\@spaces}{%
      Package graphicx or graphics not loaded%
    }{See the gnuplot documentation for explanation.%
    }{The gnuplot epslatex terminal needs graphicx.sty or graphics.sty.}%
    \renewcommand\includegraphics[2][]{}%
  }%
  \providecommand\rotatebox[2]{#2}%
  \@ifundefined{ifGPcolor}{%
    \newif\ifGPcolor
    \GPcolortrue
  }{}%
  \@ifundefined{ifGPblacktext}{%
    \newif\ifGPblacktext
    \GPblacktextfalse
  }{}%
  \let\gplgaddtomacro\g@addto@macro
  \gdef\gplbacktext{}%
  \gdef\gplfronttext{}%
  \makeatother
  \ifGPblacktext
    \def\colorrgb#1{}%
    \def\colorgray#1{}%
  \else
    \ifGPcolor
      \def\colorrgb#1{\color[rgb]{#1}}%
      \def\colorgray#1{\color[gray]{#1}}%
      \expandafter\def\csname LTw\endcsname{\color{white}}%
      \expandafter\def\csname LTb\endcsname{\color{black}}%
      \expandafter\def\csname LTa\endcsname{\color{black}}%
      \expandafter\def\csname LT0\endcsname{\color[rgb]{1,0,0}}%
      \expandafter\def\csname LT1\endcsname{\color[rgb]{0,1,0}}%
      \expandafter\def\csname LT2\endcsname{\color[rgb]{0,0,1}}%
      \expandafter\def\csname LT3\endcsname{\color[rgb]{1,0,1}}%
      \expandafter\def\csname LT4\endcsname{\color[rgb]{0,1,1}}%
      \expandafter\def\csname LT5\endcsname{\color[rgb]{1,1,0}}%
      \expandafter\def\csname LT6\endcsname{\color[rgb]{0,0,0}}%
      \expandafter\def\csname LT7\endcsname{\color[rgb]{1,0.3,0}}%
      \expandafter\def\csname LT8\endcsname{\color[rgb]{0.5,0.5,0.5}}%
    \else
      \def\colorrgb#1{\color{black}}%
      \def\colorgray#1{\color[gray]{#1}}%
      \expandafter\def\csname LTw\endcsname{\color{white}}%
      \expandafter\def\csname LTb\endcsname{\color{black}}%
      \expandafter\def\csname LTa\endcsname{\color{black}}%
      \expandafter\def\csname LT0\endcsname{\color{black}}%
      \expandafter\def\csname LT1\endcsname{\color{black}}%
      \expandafter\def\csname LT2\endcsname{\color{black}}%
      \expandafter\def\csname LT3\endcsname{\color{black}}%
      \expandafter\def\csname LT4\endcsname{\color{black}}%
      \expandafter\def\csname LT5\endcsname{\color{black}}%
      \expandafter\def\csname LT6\endcsname{\color{black}}%
      \expandafter\def\csname LT7\endcsname{\color{black}}%
      \expandafter\def\csname LT8\endcsname{\color{black}}%
    \fi
  \fi
    \setlength{\unitlength}{0.0500bp}%
    \ifx\gptboxheight\undefined%
      \newlength{\gptboxheight}%
      \newlength{\gptboxwidth}%
      \newsavebox{\gptboxtext}%
    \fi%
    \setlength{\fboxrule}{0.5pt}%
    \setlength{\fboxsep}{1pt}%
    \definecolor{tbcol}{rgb}{1,1,1}%
\begin{picture}(10204.00,6802.00)%
    \gplgaddtomacro\gplbacktext{%
    }%
    \gplgaddtomacro\gplfronttext{%
      \csname LTb\endcsname
      \put(209,3862){\rotatebox{-270}{\makebox(0,0){\strut{}Inferred event rate $[\mathrm{yr}^{-1}]$}}}%
      \put(5442,594){\makebox(0,0){\strut{}$D$ [kpc]}}%
      \csname LTb\endcsname
      \put(2794,393){\makebox(0,0)[r]{\strut{}US 708}}%
      \csname LTb\endcsname
      \put(2794,173){\makebox(0,0)[r]{\strut{}candidates}}%
      \csname LTb\endcsname
      \put(5365,393){\makebox(0,0)[r]{\strut{}10 assumed}}%
      \csname LTb\endcsname
      \put(5365,173){\makebox(0,0)[r]{\strut{}100 assumed}}%
      \csname LTb\endcsname
      \put(7936,393){\makebox(0,0)[r]{\strut{}1000 assumed}}%
      \csname LTb\endcsname
      \put(7936,173){\makebox(0,0)[r]{\strut{}10000 assumed}}%
      \csname LTb\endcsname
      \put(946,1144){\makebox(0,0)[r]{\strut{}$10^{-7}$}}%
      \put(946,2231){\makebox(0,0)[r]{\strut{}$10^{-6}$}}%
      \put(946,3319){\makebox(0,0)[r]{\strut{}$10^{-5}$}}%
      \put(946,4406){\makebox(0,0)[r]{\strut{}$10^{-4}$}}%
      \put(946,5494){\makebox(0,0)[r]{\strut{}$10^{-3}$}}%
      \put(946,6581){\makebox(0,0)[r]{\strut{}$10^{-2}$}}%
      \put(1078,924){\makebox(0,0){\strut{}$0$}}%
      \put(1776,924){\makebox(0,0){\strut{}$2$}}%
      \put(2475,924){\makebox(0,0){\strut{}$4$}}%
      \put(3173,924){\makebox(0,0){\strut{}$6$}}%
      \put(3871,924){\makebox(0,0){\strut{}$8$}}%
      \put(4570,924){\makebox(0,0){\strut{}$10$}}%
      \put(5268,924){\makebox(0,0){\strut{}$12$}}%
      \put(5966,924){\makebox(0,0){\strut{}$14$}}%
      \put(6665,924){\makebox(0,0){\strut{}$16$}}%
      \put(7363,924){\makebox(0,0){\strut{}$18$}}%
      \put(8061,924){\makebox(0,0){\strut{}$20$}}%
      \put(8760,924){\makebox(0,0){\strut{}$22$}}%
      \put(9458,924){\makebox(0,0){\strut{}$24$}}%
      \colorrgb{0.00,0.00,0.00}
      \put(7014,6263){\makebox(0,0)[l]{\strut{}Inferred Gal. SN Ia rate}}%
      \csname LTb\endcsname
      \put(3871,1828){\makebox(0,0)[r]{\strut{}US708}}%
      \put(3348,2207){\makebox(0,0)[r]{\strut{}J2050}}%
      \put(1567,2415){\makebox(0,0)[l]{\strut{}LP 398-9}}%
      \colorrgb{0.00,0.00,0.00}
      \put(2475,1471){\makebox(0,0)[l]{\strut{}US708-channel rate currently supported}}%
      \put(6315,5325){\makebox(0,0)[l]{\strut{}predicted progenitor formation rate}}%
    }%
    \gplbacktext
    \put(0,0){\includegraphics[width={510.20bp},height={340.10bp}]{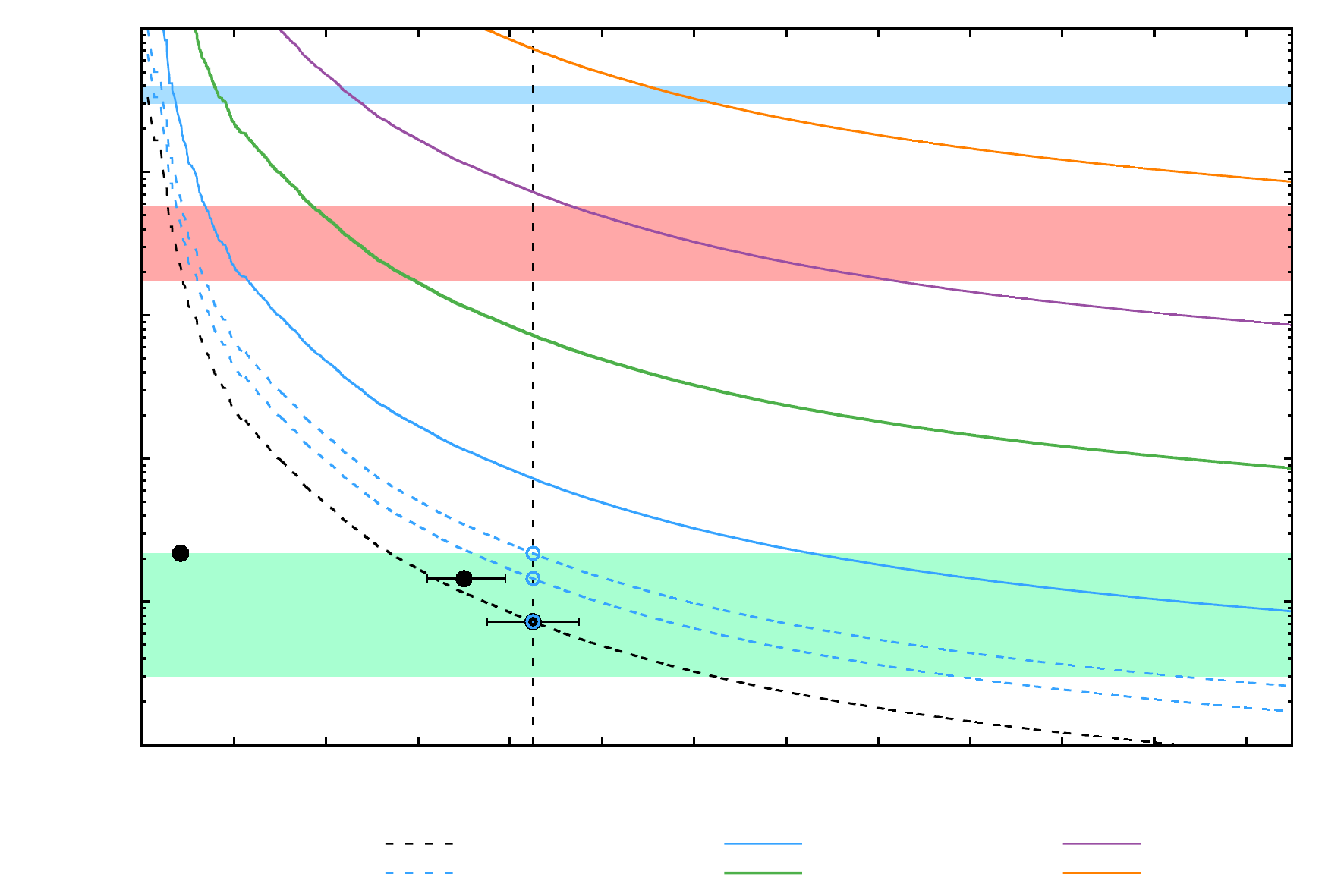}}%
    \gplfronttext
  \end{picture}%
\endgroup